\documentclass[twocolumn,showpacs,preprintnumbers,amsmath,amssymb,nofootinbib]{revtex4-2}
\usepackage{epsfig}

\usepackage{comment}
\usepackage[table]{xcolor}
\usepackage{graphicx}
\usepackage{dcolumn}
\usepackage{bm}

\usepackage{hyperref}
\hypersetup{colorlinks=true,linkcolor=magenta,anchorcolor=green,citecolor=cyan,filecolor=black,menucolor=black,urlcolor=brown}
\usepackage{slashed}
\newcommand{\ba}{\begin{eqnarray}}
\newcommand{\ea}{\end{eqnarray}}
\newcommand{\be}{\begin{equation}}
\newcommand{\ee}{\end{equation}}

\begin{document}

\title{Formation of solitons and their transitions in scalar-field dark matter models with a non-polynomial self-interaction potential}

\author{Raquel Galazo Garc\'ia}
\affiliation{Aix-Marseille Universit\'{e}, CNRS, CNES, Laboratoire d'Astrophysique de Marseille, France}
\author{Philippe Brax}
\affiliation{Universit\'{e} Paris-Saclay, CNRS, CEA, Institut de physique th\'{e}orique, 91191, Gif-sur-Yvette, France}
\author{Patrick Valageas}
\affiliation{Universit\'{e} Paris-Saclay, CNRS, CEA, Institut de physique th\'{e}orique, 91191, Gif-sur-Yvette, France}

\begin{abstract}

We study the formation of solitons inside scalar-field
dark matter halos with a non-polynomial self-interaction potential. We consider
a self-interaction potential that is quartic in  the scalar field in the low-density regime
but saturates at large densities. This mimics the behaviour of axion monodromy potentials.
We concentrate on the semi-classical regime, where the de Broglie wavelength is much smaller than the
size of the system. We find that depending on the strength and scale of the self-interactions, the system
can form solitons of the Thomas-Fermi type (dominated by self-interactions) or of the Fuzzy
Dark Matter type (dominated by the quantum pressure). The system can also display transitions
from a Thomas-Fermi soliton to a Fuzzy Dark Matter soliton as the former becomes unstable.
We show that these behaviours can be understood from a simple Gaussian ansatz.
We  find that even in cases where the self-interactions are always subdominant they can play
a critical role, by providing a small density boost that is enough to generate the seed for the
formation of a Fuzzy Dark Matter soliton at much later times.
We also point out that the intuition derived from a hydrodynamical picture can be misleading
in regimes where wave effects are important.

\end{abstract}

\date{\today}

\maketitle

\section{Introduction}

The most widely accepted model for describing Dark Matter (DM) in our Universe is the Cold Dark Matter 
(CDM) model, with  weakly interacting massive particles (WIMPs)  being the preferred choice for both 
theoretical and experimental reasons \citep{Jungman:1995df,Drees:2004jm,Steigman:1984ac}. 
However, despite numerous attempts, WIMPs remain undetected 
\citep{Schumann:2019eaa,Conrad:2014tla,Arcadi:2017kky}. 
Furthermore, as observations and simulations at galactic scales have improved, several small-scale 
tensions have emerged \citep{Weinberg:2013aya,DelPopolo:2016emo,Nakama:2017ohe} and are still under debate 
\citep{DiLuzio:2020wdo}. 
These discrepancies could potentially indicate the need for new physics  \citep{Weinberg:2013aya}. 
In this context, alternative scenarios have emerged, including the possibility that DM might exist as 
a scalar field (SFDM) with masses spanning from $10^{-22}$ eV to 1 eV  \citep{Hu:2000ke,Hui:2016ltb,Goodman:2000tg,Matos:1999et,Schive:2014dra,Ferreira:2020fam,
Li:2013nal}.
This range of masses corresponds to a de Broglie wavelength at small scales, galactic or sub-galactic scales
\begin{equation}
\lambda_{\rm dB} = \frac{2\pi}{mv}= 0.48 \, {\rm kpc} \, \left(\frac{10^{-22} {\rm eV}}{m}\right)
\left(\frac{250 \, {\rm km/s}}{v}\right) ,
\end{equation}
where $v$ is the typical velocity of the object (e.g., its virial velocity).
Hence, on scales smaller than $\lambda_{\rm dB}$ the field exhibits a wave-like behaviour.
This corresponds to the so-called``Fuzzy Dark Matter'' (FDM) models.
These wave-like effects arise from an extra pressure-like component in the equation of motion, referred to as 
quantum pressure, stemming from the gradient of the scalar field. When this quantum pressure balances  
gravity, it gives rise to equilibrium solutions known as solitons 
\citep{Lee:1991ax, Guth:2014hsa, Sikivie:2009qn,Schive:2014dra, Marsh:2015xka,Veltmaat:2018dfz}. 
These solitons appear at the center of halos and result in a flat radial density profile, i.e. a flat
core. 

If the scalar field also displays self-interactions, another way of achieving equilibrium involves a balance 
between the self-interaction pressure and gravity.
This corresponds to the``Thomas-Fermi'' (TF) regime 
\cite{Chavanis:2011zi,Chavanis:2017loo,Dawoodbhoy:2021beb,Shapiro:2021hjp}.
Axions provide a classical example for such very low mass (pseudo-) scalar fields. 
In the case of the QCD axion \citep{Peccei:1977hh,Weinberg:1977ma,Wilczek:1977pj}, which emerges from the 
Peccei-Quinn symmetry breaking, the associated potential term is non-perturbative and displays periodic 
behaviour. This periodicity is a fundamental characteristic shared by axions and axion-like particles. 
In this context, the axion field can be seen as a Goldstone mode and  its potential term  results from either 
non-perturbative effects \citep{Kim:1986ax} or from softly breaking terms before the symmetry-
breaking phase. In most cases, these potential terms generate cosine-like potentials,
which correspond to an attractive quartic self-interaction instead of a repulsive self-interaction
at low density.
In this paper, we consider instead scenarios where the quartic term leads to a repulsive self-interaction,
which allows the formation of stable solitons or boson stars supported by the associated effective
pressure. Therefore, this does not correspond to the QCD axion but to other axion-like particles
which could originate for instance from string scenarios, such as axion monodromy models 
\citep{Silverstein:2008sg,McAllister:2008hb}.
Then, the scalar potential can combine power-law terms as well as periodic terms, which allows
for both signs of the quartic coupling.
Inspired by this general framework, we investigate in this paper models where the scalar potential 
contains both a quadratic term and a cosine term, which generates a repulsive self-interaction 
at the quartic order that saturates and vanishes at high density.

Our aim is to compare the gravitational dynamics, and more specifically the formation and evolution of 
solitons, associated with such non-polynomial potentials with the well-studied cases where the
self-interaction is negligible (FDM) or quartic. 
In particular, we shall study the impact of the density threshold associated with the saturation of the
non-linear self-interactions on the formation and stability of the solitons.
Thus, we aim to determine how solitons form and whether transitions between FDM-type and
TF-type solitons can take place.
We find that this is indeed the case, both numerically and with the help of a simple Gaussian ansatz. 
Our results may be of astrophysical interest, although a detailed study is left for further investigation,
as different objects may harbour solitons of different types. 
A related study for polynomial potentials with both attractive and repulsive terms was presented
in \cite{Chavanis:2017loo}. In this paper we restrict to the non-relativistic regime. 

This paper is structured as follows. 
In Sec.~\ref{sec:cosine-model-ch6}, we introduce the axion monodromy potential and outline the simplified 
nonrelativistic case under study. 
In Sec.~\ref{sec:solitons} we recall the behaviours of solitons in the FDM and TF regimes and we introduce
a simple Gaussian ansatz that captures their main properties.
In Sec.~\ref{sec:numerical-method} we detail the numerical method employed to compute the dynamics 
of the dark matter clouds and the initial conditions. 
Following this, Sec.~\ref{sec:model-a-r-0.5} presents our results for the formation of solitons 
inside flat virialized halos for the case where the radius of TF solitons is half the one of the initial halo
(i.e., the self-interactions are large in the low-density regime).
In Sec.~\ref{sec:model-a-r-0.1} we focus on the case where the TF radius is ten percent of the initial halo
(i.e., smaller self-interactions in the low-density regime). 
Finally, in Sec.~\ref{sec:conclusion-cosine} we summarize our findings and discuss their implications.

\section{Equations of motion}
\label{sec:cosine-model-ch6}

\subsection{Nonrelativistic equation of motion}

We consider the following Lagrangian to describe the scalar field dark matter,
\be
\mathcal{L}_{\phi} = -\frac{1}{2}g^{\mu\nu}\partial_{\mu}\phi\partial_{\nu}\phi - V(\phi),
\label{eq:lagrangian-sfdm}
\ee
where $g^{\mu\nu}$ is the inverse metric, the first term is the standard kinetic term and $V(\phi)$
is the scalar potential given by
\be
V(\phi) = \frac{m^2}{2}\phi^2 + V_I(\phi),
\label{eq:potential-v-sfdm}
\ee
where $V_I(\phi)$ is the self-interaction potential.
Throughout this paper, we work in natural units, $c=\hbar=1$, and the dimensions of the scalar field
and its mass are $[\phi] =[m]=$ energy.

In this paper we focus on potentials that can be associated with axion monodromy models,
\be
V_I(\phi) = M_I^4\left[\cos(\phi/\Lambda)-1 +\frac{\phi^2}{2\Lambda^2}\right] ,
\label{eq:V_i-cosine}
\ee
which combine a quadratic term and a nonlinear periodic term.
Here the quadratic term $\phi^2/2\Lambda^2$ has been split from the mass term (\ref{eq:potential-v-sfdm}),
with which it is degenerate, to highlight that for  $\phi\ll\Lambda$ we recover the usual quartic potential
$\lambda_4 \phi^4/4 = (M_I^4/6\Lambda^4) \phi^4/4$.
The goal of this paper is to study some of the effects that arise because of the nonlinear completion
of the quartic potential, when the scalar field is large.

In the weak gravity regime and neglecting the Hubble expansion, which is appropriate for galactic
and subgalactic scales, the scalar field obeys a nonlinear Klein-Gordon equation,
\be
\ddot{\phi} -\nabla^2\phi+m^2(1+2\Phi_N)\phi  +\frac{dV_I}{d\phi}=0 ,
\label{eq:phi-without-expansion-sfdm}
\ee
where $\Phi_N$ is the gravitational potential.
Here we consider scales that are much greater than the Compton length,
i.e. wavenumbers $k \ll m$. Using $V_I \ll m^2\phi^2/2$ and the fact that the scalar field is nonrelativistic,
we can see that at leading order the field oscillates at the very high frequency $m$.
To average over these fast oscillations it is convenient to introduce a complex scalar field $\psi$
by \citep{Hu:2000ke,Hui:2016ltb},
\be
\phi= \frac{1}{\sqrt{2m}} ( \psi e^{-imt} +  \psi^* e^{imt}) .
\label{eq:phi-psi}
\ee
Substituting into the action or the equation of motion and averaging over these fast oscillations
gives the nonrelativistic equation of motion \citep{Brax:2019fzb}
\be
i \frac{\partial\psi}{\partial t} =
- \frac{\nabla^2\psi}{2m} + m ( \Phi_N + \Phi_I ) \psi ,
\label{eq:Schrod}
\ee
which has the form of a Gross-Pitaevskii equation, except that the Newtonian potential $\Phi_N$
is not external but given by the self-gravity of the scalar field.
The self-interaction factor $\Phi_I(\rho)$ is related to the scalar potential $V_I(\phi)$ by
\be
\Phi_I(\rho) = \frac{d {\cal V}_I}{d\rho} , \;\;\; \mbox{with} \;\;\; \rho = m | \psi |^2 ,
\ee
and
\be
{\cal V}_I(\rho) = \Lambda^4 \sum_{n=2}^{\infty} \frac{\lambda_{2n}}{2n} \frac{(2n)!}{(n!)^2}
\left( \frac{\rho}{2 m^2 \Lambda^2} \right)^n ,
\label{V_I-rho-def}
\ee
where the coefficients $\lambda_n$ are defined by the expansion of the scalar potential
\be
V_I(\phi) = \Lambda^4 \sum_{n \geq 3} \frac{\lambda_n}{n} \frac{\phi^n}{\Lambda^n} .
\ee
Because of the averaging over the fast oscillations, only the even terms remain in the nonrelativistic
density potential (\ref{V_I-rho-def}).

\subsection{Simplified nonrelativistic potential}
\label{sec:simplified-potential}

In the case of the axion monodromy potential (\ref{eq:V_i-cosine}), we obtain the nonrelativistic
expression
\be
\Phi_I(\rho) = \frac{8 \rho_b}{\rho_a} \left[ 1 - \frac{2 J_1(\sqrt{\rho/\rho_b})}{\sqrt{\rho/\rho_b}} \right] ,
\label{Phi_I-J1}
\ee
where $J_1$ is the Bessel function of the first kind and the characteristic densities $\rho_a$ and
$\rho_b$ are given by
\be
\rho_a = \frac{8 m^4 \Lambda^4}{M_I^4} , \;\;\; \rho_b = \frac{m^2\Lambda^2}{2} , \;\;\;
\rho_b \ll \rho_a .
\ee
At low density we recover the linear rise associated with a standard $\phi^4$ self-interaction,
\be
\rho \ll \rho_b : \;\;\; \Phi_I(\rho) \simeq \frac{\rho}{\rho_a} ,
\ee
whereas at high density the nonrelativistic potential $\Phi_I$ goes to a constant,
\be
\rho \gg \rho_b : \;\;\; \Phi_I(\rho) \simeq \frac{8 \rho_b}{\rho_a} \ll 1 .
\ee
This behavior directly follows from the bounded character of the cosine term in the axion monodromy
potential (\ref{eq:V_i-cosine}).
In this article, we investigate the effects that can arise from such a saturation at high density.
Neglecting the decaying oscillations of the Bessel function, we  consider the simplified model
\be
\Phi_I (\rho) = \begin{cases}
          \rho/\rho_a \quad &\text{if} \;\;\; \rho < \rho_c \\
          \rho_c/\rho_a \quad &\text{if} \;\;\; \rho > \rho_c  \\
     \end{cases}
     \label{eq:model-1}
\ee
which features the same linear rise at low density and saturation at high density.

\subsection{Dimensionless quantities}

As usual, it is convenient to work with dimensionless quantities, which we define by
\be
\psi = \psi_\star \tilde\psi, \;\;\; t = T_\star \tilde t, \;\;\;
\vec x = L_\star \tilde{\vec x} , \;\;\;
\Phi = V_\star^2 \tilde\Phi,
\label{eq:dimesionless-def}
\ee
where $T_\star$, $L_\star$ and $V_\star = L_\star/T_\star$ are the characteristic time, length and
velocity scales of the system. Under this rescaling, we obtain the dimensionless Schr\"odinger--Poisson
equations that govern the dynamics of the small-scale structures that we are interested in studying,
\be
i \epsilon \frac{\partial\tilde\psi}{\partial\tilde t} =
- \frac{\epsilon^2}{2} \tilde\nabla^2\tilde\psi
+ (\tilde\Phi_{N}+\tilde\Phi_{I}) \tilde\psi ,
\label{eq:Schrod-eps}
\ee
\be
\tilde\nabla^2 \tilde\Phi_{N} = 4 \pi \tilde\rho, \;\;\; \mbox{with}  \;\;\; \tilde\rho = | \tilde\psi|^2 ,
\label{eq:Poisson-eps}
\ee
\be
\tilde\Phi_I (\tilde\rho) = \begin{cases}
          \lambda \tilde\rho \quad &\text{if} \;\;\; \tilde\rho < \tilde\rho_c \\
          \lambda \tilde\rho_c \quad &\text{if} \;\;\; \tilde\rho > \tilde\rho_c  \\
     \end{cases}
     \label{eq:model-1-tilde}
\ee
and $T_\star = 1/\sqrt{{\cal G}_N m \psi_\star^2}$.
The coefficient $\epsilon$ is given by
\be
\epsilon = \frac{T_\star}{m L_\star^2} .
\label{eq:scaling-eps}
\ee
If we compare this quantity with the typical de Broglie wavelength $\lambda_{\rm dB} = 2\pi/(m V_\star)$,
we have
\be
\epsilon \sim \frac{\lambda_{\rm dB}}{L_\star} .
\label{eq:epsilon-de-Broglie}
\ee
Therefore, the parameter $\epsilon$, which appears in the dimensionless Schr\"odinger equation
(\ref{eq:Schrod-eps}) plays the role of $\hbar$ in quantum mechanics. This parameter measures the
relevance of wave effects in the system, such as interferences or the effect of the quantum pressure.
More precisely, we have for the rescaled de Broglie wave length
\be
\tilde\lambda_{\rm dB} = \frac{2\pi \epsilon}{\tilde v} ,
\label{eq:de-Broglie-rescaled}
\ee
where $\tilde v=v/V_\star$ is the rescaled velocity.
In the following we omit the tilde to simplify the notations.

\section{Solitons and Gaussian ansatz}\label{sec:solitons}

\subsection{Hydrodynamical picture}
\label{sec:hydro}

As in Eq.(\ref{eq:Poisson-eps}), the matter density is the square root of the amplitude of the
wave function $\psi$. Defining a velocity field $\vec v$ from the phase $S$ by \citep{Madelung:1927ksh}
\be
\psi = \sqrt{\rho} \, e^{i S} , \;\;\; \vec v = \epsilon \nabla S ,
\label{eq:Madelung}
\ee
and substituting into the equation of motion (\ref{eq:Schrod-eps}), the real and imaginary parts
give the continuity and Euler equations,
\be
\frac{\partial\rho}{\partial t} + \nabla\cdot(\rho \vec v) = 0 , \;\;\;
\frac{\partial\vec v}{\partial t} + (\vec v \cdot \nabla) \vec v = - \nabla( \Phi_Q + \Phi_N + \Phi_I ) ,
\label{eq:Hydro}
\ee
where we introduced the so-called quantum pressure defined by
\be
\Phi_Q = - \frac{\epsilon^2}{2} \frac{\nabla^2 \sqrt{\rho}}{\sqrt{\rho}} .
\label{eq:PhiQ-def}
\ee
On the other hand, the self-interaction potential can be identified with an effective polytropic
pressure $P_I(\rho)$, with $\nabla\Phi_I = (\nabla P_I)/\rho$ and
\be
P_I = \int d\rho \, \rho \frac{d\Phi_I}{d\rho} .
\label{eq-P_I}
\ee
In particular, in the low density regime where $\Phi_I=\lambda\rho$ we have
$P_I = \lambda\rho^2/2$.
Thus, the Madelung transform (\ref{eq:Madelung}) maps the nonlinear Schr\"odinger equation
(\ref{eq:Schrod-eps}) to the hydrodynamical equations (\ref{eq:Hydro}).
However, this mapping is not complete. At locations where $\psi$ vanishes the phase and the velocity
are not well defined. These singularities can lead to vortices that are not captured by
a curl-free velocity field, as assumed in (\ref{eq:Madelung}).

In fact, we shall find in Sec.~\ref{sec:R0p5-rhoc0p5} that this hydrodynamical picture can be
misleading, in regimes where the density fluctuations are large and wave effects play a significant
role.
Indeed, as the self-interactions are associated with an effective pressure (\ref{eq-P_I}) we could
expect that they make it more difficult to build structures and large density contrasts.
However, we shall see in Sec.~\ref{sec:R0p5-rhoc0p5} that this is not always the case.

\subsection{Solitons}

The Schr\"odinger equation (\ref{eq:Schrod-eps}) admits hydrostatic equilibria, also called solitons \citep{Chavanis:2011zi,Chavanis:2011zm,Harko:2011jy,Brax:2019fzb},
which correspond to the ground state of the potential $\Phi_N+\Phi_I$.
These configurations, of the form $e^{-i \mu t/\epsilon} \hat\psi(\vec x)$, are
solutions of the time-independent Schr\"odinger equation
\be
\Phi_Q + \Phi_N + \Phi_I = \mu , \;\;\; \psi_{\rm sol}(\vec x,t) = e^{-i \mu t/\epsilon}
\hat\psi_{\rm sol}(r) ,
\label{eq:hydrostatic-full}
\ee
where we considered spherically symmetric solutions and the quantum pressure $\Phi_Q$ was
already defined in Eq.(\ref{eq:PhiQ-def}).
They also correspond to the hydrostatic equilibria of the hydrodynamical equations (\ref{eq:Hydro}).

\subsubsection{Low-density regime}

In the Thomas-Fermi regime where we can neglect the quantum pressure because $\epsilon \ll 1$
and gravity is balanced by the repulsive self-interactions, the hydrostatic equilibrium is given by
\be
\mbox{TF regime} : \;\;\; \Phi_N + \Phi_I = \mu  .
\label{eq:TF-static}
\ee
At low density $\rho < \rho_c$, where $\Phi_I = \lambda \rho$, the soliton density profile reads
\cite{Chavanis:2011zi,Brax:2019fzb}
\be
\rho_{0} < \rho_c : \;\;\;
\rho_{\rm TF}(r) = \rho_{0} \frac{\sin(\pi r/R_{\rm TF})}{\pi r/R_{\rm TF}} ,
\label{eq:rho-sol-TF}
\ee
where the subscript ``TF'' stands for the Thomas-Fermi regime, and
\be
R_{\rm TF} = \frac{\sqrt{\lambda \pi}}{2} , \;\;\; M = \frac{4}{\pi} \rho_{0} R_{\rm TF}^3 ,
\;\;\; \rho_{0} = \frac{2 M}{\sqrt{\pi \lambda^3}} .
\label{eq:Rsol-lambda}
\ee
In particular, the soliton radius $R_{\rm TF}$ does not depend on its mass.
The expression (\ref{eq:rho-sol-TF}) only applies to $r < R_{\rm TF}$.
For $r \geq R_{\rm TF}$ one cannot neglect the quantum pressure and the density profile shows
a negligible exponential tail.
Because the solitons have a strictly positive density and a constant phase, which can be taken to zero,
they are also fully described by the hydrodynamical picture recalled in Sec.~\ref{sec:hydro} above.

\subsubsection{High-density regime}

At high density, $\rho \gg \rho_c$, the potential $\Phi_I$ is flat, which corresponds to vanishing
self-interactions. The Thomas-Fermi regime no longer exists and the soliton is determined by the
balance between gravity and the quantum pressure. This gives
\be
\mbox{FDM regime} : \;\;\; \Phi_Q + \Phi_N = \mu - \lambda \rho_c ,
\label{eq:Phi-FDM-static}
\ee
with the scaling
\be
\rho_{0} \gg \rho_c : \;\;\; R_{\rm FDM} \sim \frac{\epsilon^2}{M} , \;\;\;
\rho_{0} \sim \frac{M^4}{\epsilon^6} ,
\label{eq:rho-sol-FDM}
\ee
where the subscript ``FDM'' stands for the Fuzzy Dark Matter regime (i.e. negligible self-interactions).
Thus, starting from a system of mass $M \sim 1$, in the dimensionless units (\ref{eq:dimesionless-def}),
we can expect to form first a low density central soliton (\ref{eq:rho-sol-TF}) governed by the
self-interactions. Then, if the soliton density grows beyond $\rho_c$ we can expect a collapse
to a much smaller soliton (\ref{eq:rho-sol-FDM}) determined by the quantum pressure, i.e. the Laplacian
term in the Schr\"odinger equation.
The aim of this paper is to study in more details this transition.

\subsection{Gaussian ansatz for the radial profile}
\label{sec:gaussian}

\subsubsection{Energy functional}
\label{sec:energy}

To study the transitions between low and high density solitons we can use a simple analytical approach
where we use a Gaussian ansatz for the density radial profile \citep{Chavanis:2011zi,Chavanis:2017loo,Brax:2020oye},
\be
\rho(r)= \rho_0 \; e^{-(r/R)^2}, \;\;  \psi = \sqrt{\rho} , \;\; \text{with} \;\;  \rho_0 = \frac{M}{\pi^{3/2}R^3} .
\label{eq:Gaussian}
\ee
This bypasses the numerical computation of the soliton profiles and should provide the correct
scalings and behaviors.
The energy functional associated with the equation of motion (\ref{eq:Schrod-eps}) reads
\be
E_{\rm tot} = E_K + E_N + E_I = \int d{\vec x} \left( \frac{\epsilon^2}{2} | \nabla\psi |^2
+ \frac{1}{2} \rho \Phi_N + {\cal V}_I \right) .
\label{eq:E-tot-psi}
\ee
The three terms correspond to the kinetic, gravitational and self-interaction energies, where the
nonrelativistic potential ${\cal V}_I$ associated with (\ref{eq:model-1-tilde}) reads
\be
{\cal V}_I (\rho) = \begin{cases}
          \lambda \rho^2/2 \quad &\text{if} \;\;\; \rho < \rho_c \\
          \lambda \rho_c \rho-\lambda \rho_c^2/2 \quad &\text{if} \;\;\; \rho > \rho_c  \\
     \end{cases}
     \label{eq:VI-rho-bimodal}
\ee
and $\Phi_I = d {\cal V}_I/d\rho$.
The solitons are minima of the energy functional at fixed mass. Substituting the Gaussian ansatz
(\ref{eq:Gaussian}) and looking for a minimum of its total energy with respect to the parameter
$\rho_0$ thus provides an estimate of the soliton properties, for a given mass $M$.
The kinetic and gravitational energies of this Gaussian profile are given by
\be
E_K = \epsilon^2 \, \frac{3\pi}{4} M^{1/3} \rho_0^{2/3} , \;\;\;
E_N = -\frac{1}{\sqrt{2}} \, M^{5/3} \rho_0^{1/3} .
\label{eq:Gauss-EK-EN}
\ee
The self-interaction energy reads at low density
\be
\rho_0 < \rho_c : \;\;\; E_I = \frac{\lambda}{4\sqrt{2}} \, M \rho_0 ,
\label{eq:Gauss-EI1}
\ee
and at high density
\ba
&& \hspace{-0.9cm}  \rho_0 > \rho_c : \;  E_I = \frac{\lambda M \rho_0}{4\sqrt{2}}
\left[ {\rm Erfc} \left( \sqrt{2 \ln \frac{\rho_0}{\rho_c} } \right) + 4 \sqrt{2} \frac{\rho_c}{\rho_0} \right.
\nonumber \\
&& \hspace{-0.9cm}  \left. \times {\rm Erf} \left( \sqrt{\ln \frac{\rho_0}{\rho_c} } \right) -
\frac{2\sqrt{2}}{3\sqrt{\pi}} \frac{\rho_c^2}{\rho_0^2} \sqrt{\ln \frac{\rho_0}{\rho_c} }
\left( 9 + 4 \ln \frac{\rho_0}{\rho_c} \right) \right] ,
\label{eq:Gauss-EI2}
\ea
with the large-density limit
\be
\rho_0 \to \infty : \;\;\; E_I = \lambda M \rho_c - \frac{2 \lambda M \rho_c^2 \ln(\rho_0/\rho_c)^{3/2}}
{3 \sqrt{\pi} \rho_0} + \dots
\ee
The saturation of $E_I$ to a finite value means that at large densities the self-interactions can no longer
balance gravity and the gravitational collapse is only stopped by the quantum pressure.

The Gaussian ansatz (\ref{eq:Gaussian}) simplifies the problem by replacing the energy functional
$E_{\rm tot}[\psi(\vec x)]$ of Eq.(\ref{eq:E-tot-psi}) by the simple function $E_{\rm tot}(\rho_0)$
of the soliton central density defined by Eqs.(\ref{eq:Gauss-EK-EN})-(\ref{eq:Gauss-EI2}).
However, this can only provide information on the soliton and not on the full system, which
usually also contains an outer halo that is not in hydrostatic equilibrium.
Then, for a given soliton mass, the minima of the total energy are given by the condition
\be
\frac{d E_{\rm tot}}{d\rho_0} = 0 .
\label{eq:minimum-E}
\ee
This provides a simple estimate of the soliton density and radius, without the need to solve the
integro-differential equation (\ref{eq:hydrostatic-full}).

\subsubsection{Low-density regime}

In the low-density Thomas-Fermi regime (\ref{eq:rho-sol-TF}) we write
\be
\rho_0 < \rho_c : \; E_{\rm tot} \simeq E_N+E_I = -\frac{M^{5/3} \rho_0^{1/3}}{\sqrt{2}}
+ \frac{\lambda M \rho_0}{4\sqrt{2}} ,
\ee
and the saddle-point condition (\ref{eq:minimum-E}) gives
\be
\rho_0 = \left( \frac{4}{3\lambda} \right)^{3/2} M , \;\;\; R = \sqrt{ \frac{3\lambda}{4\pi} } .
\label{eq:rho0-TF}
\ee
Up to factors of order unity, we recover the exact result (\ref{eq:Rsol-lambda}).

\subsubsection{High-density regime}

In the high density regime (\ref{eq:rho-sol-FDM}) we write
\be
\rho_0 \gg \rho_c : \; E_{\rm tot} \simeq E_K + E_N  =
\frac{\epsilon^2 3\pi M^{1/3} \rho_0^{2/3}}{4}  -\frac{M^{5/3} \rho_0^{1/3}}{\sqrt{2}} ,
\ee
and the saddle-point condition (\ref{eq:minimum-E}) gives
\be
\rho_0 = \left( \frac{\sqrt{2}}{\epsilon^2 3\pi} \right)^3 M^4 , \;\;\;
R = \sqrt{ \frac{\pi}{2} } \frac{3 \epsilon^2 }{M} .
\label{eq:R-FDM}
\ee
We recover the scaling (\ref{eq:rho-sol-FDM}).
For a virial velocity $v^2=M/R$, this gives for the dimensionless de Broglie wave length
(\ref{eq:de-Broglie-rescaled})
\be
\lambda_{\rm dB} = \left( \frac{\pi}{2} \right)^{1/4} \frac{2\pi \sqrt{3} \epsilon^2}{M} ,
\ee
which as expected is of the same order as the radius $R$ in Eq.(\ref{eq:R-FDM}).

\section{Numerical simulations}\label{sec:numerical-method}

\subsection{Initial conditions and simulation set up}

\begin{figure}
\centering
\includegraphics[height=6.5cm,width=0.49\textwidth]{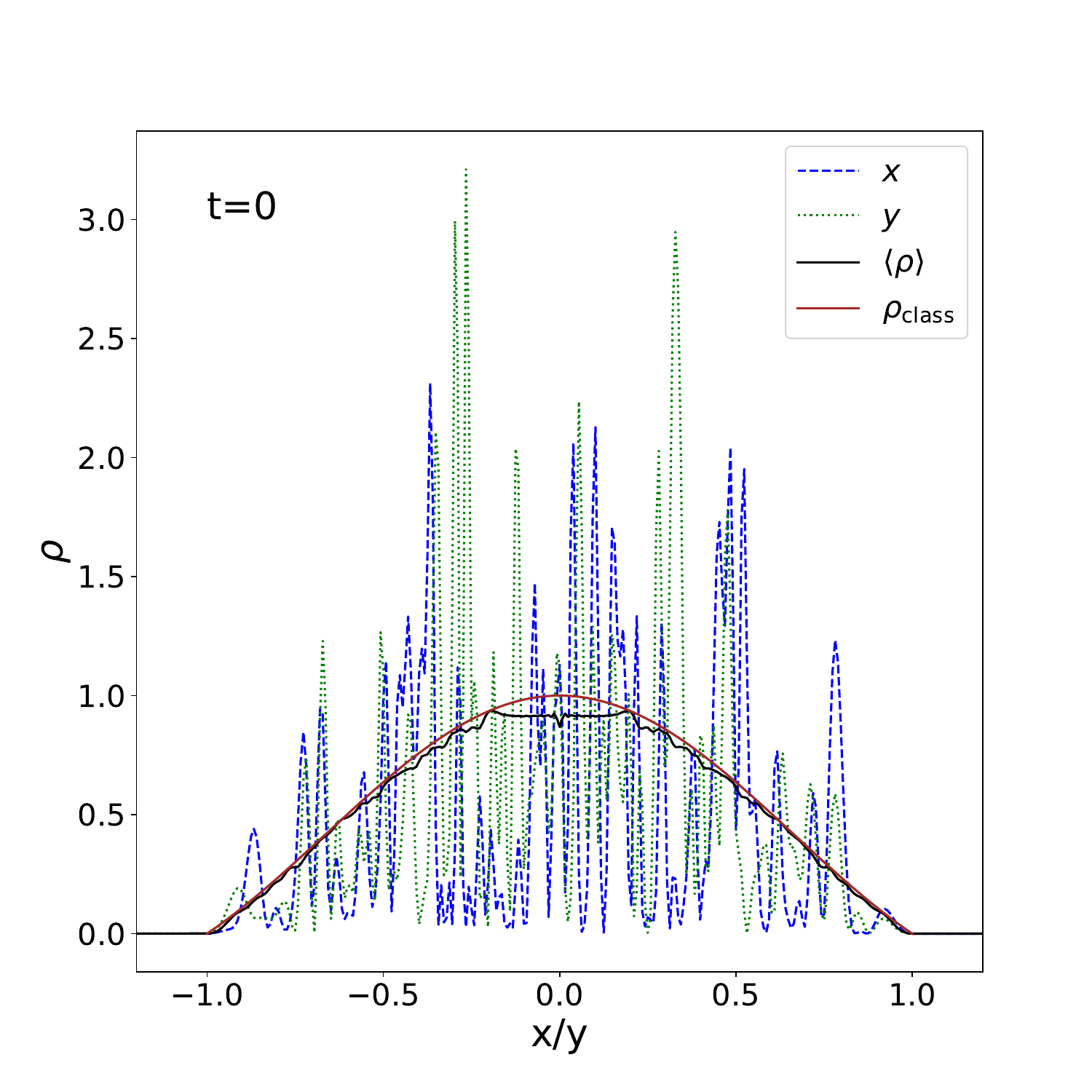}
\includegraphics[height=6.5cm,width=0.49\textwidth]{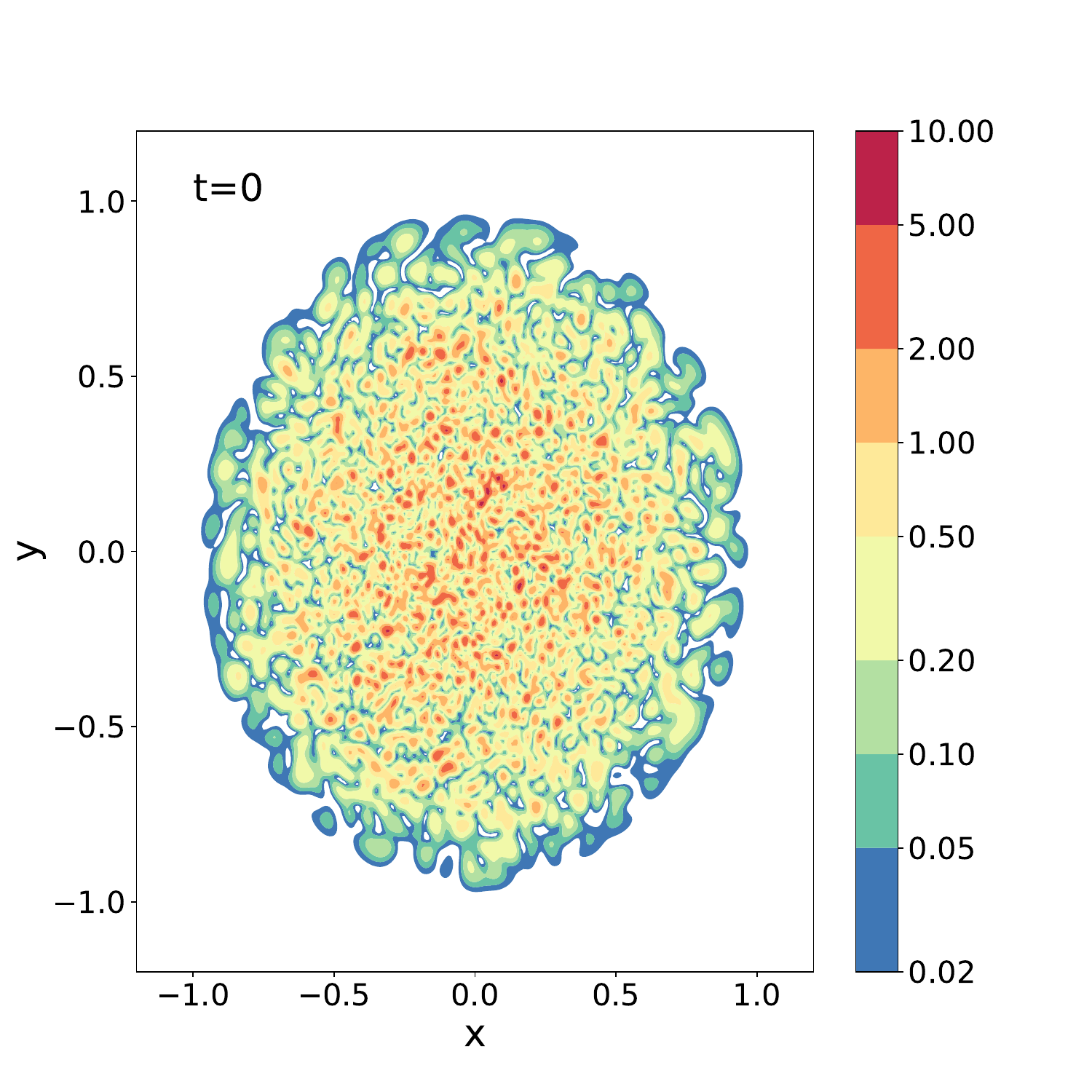}
\caption{Initial condition of our simulations.
{\it Upper panel:} density profiles along the x and y axis (dashed and dotted lines with large
fluctuations), mean density $\langle\rho\rangle$ of Eq.(\ref{eq:rho-average}) (black line with
small wiggles) and classical density profile $\rho_{\rm class}$ (smooth red line).
{\it Lower panel:} 2D density map on the $(x,y)$ plane at $z=0$.}
\label{fig:initial}
\end{figure}

As we are interested in the formation of possible solitons, we start our simulations without a central
soliton. Instead, our initial conditions correspond in a semi-classical limit to an equilibrium
phase space distribution of classical collisionless particles.
We choose for simplicity a Lane-Emden profile of polytropic index $n=1$ for this initial classical density profile,
\be
0 < r < 1 : \;\;\; \rho_{\rm class}(r) = \rho_0 \frac{\sin(\pi r)}{\pi r} .
\label{eq:rho-class}
\ee
This corresponds to a classical system of particles with the isotropic phase-space distribution
\be
- \frac{4\rho_0}{\pi} < E < 0 : \;\;\; f_{\rm class}(E) = \frac{1}{8\pi\sqrt{-2E}} , \;\;\;  E = \frac{v^2}{2} + \Phi_N .
\label{eq:fE-class}
\ee
Within our framework, this classical profile can be recovered as a weak limit $\epsilon \to 0$
of the Schr\"odinger equation. Then, one considers for the wave function \citep{Lin:2018whl,Yavetz:2021pbc} a sum over the eigenmodes
$\hat \psi_{n\ell m}(\vec x)$ of the Schr\"odinger equation defined by the target classical gravitational
potential $\Phi_{N\rm class}$,
\be
\psi(\vec x) = \sum_{n\ell m} a_{n\ell m} \hat\psi_{n \ell m}(\vec x) ,
\label{eq:psi-halo-a_nlm}
\ee
with
\be
- \frac{\epsilon^2}{2} \nabla^2 \hat\psi_{n\ell m} + \Phi_{N\rm class} \hat\psi_{n\ell m} = E_{n\ell m} \hat \psi_{n\ell m} .
\ee
For spherically symmetric density and gravitational potential profiles, these modes can be expanded as usual
in spherical harmonics, $\hat\psi_{n\ell m}(\vec x) = {\cal R}_{n\ell}(r) Y^m_\ell(\theta,\varphi)$.
Then, one can check that by choosing for the coefficients $a_{n\ell m}$ in the expansion
(\ref{eq:psi-halo-a_nlm}) the stochastic values
\be
a_{n\ell m} = a(E_{n\ell}) e^{i\Theta_{n\ell m}} , \;\;\; a(E) = (2\pi \epsilon)^3 f_{\rm class}(E) ,
\ee
where the phases $\Theta_{n\ell m}$ are uncorrelated random variables with a uniform distribution over
$0\leq \Theta < 2\pi$, the mean density profile
\be
\langle\rho\rangle = \langle | \psi |^2 \rangle = \sum_{n\ell m} a(E_{n\ell})^2 | \hat\psi_{n \ell m}|^2
\label{eq:rho-average}
\ee
converges to the classical profile (\ref{eq:rho-class}) in the semi-classical limit \citep{Binney2008}
\be
\epsilon \to 0 : \;\;\; \langle\rho(r)\rangle \to \rho_{\rm class}(r) .
\ee
Here the statistical average $\langle\dots\rangle$ is taken over the random phases $\Theta_{n\ell m}$.
This provides a realization of the classical profile (\ref{eq:rho-class})-(\ref{eq:fE-class}) within the
framework of the wave function $\psi$.
However, this only holds in a weak sense as each realization shows density fluctuations of order unity
on the de Broglie scale (\ref{eq:de-Broglie-rescaled}).
In particular, one has
\be
\langle \rho^2 \rangle_c \equiv \langle ( \rho - \langle\rho\rangle )^2 \rangle = \langle \rho \rangle .
\label{eq:rho-variance}
\ee
Moreover, here we neglected the self-interactions and the formation of solitons, which are absent in the
classical collisionless system.

Throughout this paper we take $\epsilon=0.01$, to focus on the semi-classical regime, and
$R=1$ for the initial radius of the halo, as in Eq.(\ref{eq:rho-class}).
We also take for the initial target central density $\rho_0=1$ in Eq.(\ref{eq:rho-class}),
which gives $\langle M \rangle = 4/\pi$.
Thus, our initial condition is a halo of mass, density and radius of order unity, in the dimensionless
coordinates (\ref{eq:dimesionless-def}), which would be an equilibrium configuration supported by its
velocity dispersion in the classical limit.
This initial condition, which is common to all the runs performed in this paper, is shown in
Fig.~\ref{fig:initial}. We can see that the density profile of a given realization shows large
fluctuations, in agreement with (\ref{eq:rho-variance}), whereas the mean $\langle\rho\rangle$
closely follows the classical profile $\rho_{\rm class}$, with small deviations due to the finite
value of $\epsilon$.

In addition to $\epsilon$, the equations of motion are determined by the self-interaction potential
(\ref{eq:model-1-tilde}).
Instead of the parameters $(\lambda,\rho_c)$ we use the parameters $(R_{\rm TF},\rho_c)$,
where $R_{\rm TF}$ is the soliton radius in the Thomas-Fermi regime (\ref{eq:Rsol-lambda}).
This provides a more immediate sense of the strength of the self-interactions in the low-density
regime. In this paper we consider the two cases $R_{\rm TF} = 0.5$ and $R_{\rm TF} = 0.1$,
associated with large and small self-interactions as compared with the system size and density.
We consider three cases for $\rho_c$, with a small, intermediate and large value.
These different choices for $\rho_c$ allow us to examine how changes in the density threshold
affect the dynamics.

\subsection{Numerical algorithm}

To solve the nonlinear Schrödinger--Poisson system (\ref{eq:Schrod-eps})-(\ref{eq:model-1-tilde})
we use a pseudo-spectral method with a split-step algorithm
\citep{1990JCoPh..87..108P,2008JLwT...26..302Z,2018JCAP...10..027E}.
The wave function after a time step $\Delta t$ is obtained as
\ba
&&\psi(\vec{x},t+ \Delta t) = \exp\left[-\frac{i \Delta t}{2\epsilon}
\Phi(\vec{x},t+ \Delta t)\right] \times \nonumber \\
&& \mathcal{F}^{-1} \exp\left[-\frac{i \epsilon\Delta t}{2} k^2\right]\mathcal{F}
\exp\left[-\frac{i \Delta t }{2\epsilon}\Phi(\vec{x},t)\right]\psi(\vec{x},t) \qquad .
\label{eq:step}
\ea
where $\mathcal{F}$ and $\mathcal{F}^{-1}$ are the discrete Fourier
transform and its inverse, $k$ is the wavenumber in Fourier space and $\Phi=\Phi_N+\Phi_I$.
The sequence of the operations is from right to left, and in a nutshell the algorithm can be explained as
follows: first, a half time step is taken where only the potential operator is applied, followed by a
complete time step where the Laplacian operator is applied. Afterwards, the potential field is updated,
and a final half time step is performed in the potential operator.
This algorithm makes use of the fact that the potential term in the Schrödinger equation
(\ref{eq:Schrod-eps}) is a local multiplication in configuration space whereas the Laplacian term is a local
multiplication in Fourier space.

We obtain the gravitational potential using Fourier transforms,
\be
\Phi_{N}(\vec{x}) = \mathcal{F}^{-1}\left(-\frac{4\pi}{k^2}\right)\mathcal{F} |\psi|^2 ,
\ee
and we apply periodic boundary conditions to our simulation box.

\section{Results for $R_{\rm TF}$ = 0.5}\label{sec:model-a-r-0.5}

We first consider in this section the case $R_{\rm TF}=0.5$, where the Thomas-Fermi radius
(\ref{eq:Rsol-lambda}) defined in the low-density regime is of the order of the system size.

\begin{figure*}[ht]
\centering
\includegraphics[height=4.cm,width=0.28\textwidth]{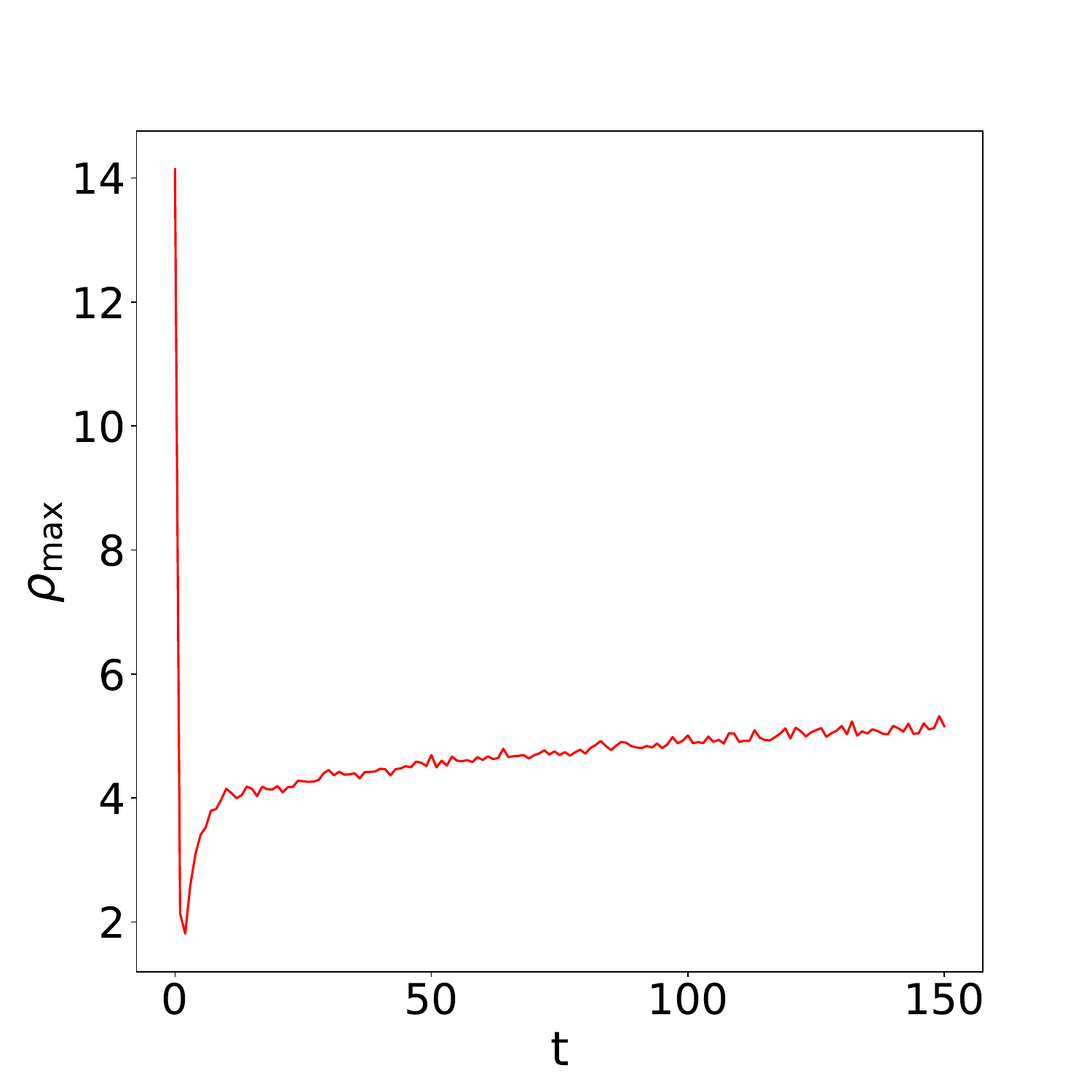}
\includegraphics[height=4.cm,width=0.28\textwidth]{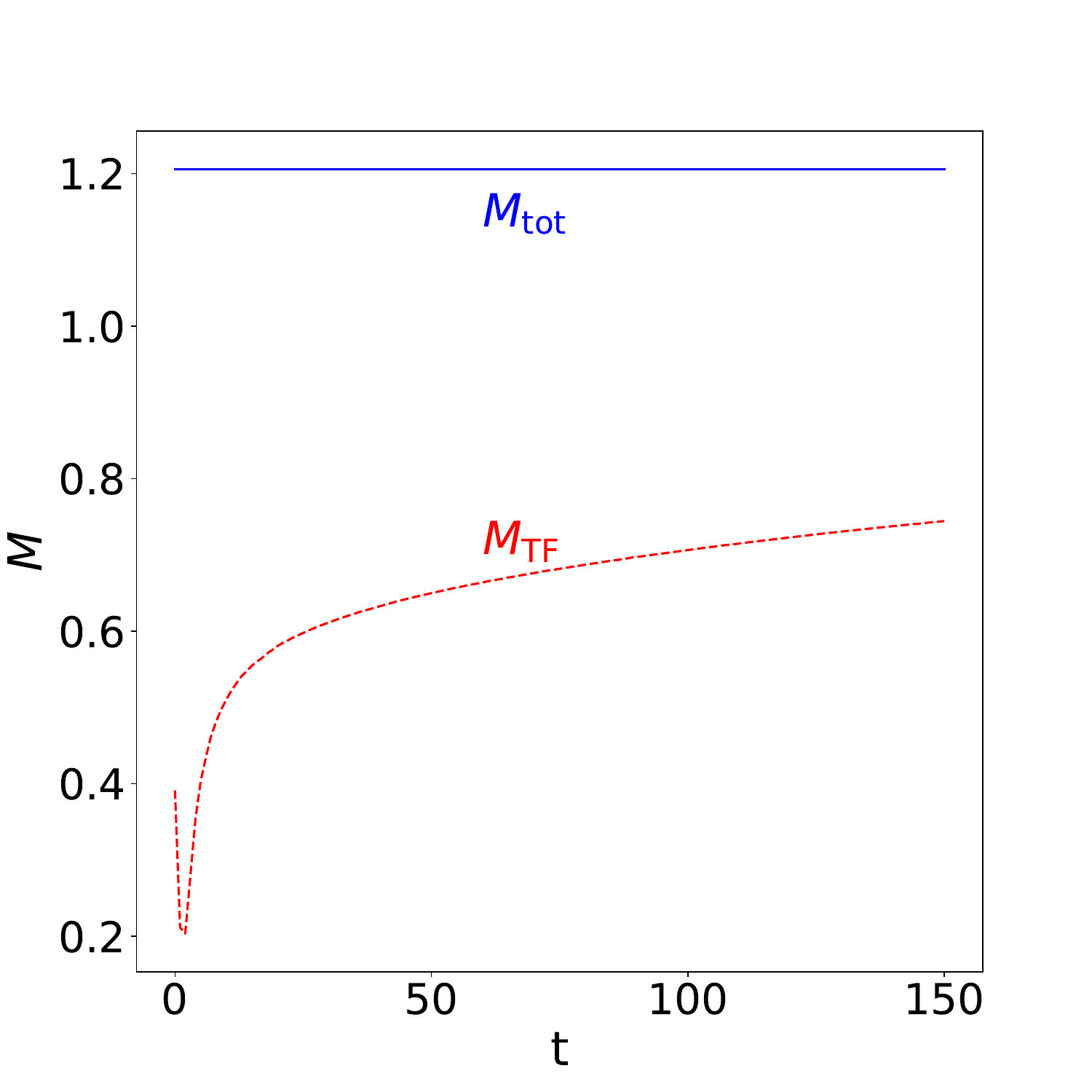}
\includegraphics[height=4.cm,width=0.28\textwidth]{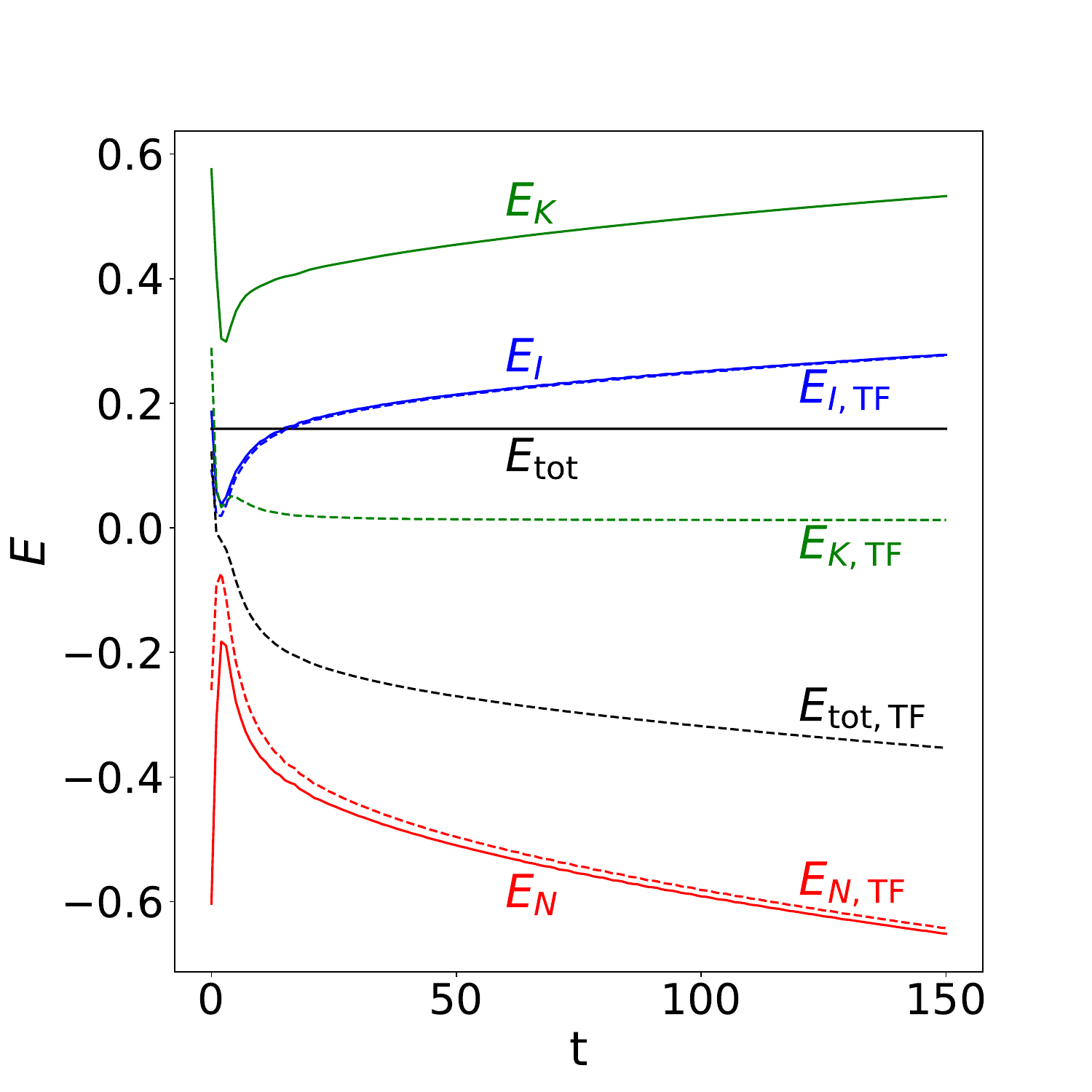}\\
\includegraphics[height=4.cm,width=0.28\textwidth]{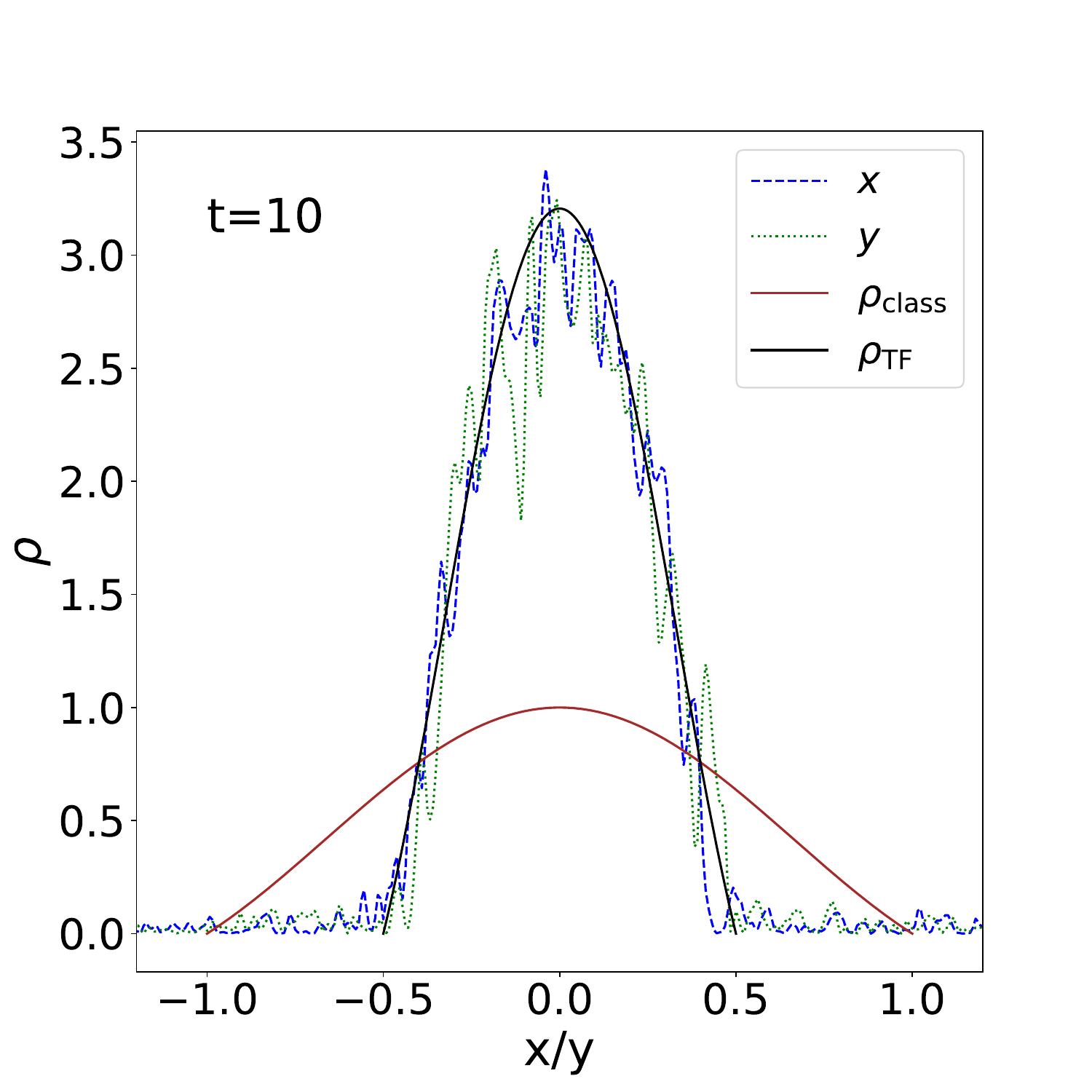}
\includegraphics[height=4.cm,width=0.28\textwidth]{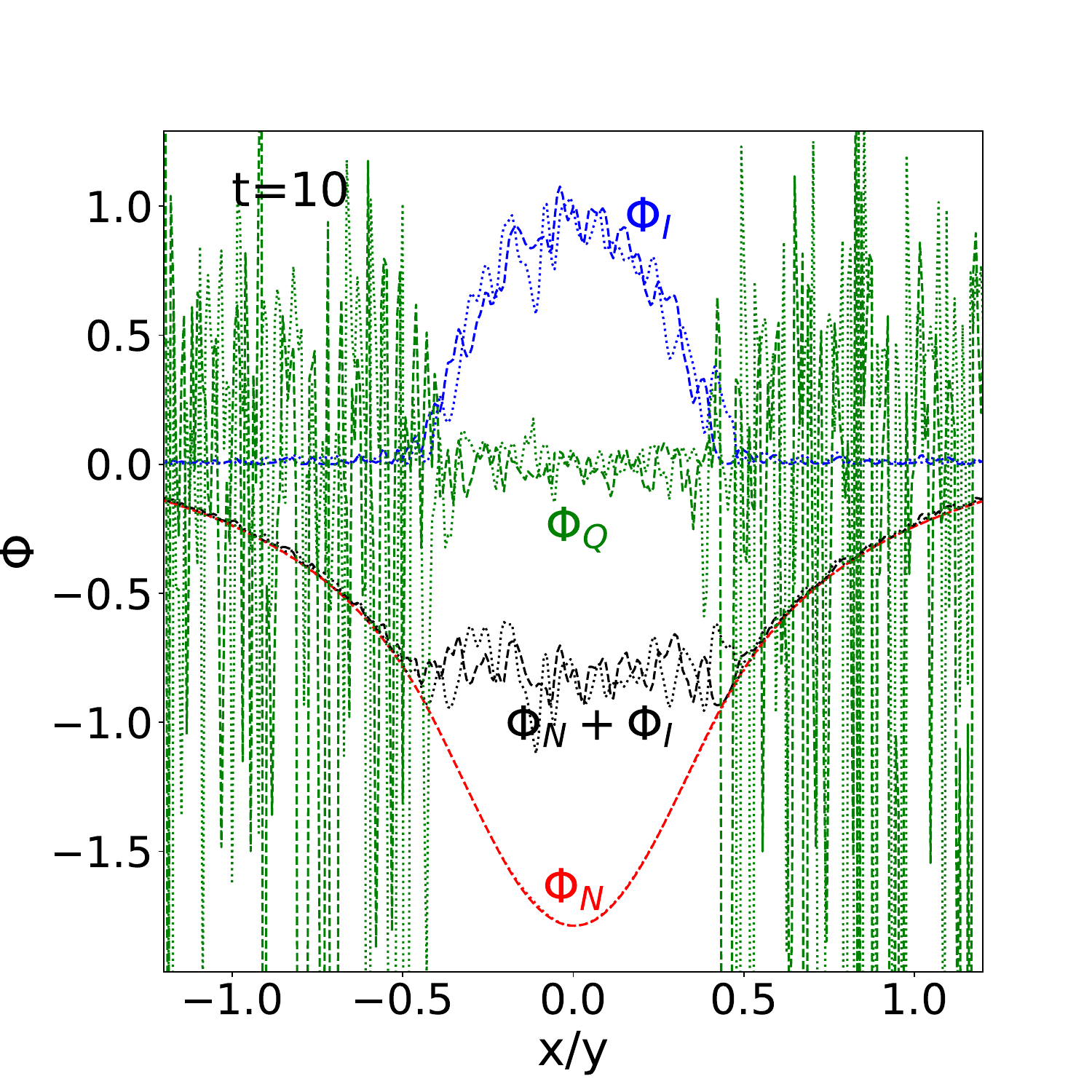}
\includegraphics[height=4.cm,width=0.29\textwidth]{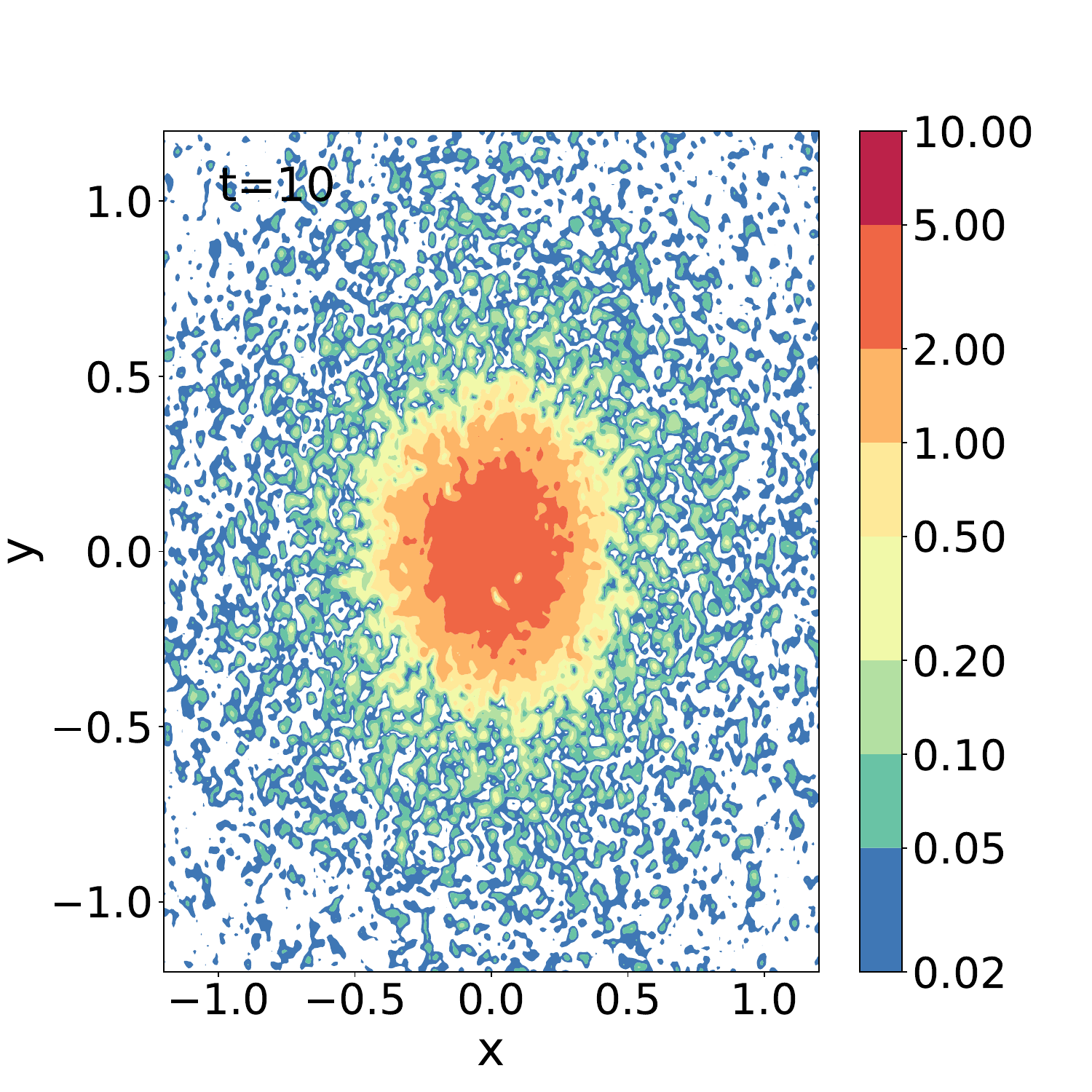}\\
\includegraphics[height=4.cm,width=0.28\textwidth]{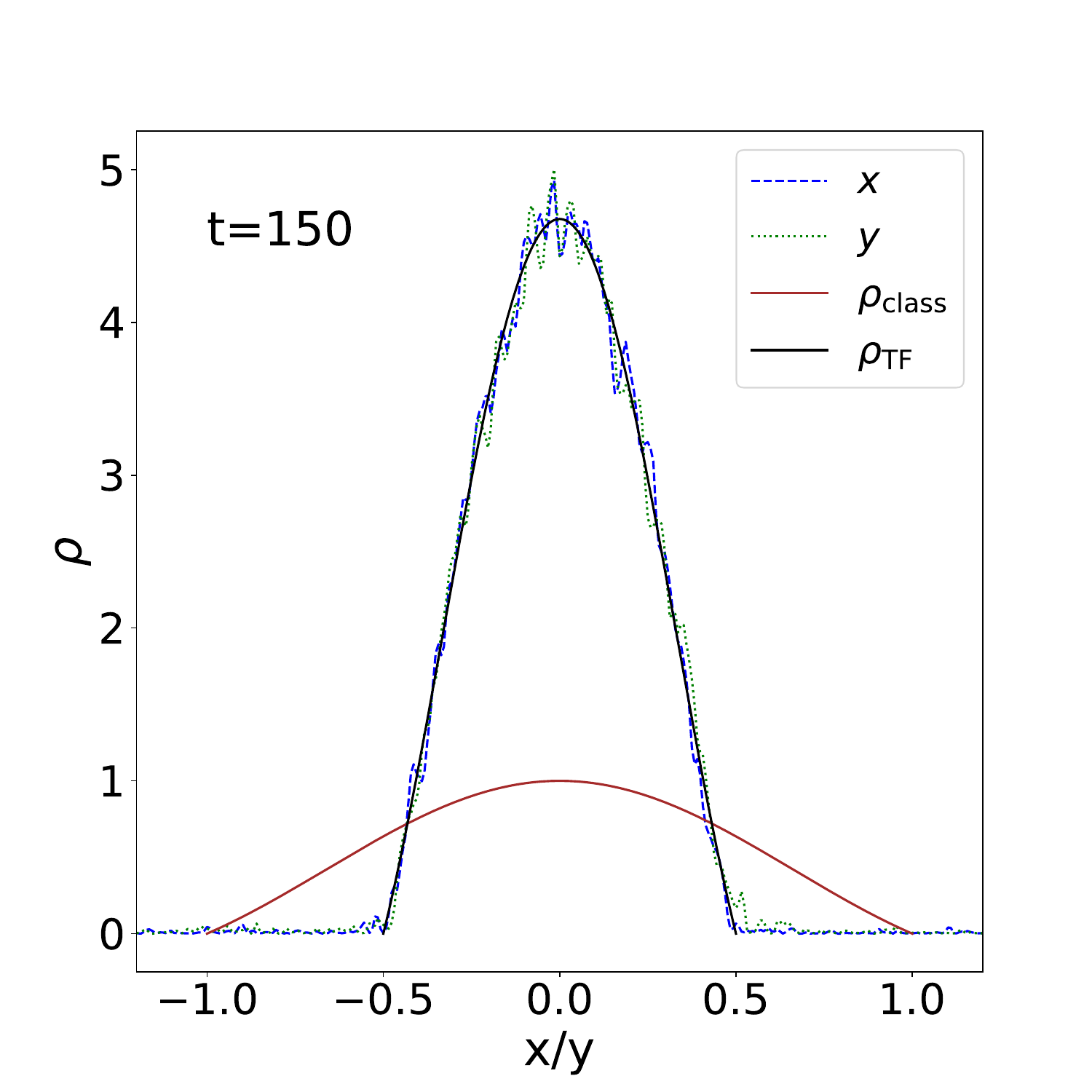}
\includegraphics[height=4.cm,width=0.28\textwidth]{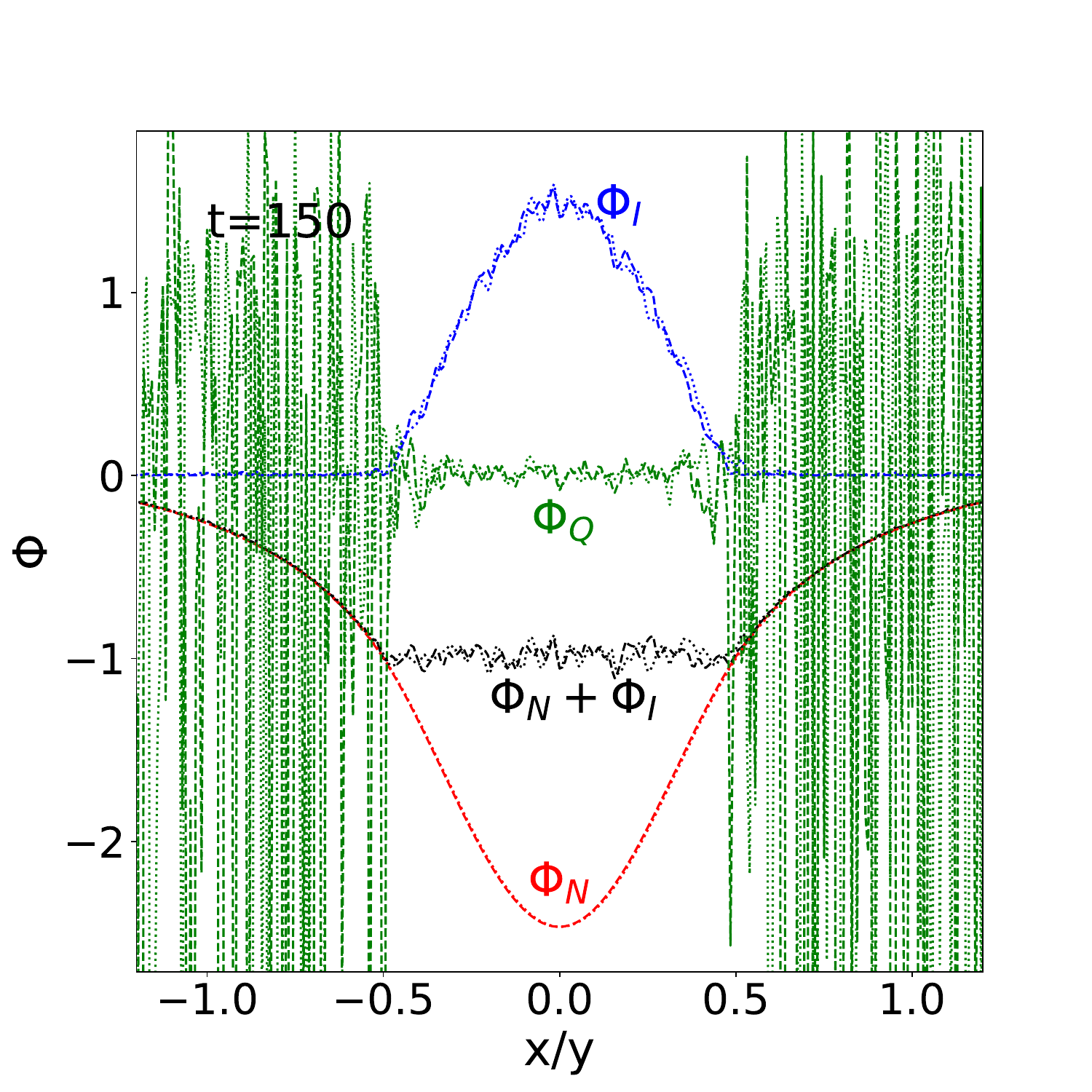}
\includegraphics[height=4.cm,width=0.29\textwidth]{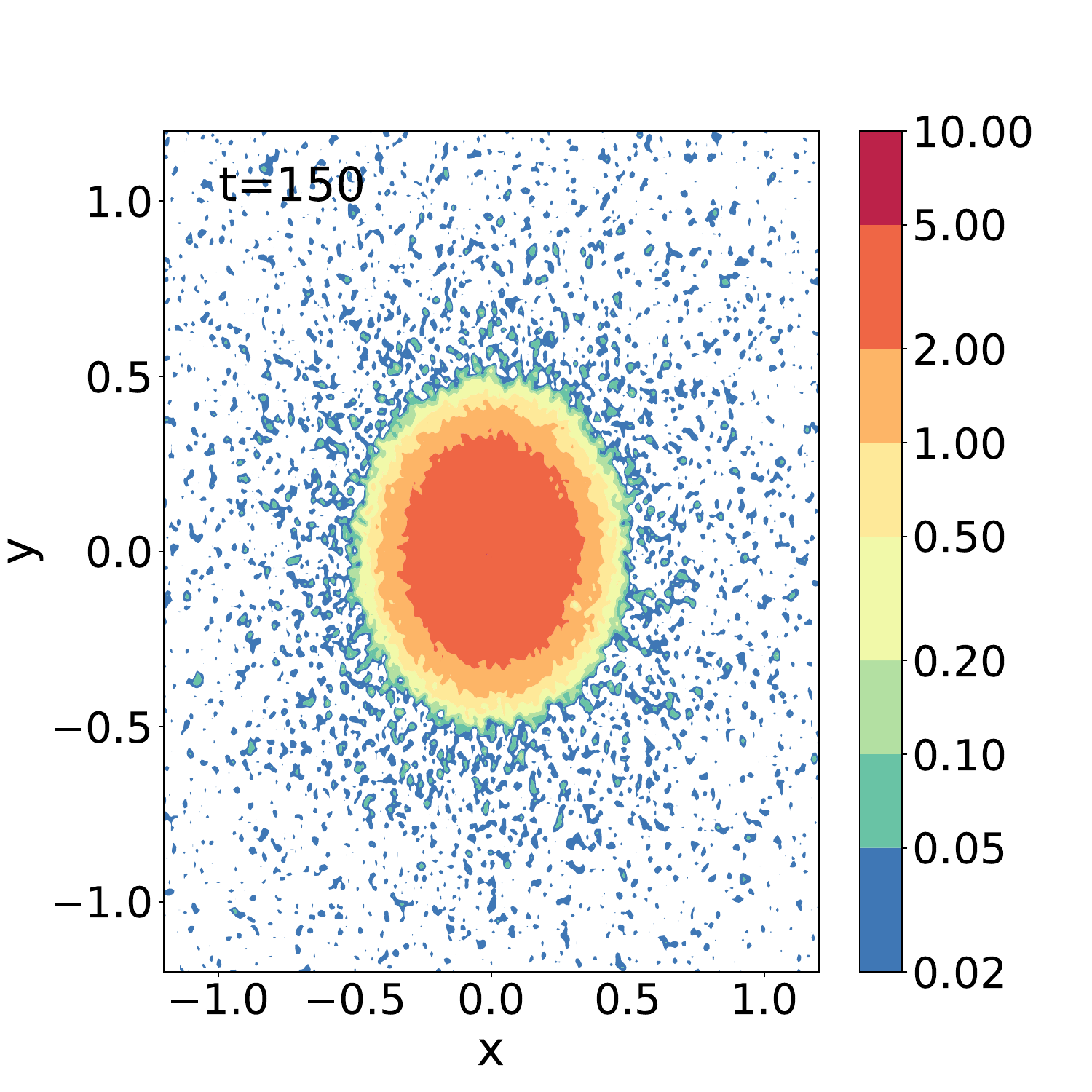}
\caption{
[$R_{\rm TF}=0.5$, $\rho_c=100$.]
(a) Evolution with time of the maximum density $\rho_{\max}$.
(b) Total mass $M_{\rm tot}$ in the system and mass $M_{\rm TF}$ enclosed within the Thomas-Fermi
radius (\ref{eq:rho0-TF}).
(c) Energies $E_K$, $E_I$, $E_N$, $E_{\rm tot}$ of the total system (solid lines), and
energies $E_{K,\rm TF}$, $E_{I,\rm TF}$, $E_{N,\rm TF}$, $E_{\rm tot, \rm TF}$ (dashed lines)
enclosed within the Thomas-Fermi radius (\ref{eq:rho0-TF}).
(d) Density $\rho$ along the $x$ (blue dashed line) and $y$ (green dotted line) axis running through
the center of the halo, at time $t=10$.
The smooth brown solid line is the initial classical density profile (\ref{eq:rho-class})
as in Fig.~\ref{fig:initial}. The smooth black solid line is the Thomas-Fermi soliton profile
(\ref{eq:rho-sol-TF}), normalized to the mass within radius $R_{\rm TF}$.
(e) Potentials $\Phi_N$ (red lines), $\Phi_I$ (blue lines), $\Phi_Q$ (green lines),
and the sum $\Phi_N+\Phi_Q$ (black lines), along the $x$ (dashed lines) and
$y$ (dotted lines) axis running through the center of the halo, at time $t=10$.
(f) 2D density map at time $t=10$ on the $(x,y)$ plane.
(g) Density $\rho$ at time $t=150$.
(h) Potentials at time $t=150$.
(i) 2D density map at time $t=150$.
}
\label{fig:R0p5-rhoc-100}
\end{figure*}

\begin{figure*}[ht]
\centering
\includegraphics[height=4.cm,width=0.28\textwidth]{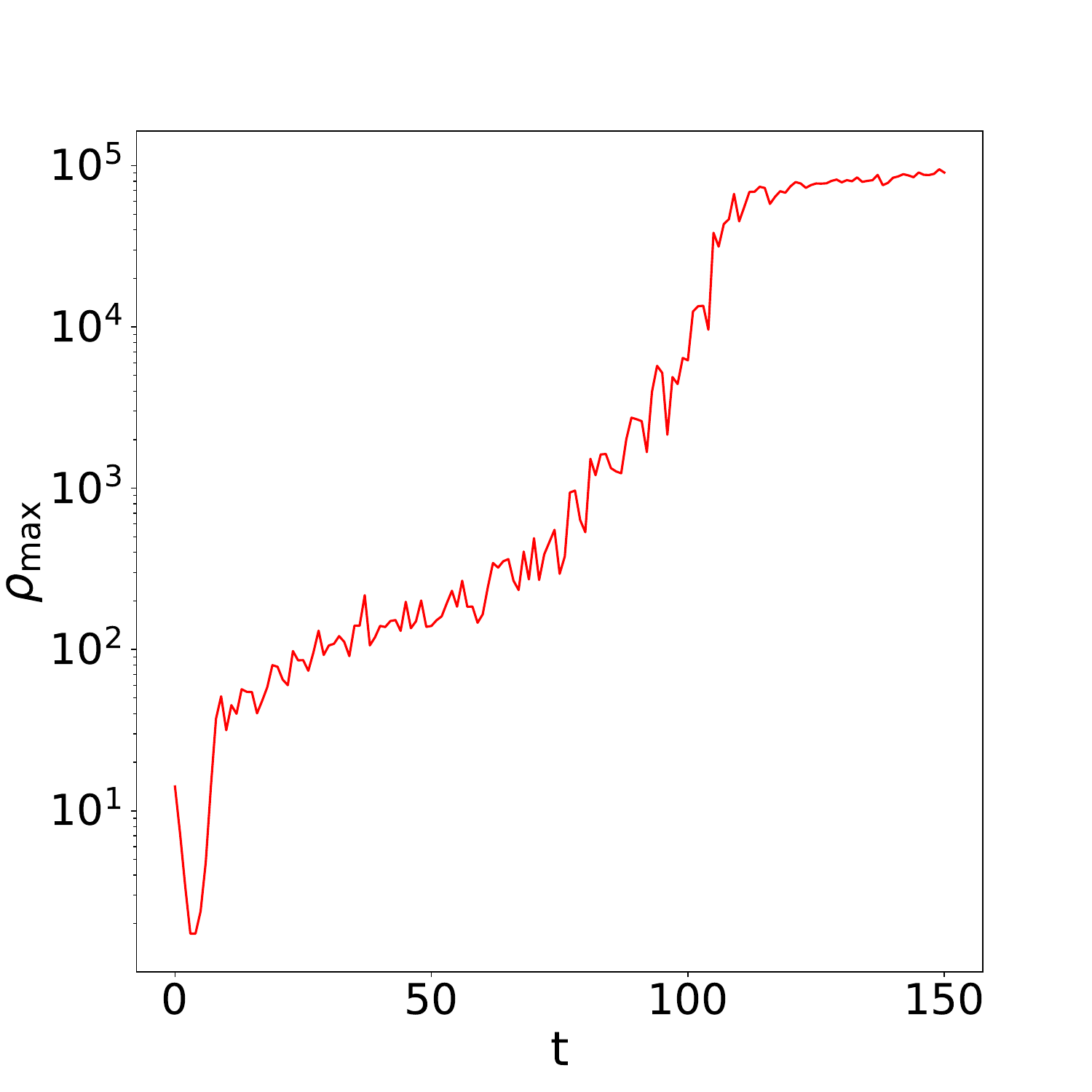}
\includegraphics[height=4.cm,width=0.28\textwidth]{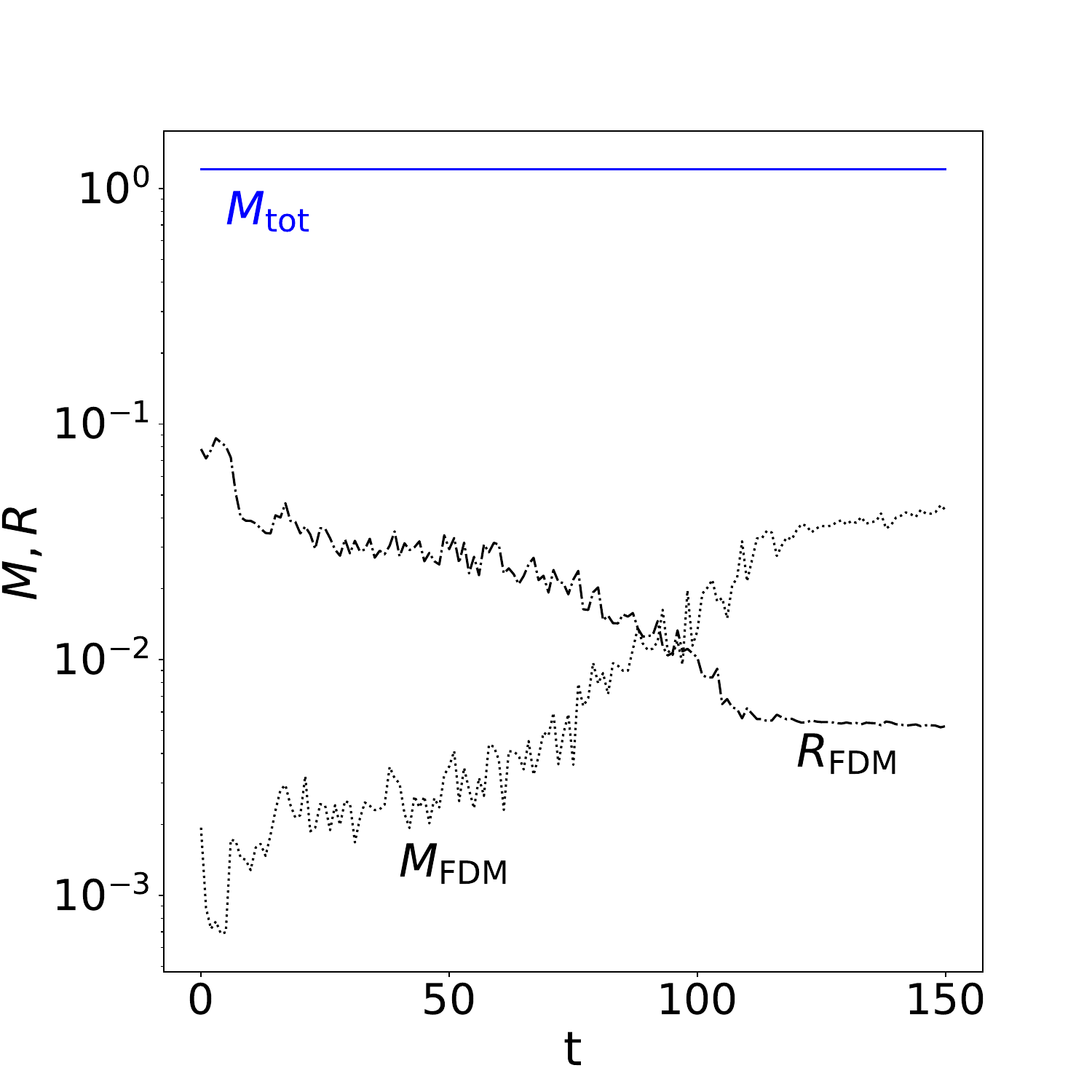}
\includegraphics[height=4.cm,width=0.28\textwidth]{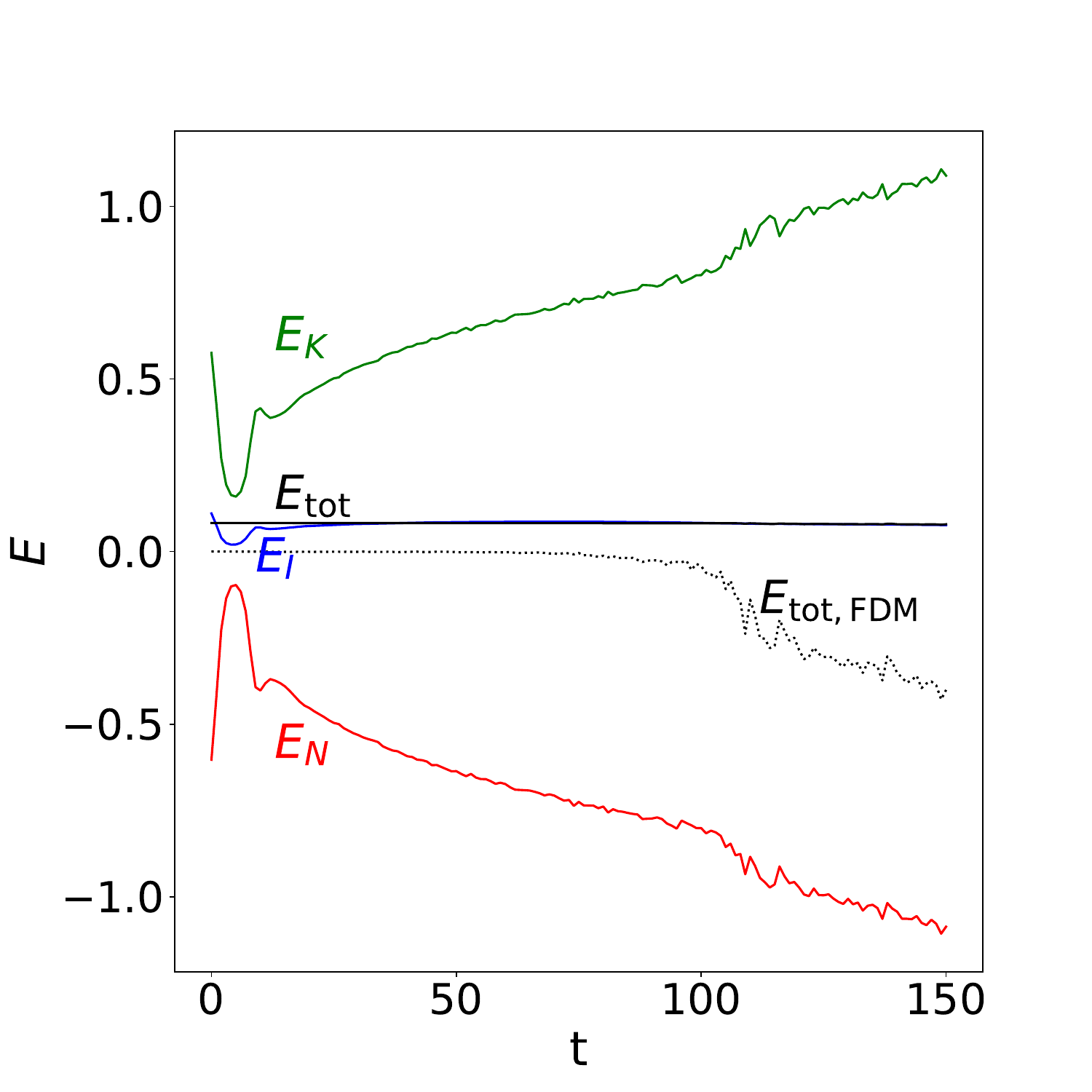}\\
\includegraphics[height=4.cm,width=0.28\textwidth]{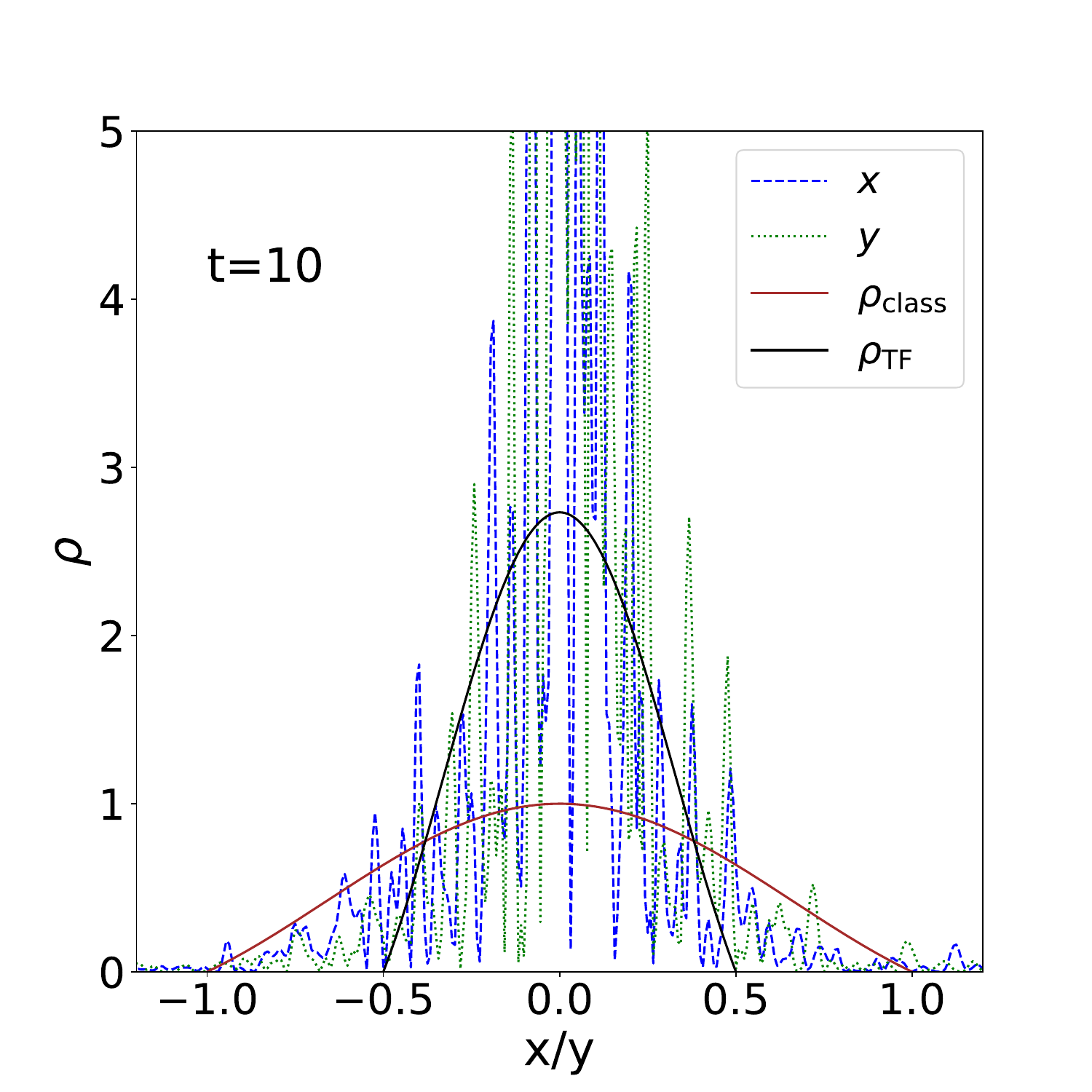}
\includegraphics[height=4.cm,width=0.28\textwidth]{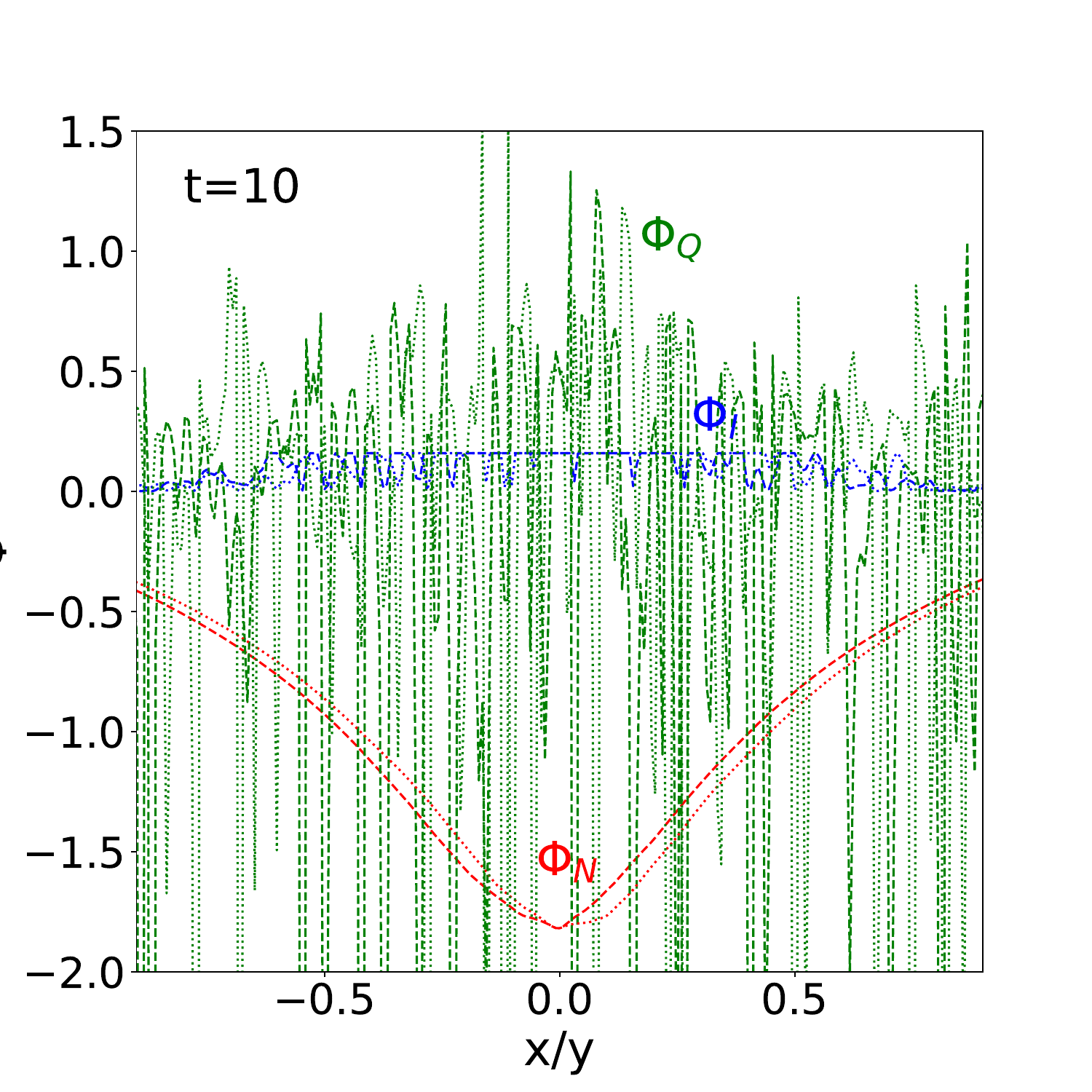}
\includegraphics[height=4.cm,width=0.29\textwidth]{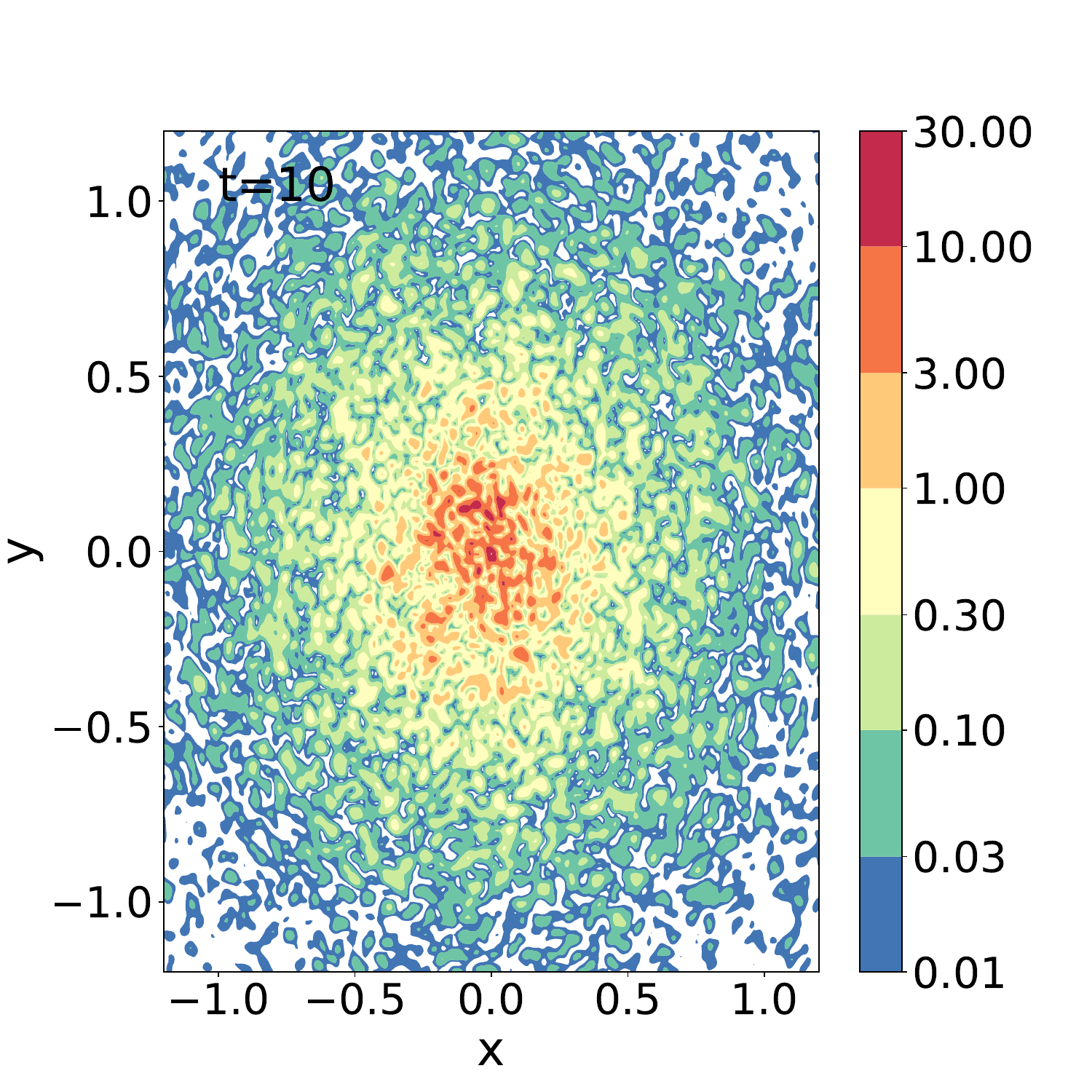}\\
\includegraphics[height=4.cm,width=0.28\textwidth]{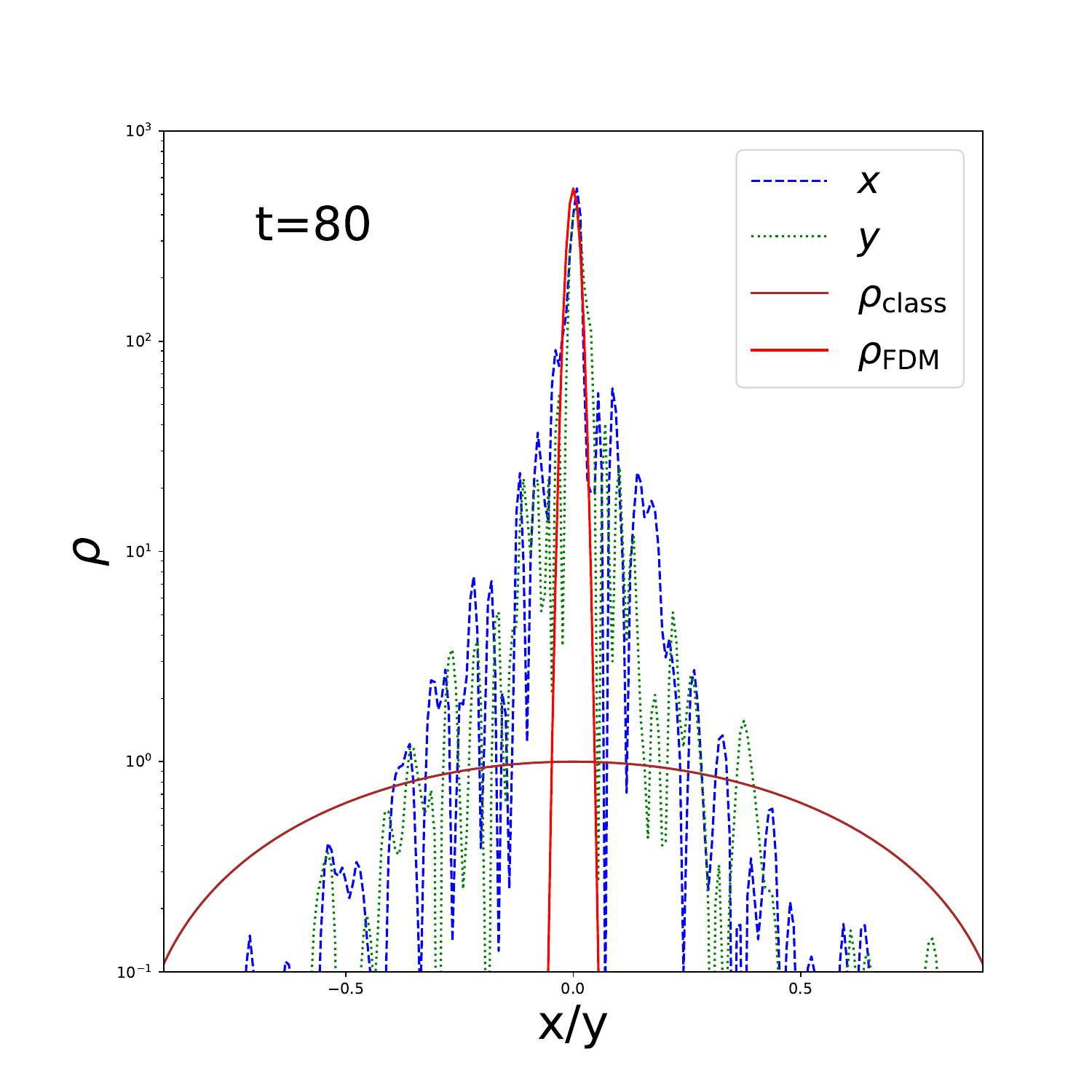}
\includegraphics[height=4.cm,width=0.28\textwidth]{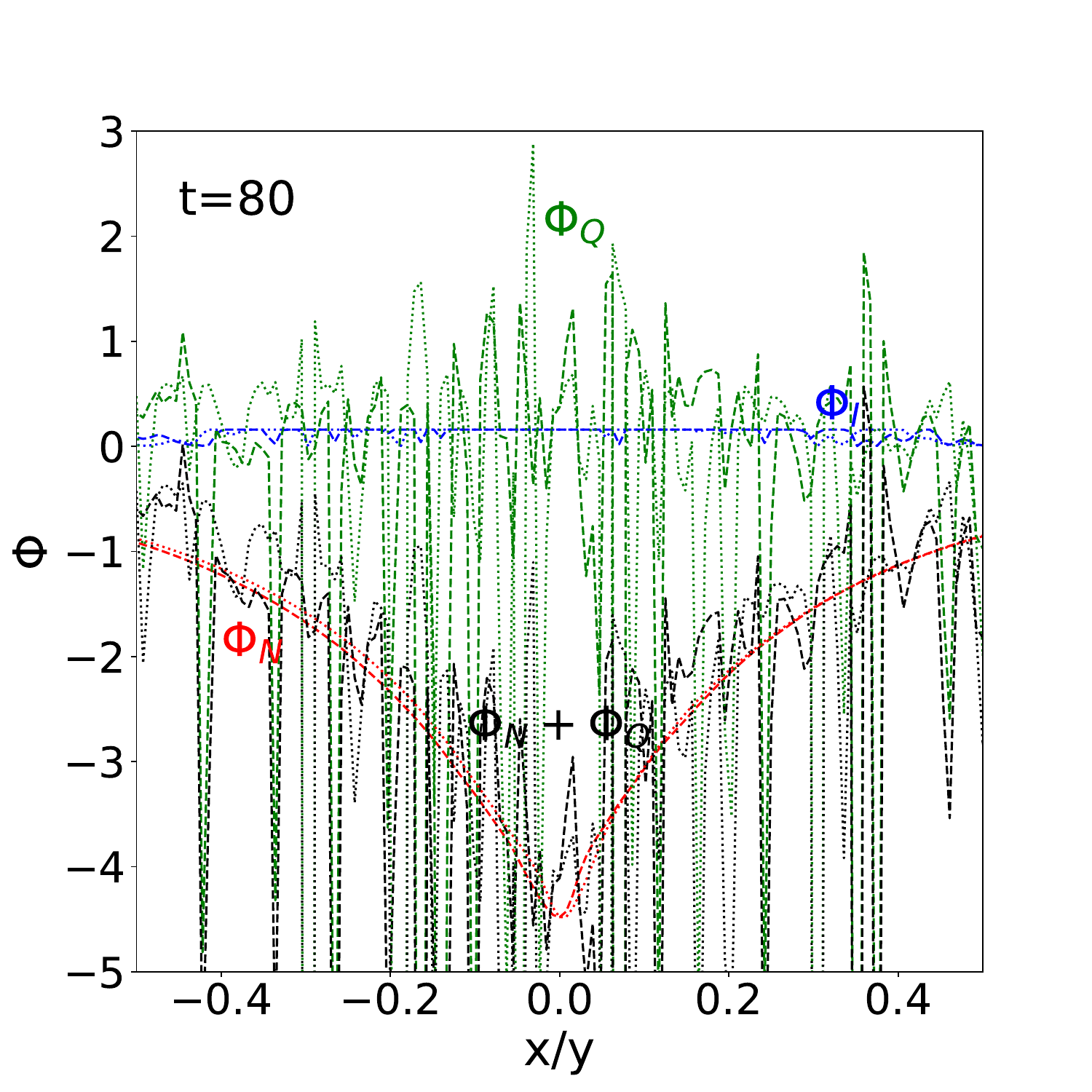}
\includegraphics[height=4.cm,width=0.29\textwidth]{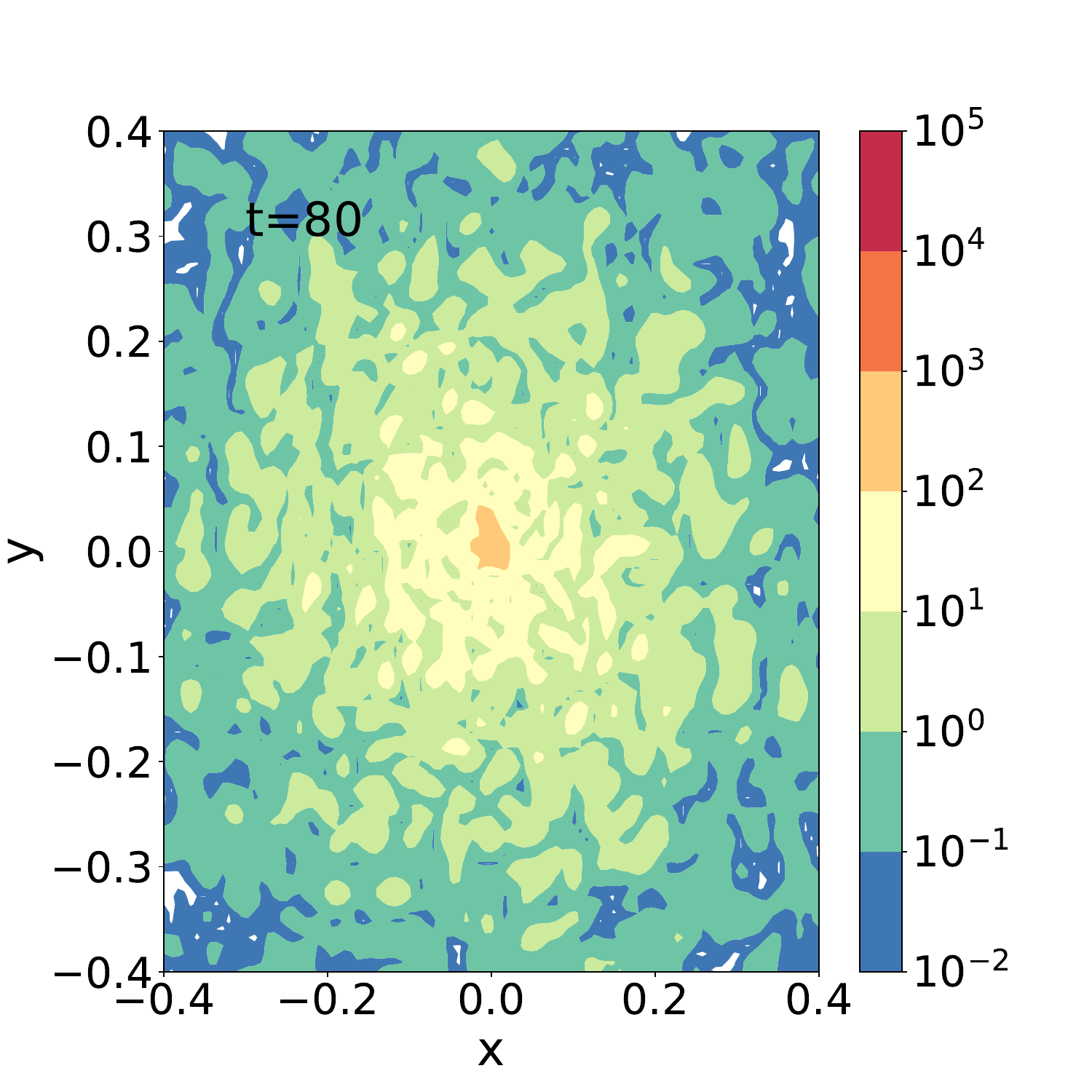}\\
\includegraphics[height=4.cm,width=0.28\textwidth]{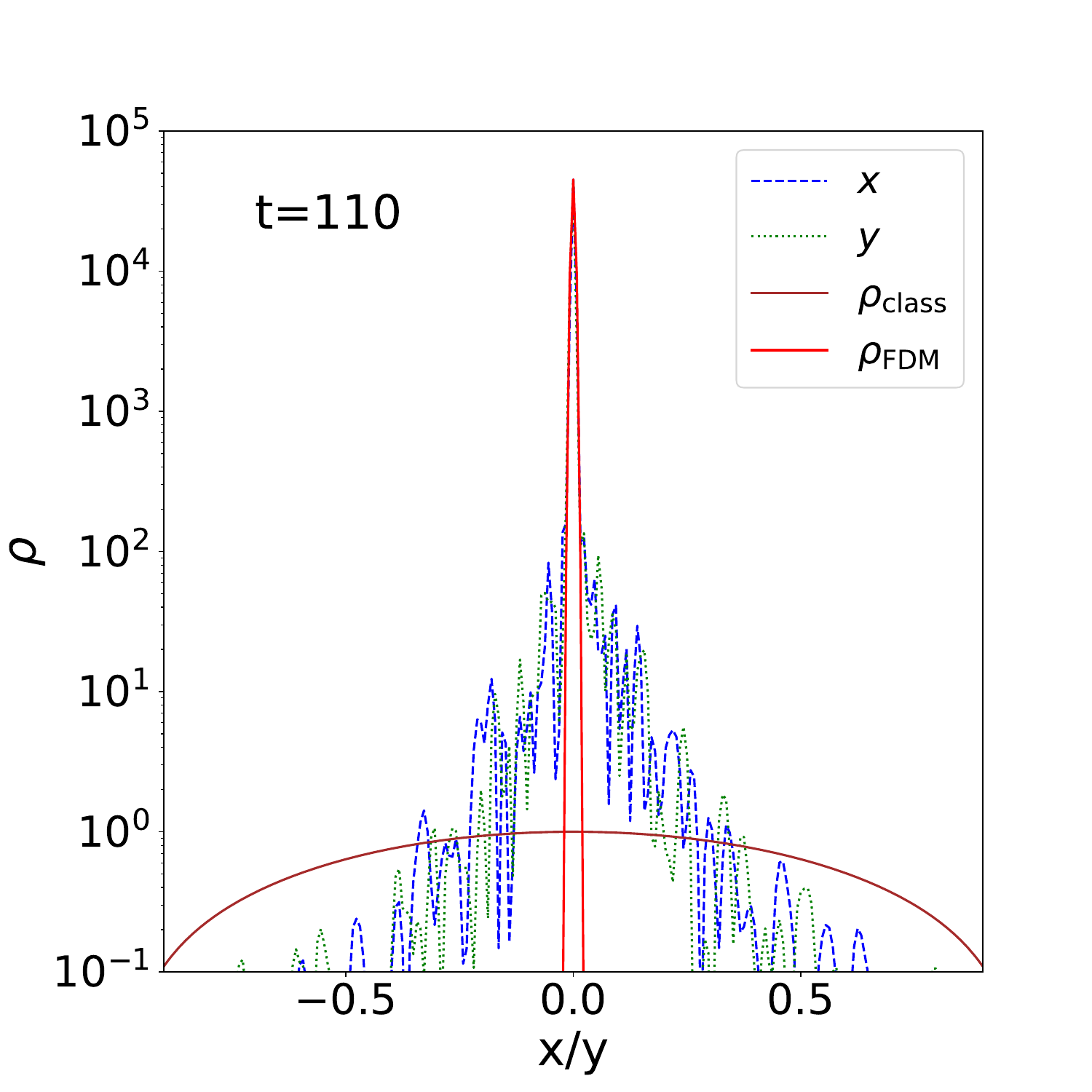}
\includegraphics[height=4.cm,width=0.28\textwidth]{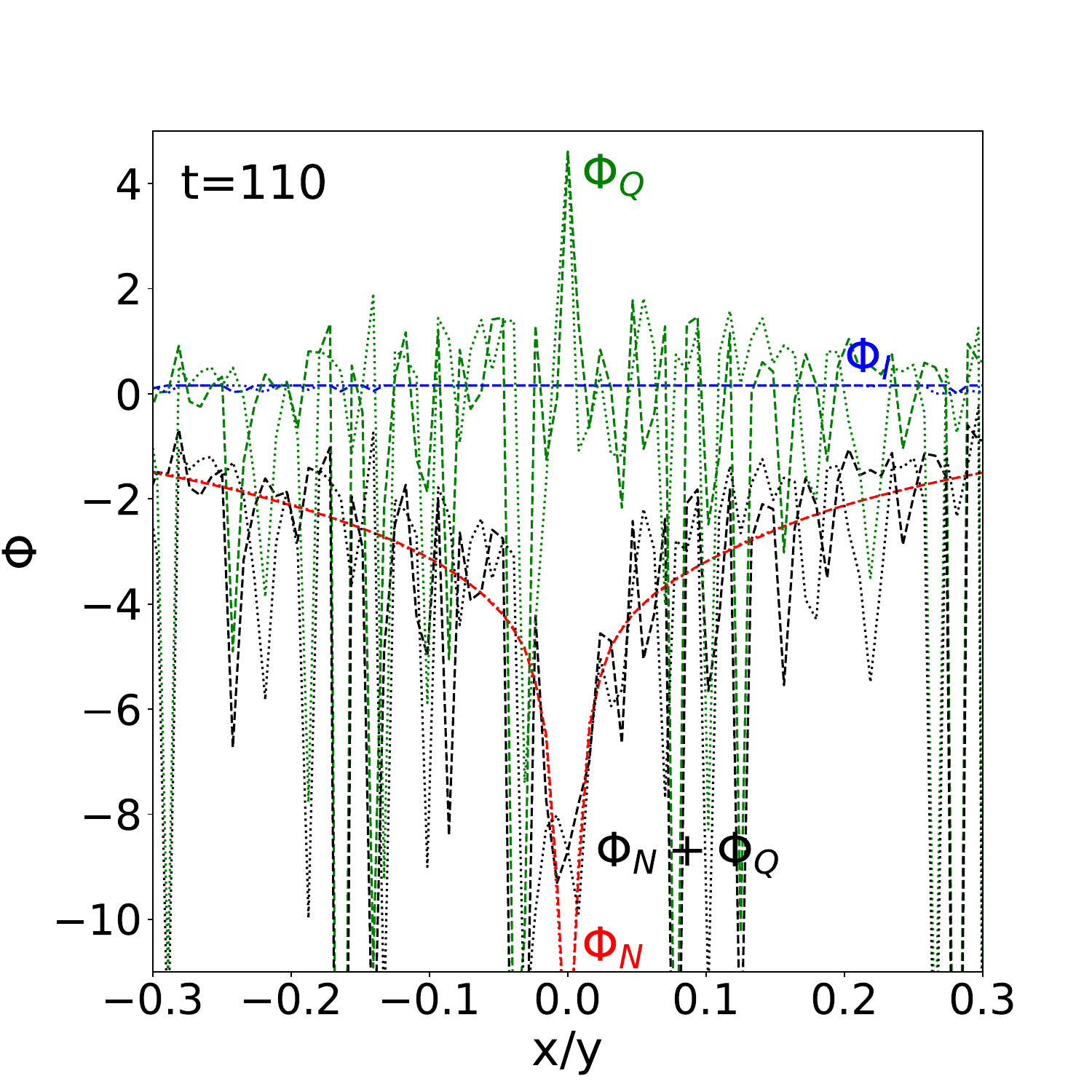}
\includegraphics[height=4.cm,width=0.29\textwidth]{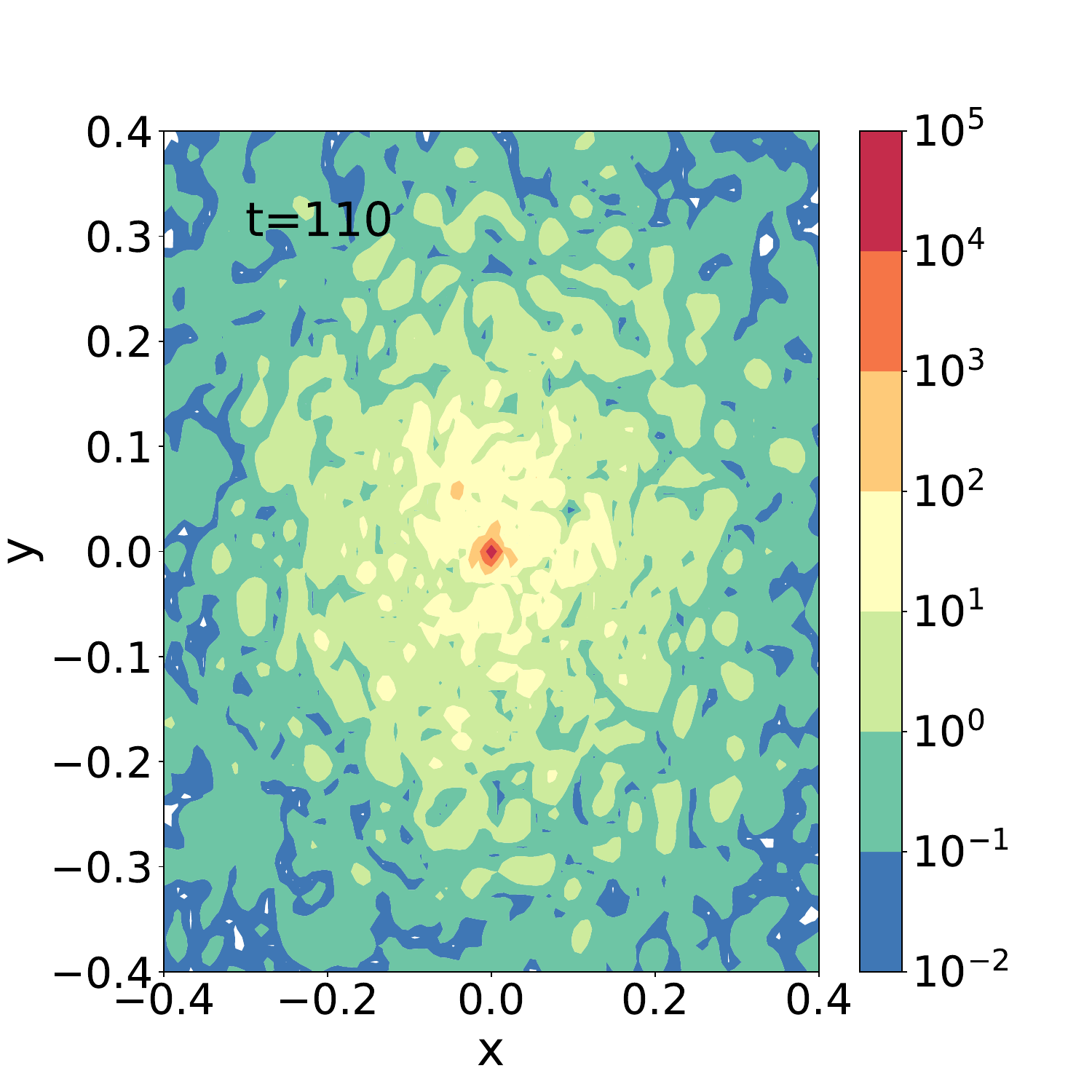}\\
\caption{
[$R_{\rm TF}=0.5$, $\rho_c=0.5$.]
Same panels as in Fig.~\ref{fig:R0p5-rhoc-100}.
However, in panels (g) and (j) we also show the Gaussian FDM profile (\ref{eq:R-FDM}) (red solid curve),
normalized to the mass $M_{\rm FDM}$ enclosed within radius $R_{\rm FDM}$.
In panel (b) we also show the radius $R_{\rm FDM}$ and the enclosed mass $M_{\rm FDM}$.
In panel (c) we also show the total energy $E_{\rm tot, FDM}$ within radius $R_{\rm FDM}$.
In panels (h) and (k) we show the sum $\Phi_N+\Phi_Q$ instead of $\Phi_N+\Phi_I$.
}
\label{fig:R0p5-rhoc-0p5}
\end{figure*}

\begin{figure*}[ht]
\centering
\includegraphics[height=4.cm,width=0.28\textwidth]{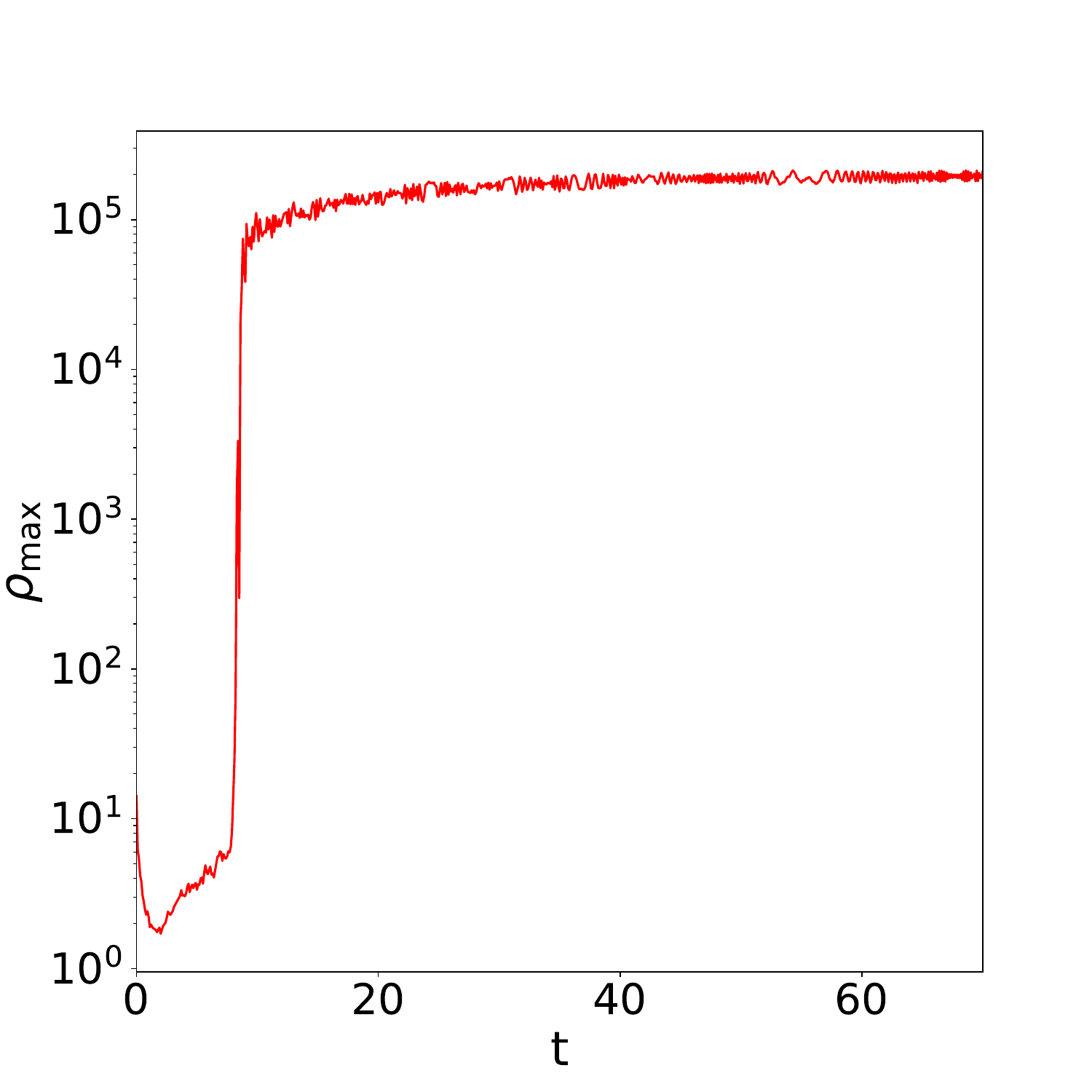}
\includegraphics[height=4.cm,width=0.28\textwidth]{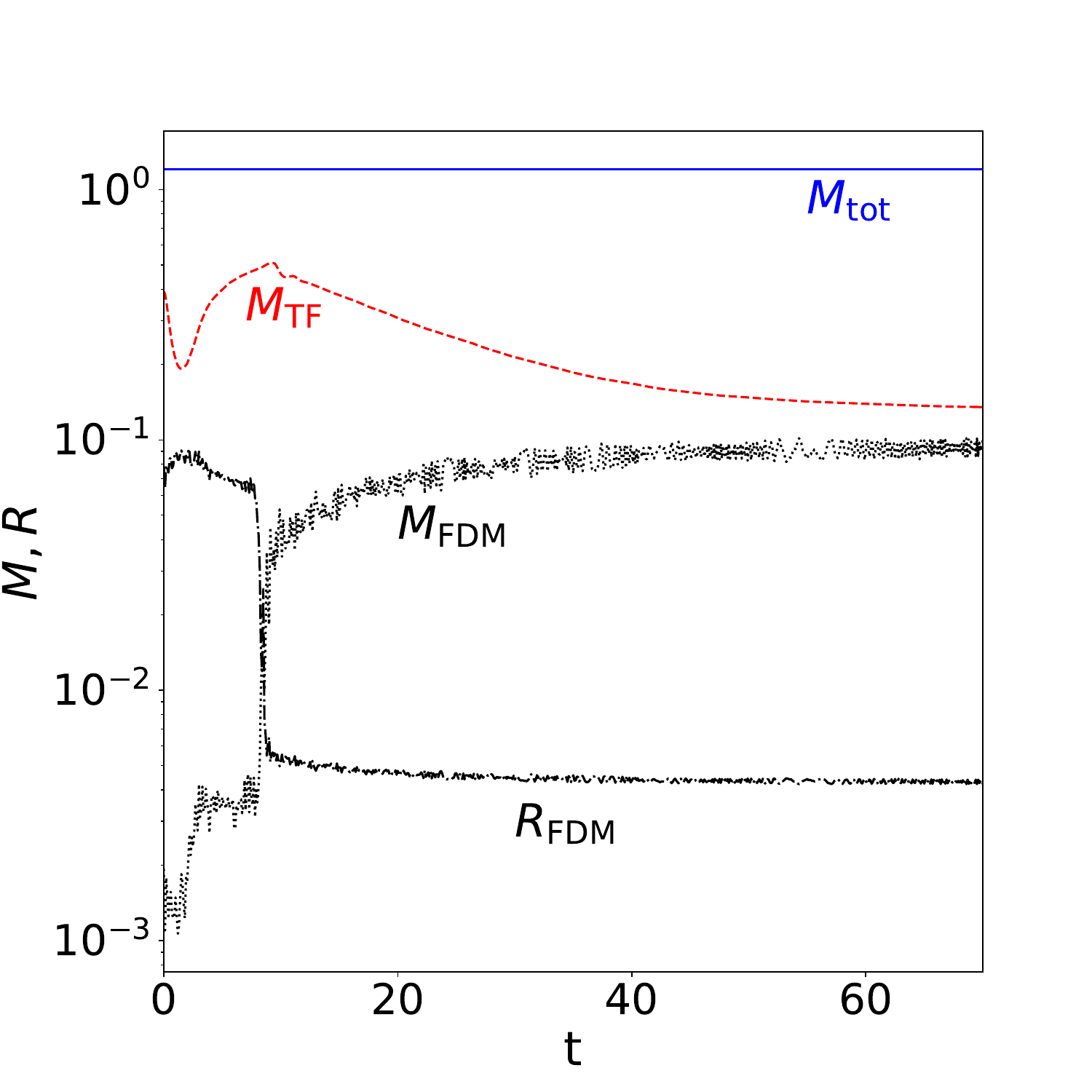}
\includegraphics[height=4.cm,width=0.28\textwidth]{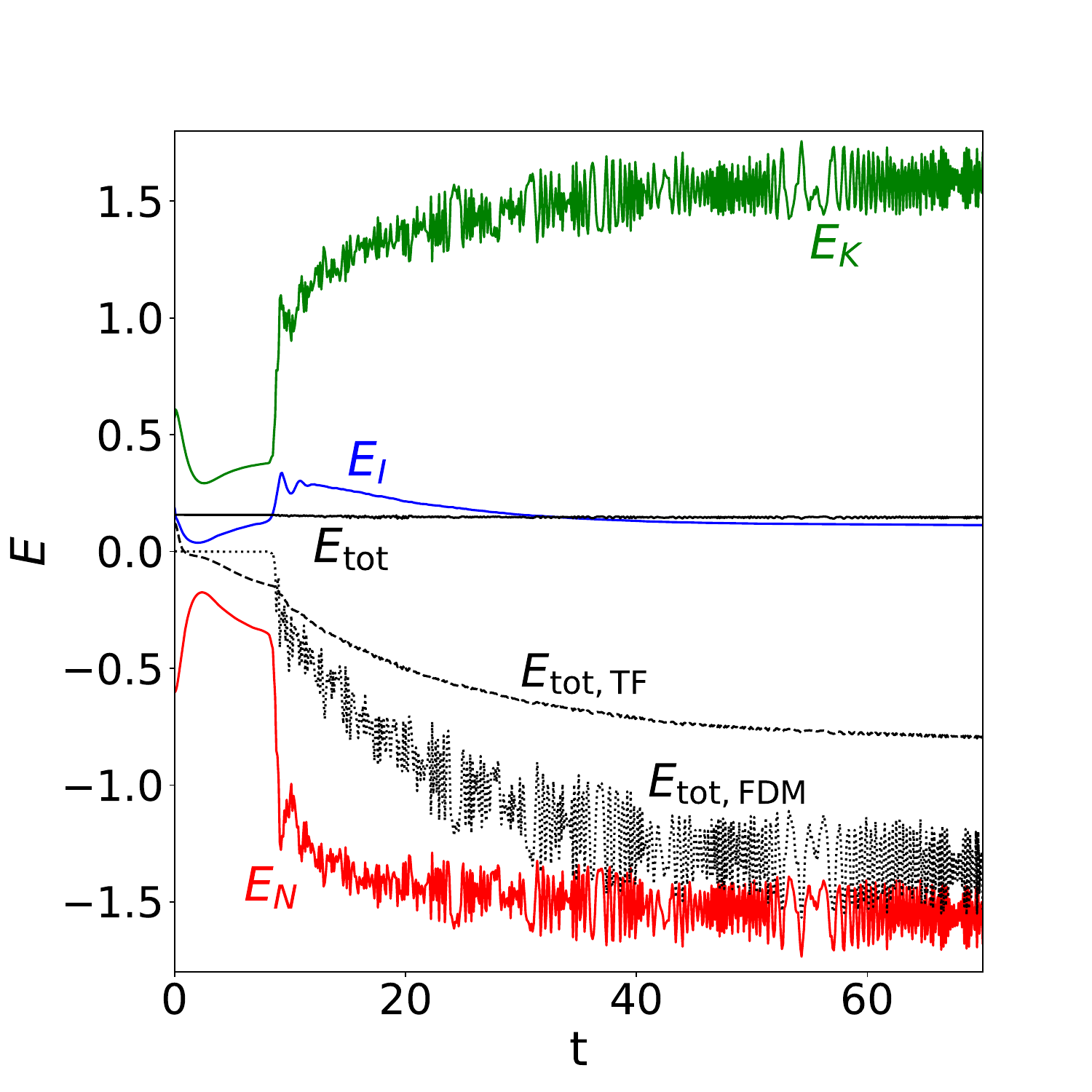}\\
\includegraphics[height=4.cm,width=0.28\textwidth]{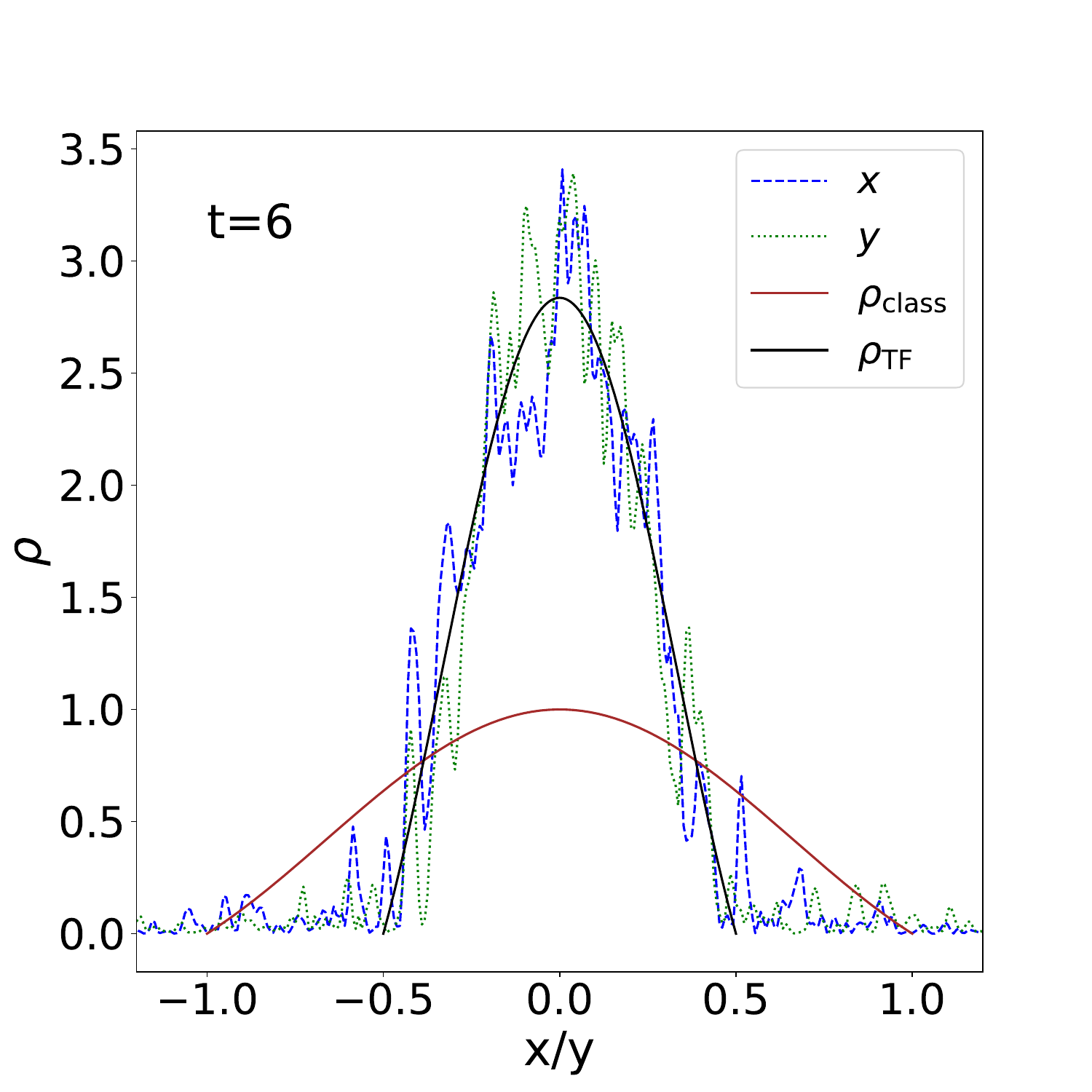}
\includegraphics[height=4.cm,width=0.28\textwidth]{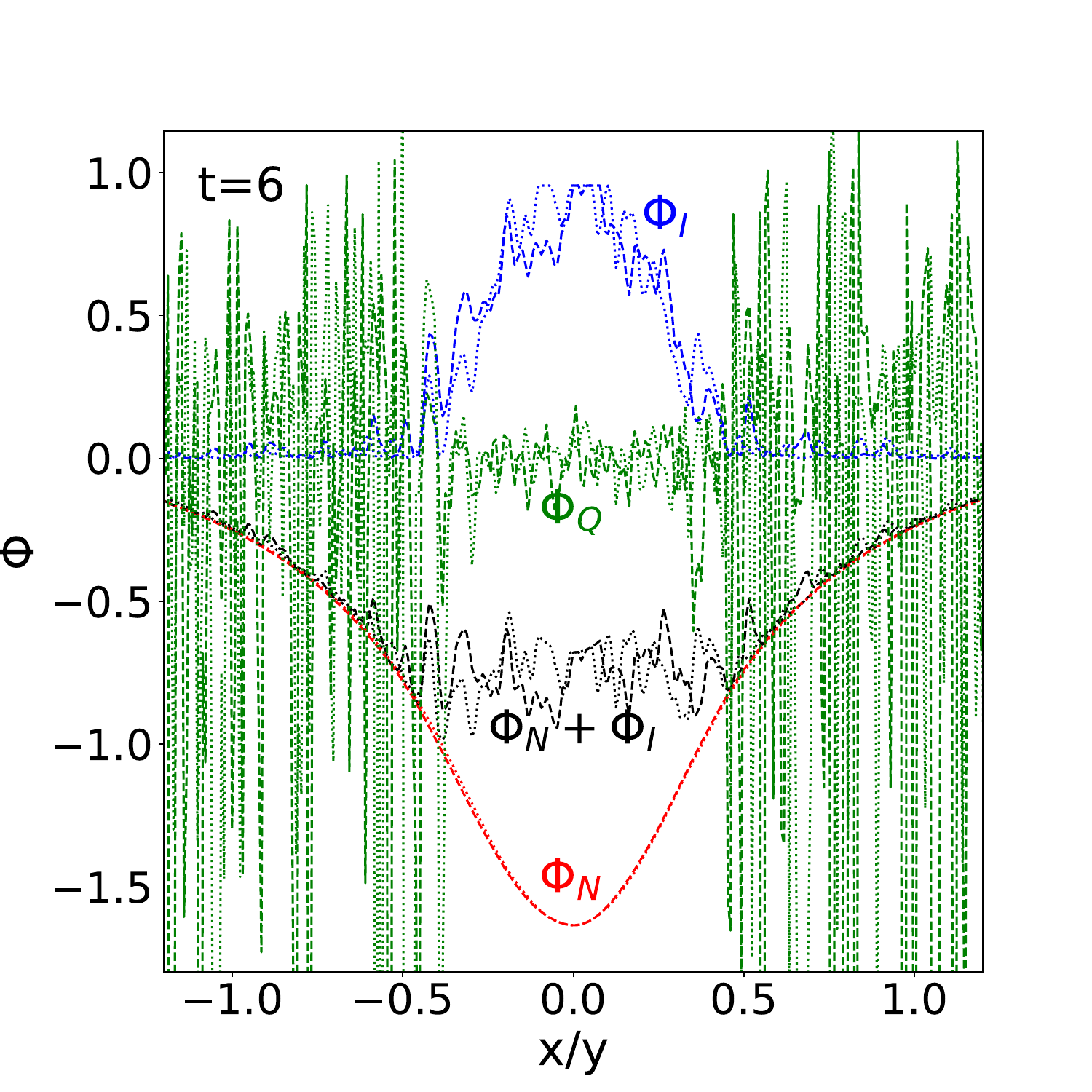}
\includegraphics[height=4.cm,width=0.29\textwidth]{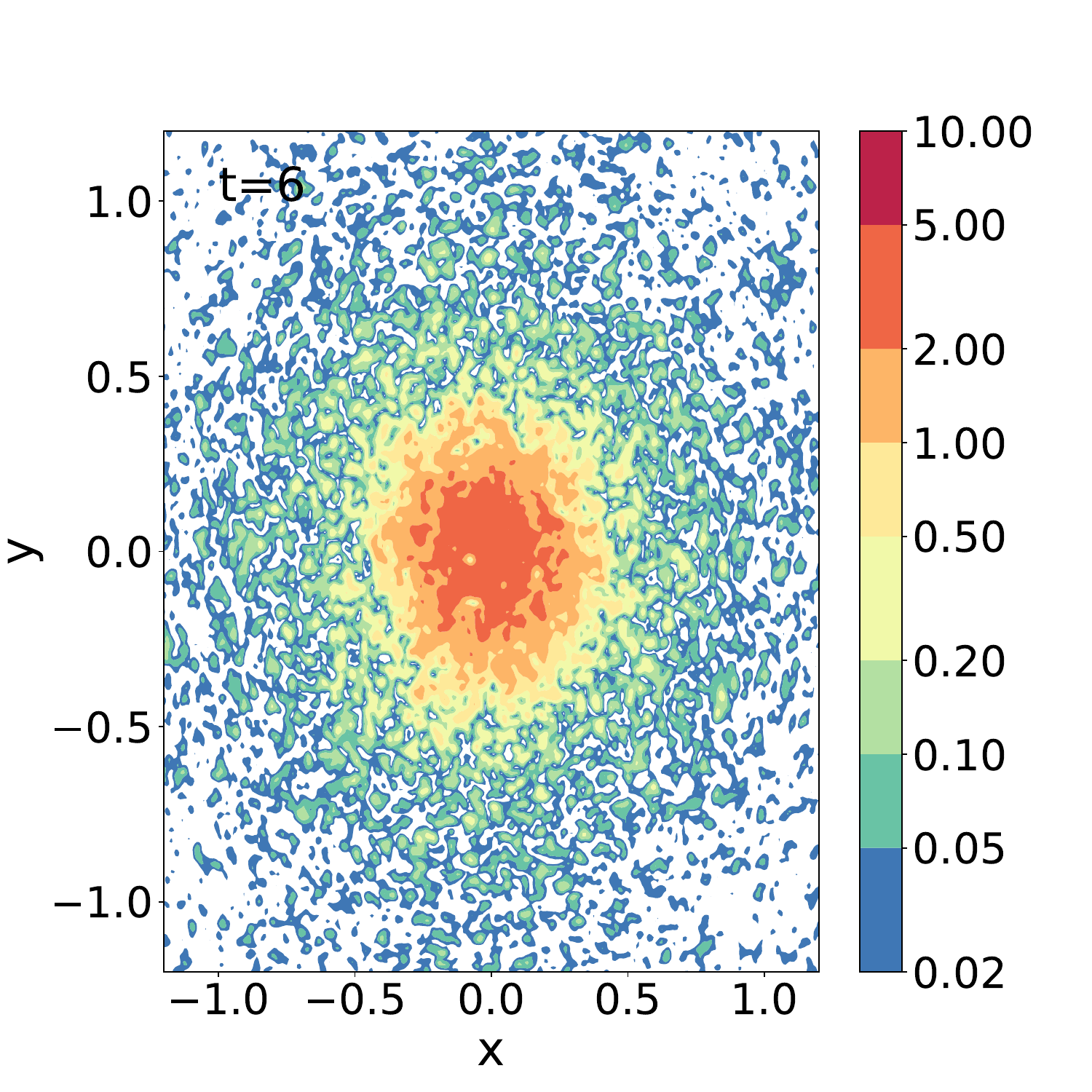}\\
\includegraphics[height=4.cm,width=0.28\textwidth]{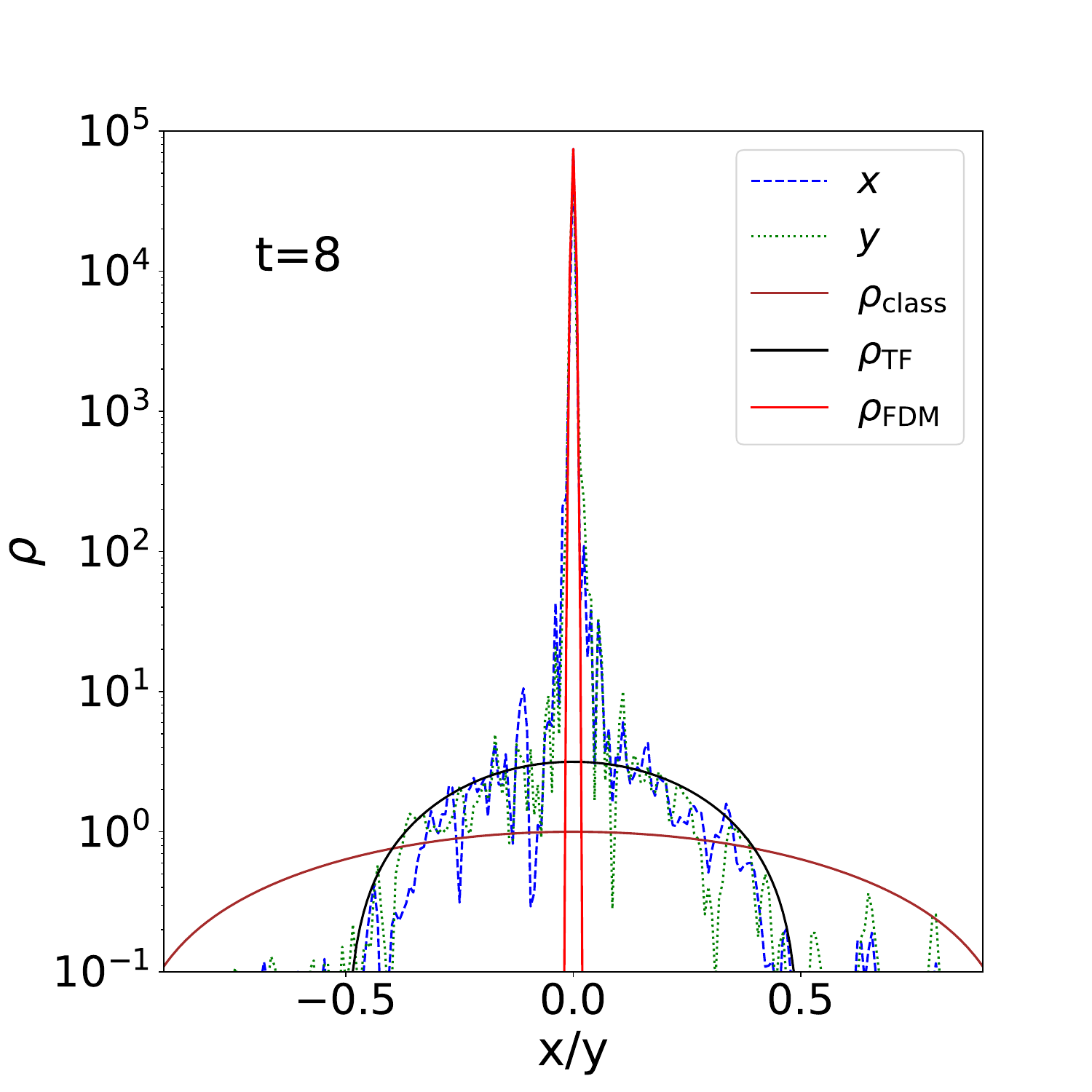}
\includegraphics[height=4.cm,width=0.28\textwidth]{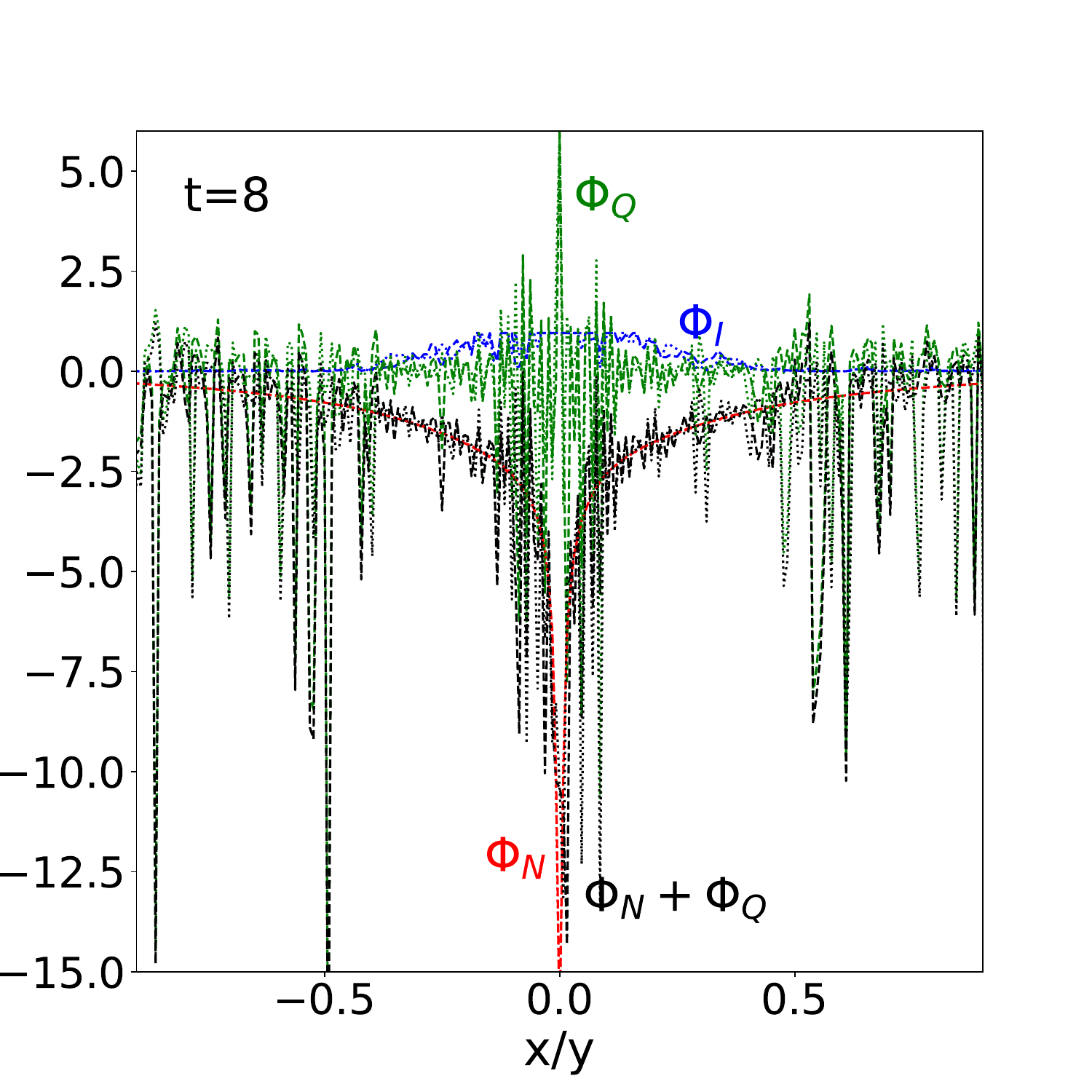}
\includegraphics[height=4.cm,width=0.29\textwidth]{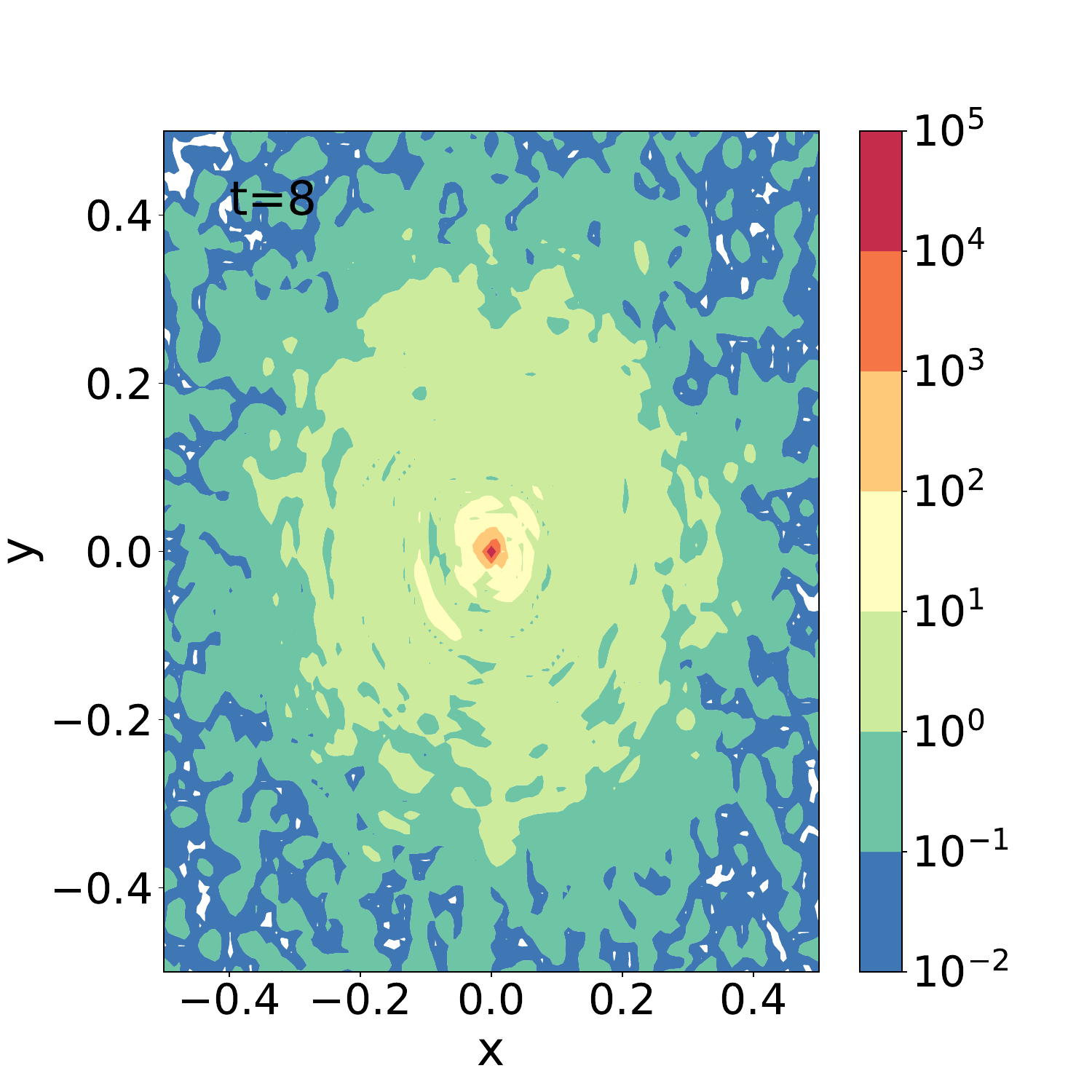}\\
\includegraphics[height=4.cm,width=0.28\textwidth]{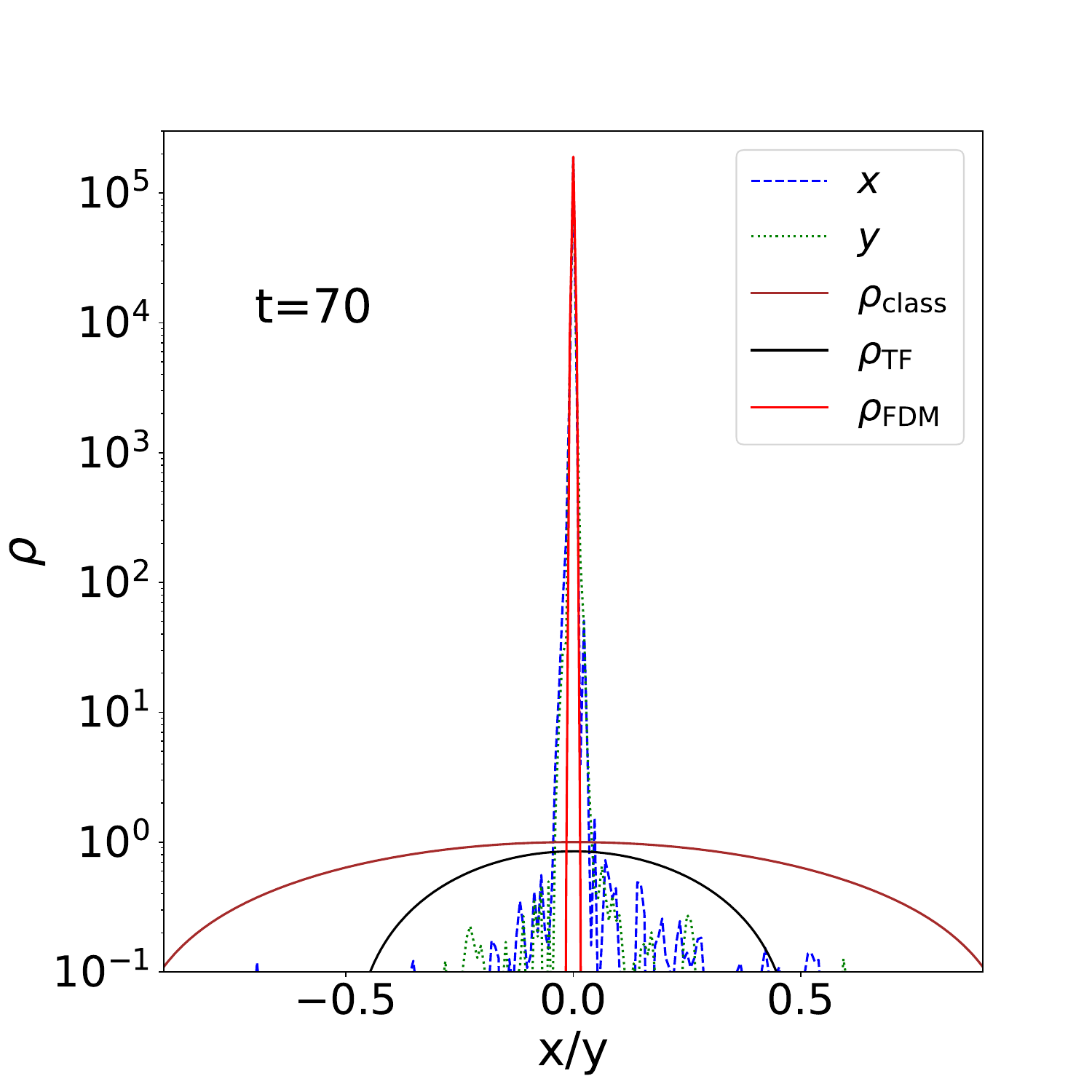}
\includegraphics[height=4.cm,width=0.28\textwidth]{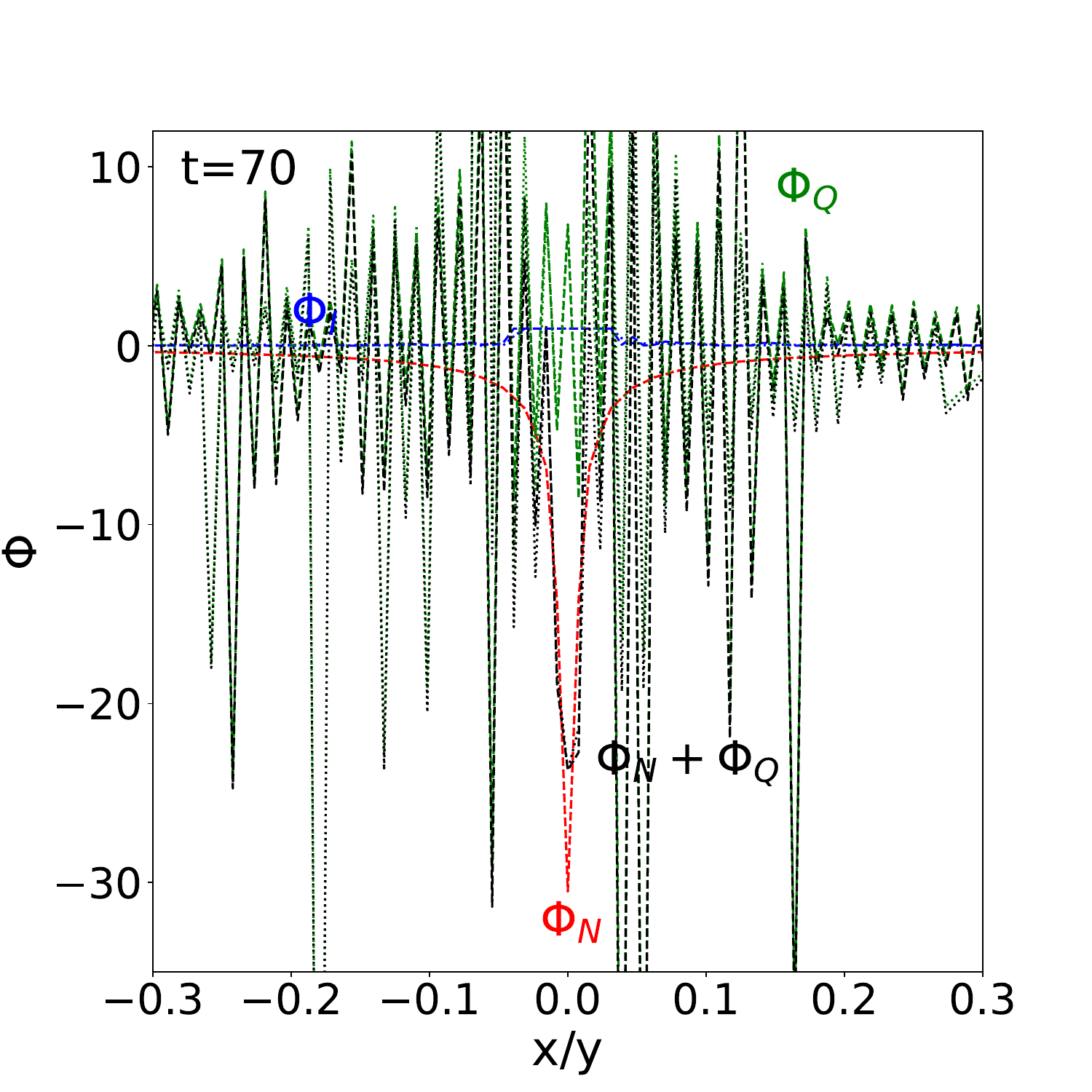}
\includegraphics[height=4.cm,width=0.29\textwidth]{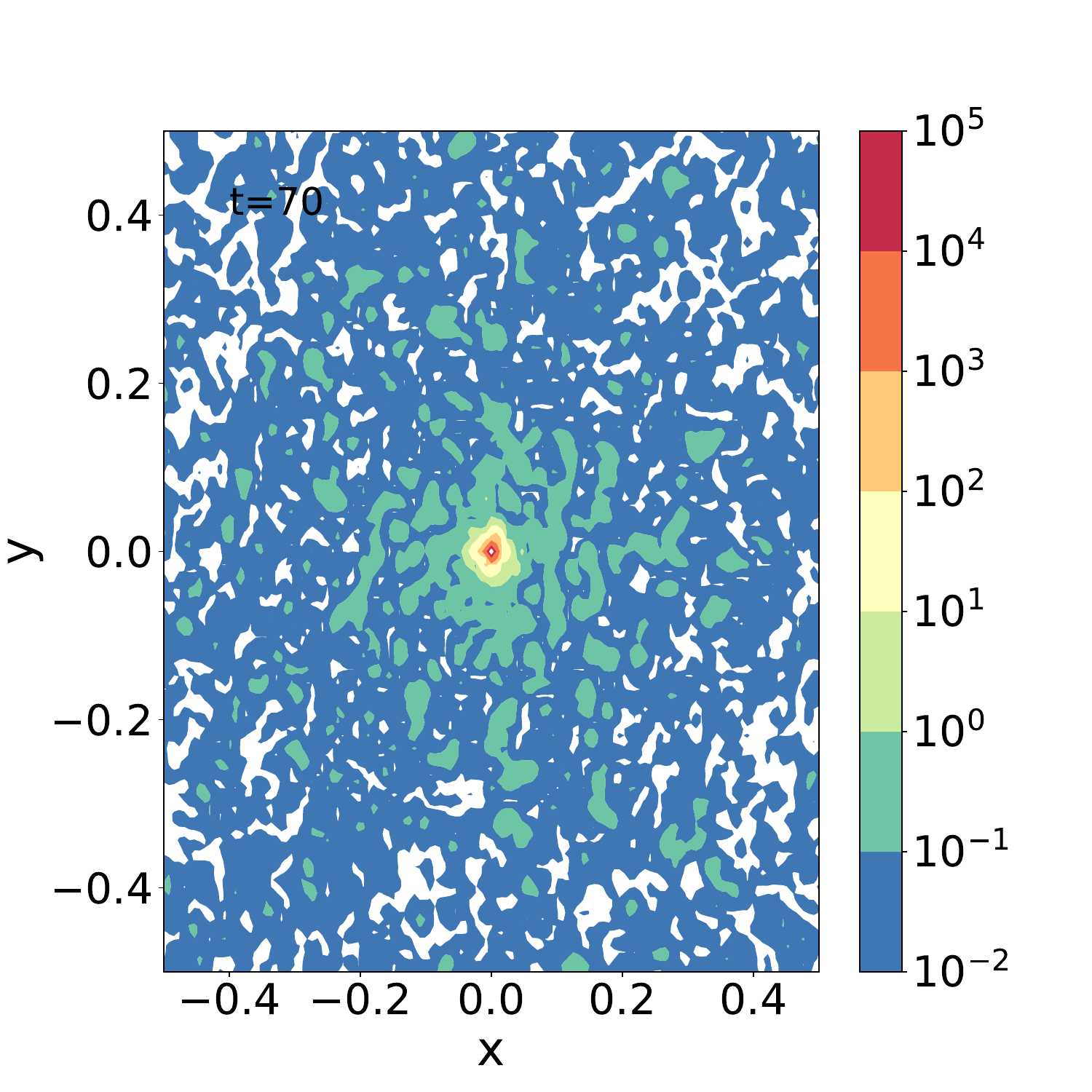}
\caption{Evolution of a halo with $R_{\rm TF}=0.5$ and $\rho_c=3$.}
\label{fig:R0p5-rhoc-3}
\end{figure*}

\subsection{Numerical simulations}
\label{sec:numerical-R_TF=0p5}

\subsubsection{Large density threshold, $\rho_c=100$}
\label{sec:R0p5-rhoc-100}

We first set $\rho_c=100$ in Eq. (\ref{eq:model-1}). This high density threshold means that the central
density will never reach the threshold $\rho_c$ and the self-interaction potential always takes the
form $\Phi_I = \lambda \rho$, as for a quartic self-interaction $\lambda_4 \phi^4/4$.
Thus, as seen in Fig.~\ref{fig:R0p5-rhoc-100}
we recover the results obtained in Fig.~3 of \cite{Garcia:2023abs}, where we studied
a simple quartic self-interaction model.
Again, within a few dynamical times, typically by $t \sim 8$, the system reaches a quasi-stationary state
with a central soliton of radius $R_{\rm TF}$, which contains about half of the mass of the system.
This soliton remains relatively stable over time, with only a slow growth of its mass,
shown by the dashed line in panel (b), and of the maximum density.

The presence of the central equilibrium soliton is clearly seen in the density profiles shown in panels
(d) and (g).
Its radius is fixed at $R_{\rm TF}=0.5$ and does not change as the soliton mass grows.
Panels (e) and (h) provide a visual representation of the hydrostatic equilibrium, which
corresponds to the constant plateau observed in the total potential $\Phi \simeq \Phi_N+\Phi_I$
over the soliton extent (i.e., $r \leq 0.5$), in agreement with Eq.(\ref{eq:TF-static}).

Accompanying the soliton is a depleted halo, retaining the remaining half of the initial mass.
Its density is reduced as compared with the initial conditions but it extends somewhat beyond
the initial radius $R_{\rm init}=1$, because of the violent dynamics associated with the formation
of the central soliton, which give rise to a time-dependent potential in the transition phase.
The excited modes associated with the halo also extend over the central region and produce small
fluctuations on top of the smooth soliton density profile.
As seen in panels (e) and (h), in the Thomas-Fermi regime achieved here the quantum pressure
$\Phi_Q$ is negligible in the soliton, as compared with the self-interaction and gravitational potentials
$\Phi_I$ and $\Phi_N$. However, in the halo the self-interaction potential $\Phi_I$ is negligible
as compared with the quantum pressure $\Phi_Q$. Thus, the Thomas-Fermi regime does not
apply to the halo, which is governed by the balance between gravitational and kinetic energies,
as in the semiclassical limit of collisionless particles.
The large fluctuations of $\Phi_Q$ show that this halo does not correspond to a FDM soliton
(\ref{eq:Phi-FDM-static}). This is not a hydrostatic equilibrium but a quasi-stationnary state
made of a superposition of a large number of excited eigenmodes.
In the semiclassical limit this corresponds to a virialized halo supported by its velocity dispersion

The comparison between the 2D density maps in panels (f) and (i) shows that over time the soliton
slowly gains mass and the system becomes smoother and more spherically symmetric.
We can see in panel (c) that while the total energy of the system is constant, the energy within
radius $R_{\rm TF}=0.5$ slowly decreases. This is the signature of a gravitational cooling process,
where energy is transferred towards the outer halo. This allows the central region to decrease its total energy
and build an equilibrium soliton with a slowly growing mass.
Indeed, the virial theorem gives for the equilibrium soliton
$2 E_{K,\rm TF} + E_{N,\rm TF} + 3 E_{I,\rm TF}=0$. As the kinetic energy of the soliton is negligible,
we have $E_{I, \rm TF} \simeq - E_{N, \rm TF}/3$ and $E_{\rm tot, TF} \simeq 2 E_{N, \rm TF}/3 < 0$.
As usual for gravitational systems, a higher mass is associated with a lower energy.
These behaviors agree with the results found in panel (c).
We can also see in panel (c) that whereas most of the gravitational and self-interaction energies
are located in the central soliton, most of the kinetic energy is located in the excited halo.

\subsubsection{Small density threshold, $\rho_{{\rm c}}=0.5$}
\label{sec:R0p5-rhoc0p5}

We now consider the opposite case, $\rho_c=0.5$, in Fig.~\ref{fig:R0p5-rhoc-0p5}.
This low density threshold means that the self-interaction potential is mostly constant and small
and it is unable to form a TF soliton, as we can see in the figure.
Nevertheless, we can see in the second row, by comparison with the initial condition in
Fig.~\ref{fig:initial}, that by $t=10$ the small self-interaction has been able to increase the mass inside
the radius $R_{\rm TF}=0.5$. This is most apparent in the comparison of the 2D maps at times
$t=0$ and $t=10$, as well as in the comparison with Fig.~\ref{fig:R0p1-rhoc-0p5}, where we also
have $\rho_c=0.5$ but $R_{\rm TF}=0.1$. There, the self-interactions are fully negligible and the
system remains close to the initial state shown in Fig.~\ref{fig:initial} for a long time $t \lesssim 2500$.
In contrast, despite the small value of $\rho_c$, in Fig.~\ref{fig:R0p5-rhoc-0p5}
the self-interactions are able to increase the central
densities by a factor of a few in the early stages. However, they are too weak to form a smooth
TF soliton, as shown in the second row by the large fluctuations of the density field on the de Broglie
scale and by the low value of $\Phi_I$ as compared with $\Phi_Q$.

In a sense, this behavior is somewhat counter-intuitive.
Indeed, as recalled in Sec.~\ref{sec:hydro}, in the hydrodynamical formulation of the equations of motion
the self-interactions are akin to an effective pressure.
Therefore, one could expect them to make the system expand rather than collapse, and to make it
even more difficult to reach large densities, as compared with the case without self-interactions.
However, the comparison of Figs.~\ref{fig:R0p5-rhoc-0p5} and \ref{fig:R0p1-rhoc-0p5} shows
that this is not always the case.
This is because in the regime considered here, the hydrodynamical picture breaks down.
Because of the large density fluctuations the density field vanishes in many places, where
the hydrodynamical mapping (\ref{eq:Madelung}) is ill-defined.
For instance, this can lead to singularities in the velocity field and vortices, which are not captured
by a curl-free velocity field.
Then, we must go back to the Gross-Pitaevskii equation (\ref{eq:Schrod}).
Here the physics at play does not seem related to small-scale vortices but to a trace of the
large-scale failed TF soliton and its characteristic radius $R_{\rm TF}$.
Somehow, there is a partly successful collapse to form the TF soliton, even though the density profile
cannot fully relax to a smooth TF soliton.
Thus, it appears that this nonlinear Sch\"rodinger equation can display intricate behaviors.
This shows that the hydrodynamical picture can be misleading in regimes where the
wave effects are large.

Until $t \lesssim 100$, the system shows a slow evolution, with a slow rise of the maximum density
$\rho_{\max}$. As seen in the third row, until $t \lesssim 80$ the density field shows strong fluctuations
with many peaks in the central region that have similar amplitudes.
In the semiclassical limit, this corresponds to a virialized halo supported by its velocity dispersion.
Here, the field $\psi$ is made of a superposition of many excited eigenmodes, as for the outer halo
found in Fig.~\ref{fig:R0p5-rhoc-100}.
Around $t\sim 100$, the merging of some of these peaks builds a density peak that rises above the rest.
As seen in the fourth row, this gives rise to a central soliton which is now supported by the quantum
pressure rather than by the self-interactions. In agreement with the scalings (\ref{eq:R-FDM}),
the rise of the mass $M_{\rm FDM}$ is associated with a decrease of the radius $R_{\rm FDM}$.
Again, we can see in panel (c) a gravitational cooling process that transfers energy from the central
region to the outer halo, which allows the formation of a massive gravitational equilibrium at the center.

\subsubsection{Intermediate density threshold, $\rho_c=3$}
\label{sec:sim-R0p5-rhoc-3}

We now discuss the intermediate case, $\rho_c=3$, shown in
Fig.~\ref{fig:R0p5-rhoc-3}.
At early times the dynamics are identical to those of the high-density threshold case shown in
Fig.~\ref{fig:R0p5-rhoc-100}, with the fast formation by $t \sim 6$ of a soliton of radius
$R_{\rm TF}=0.5$ supported by the pressure associated with the self-interactions.
This is because the density remains below $\rho_c$ during this stage.

Later, at $t \simeq 7$, the soliton reaches densities where the self-interaction potential saturates to a constant.
As a consequence, the self-interactions can no longer support the soliton, which collapses
to form of a much narrower and higher density peak supported by its kinetic energy.
Its radius is determined by the de Broglie wave length (\ref{eq:epsilon-de-Broglie}), which is of the order of
$\epsilon$.
This sharp transition can be clearly seen in the first row. The sudden rise of the maximum density
occurs along with the rise of $M_{\rm FDM}$ and the decrease of $E_{\rm tot, FDM}$.

After this transition, the central density peak slowly grows but the main evolution is the slow decrease
of the mass $M_{\rm TF}$ and of the density peaks within the intermediate radius $R_{\rm TF}=0.5$,
which leads to $M_{\rm FDM} \simeq M_{\rm TF}$.
As seen in the bottom row, this leaves an isolated and narrow FDM soliton without much trace of
the TF soliton that had formed at $t \lesssim 7$.

However, as compared with the case shown in Fig.~\ref{fig:R0p5-rhoc-0p5}, we can see that the
formation of an earlier TF soliton has greatly accelerated the dynamics.
The narrow FDM soliton now forms at $t \simeq 7$ instead of $t\sim 100$.
Indeed, the initial halo first efficiently forms a TF soliton because $R_{\rm TF}$ is of the
same order as the size of the system (i.e. the self-interactions are rather strong).
Next, because this static soliton has negligible kinetic support when it leaves the Thomas-Fermi regime
(the self-interactions vanish) it quickly undergoes a collective gravitational collapse and leads to the
formation of a small FDM soliton of size $R_{\rm FDM}$.
However, we can see that only about $10\%$ of the TF soliton ends up in the narrower FDM soliton.
Nevertheless, this gives a final FDM soliton mass $M_{\rm FDM} \simeq 0.09$ that is larger than
the final mass $M_{\rm FDM} \simeq 0.04$ found in Fig.~\ref{fig:R0p5-rhoc-0p5}.

Thus, this two-stages process accelerates the formation of the final FDM peak.
Unlike the case displayed in Fig.~\ref{fig:R0p5-rhoc-0p5}, where the central peak gradually emerges
from strong fluctuations, the presence of the coherent state associated with the early
TF soliton provides a more favourable initial condition for the later formation of the FDM soliton,
which appears much earlier and is able to gather a larger mass.

\subsection{Gaussian ansatz}

The dynamics found in the numerical simulations displayed in Sec.~\ref{sec:numerical-R_TF=0p5}
can be understood from the Gaussian ansatz presented in Sec.~\ref{sec:gaussian}.

\subsubsection{Large density threshold, $\rho_c=100$}

\begin{figure}[ht]
\centering
\includegraphics[height=4.1cm,width=0.235\textwidth]{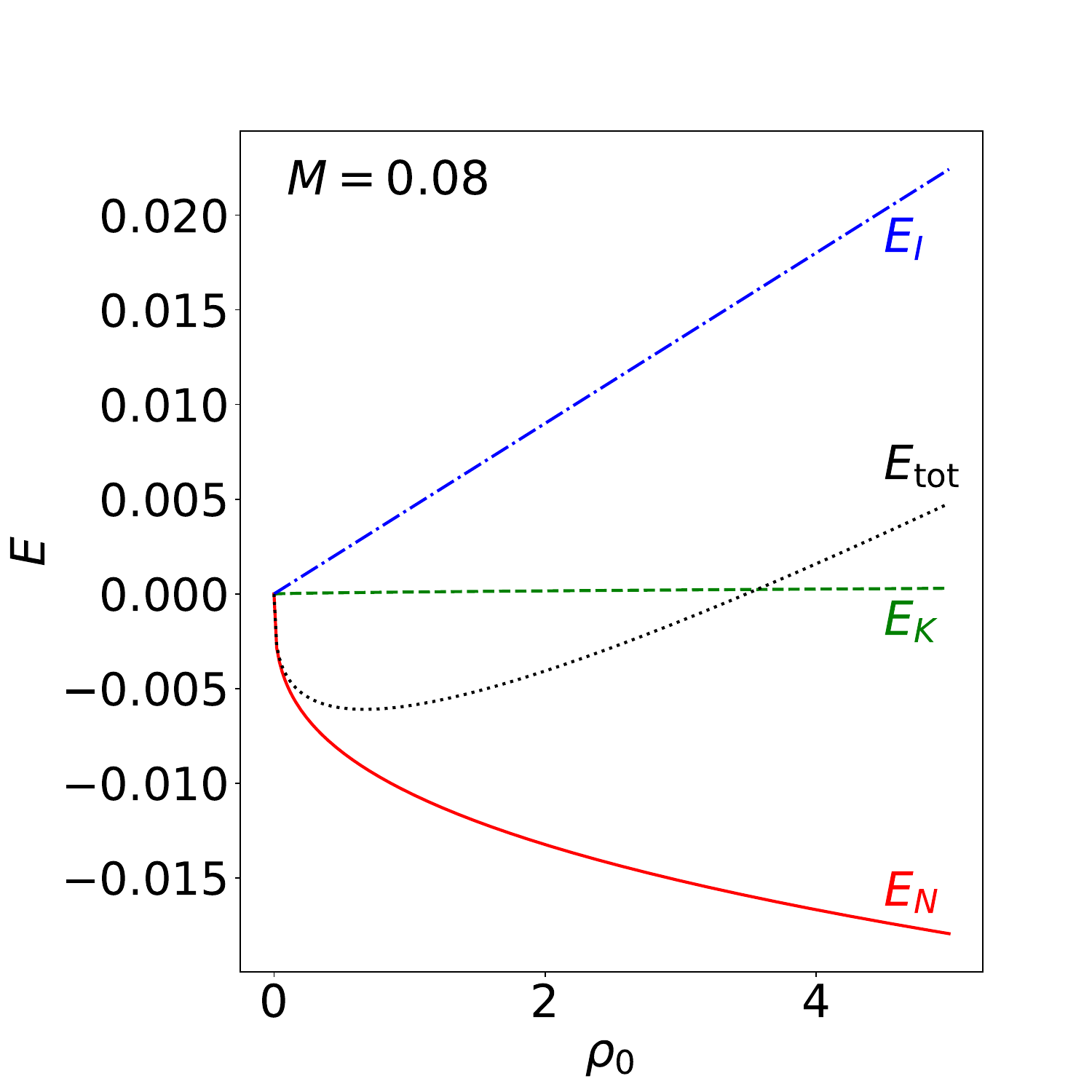}
\includegraphics[height=4.1cm,width=0.235\textwidth]{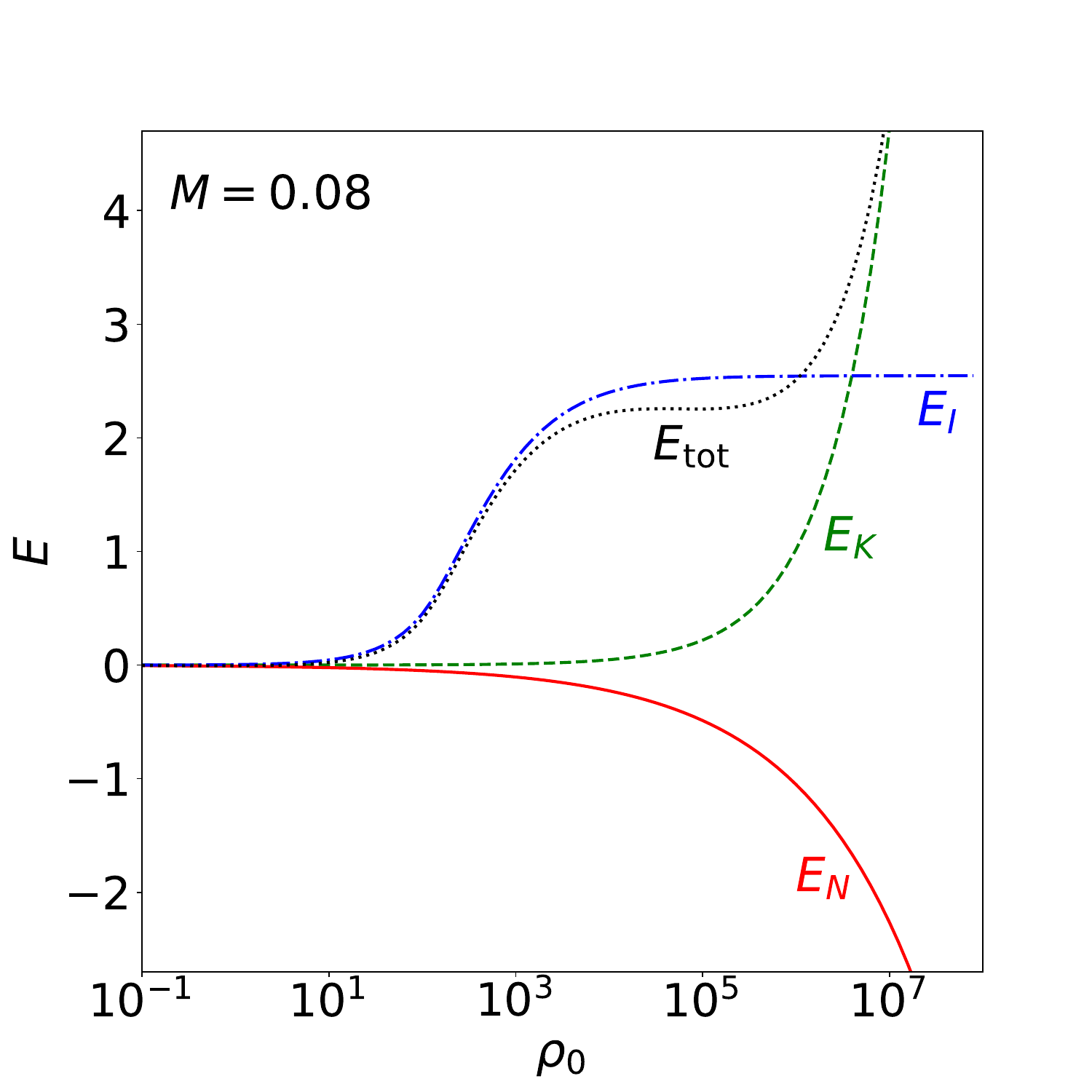}\\
\includegraphics[height=4.1cm,width=0.235\textwidth]{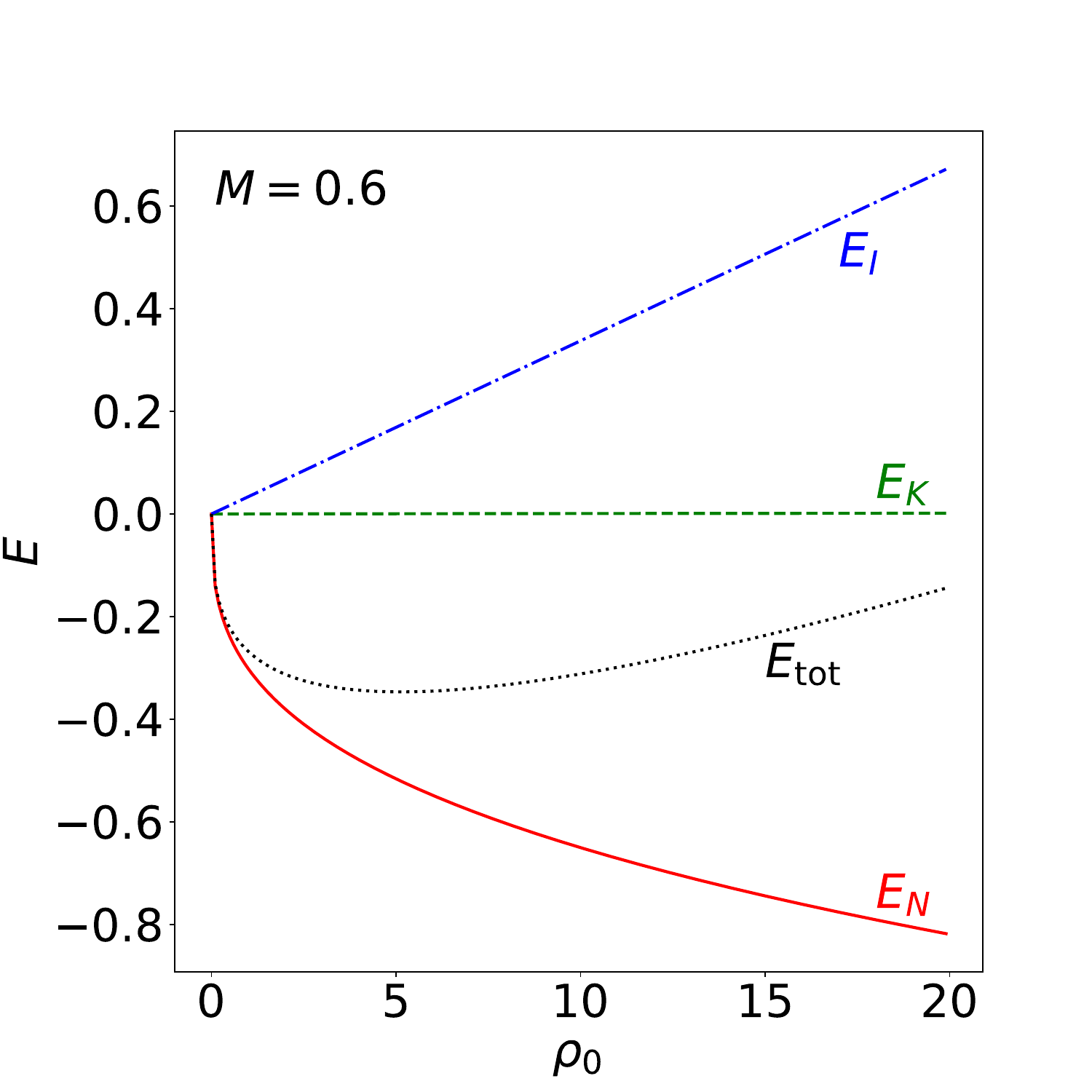}
\includegraphics[height=4.1cm,width=0.235\textwidth]{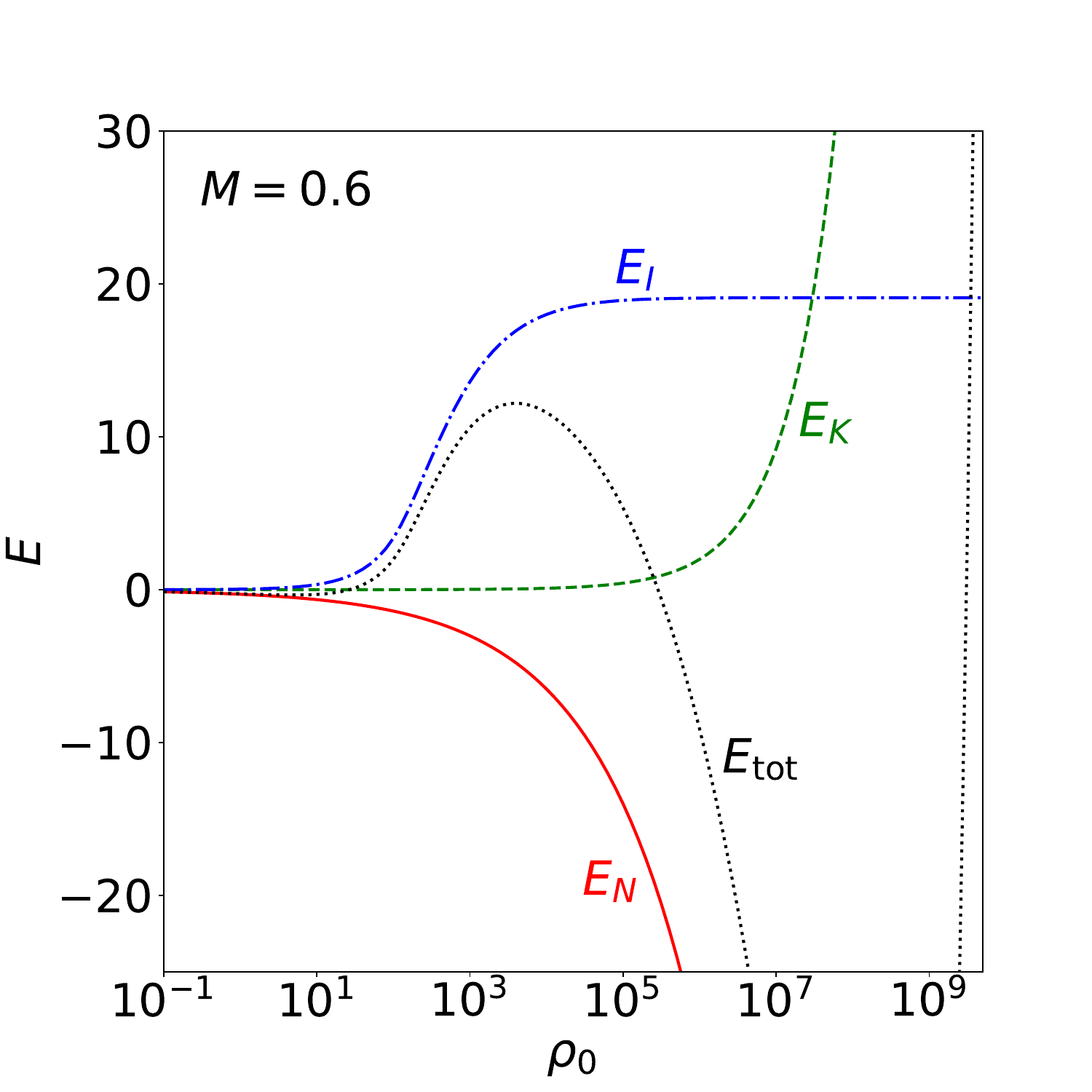}\\
\includegraphics[height=4.1cm,width=0.235\textwidth]{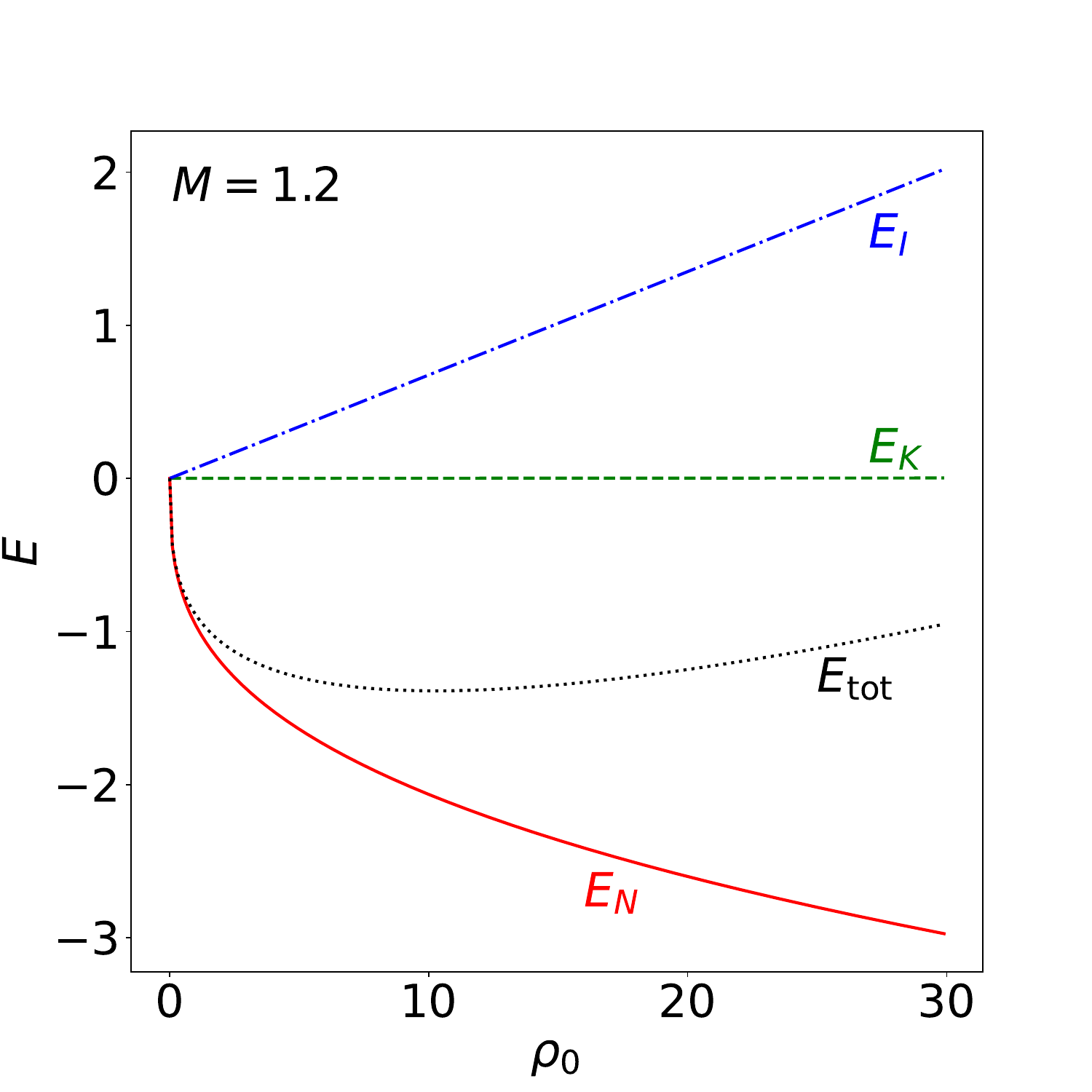}
\includegraphics[height=4.1cm,width=0.235\textwidth]{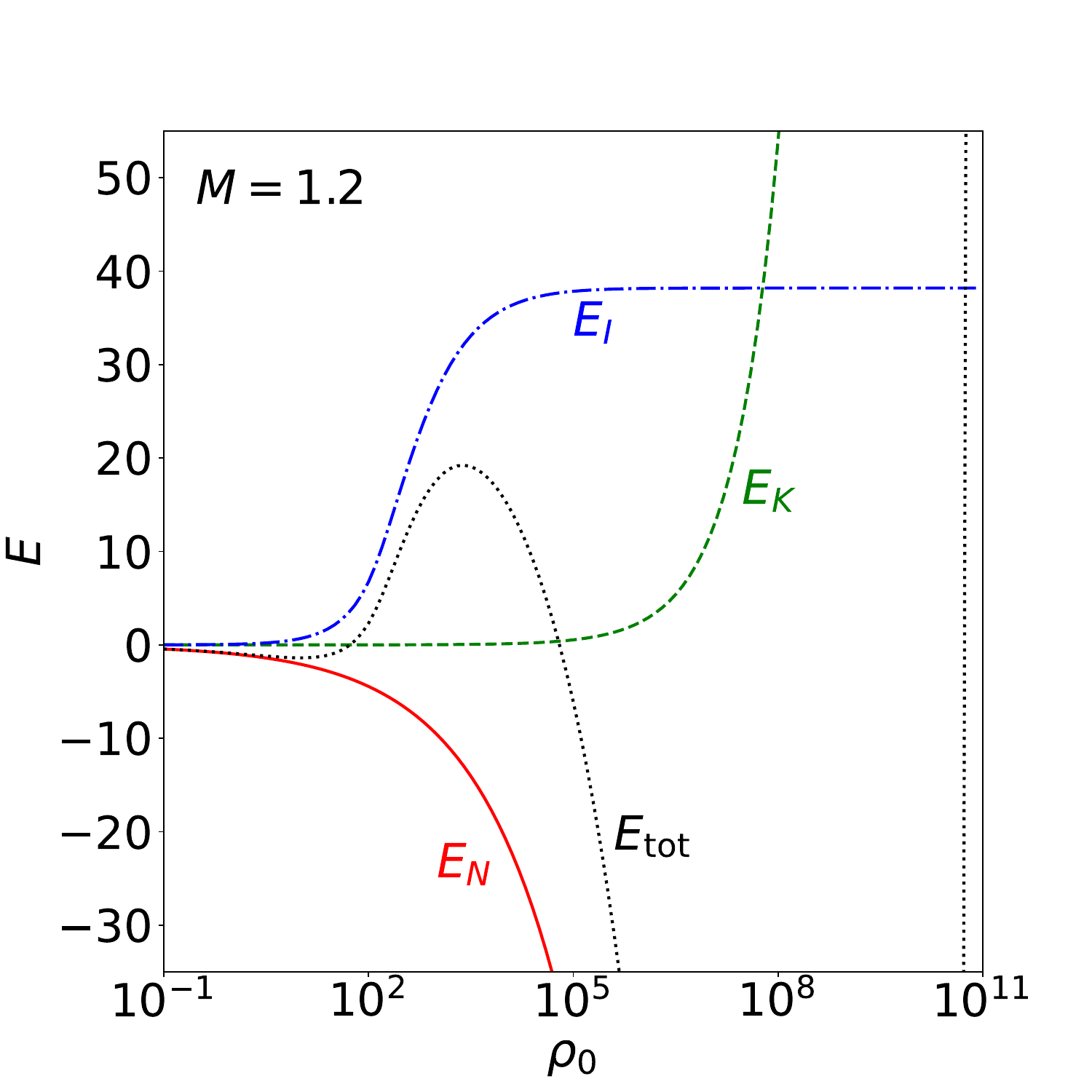}\\
\caption{
[$R_{\rm TF}=0.5$, $\rho_c=100$.]
Energy of a Gaussian density profile with total mass $M=0.08$ ({\it first row}),
$M=0.6$ ({\it second row}) and $M=1.2$ ({\it third row}).
The {\it left column} shows a zoom on low densities whereas the {\it right column} shows a large range
of densities on a logarithmic scale.
We display the energies $E_K$ (green dashed lines), $E_I$ (blue dot-dashed lines),
$E_N$ (red solid lines), and the total energy $E_{\rm tot}$ (black dotted lines).}
\label{fig:R0p5-rhoc-100-Gauss}
\end{figure}

We show in Fig.~\ref{fig:R0p5-rhoc-100-Gauss} the energies $E_K$, $E_I$, $E_N$ and the total energy
$E_{\rm tot}$ obtained from Eqs.(\ref{eq:Gauss-EK-EN})-(\ref{eq:Gauss-EI2}), for the
high density threshold case $\rho_c=100$ displayed in Fig.~\ref{fig:R0p5-rhoc-100} and three
different masses.
We can see that for all masses $M < 1.2$, which is the total mass of the system in the numerical
simulation of Fig.~\ref{fig:R0p5-rhoc-100}, there is a local minimum of the energy at a central
density $\rho_0$ of the order of unity. This corresponds to the TF solution (\ref{eq:rho0-TF}),
which approximates the TF soliton (\ref{eq:rho-sol-TF}) supported by the repulsive self-interactions.
For $0.08 < M < 1.2$, thanks to the saturation of the self-interactions at high densities,
there appears a second minimum at a very high density, $\rho_0 > 10^5$.
This is the FDM solution (\ref{eq:R-FDM}), which approximates the FDM soliton (\ref{eq:Phi-FDM-static})
supported by the quantum pressure.
These two minima are widely separated because we have $\epsilon \ll 1$.

 Because the TF radius (\ref{eq:Rsol-lambda}) is of the order of the system size (i.e. the self-interactions
 are rather strong), the system quickly forms a TF soliton as found in the simulation in
 Fig.~\ref{fig:R0p5-rhoc-100}, wich contains about half of the total mass.
 This is possible because the energy and central density of this soliton are not too
 far from those of the initial conditions.
 On the other hand, we can see in Fig.~\ref{fig:R0p5-rhoc-100-Gauss} that for all possible
 masses of the soliton, $M < 1.2$, the high-density FDM soliton is separated from the TF soliton
 by a large energy barrier.
 This explains why the TF soliton remains stable and no high-density FDM soliton is formed
 in the simulation.

\subsubsection{Small density threshold, $\rho_c=0.5$}

\begin{figure}[ht]
\centering
\includegraphics[height=4.1cm,width=0.235\textwidth]{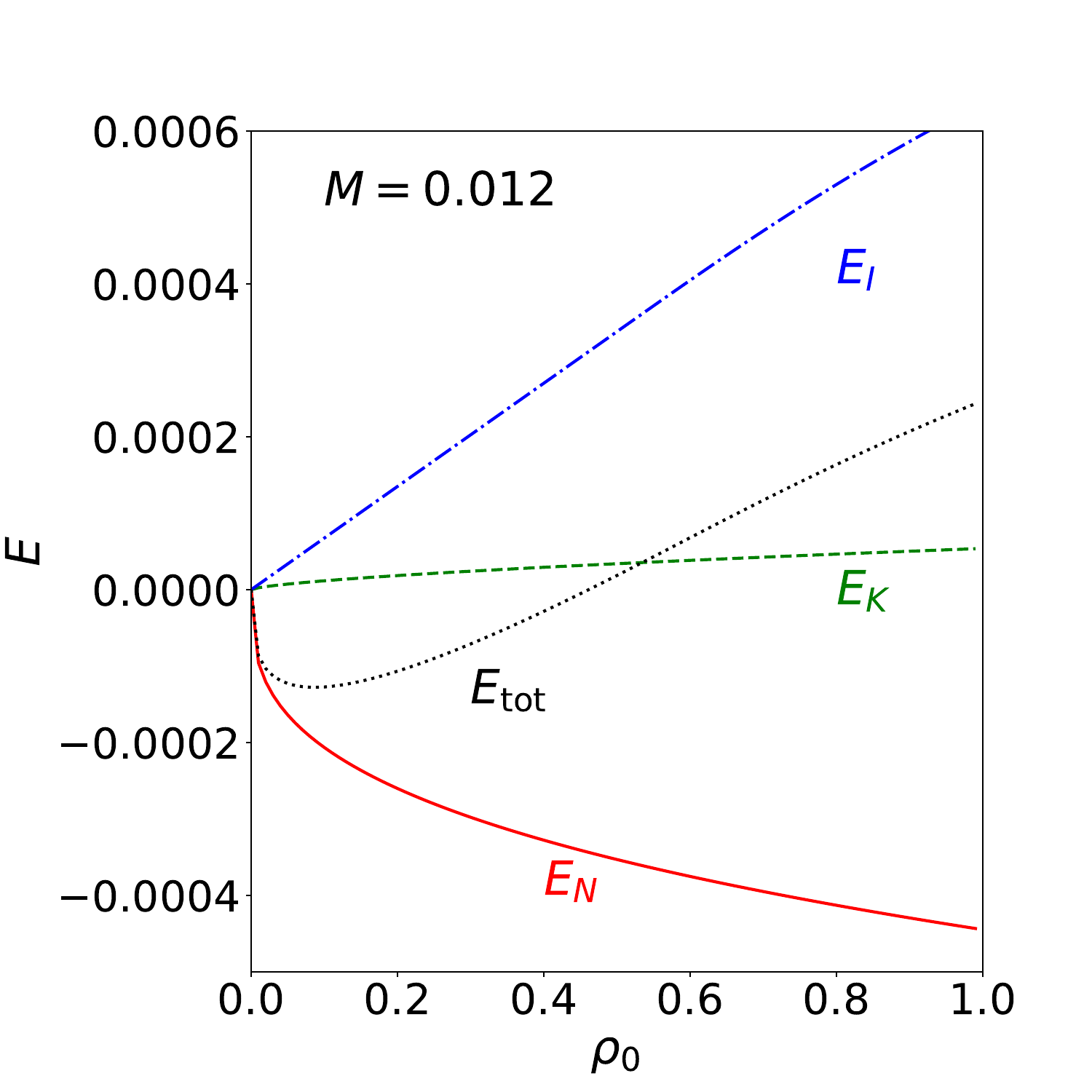}
\includegraphics[height=4.1cm,width=0.235\textwidth]{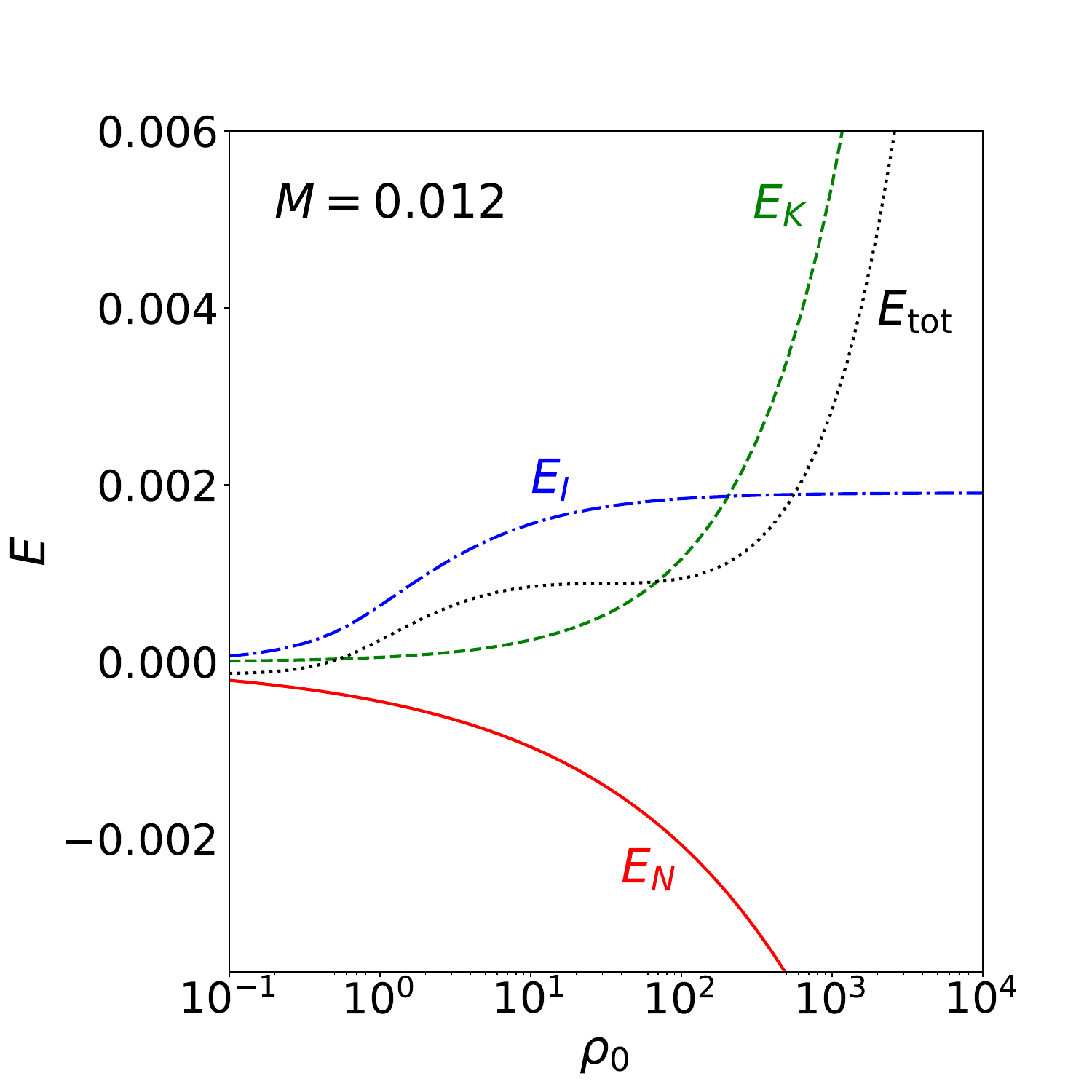}\\
\includegraphics[height=4.1cm,width=0.235\textwidth]{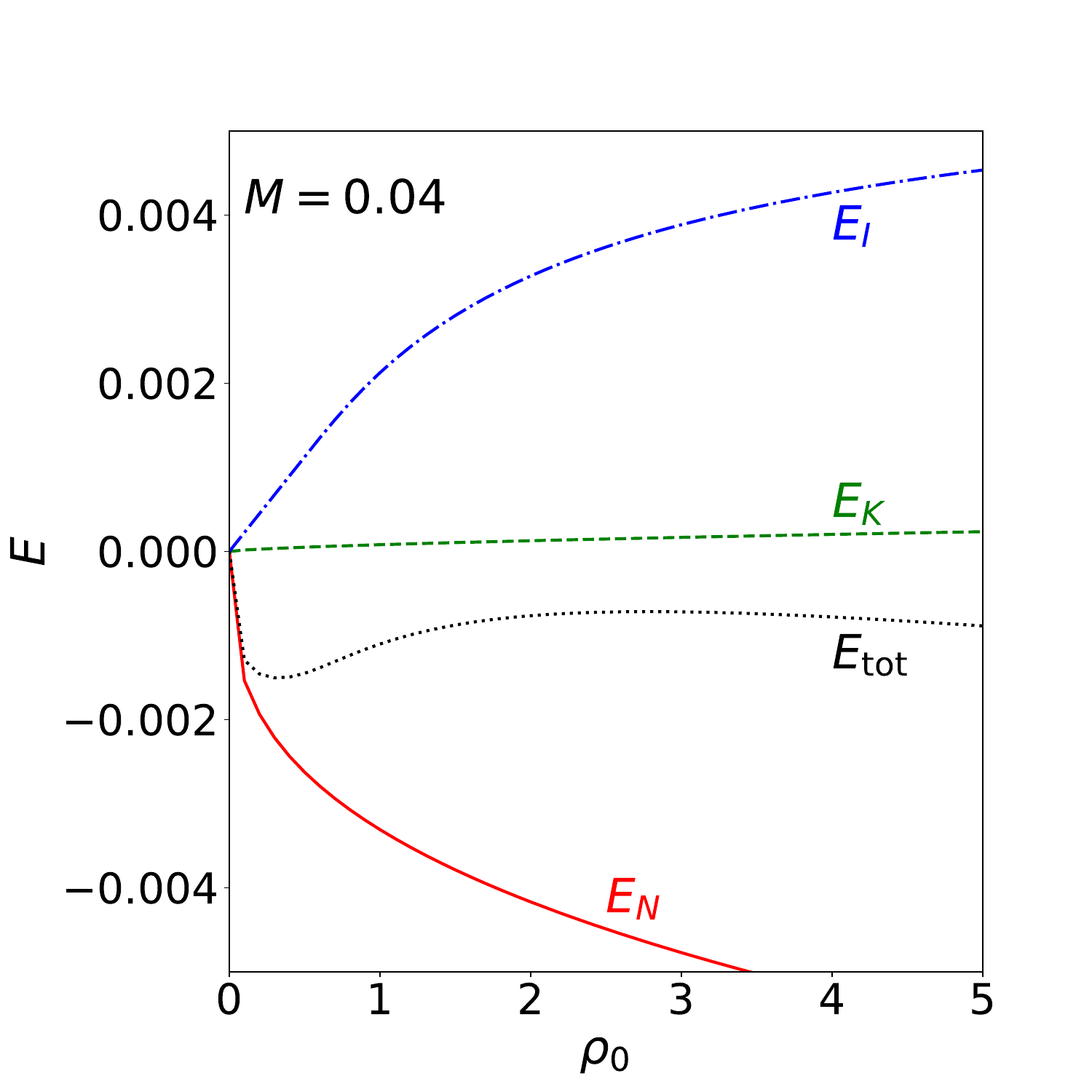}
\includegraphics[height=4.1cm,width=0.235\textwidth]{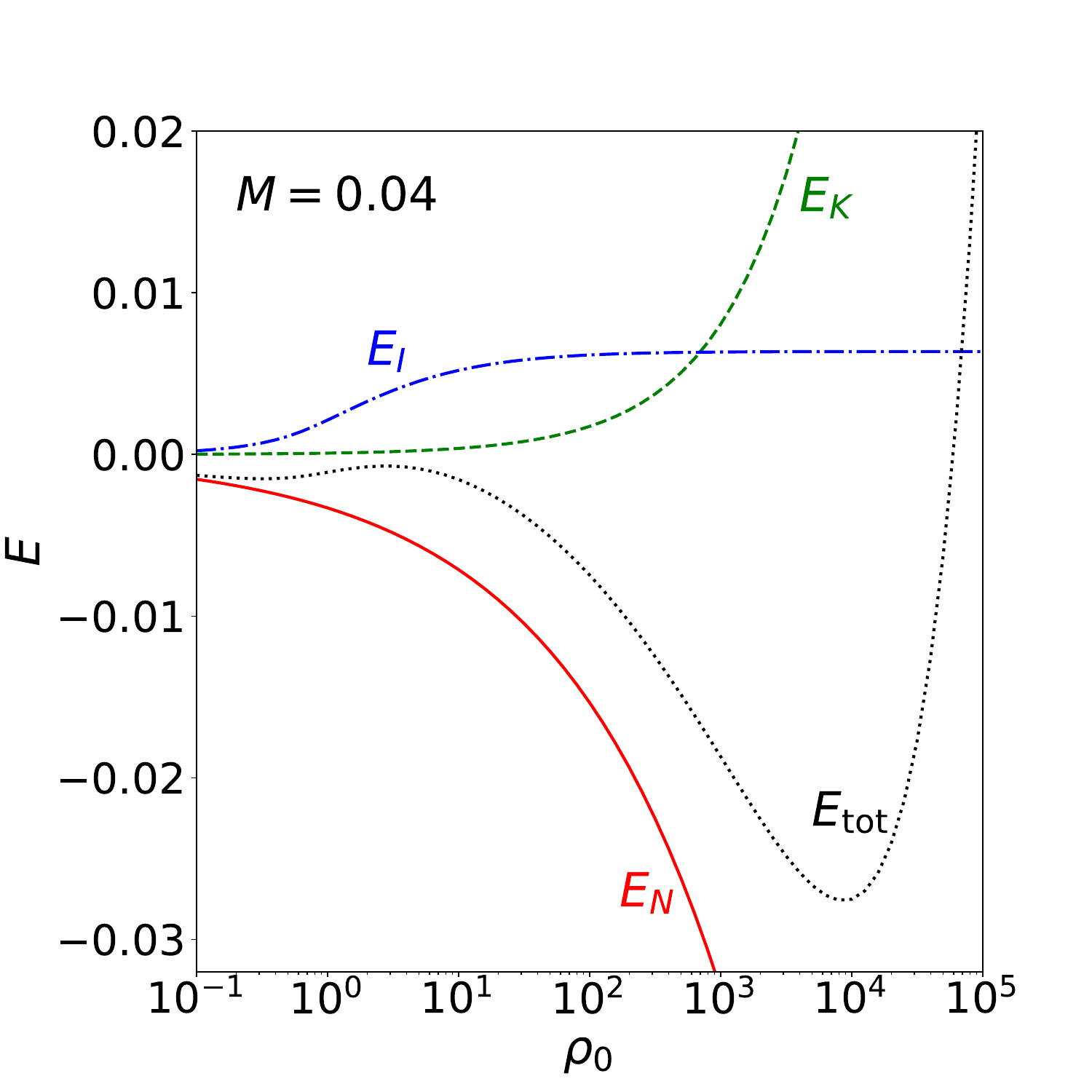}\\
\caption{
[$R_{\rm TF}=0.5$, $\rho_c=0.5$.]
Energy of a Gaussian density profile.}
\label{fig:R0p5-rhoc-0p5-Gauss}
\end{figure}

We show in Fig.~\ref{fig:R0p5-rhoc-0p5-Gauss} the case $\rho_c=0.5$, which corresponds to the
simulation of Fig.~\ref{fig:R0p5-rhoc-0p5}.
For low masses, $M < 0.09$, there exists a TF soliton supported by the self-interactions with a low
central density $\rho_0 <0.5$. However, this configuration is irrelevant as the intrinsic
density fluctuations are already of the order of unity in the initial state,
as seen in Fig.~\ref{fig:initial}, and do not allow such a small soliton to be formed.
On the other hand, even though the threshold $\rho_c$ is too low to permit the formation of
a massive TF soliton, the self-interactions still prevent the existence of a FDM soliton
for $M < 0.012$, as seen in the upper right panel.
For higher masses, there exists a FDM soliton (\ref{eq:R-FDM}) with a very high central density,
$\rho_0 \gg 1$.
This explains why the simulation shown in Fig.~\ref{fig:R0p5-rhoc-0p5} cannot form a stable FDM
soliton at the center until $t \sim 100$, when the mass in the central region reaches
$M_{\rm FDM} \sim 0.01$.
At $t \lesssim 100$, with $M_{\rm FDM} \lesssim 0.01$, we instead have competiting density
peaks with a range of densities $1 \lesssim \rho_0 \lesssim 10^3$, in agreement with the
upper right panel in Fig.~\ref{fig:R0p5-rhoc-0p5-Gauss}.
After this critical mass is reached, a stable FDM soliton can form, which quickly obtains a mass
$M_{\rm FDM} \simeq 0.04$ and a central density $\rho_0 \simeq 10^4$, in agreement with the
lower right panel in Fig.~\ref{fig:R0p5-rhoc-0p5-Gauss}.
Because the energies and the gravitational potential well associated with this FDM soliton are rather
small as compared with the energies of the full system, the soliton growth seems to stall in the
simulation shown in Fig.~\ref{fig:R0p5-rhoc-0p5}, or to proceed on longer timescales than the
duration of our simulation.

\subsubsection{Intermediate density threshold, $\rho_c=3$}
\label{sec:Gaussian-R0p5-rhoc3}

\begin{figure}[ht]
\centering
\includegraphics[height=4.1cm,width=0.235\textwidth]{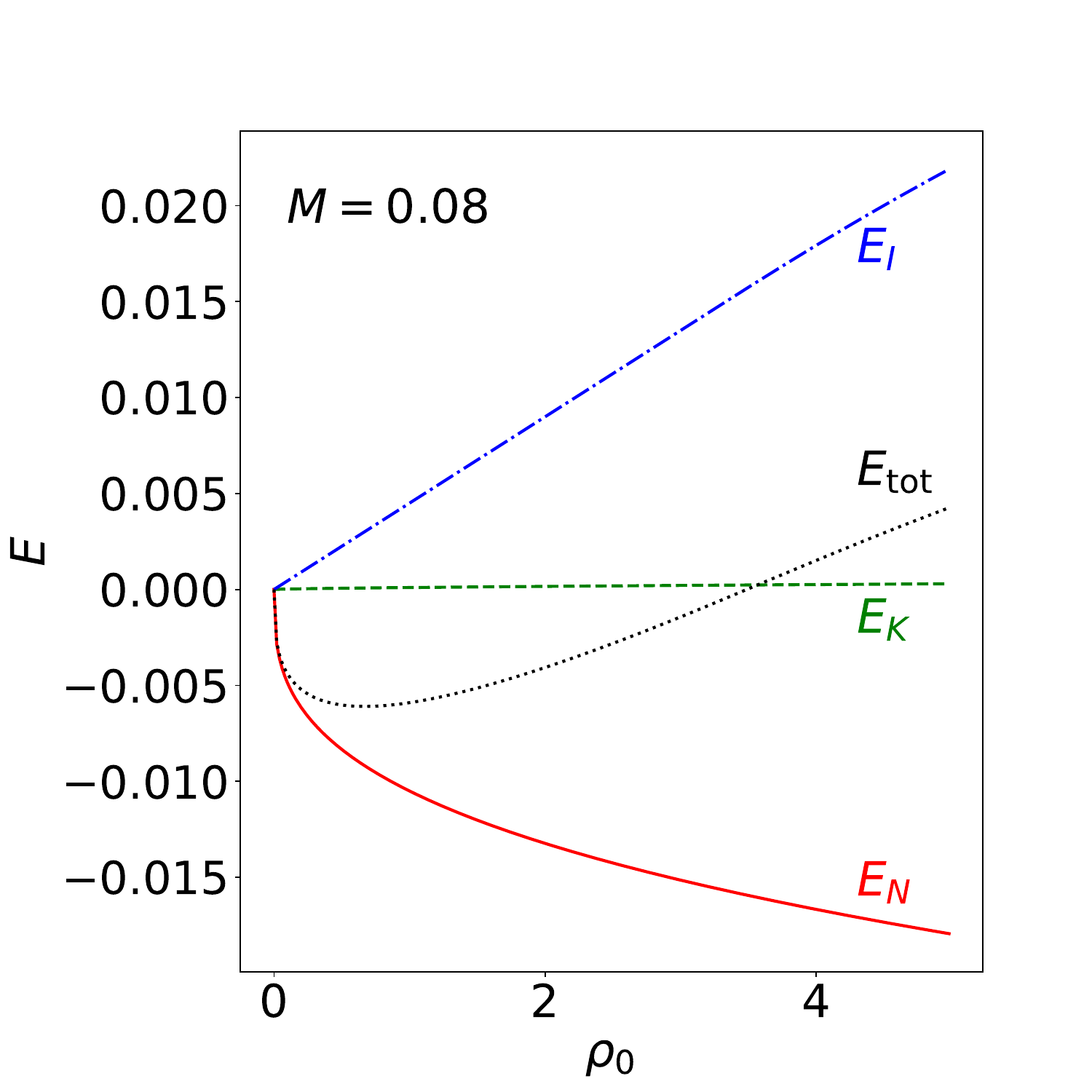}
\includegraphics[height=4.1cm,width=0.235\textwidth]{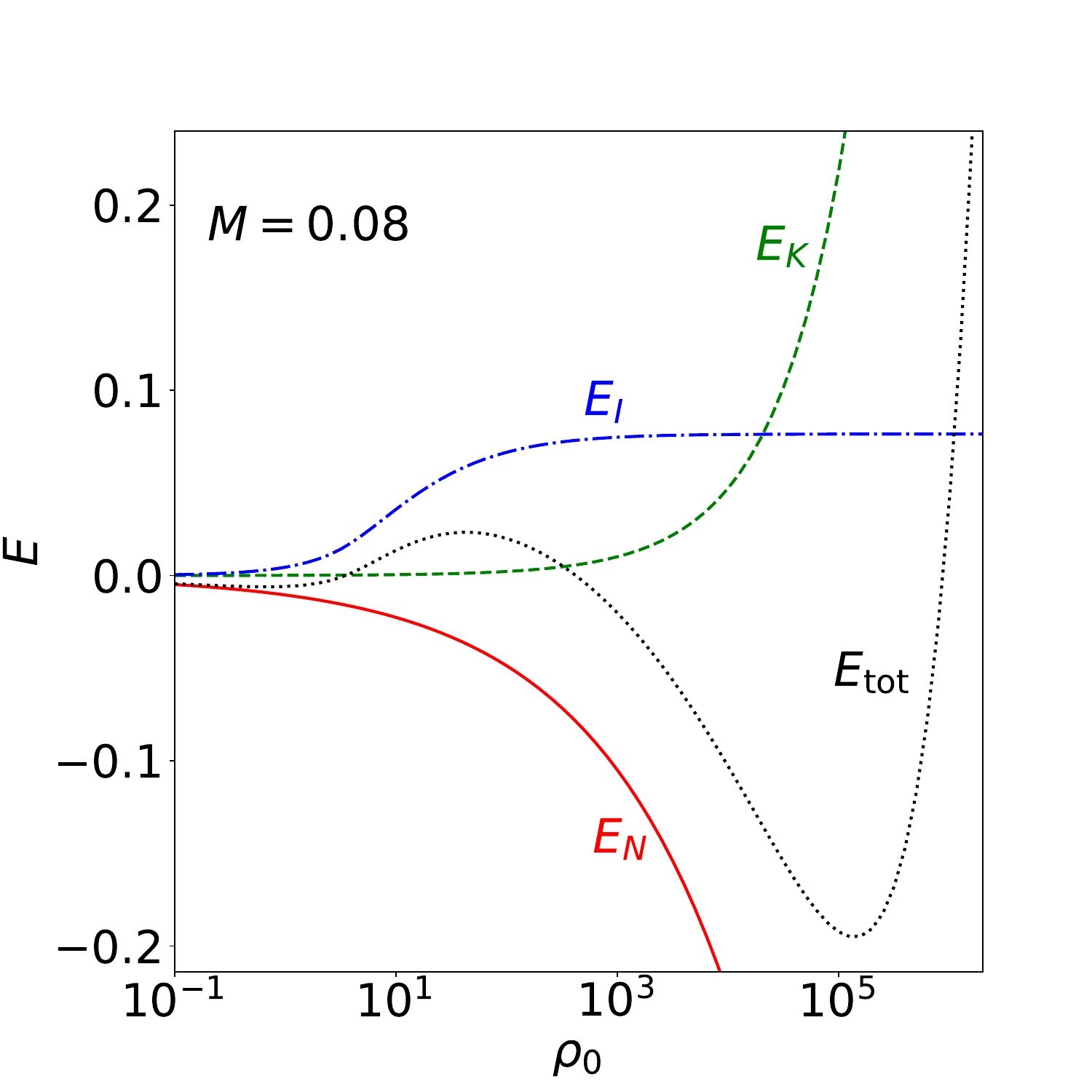}\\
\includegraphics[height=4.1cm,width=0.235\textwidth]{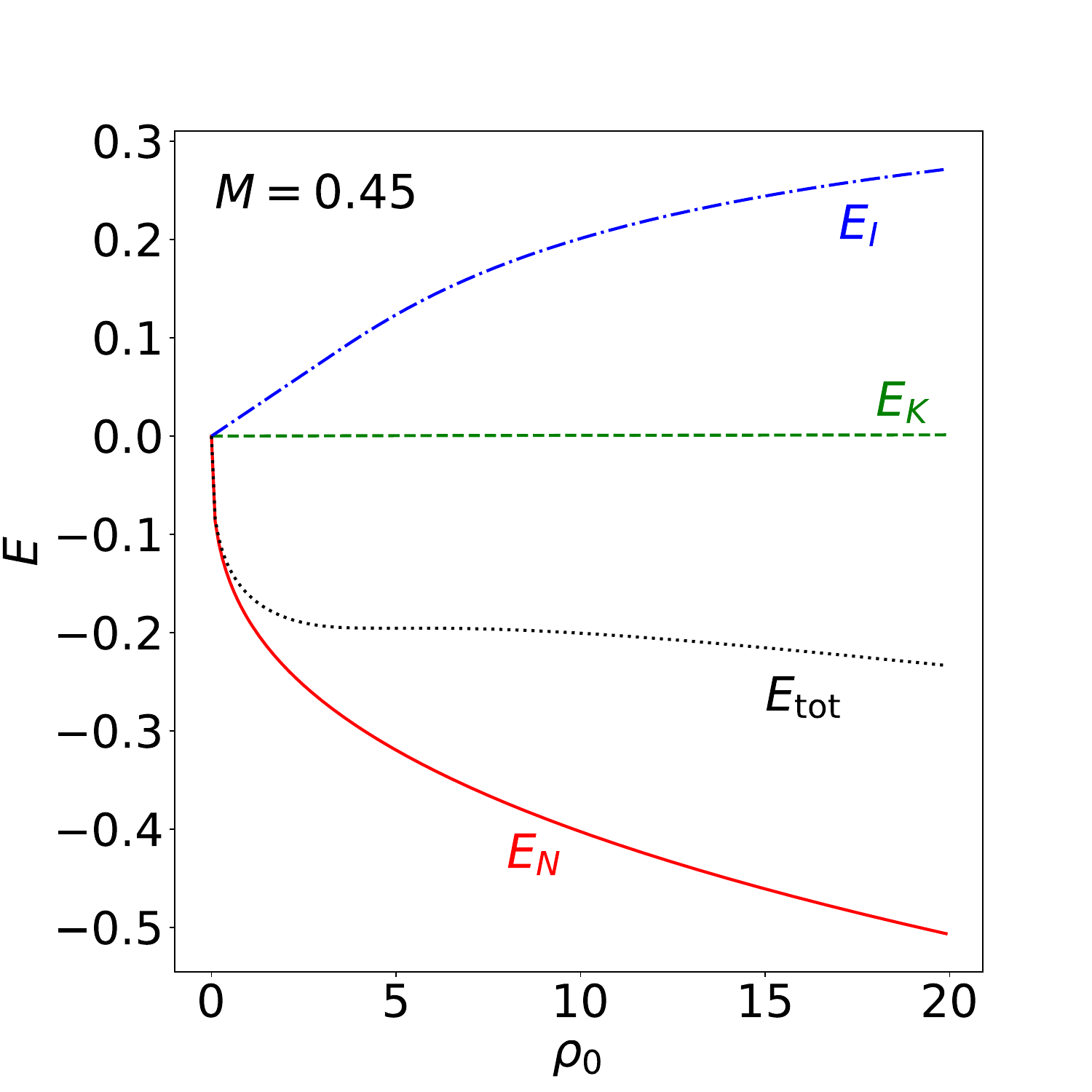}
\includegraphics[height=4.1cm,width=0.235\textwidth]{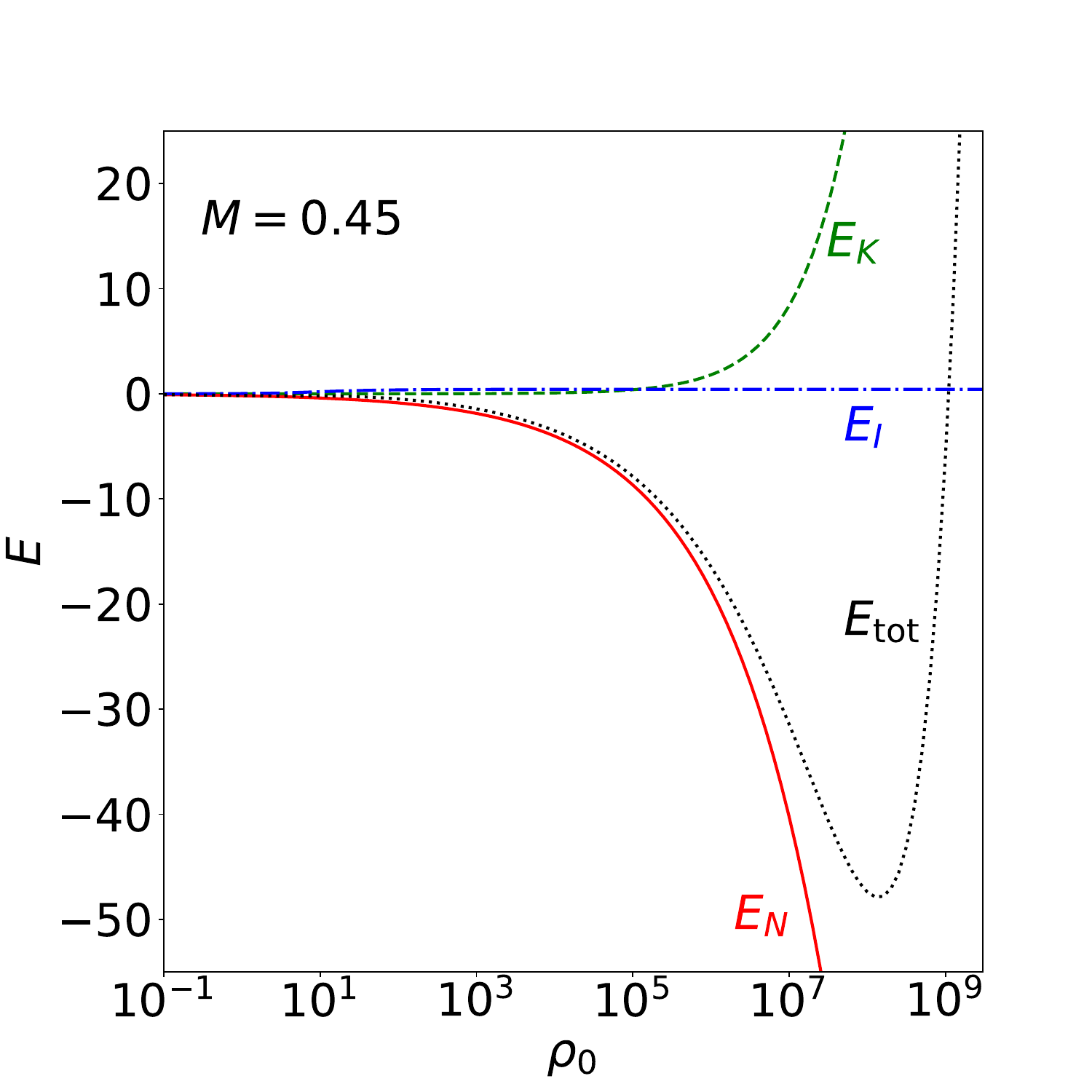}\\
\caption{
[$R_{\rm TF}=0.5$, $\rho_c=3$.]
Energy of a Gaussian density profile.}
\label{fig:R0p5-rhoc-3-Gauss}
\end{figure}

We show in Fig.~\ref{fig:R0p5-rhoc-3-Gauss} the case $\rho_c=3$, which corresponds to the
simulation of Fig.~\ref{fig:R0p5-rhoc-3}.
We can see that for a low soliton mass, $M<0.45$, there is a local energy minimum at a density $\rho_0$
of the order of unity and another local energy minimum at a much higher density $\rho_0 \gg 1$.
Again, the low-density minimum corresponds to the TF soliton (\ref{eq:rho0-TF}) whereas the
high-density minimum corresponds to the FDM soliton (\ref{eq:R-FDM}).
We can check in Fig.~\ref{fig:R0p5-rhoc-3-Gauss} that the low-density equilibrium is governed by the
balance between the gravitational and self-interaction energies ($E_K$ is small and flat),
whereas the high-density equilibrium is governed by the balance between the gravitational and
kinetic energies ($E_I$ is small and flat).

Because of the saturation of the self-interaction energy at high density, the low-density minimum disappears
at $\rho_0 \simeq 5$ and $M \simeq 0.45$.
This agrees with the sudden jump in the maximum density $\rho_{\max}$ found in Fig.~\ref{fig:R0p5-rhoc-3}
after $\rho_{\max}$ reaches a value of about 6.
It is expected that this critical value for $\rho_{\max}$ is somewhat above the threshold of 5 obtained in
Fig.~\ref{fig:R0p5-rhoc-3-Gauss}, because the density field in the simulations shows fluctuations of the
order of unity which add up to the smooth underlying soliton profile in the definition of $\rho_{\max}$,
as clearly seen in panel (d) in Fig.~\ref{fig:R0p5-rhoc-3}.
There is also a good agreement for the values of $M_{\rm TF} \simeq 0.45$ and $E_{\rm tot,TF} \simeq -0.2$
at the transition with the simulation results in Fig.~\ref{fig:R0p5-rhoc-3}.

As seen in the lower right panel, the high-density minimum is then located at the much higher density
$\rho_0 \sim 10^8$ with a significantly lower energy $E_{\rm tot} \simeq -50$ than the vanishing
TF soliton which had $E_{\rm tot,TF} \simeq -0.2$.
This large energy mismatch implies that the TF soliton at $\rho_0 \simeq 5$
cannot quickly settle down to the higher-density FDM soliton while keeping its mass and energy.
In fact, as seen in panel (b) in Fig.~\ref{fig:R0p5-rhoc-3}, only a small fraction of the TF soliton
mass ends up in the smaller FDM soliton at the transition, as the simulation gives
$M_{\rm FDM} \simeq 0.05$ after the transition, that is, a ratio of $10\%$.
Let us assume that the energy of the new FDM soliton is about the same as the energy of the previous
TF soliton, $E_{\rm tot, FDM} \simeq -0.2$, with the remaining TF matter having a zero energy
(in a classical picture, they correspond to barely unbound particles in the gravitational potential well
now set by the FDM soliton). Then, we can see in the upper right panel in
Fig.~\ref{fig:R0p5-rhoc-3-Gauss} that we predict $M_{\rm FDM} \simeq 0.08$.
This is not too far from the value $M_{\rm FDM} \simeq 0.05$ obtained in the simulation.
We can also see in panel (c) in Fig.~\ref{fig:R0p5-rhoc-3} that the energy of the FDM soliton
is of the same order as that of the previous TF soliton at the transition.
We can see in the simulation that this transition is irreversible. Indeed, as the FDM soliton gains some
mass and decreases its energy it cannot go back to the TF configurations, which all have higher energy.

\section{Results for $R_{\rm TF}$ = 0.1}\label{sec:model-a-r-0.1}

We explore in this section the case $R_{\rm TF} = 0.1$, where the self-interactions are rather small
and the Thomas-Fermi radius (\ref{eq:Rsol-lambda}) defined in the low-density regime is much
smaller than the system size.
In the cosmological context, this corresponds to structures that form at later times, collapsing on a much
larger scale than the characteristic length associated with the self-interactions.

\begin{figure*}[ht!]
\centering
\includegraphics[height=4.cm,width=0.28\textwidth]{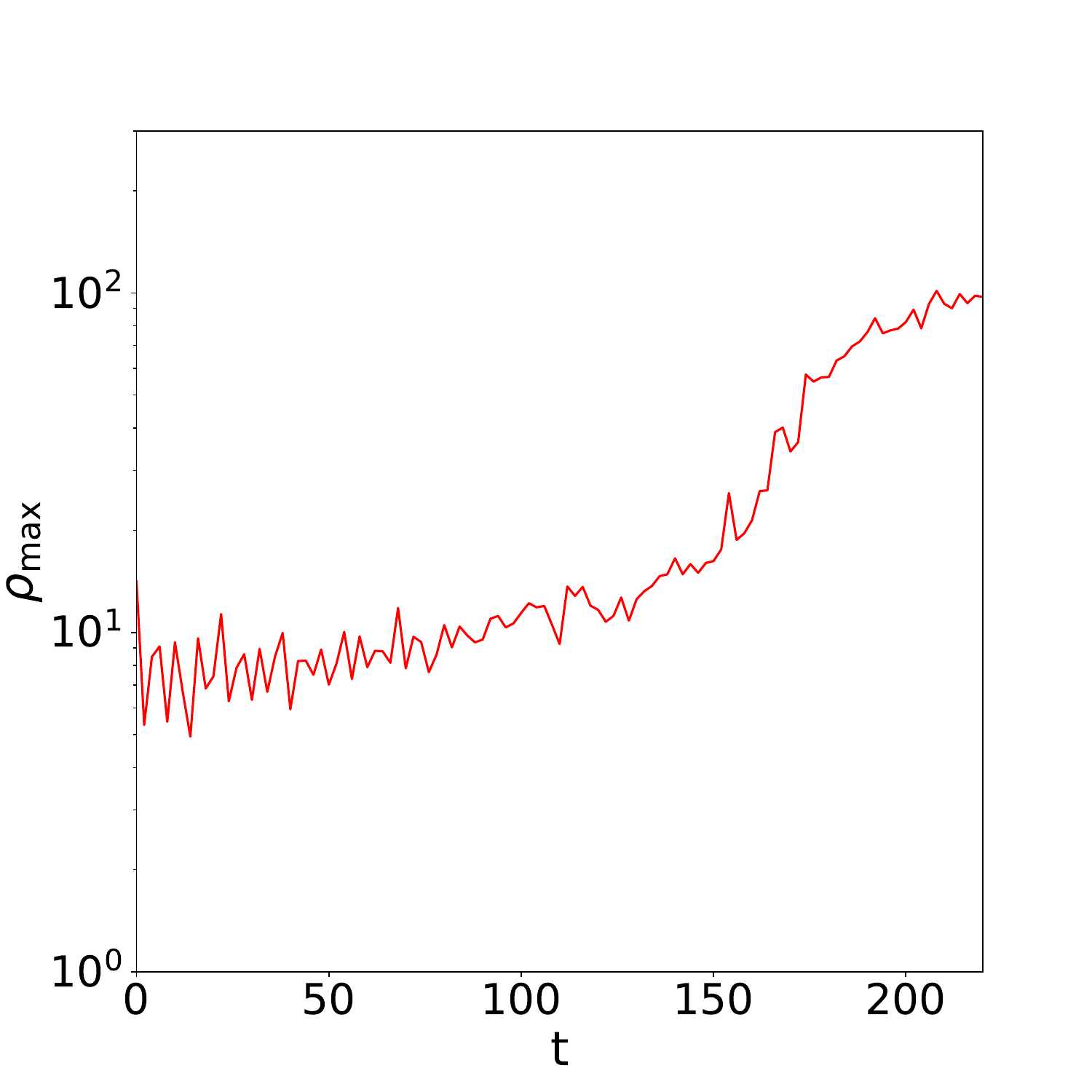}
\includegraphics[height=4.cm,width=0.28\textwidth]{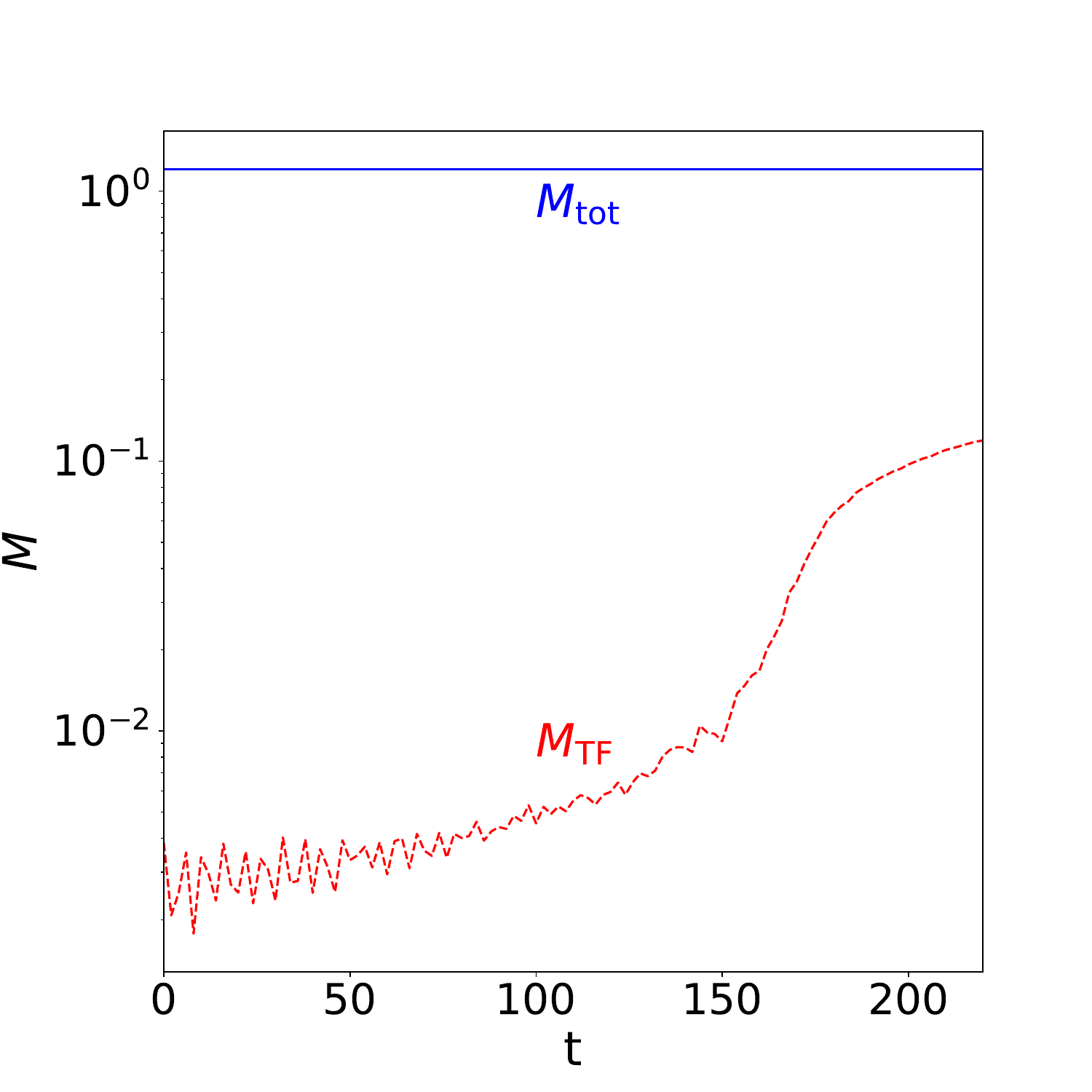}
\includegraphics[height=4.cm,width=0.28\textwidth]{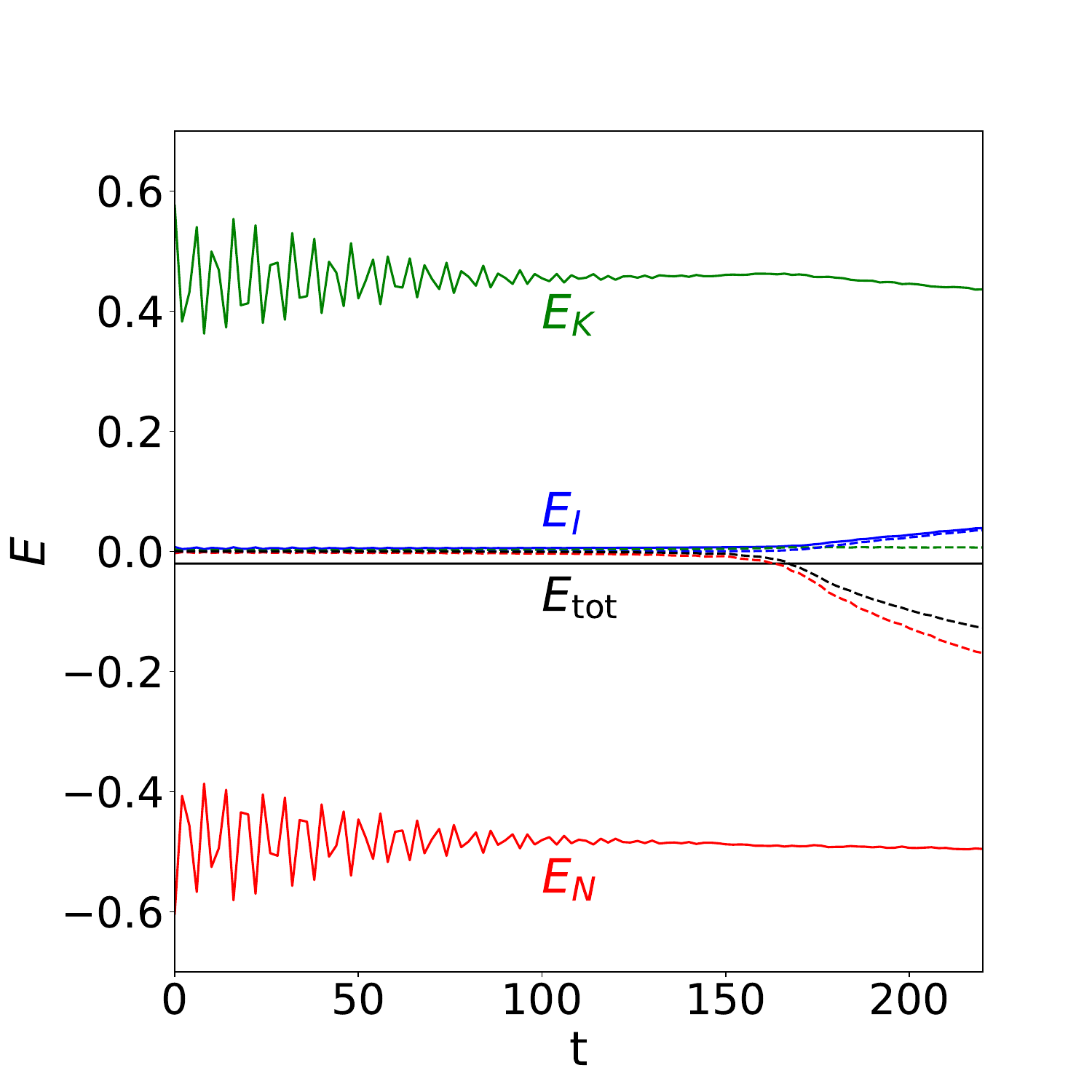}\\
\includegraphics[height=4.cm,width=0.28\textwidth]{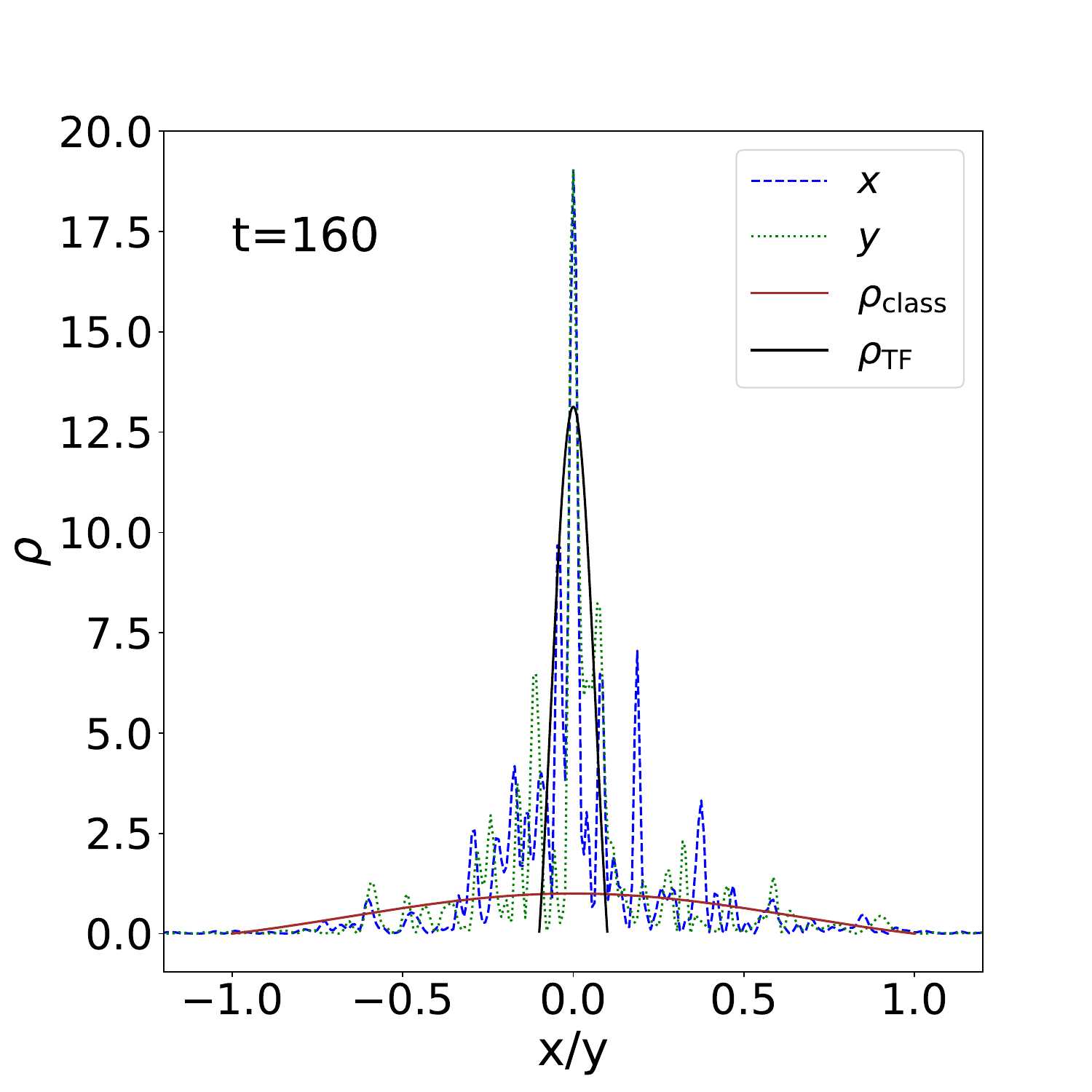}
\includegraphics[height=4.cm,width=0.28\textwidth]{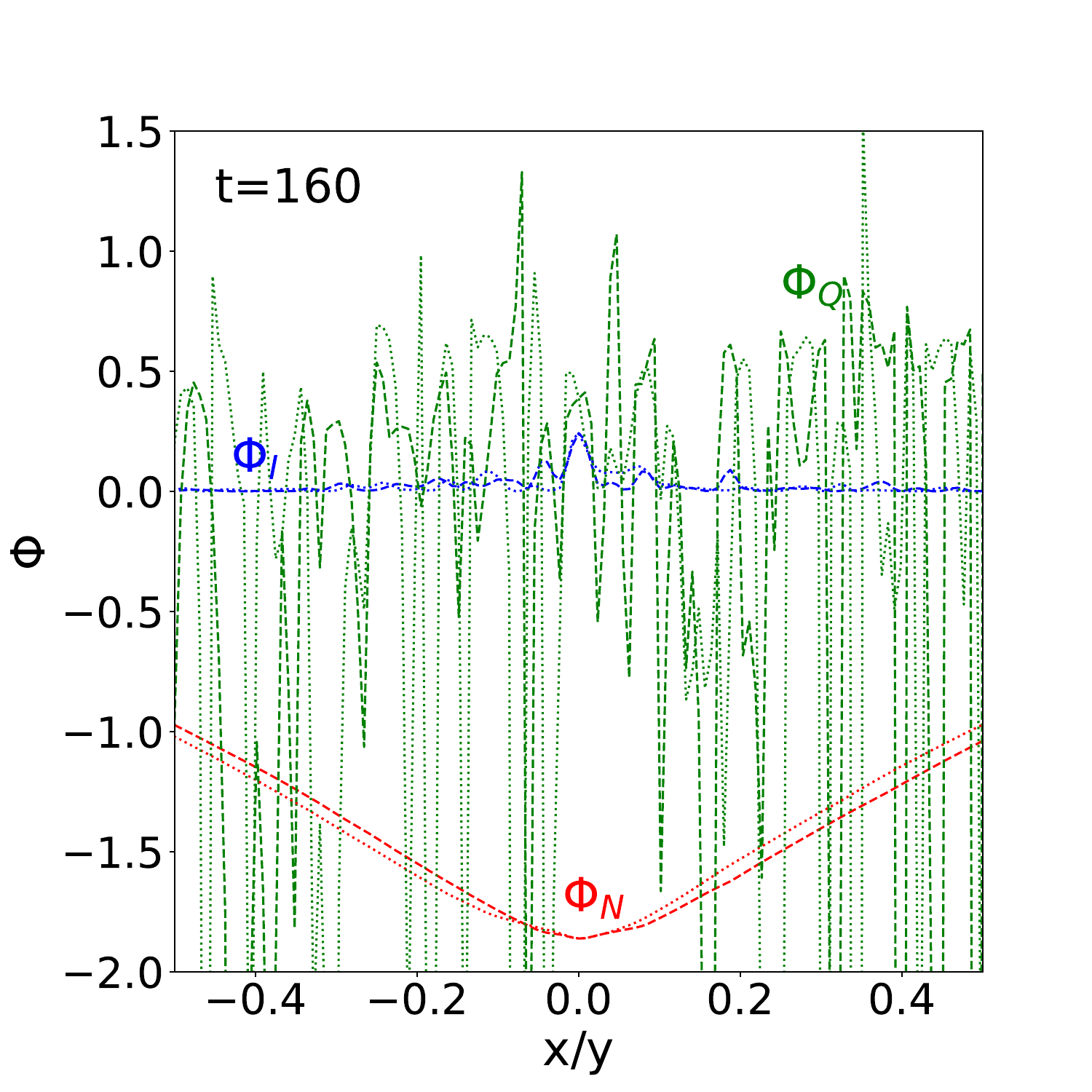}
\includegraphics[height=4.cm,width=0.29\textwidth]{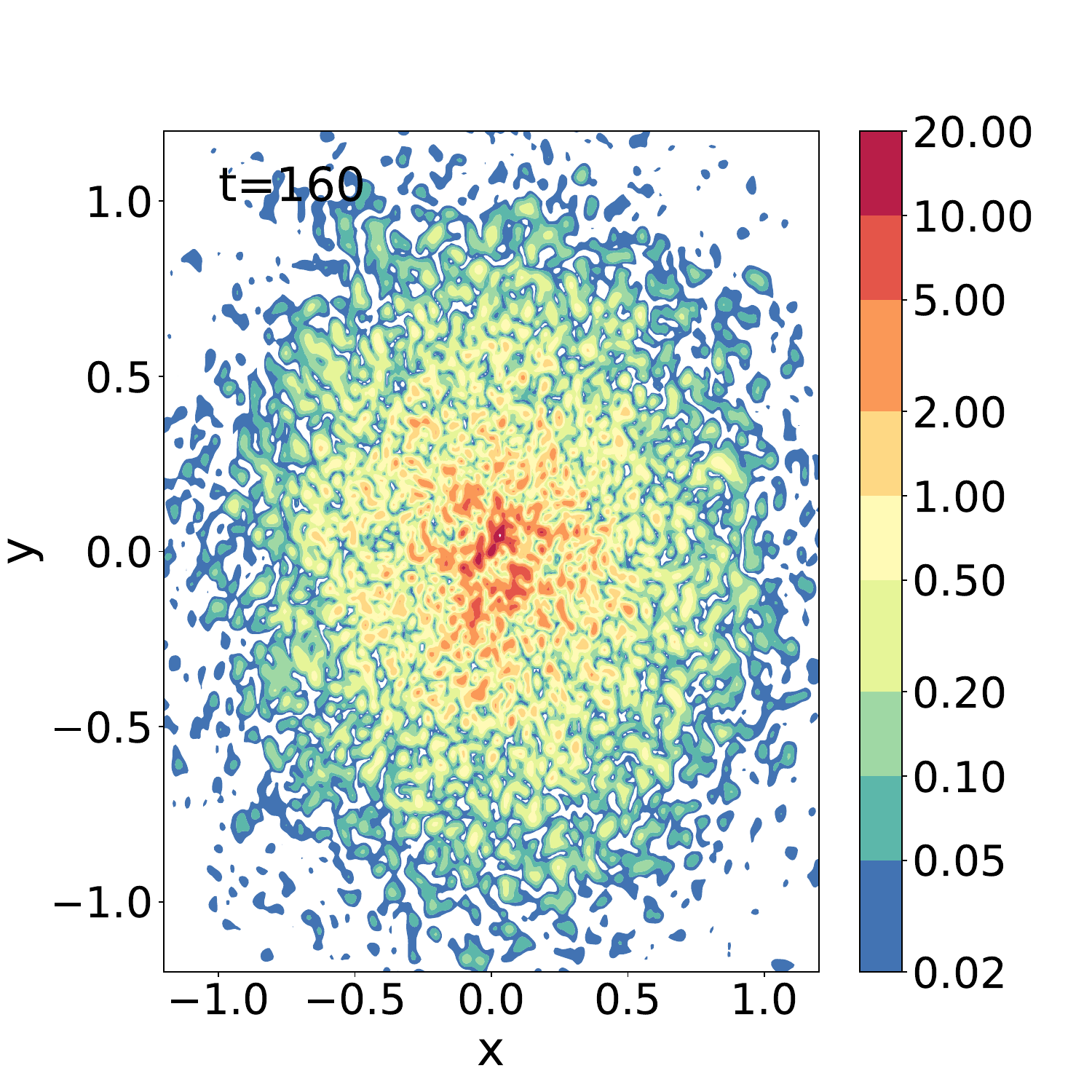}\\
\includegraphics[height=4.cm,width=0.28\textwidth]{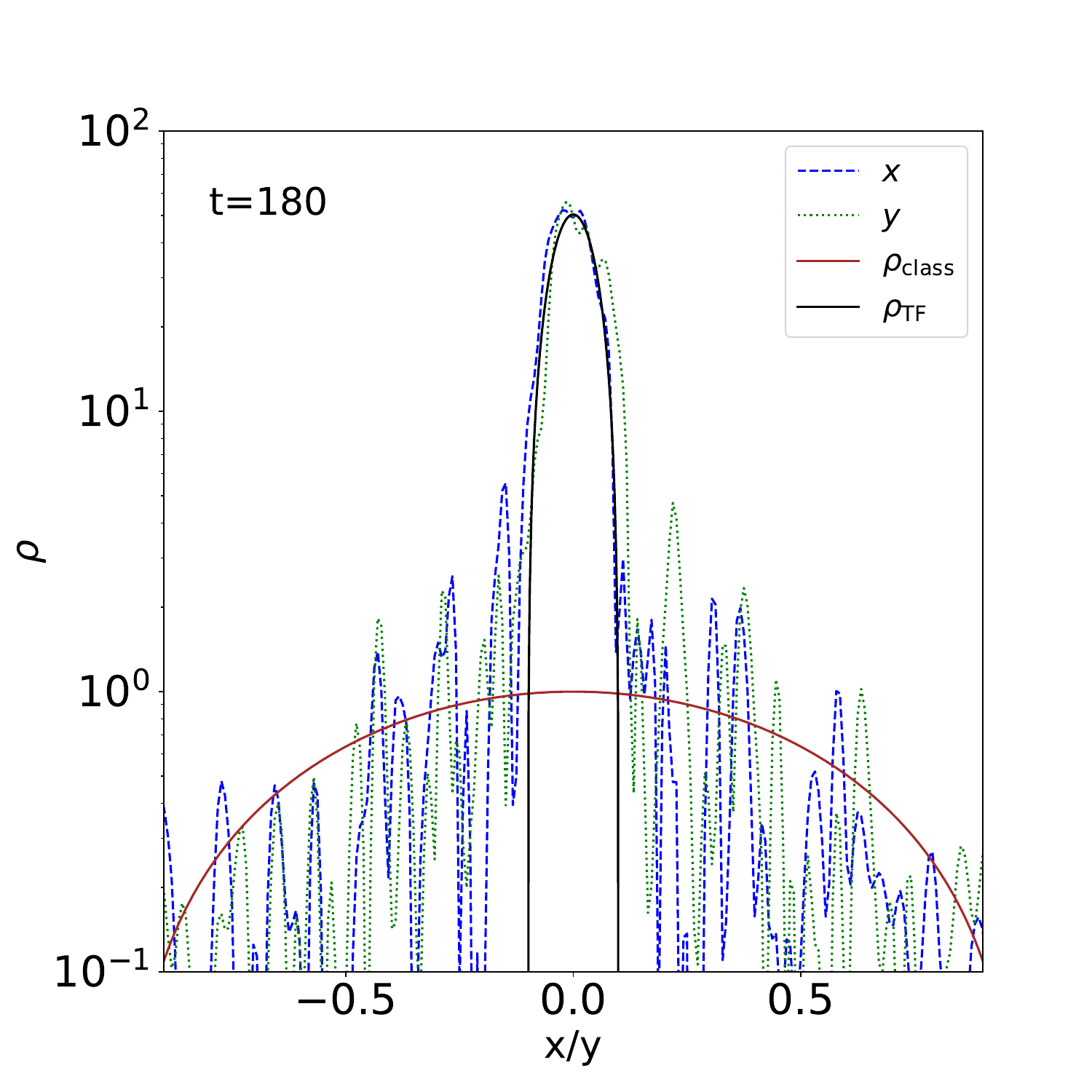}
\includegraphics[height=4.cm,width=0.28\textwidth]{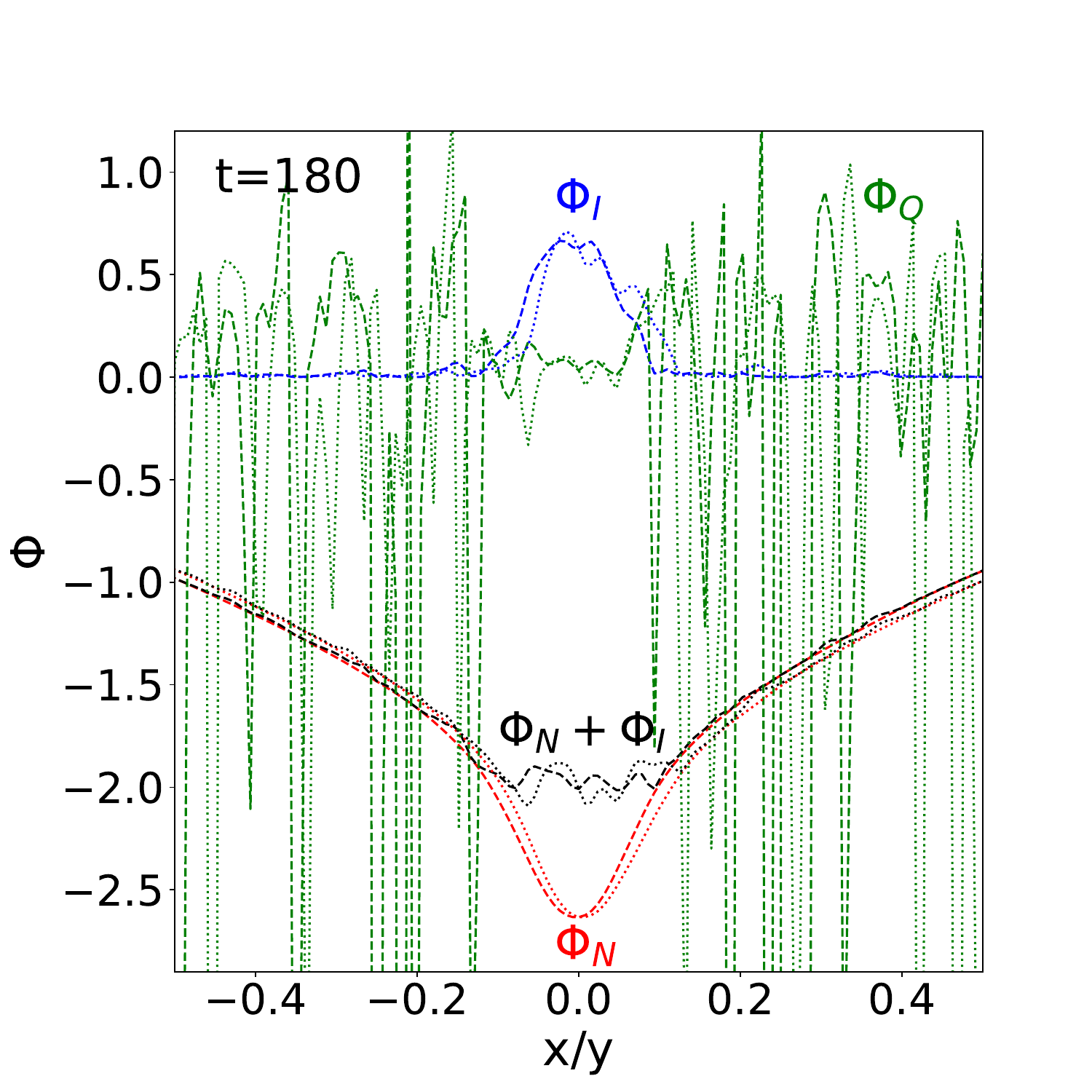}
\includegraphics[height=4.cm,width=0.29\textwidth]{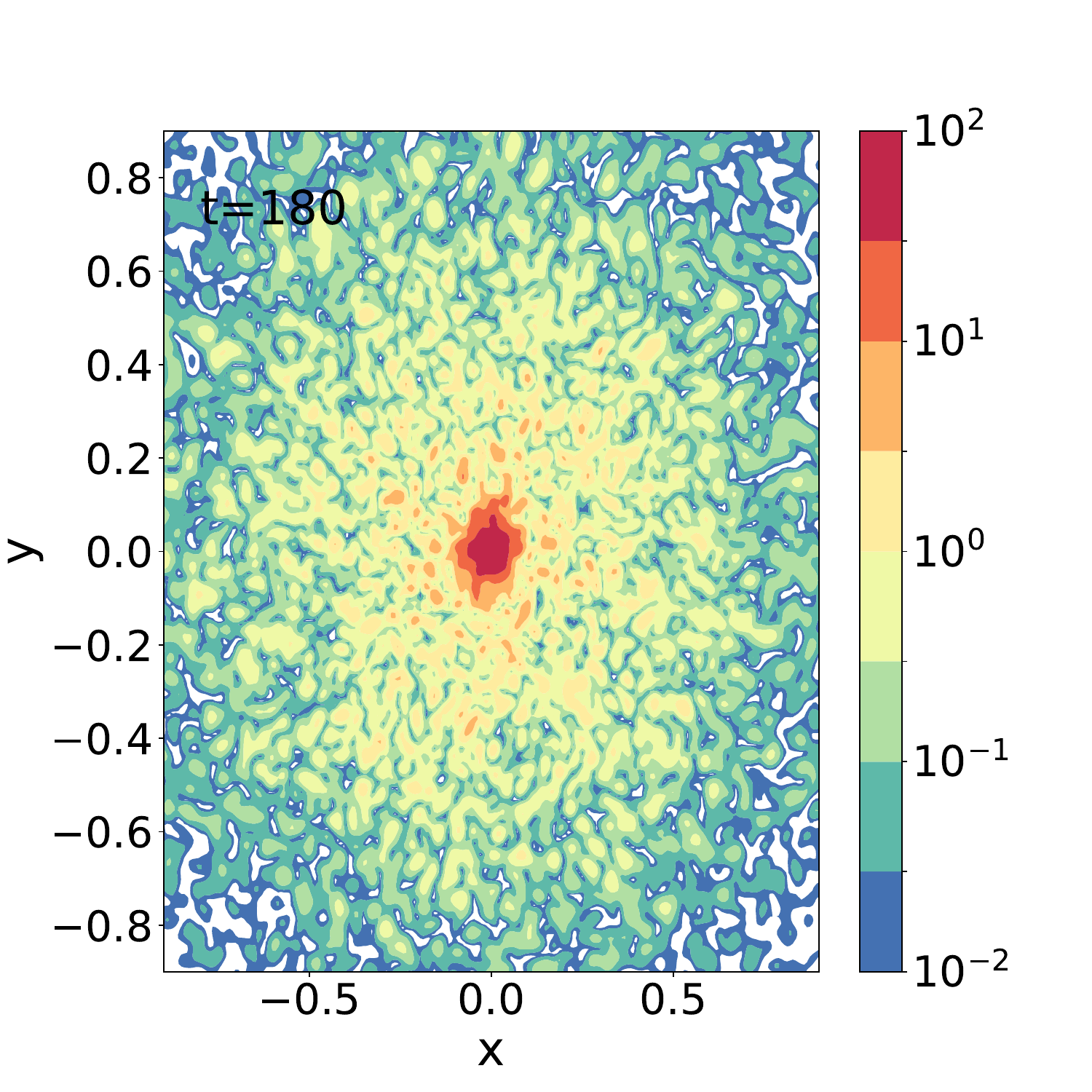}
\caption{Evolution of a halo with $R_{\rm TF}=0.1$ and $\rho_c=500$.}
\label{fig:R0p1-rhoc-500}
\end{figure*}

\begin{figure*}[ht!]
\centering
\includegraphics[height=4.cm,width=0.28\textwidth]{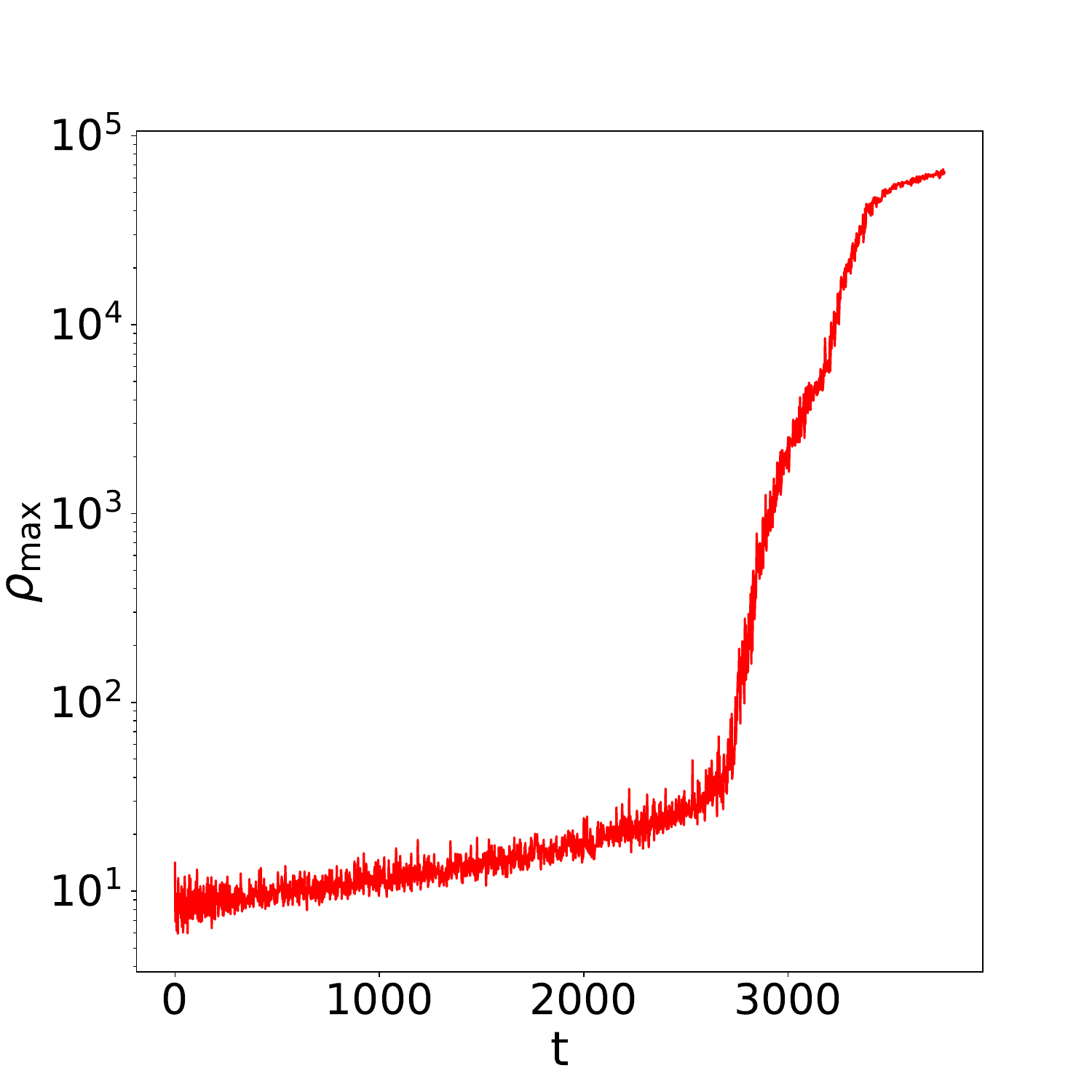}
\includegraphics[height=4.cm,width=0.28\textwidth]{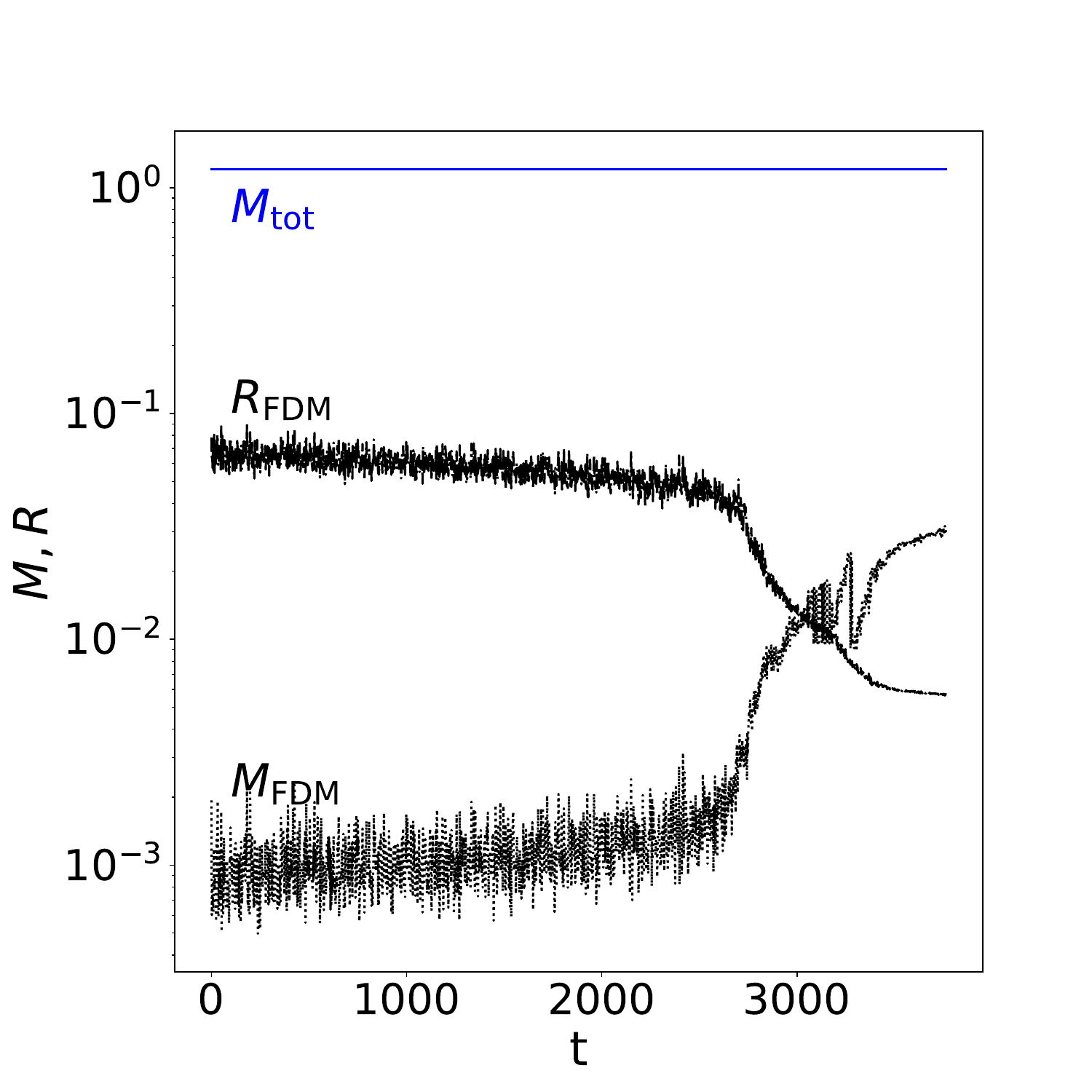}
\includegraphics[height=4.cm,width=0.28\textwidth]{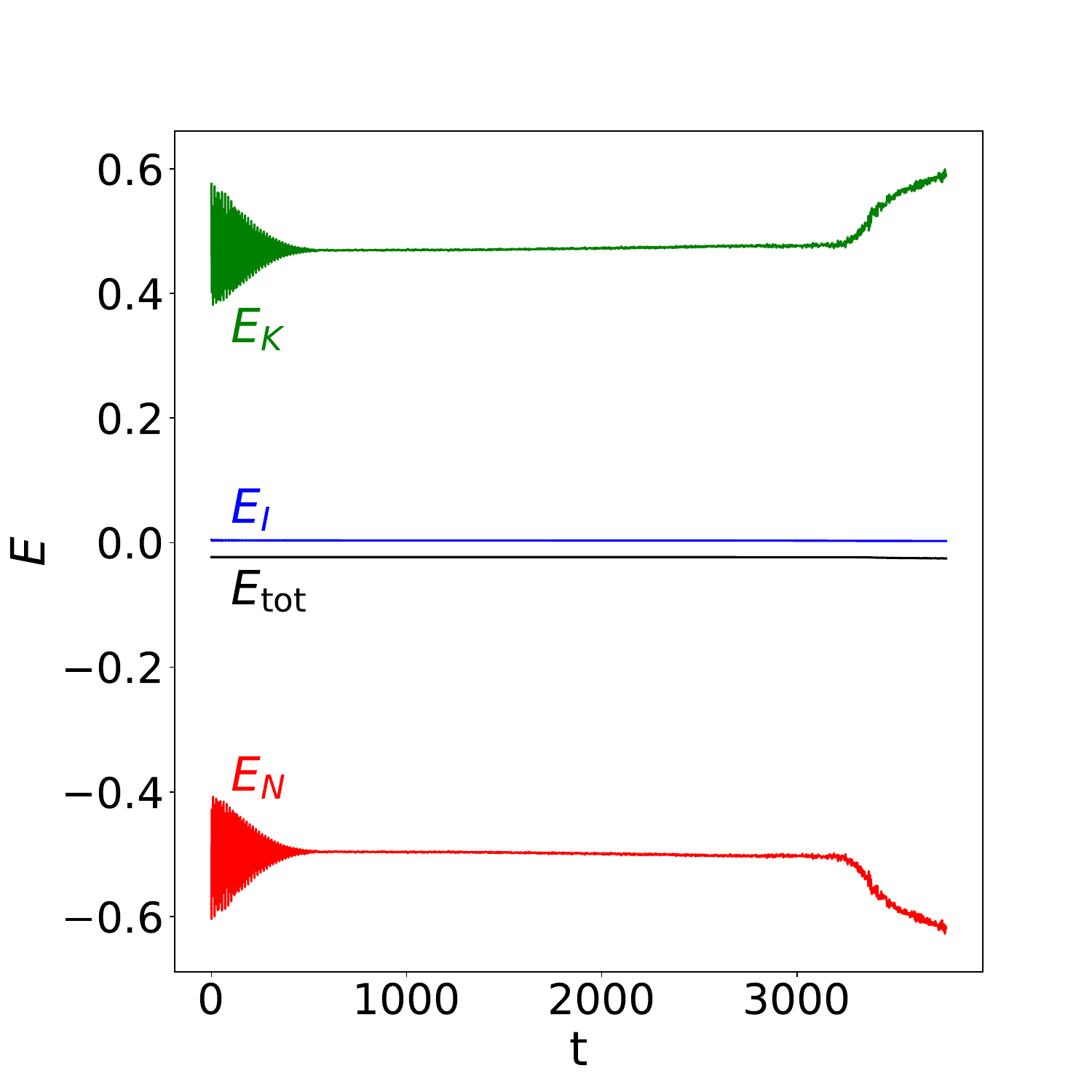}\\
\includegraphics[height=4.cm,width=0.28\textwidth]{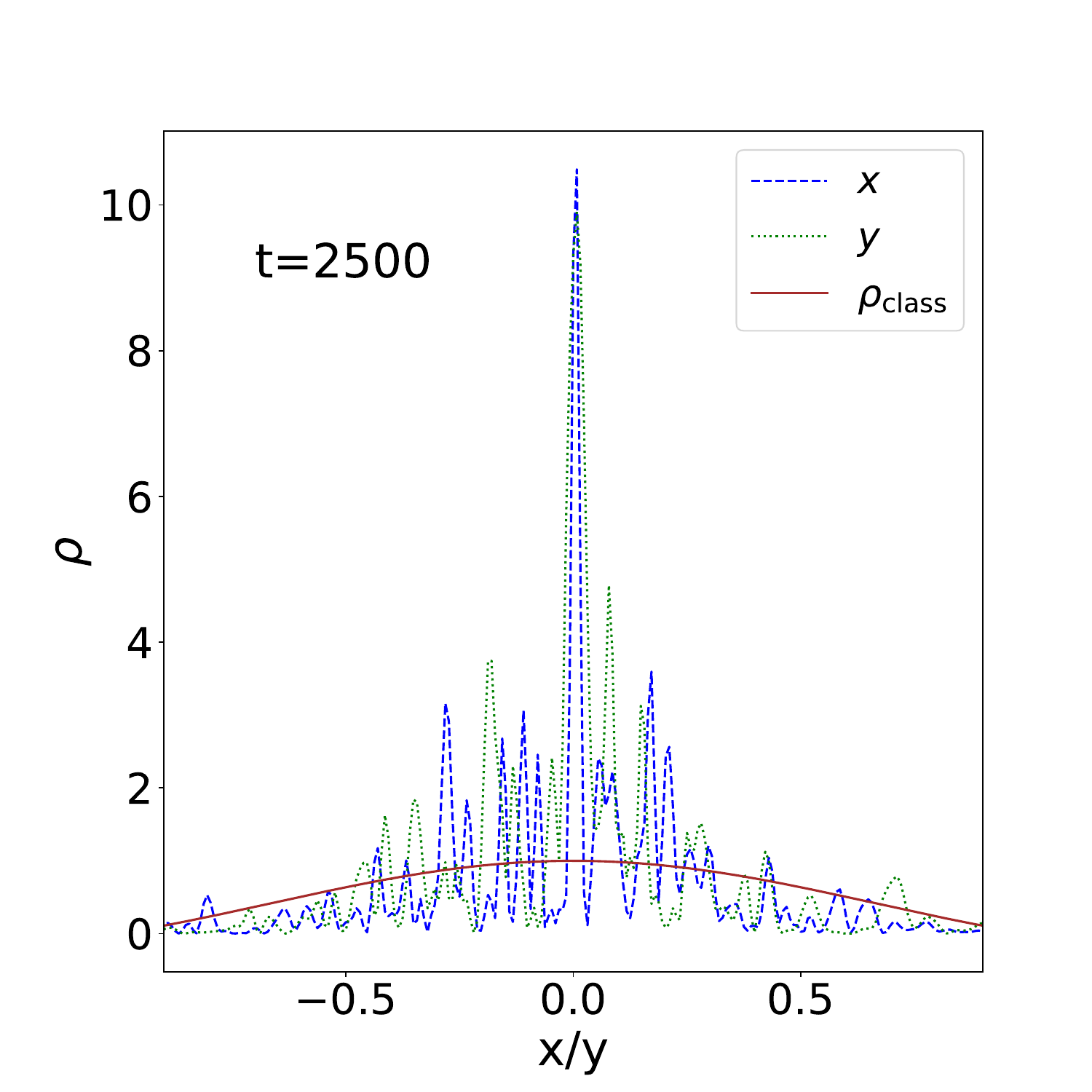}
\includegraphics[height=4.cm,width=0.28\textwidth]{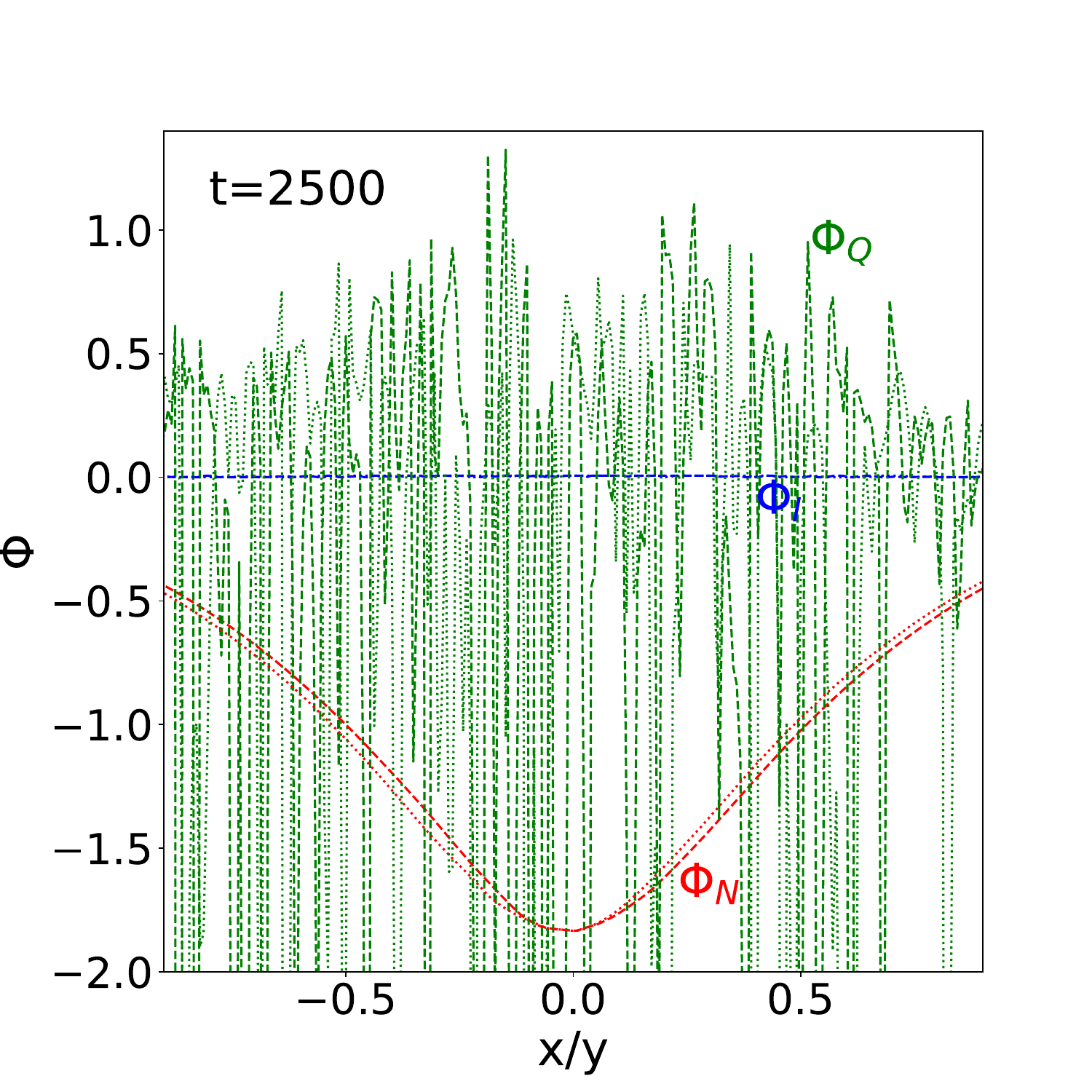}
\includegraphics[height=4.cm,width=0.29\textwidth]{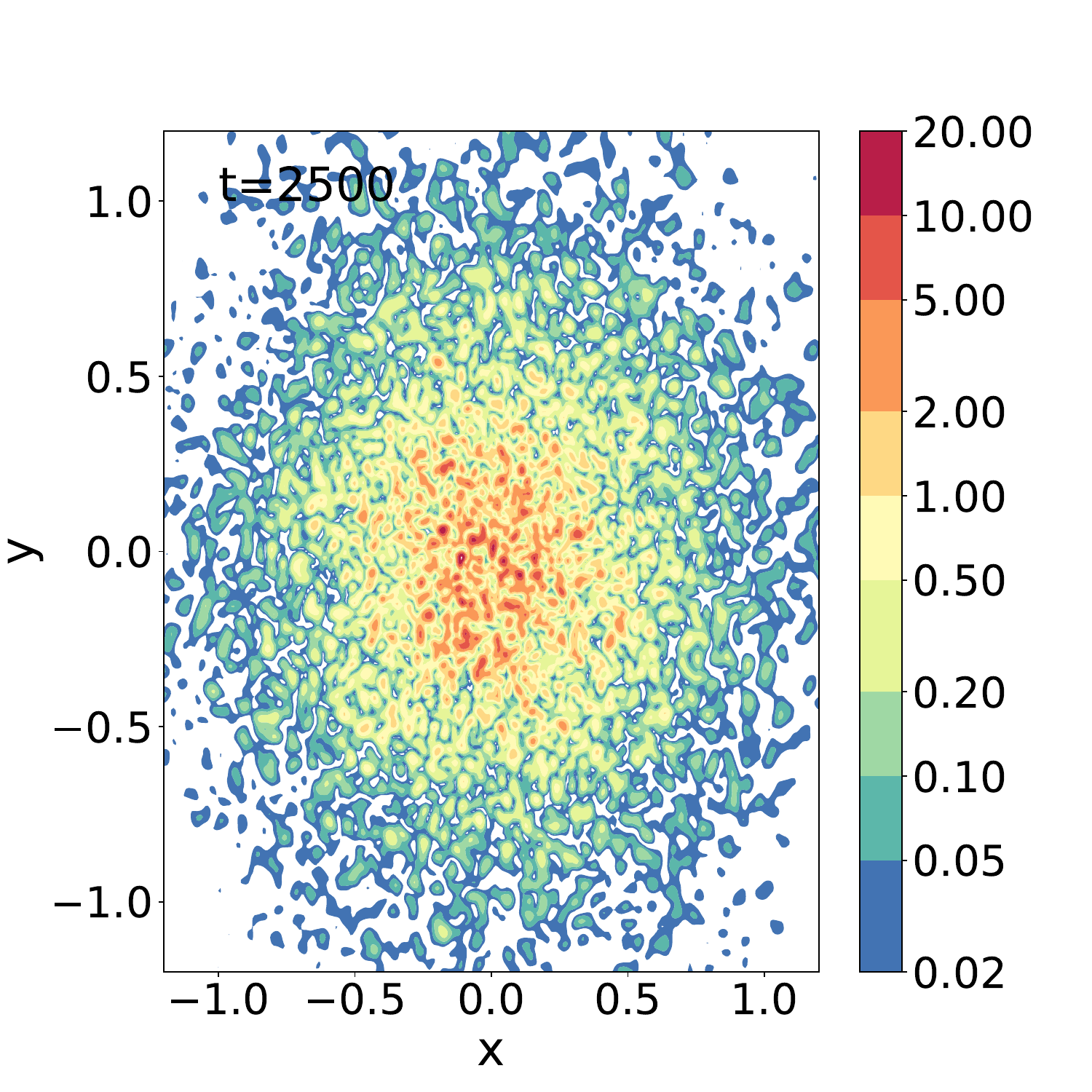}\\
\includegraphics[height=4.cm,width=0.28\textwidth]{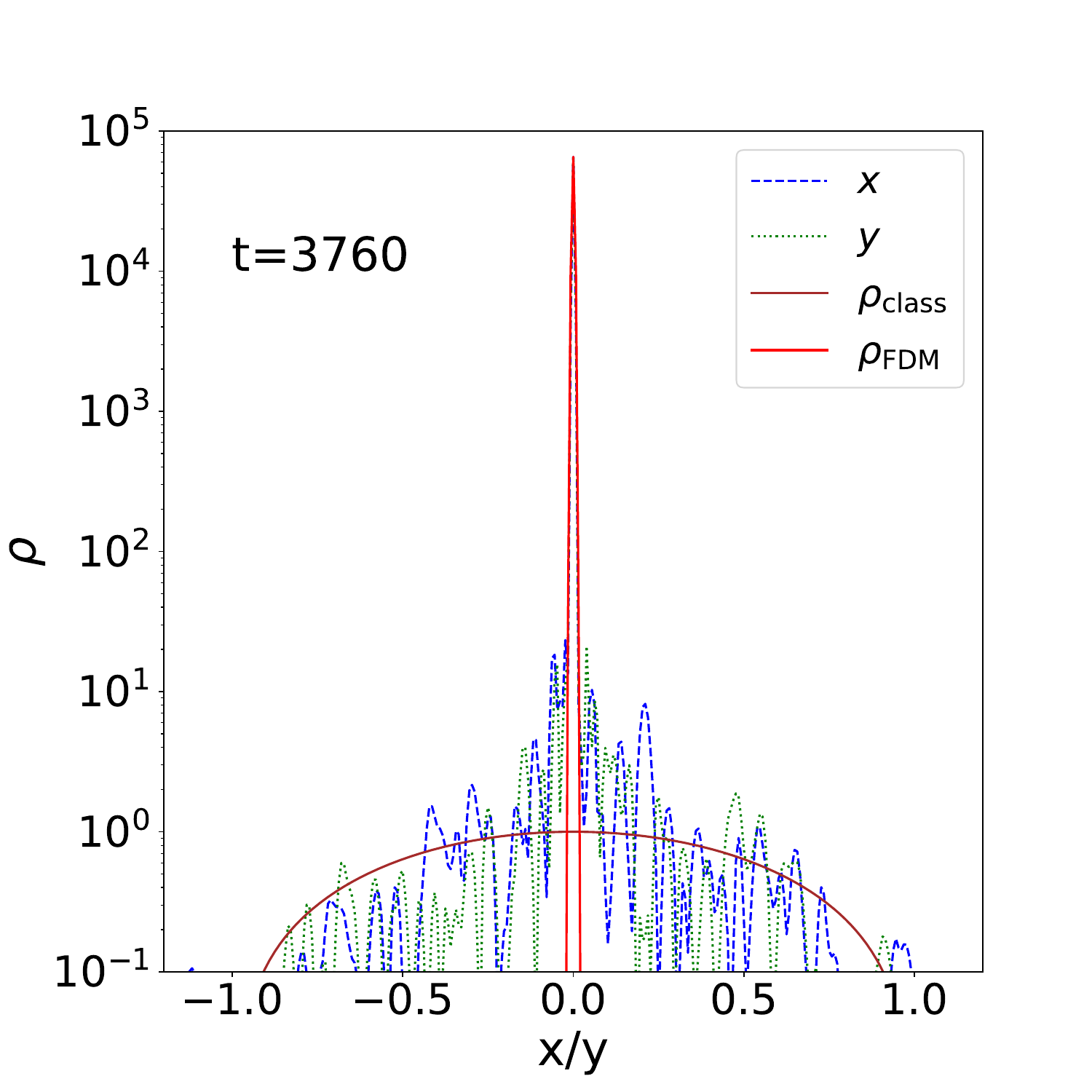}
\includegraphics[height=4.cm,width=0.28\textwidth]{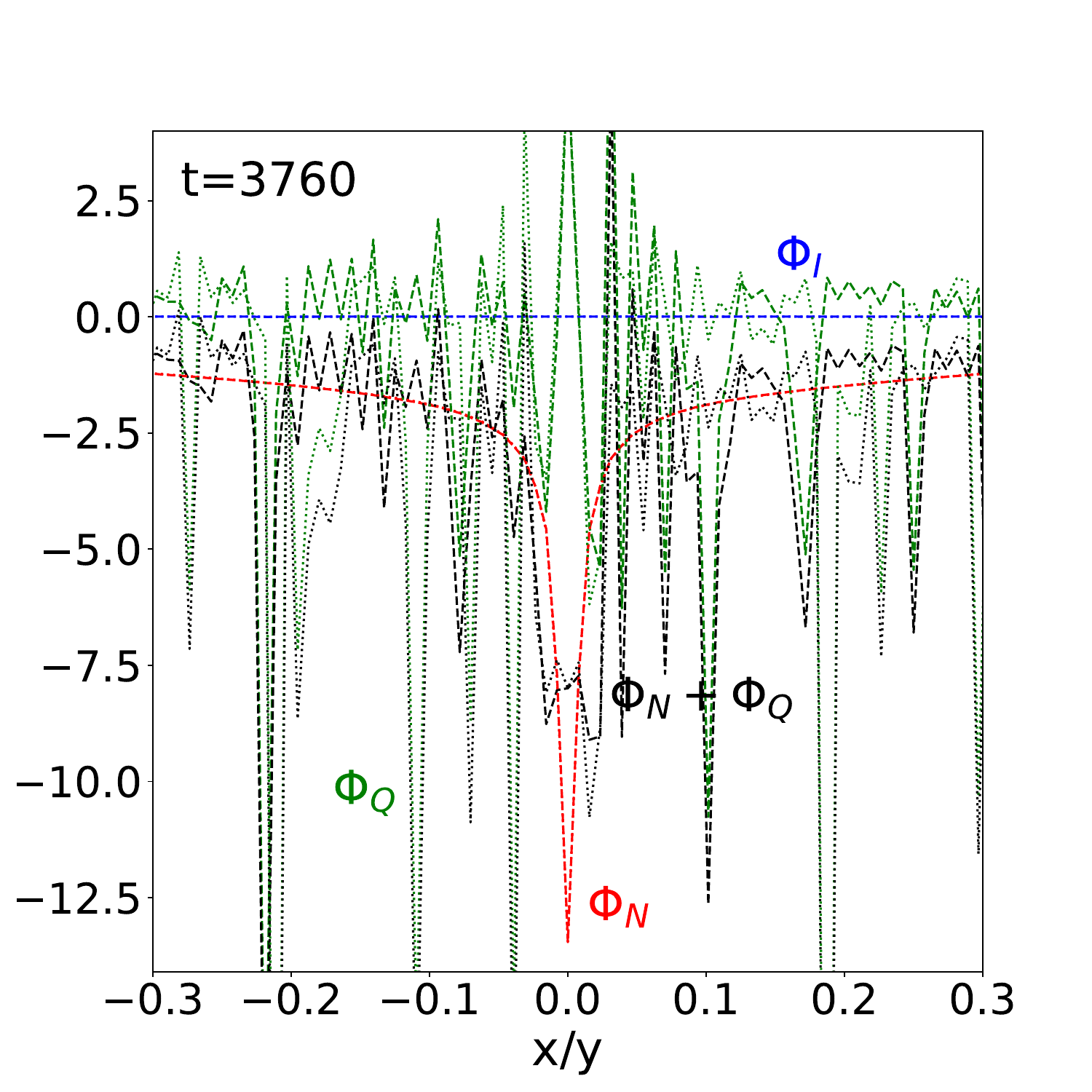}
\includegraphics[height=4.cm,width=0.29\textwidth]{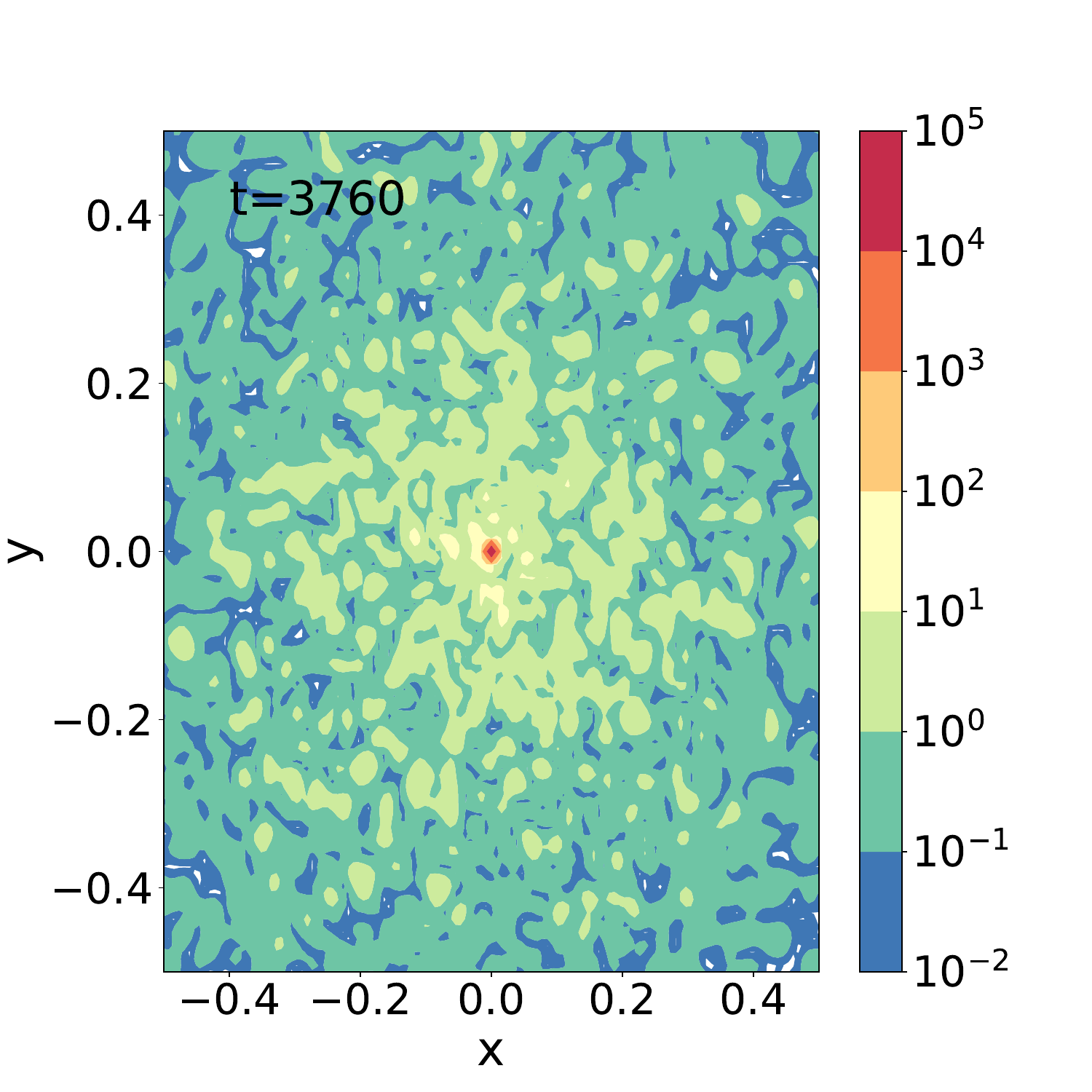}
\caption{Evolution of a halo with $R_{\rm TF}=0.1$ and $\rho_c=0.5$.}
\label{fig:R0p1-rhoc-0p5}
\end{figure*}

\begin{figure*}[ht]
\centering
\includegraphics[height=4.cm,width=0.28\textwidth]{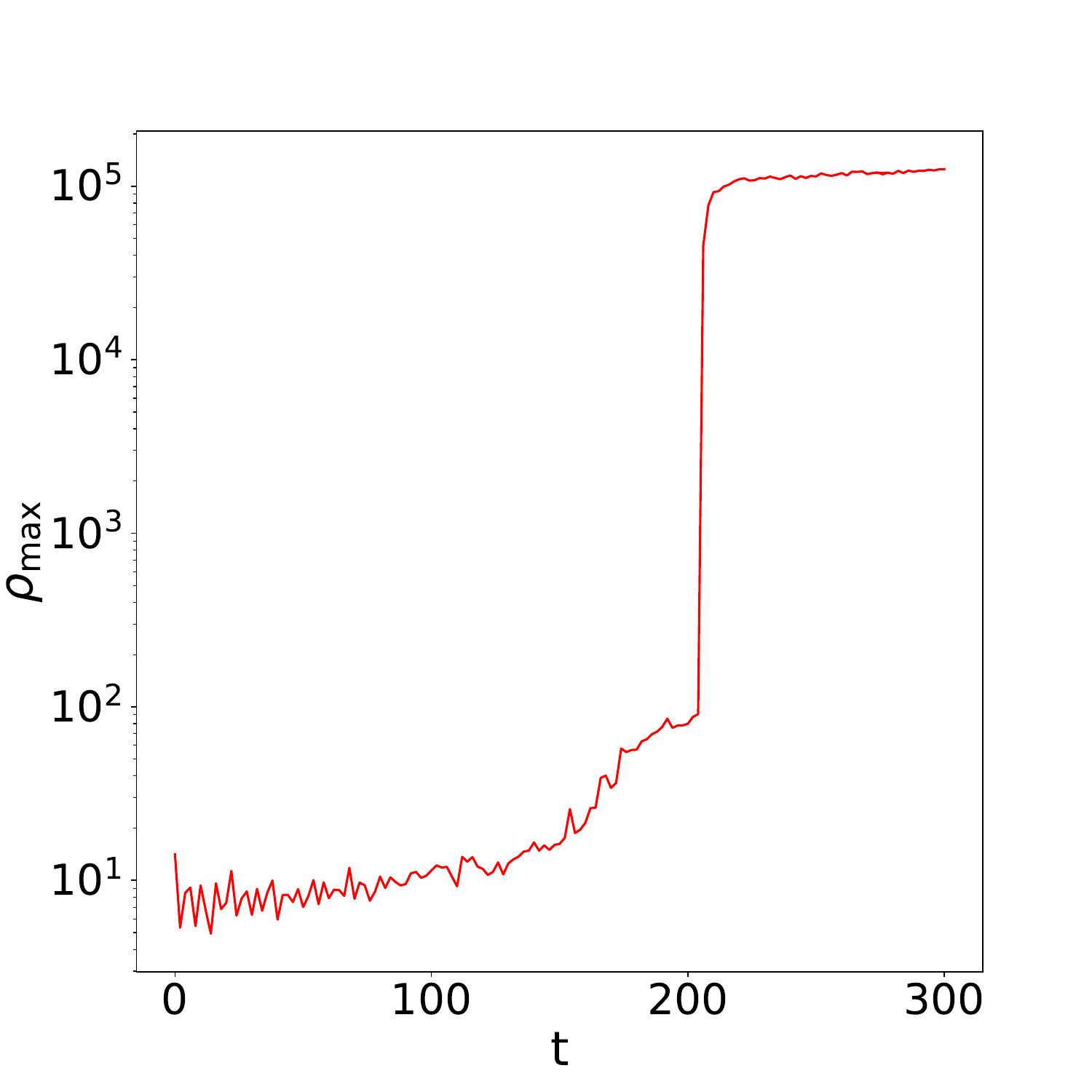}
\includegraphics[height=4.cm,width=0.28\textwidth]{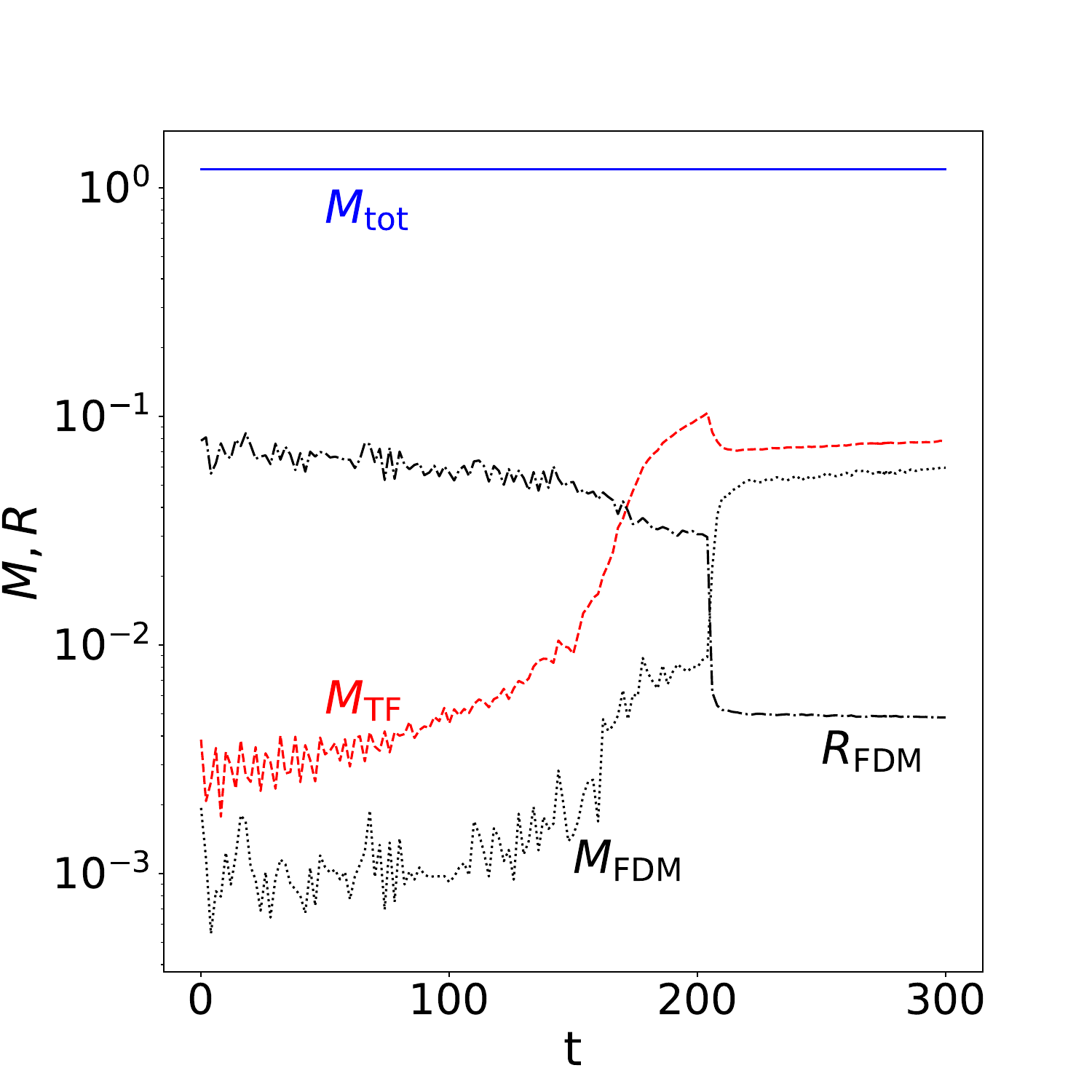}
\includegraphics[height=4.cm,width=0.28\textwidth]{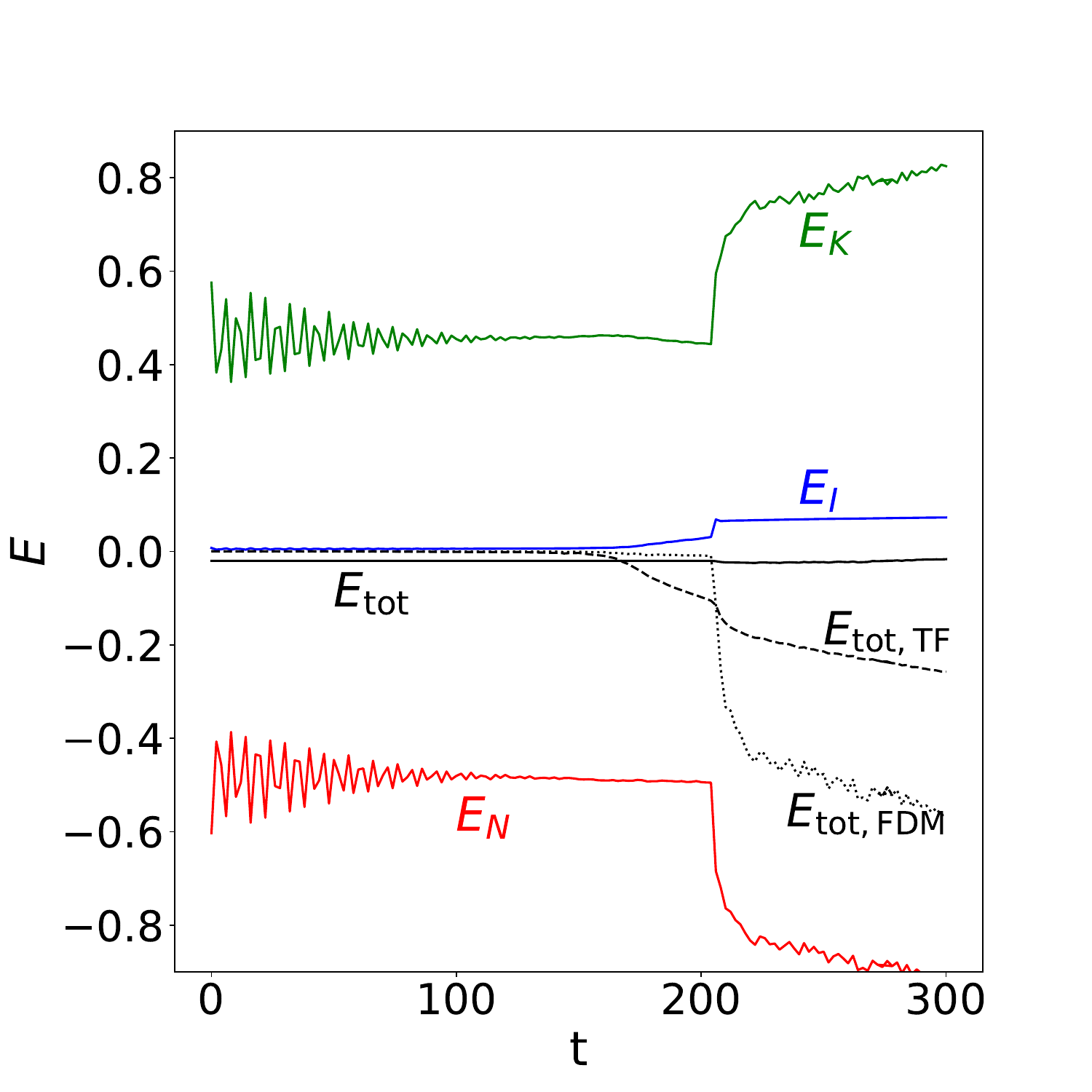}\\
\includegraphics[height=4.cm,width=0.28\textwidth]{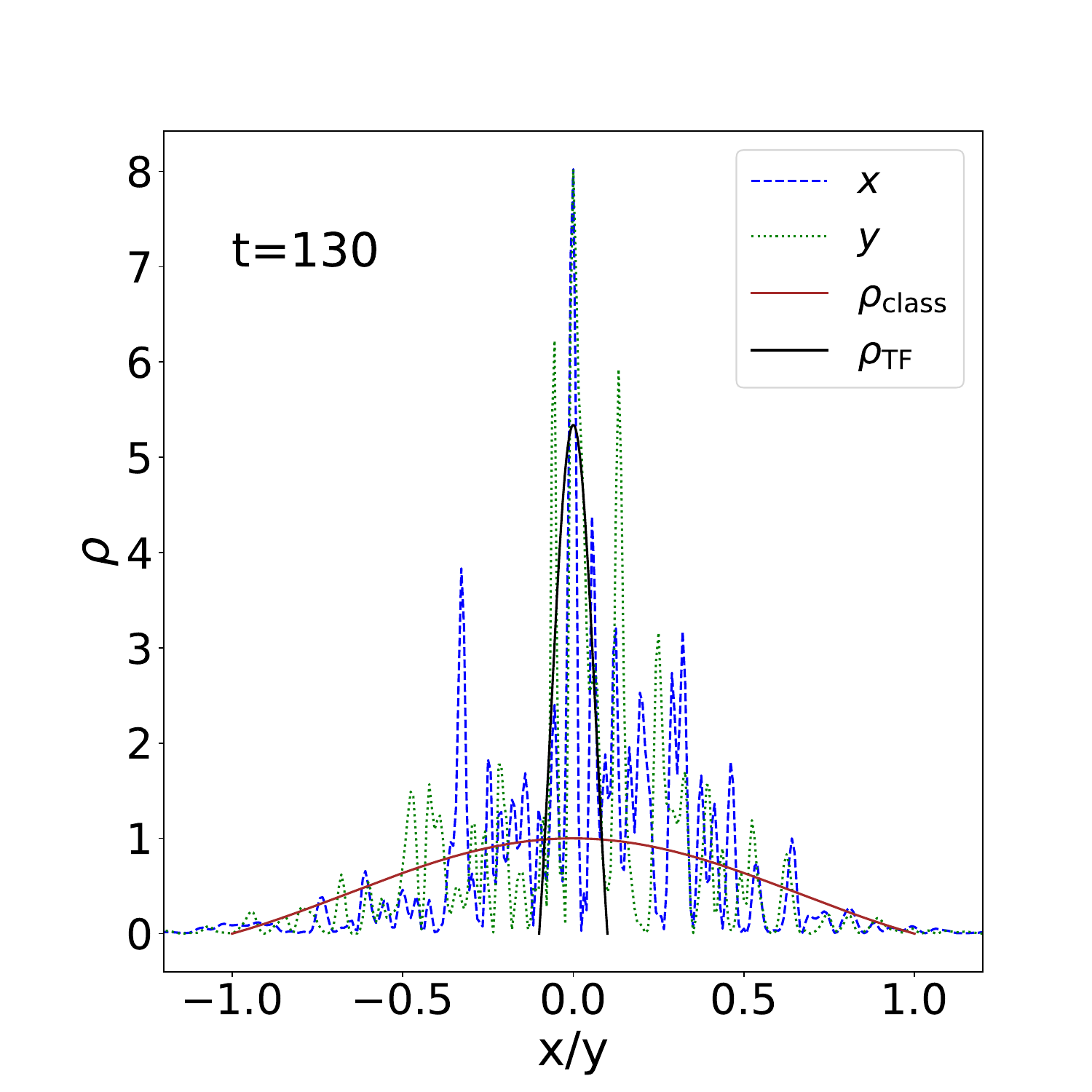}
\includegraphics[height=4.cm,width=0.28\textwidth]{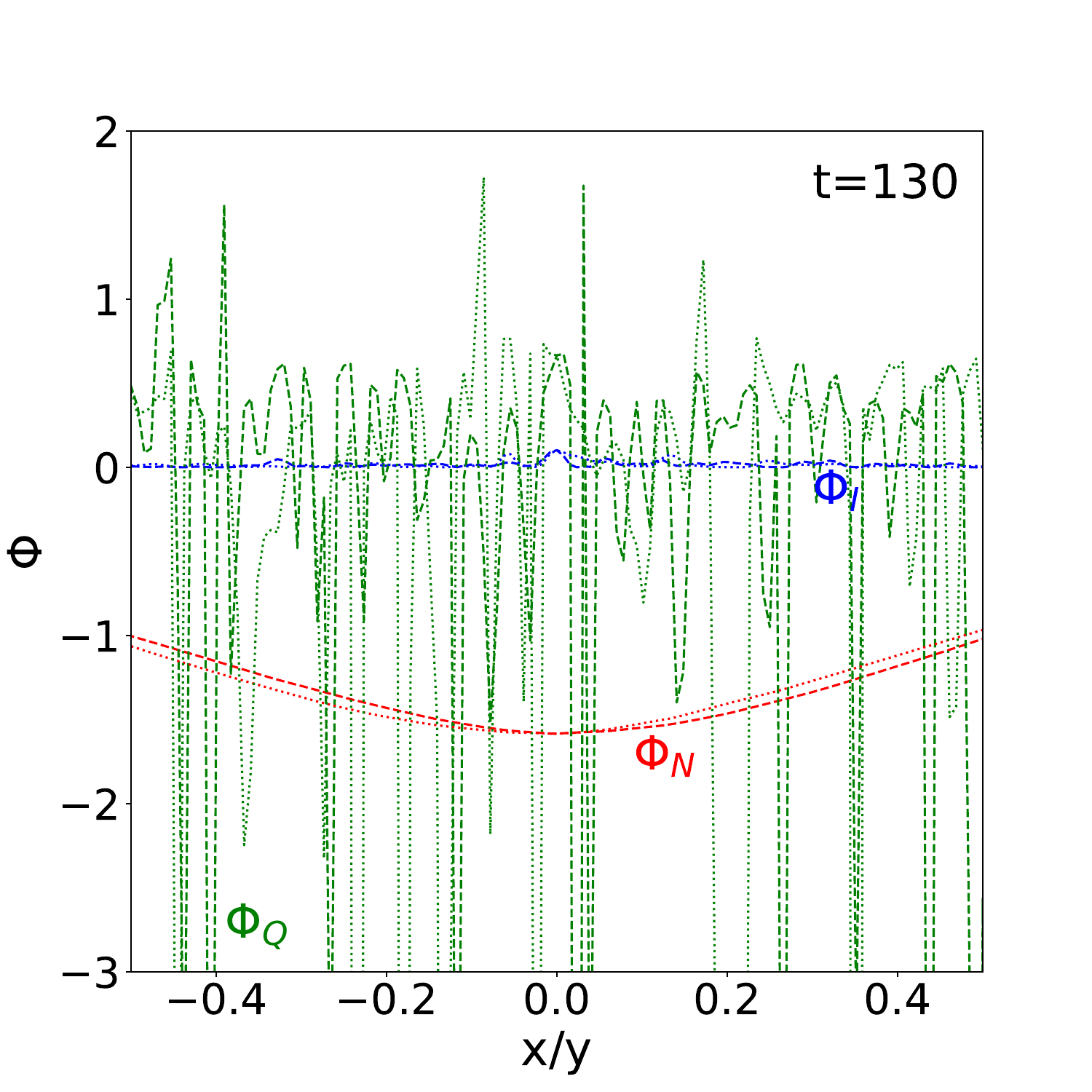}
\includegraphics[height=4.cm,width=0.29\textwidth]{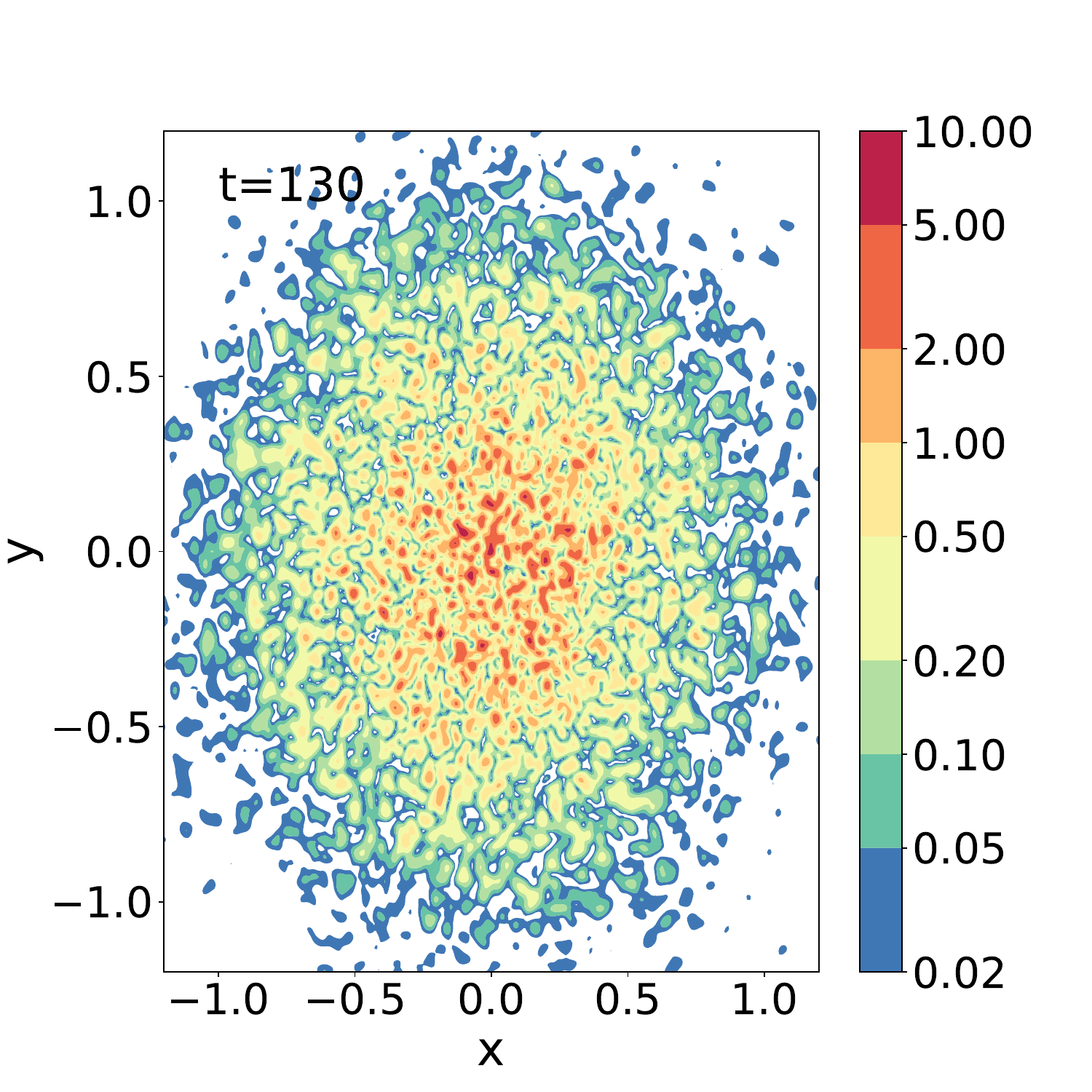}\\
\includegraphics[height=4.cm,width=0.28\textwidth]{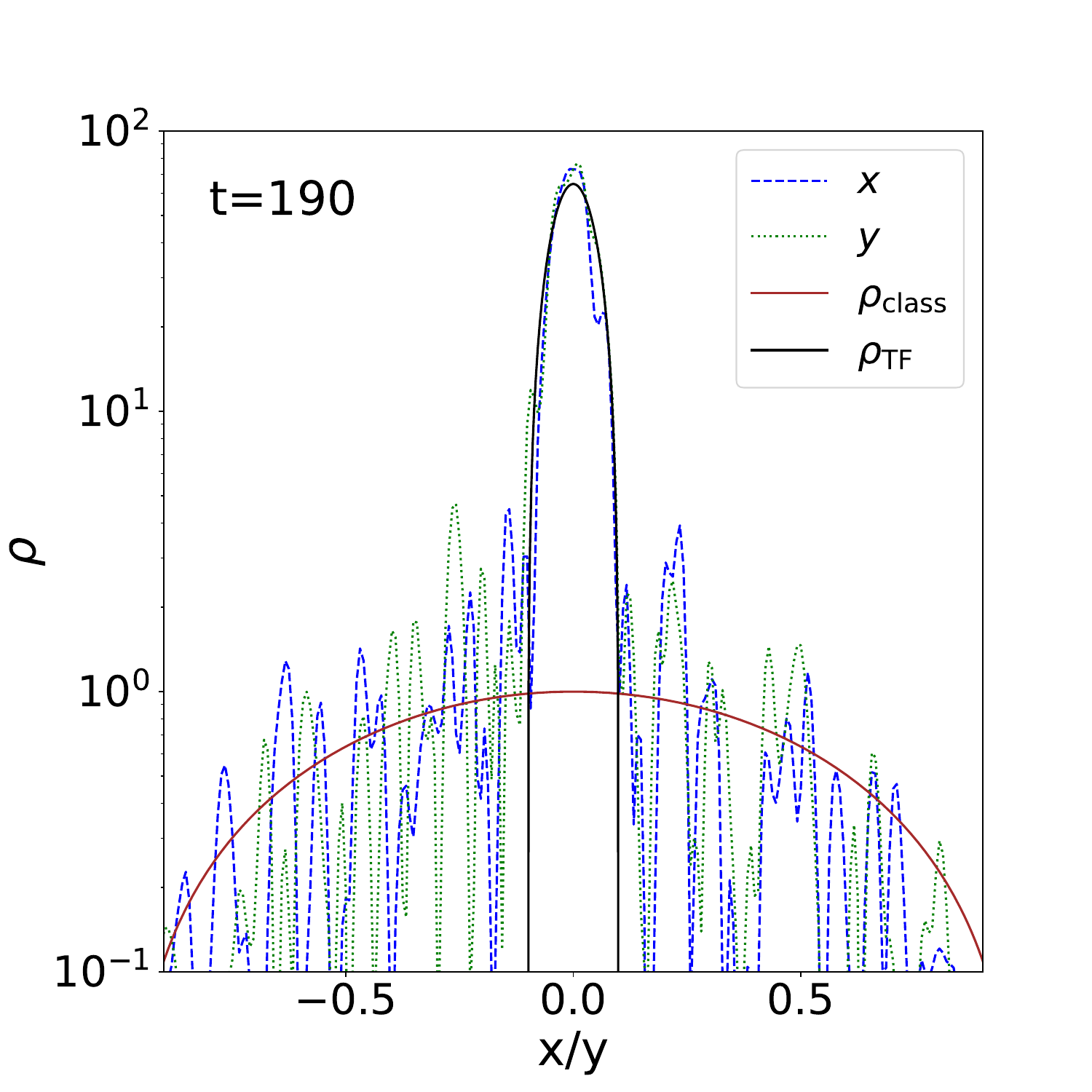}
\includegraphics[height=4.cm,width=0.28\textwidth]{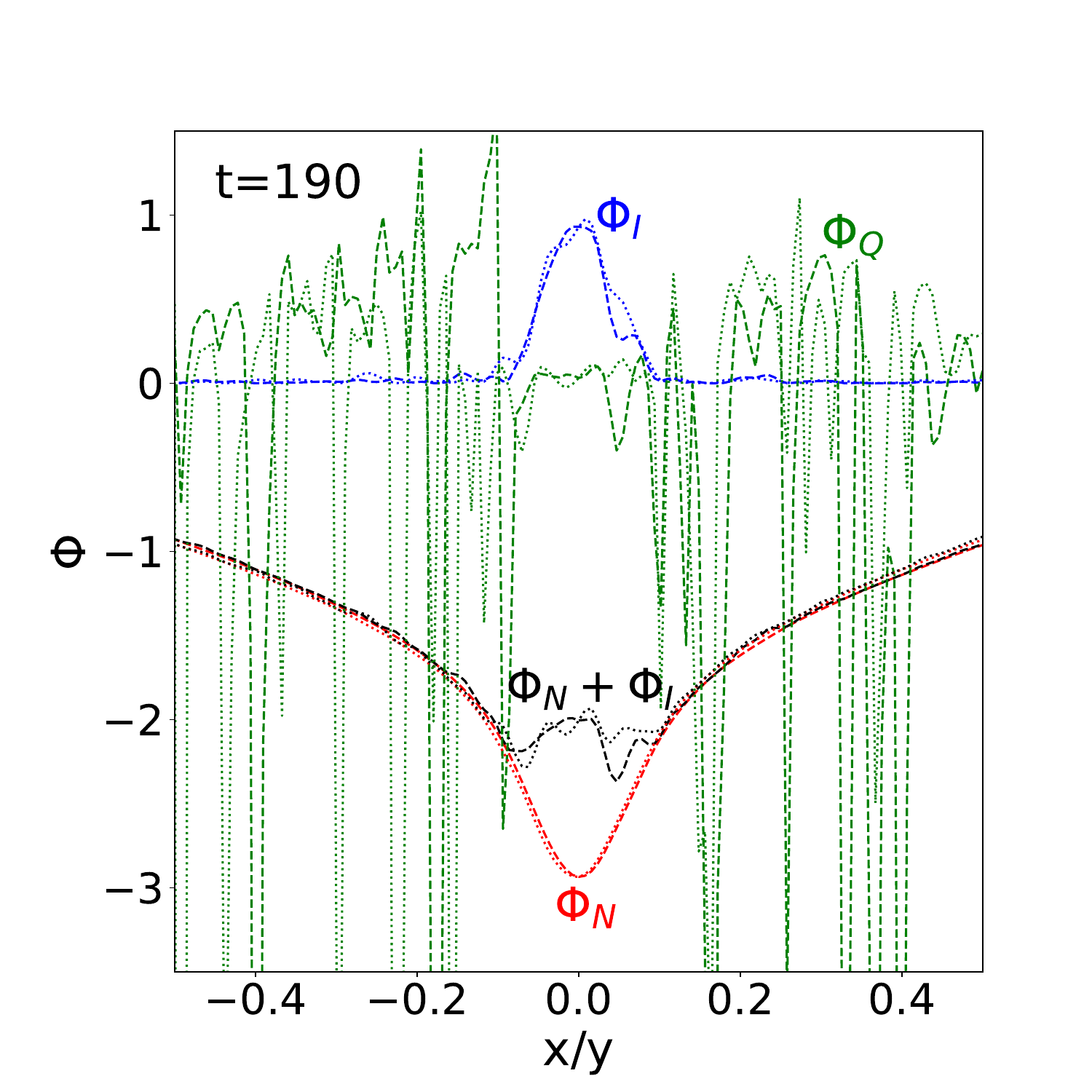}
\includegraphics[height=4.cm,width=0.29\textwidth]{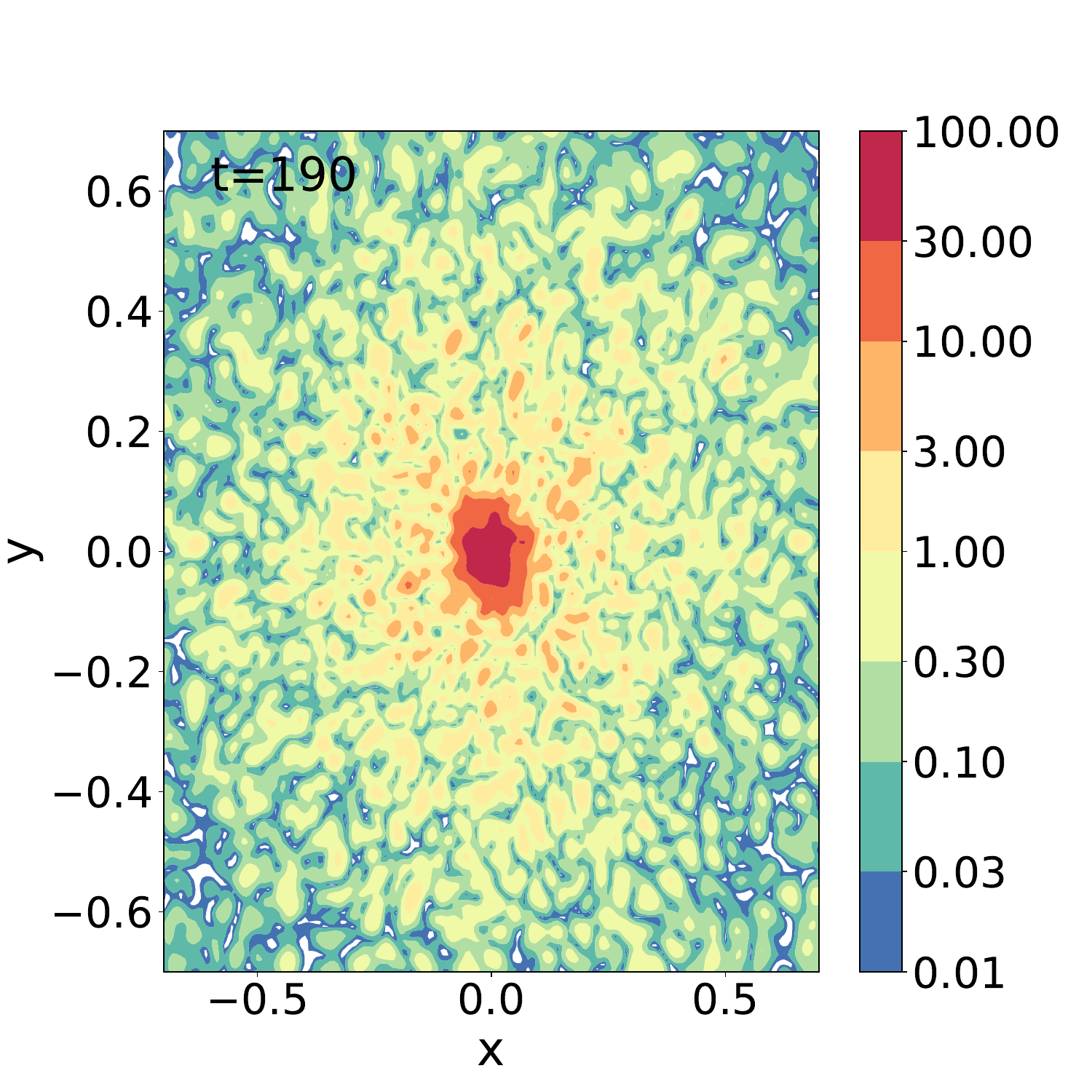}\\
\includegraphics[height=4.cm,width=0.28\textwidth]{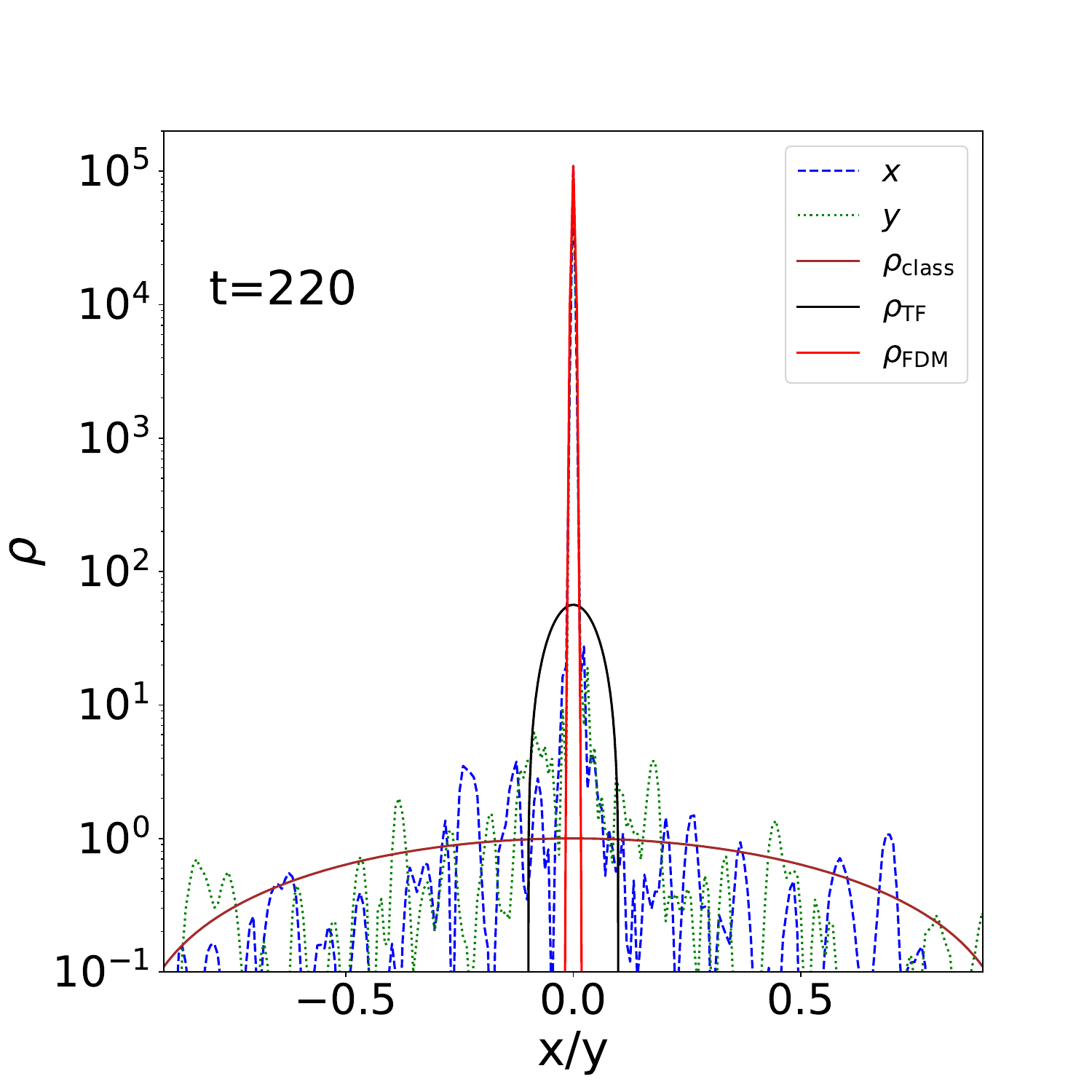}
\includegraphics[height=4.cm,width=0.28\textwidth]{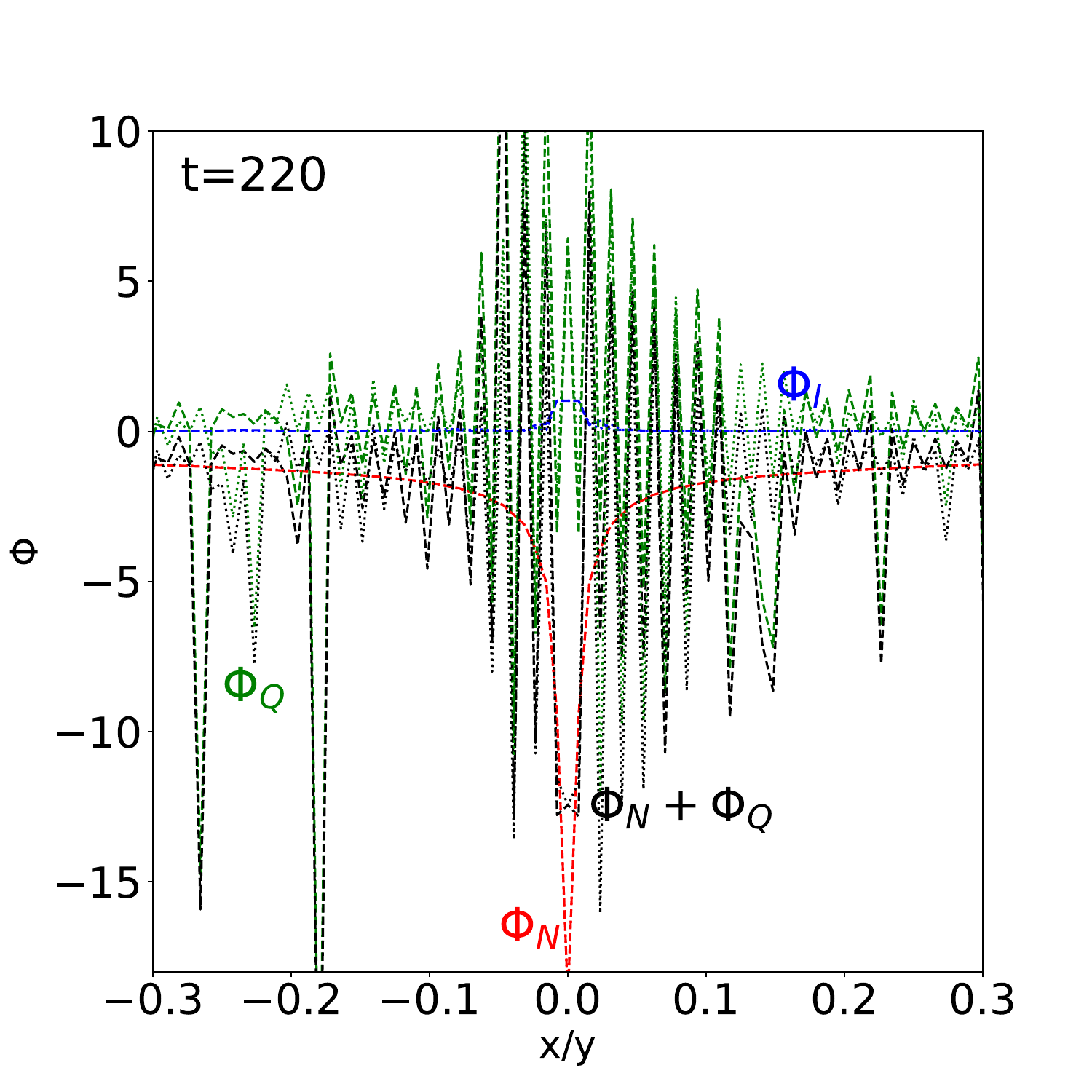}
\includegraphics[height=4.cm,width=0.29\textwidth]{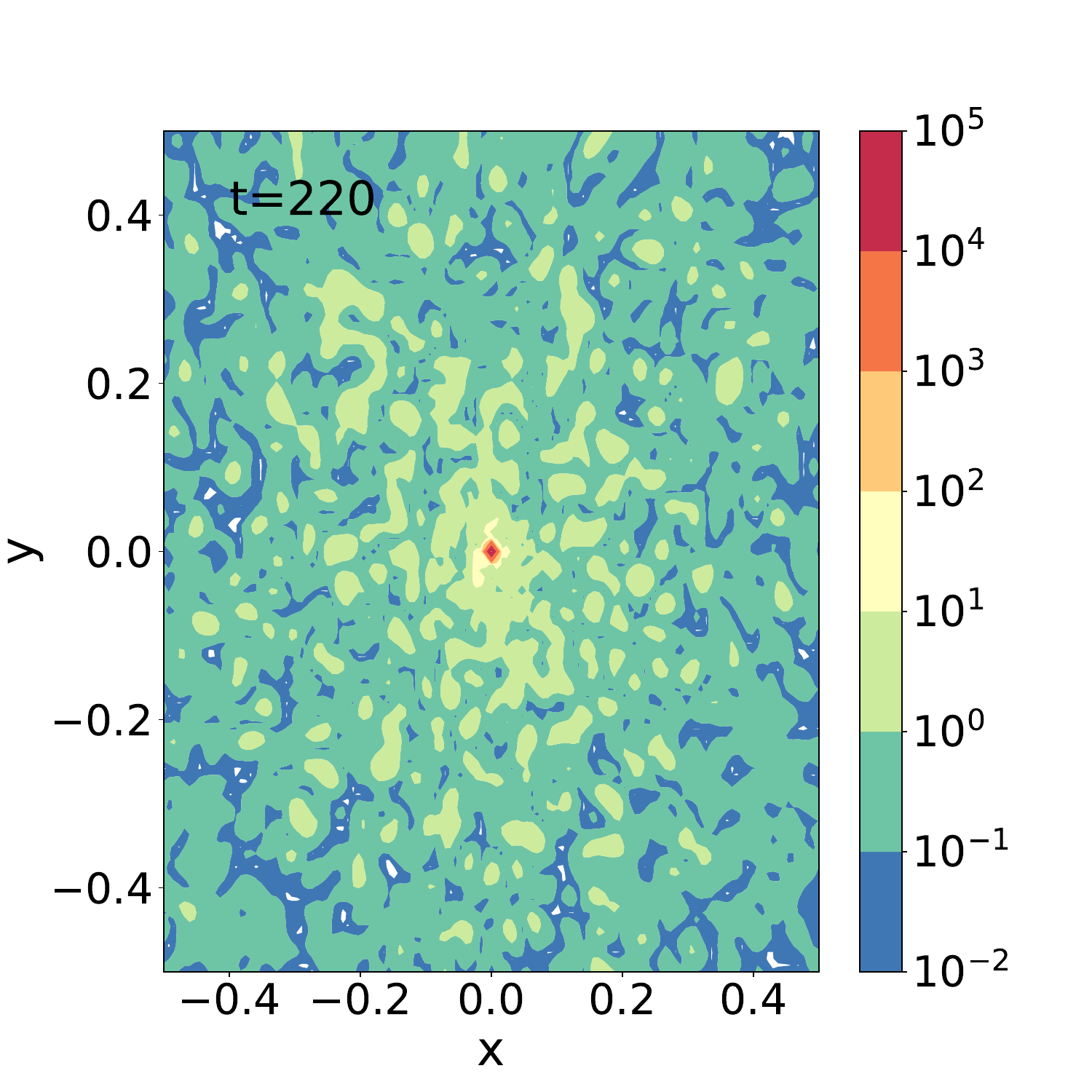}
\caption{Evolution of a halo with $R_{\rm TF}=0.1$ and $\rho_c=80$.}
\label{fig:R0p1-rhoc-80}
\end{figure*}

\subsection{Numerical simulations}

\subsubsection{Large density threshold, $\rho_c=500$}

As in Sec.~\ref{sec:R0p5-rhoc-100}, we first set the threshold $\rho_c=500$ to a very high value
so that the self-interaction potential always takes the form $\Phi_I = \lambda\rho$.
Therefore, the system behaves as in \citep{Garcia:2023abs}.
In contrast with the case $R_{\rm TF}=0.5$ shown in Fig.~\ref{fig:R0p5-rhoc-100}, the system does not
quickly form a TF soliton.
Until $t \lesssim 160$, the system remains dominated by gravity and the quantum pressure, or
wave effects, with large density fluctuations on the de Broglie scale.
However, the maximum amplitude of the competiting density peaks slowly increases until one of them
becomes large enough to overwhelm the other fluctuations and form a stable TF soliton.
Thus, at $t=180$ we find a relaxed TF soliton of radius $R_{\rm TF}=0.1$, which is governed by
the balance between gravity and the self-interactions.
Afterwards, the soliton grows at an increasingly smaller rate until the end of our simulation.
Again, the formation of the soliton is associated with a gravitational cooling process, i.e.
the transfer of energy to the outer halo, which consists of a superposition of many excited
states.

\subsubsection{Small density threshold, $\rho_c=0.5$}

We now consider in Fig.~\ref{fig:R0p1-rhoc-0p5} the case of a small threshold, $\rho_c=0.5$.
As noticed in Sec.~\ref{sec:R0p5-rhoc0p5}, as compared with the case shown in
Fig.~\ref{fig:R0p5-rhoc-0p5} which has the same threshold $\rho_c=0.5$ but a larger TF radius
$R_{\rm TF}=0.5$, the self-interactions no longer play any role, even at early times.
Until $t \lesssim 2500$, the system remains similar to the initial condition of Fig.~\ref{fig:initial},
with density fluctuations of the order of unity within the initial radius $R=1$.
In particular, the self-interactions no longer manage to boost the density fluctuations in the central
region at early times.
Thus, the dynamics are similar to those of an exact FDM system, with vanishing self-interactions.
On a very long time scale the height of the density peaks at the center slowly grows and one of them
manages to reach a density of the order of 50. Then, at $t \sim 3000$ a FDM soliton forms at the center
of the halo. 
This agrees with the gravitational condensation time $t_{\rm gr}$ obtained by a kinetic approach
in \citep{Levkov:2018kau}, which reads in our dimensionless units as
\be
t_{\rm gr} \simeq \frac{\sqrt{2} v^6} {12 \pi^3 \rho^2 \epsilon^3 \ln(v/\epsilon)} \sim 3300 ,
\label{eq:t-gr}
\ee
where we took $\rho \sim 0.5$ and $v\sim 1$.
This shows that even though the self-interactions were always subdominant in the case
$R_{\rm TF}=0.5$ displayed in Fig.~\ref{fig:R0p5-rhoc-0p5}, they nevertheless played a critical
role. Indeed, they managed to boost by a factor of order unity the density contrast in the
central region at early times, which appeared to be sufficient to initiate a slow rise of the central
density contrast until the formation of a FDM soliton at $t \sim 100$.
In Fig.~\ref{fig:R0p1-rhoc-0p5} there is no such initial boost and the system remains 
for a much longer time in a virialized configuration without significant central overdensity,
until $t \sim 3000$.

\subsubsection{Intermediate density threshold, $\rho_c=80$}

We now discuss the intermediate case, $\rho_c=80$, shown in Fig.~\ref{fig:R0p1-rhoc-80}.
This is somewhat similar to the other intermediate case shown in Fig.~\ref{fig:R0p5-rhoc-3},
in the sense that there is a transition from a TF soliton to a FDM soliton.

At early times, the dynamics are again identical to those of the high-density threshold case shown in
Fig.~\ref{fig:R0p1-rhoc-500}, with the very slow formation at $t \sim 170$ of a soliton of radius
$R_{\rm TF}=0.1$ supported by the pressure associated with the self-interactions.
This is because the density remains below $\rho_c$ during this stage.
This TF soliton is well relaxed at $t\sim 190$ and next slowly grows, as in the high-threshold
case shown in Fig.~\ref{fig:R0p1-rhoc-500}.

However, at $t \simeq 200$, the soliton reaches densities where the self-interactions saturate
and can no longer balance its self-gravity. It then collapses to form of a much narrower and higher
density peak. This event is accompanied by some gravitational cooling, with the ejection of some
mass and energy out of the central region of radius $R_{\rm TF}$.
At time $t\sim 220$ this configuration has already relaxed to a FDM soliton supported by the
quantum pressure. In contrast with the case shown in Fig.~\ref{fig:R0p5-rhoc-3}, the FDM soliton
contains as much as $60\%$ of the mass of the previous TF soliton.
Afterwards, its mass slowly grows within the duration of our simulation.

Thus, as in Fig.~\ref{fig:R0p5-rhoc-3}, this simulation illustrates a rapid transition from the TF regime
to the FDM-like regime. Initially, the TF soliton gradually forms from the fluctuations in the halo.
Once fully formed, it acts as a coherent state, accelerating the transition process and the formation
of the FDM peak. This coherent state provides a clear starting point for the FDM soliton development,
enabling it to quickly emerge and dominate the system.
In particular, as seen in Fig.~\ref{fig:R0p1-rhoc-0p5}, without this first TF stage the system would not
form any FDM soliton and would remain a superposition of many excited states.

\subsection{Gaussian ansatz}

We now discuss how the dynamics found in the simulations can be understood from the Gaussian ansatz.

\subsubsection{Large density threshold, $\rho_c=500$}

\begin{figure}[ht]
\centering
\includegraphics[height=4.1cm,width=0.235\textwidth]{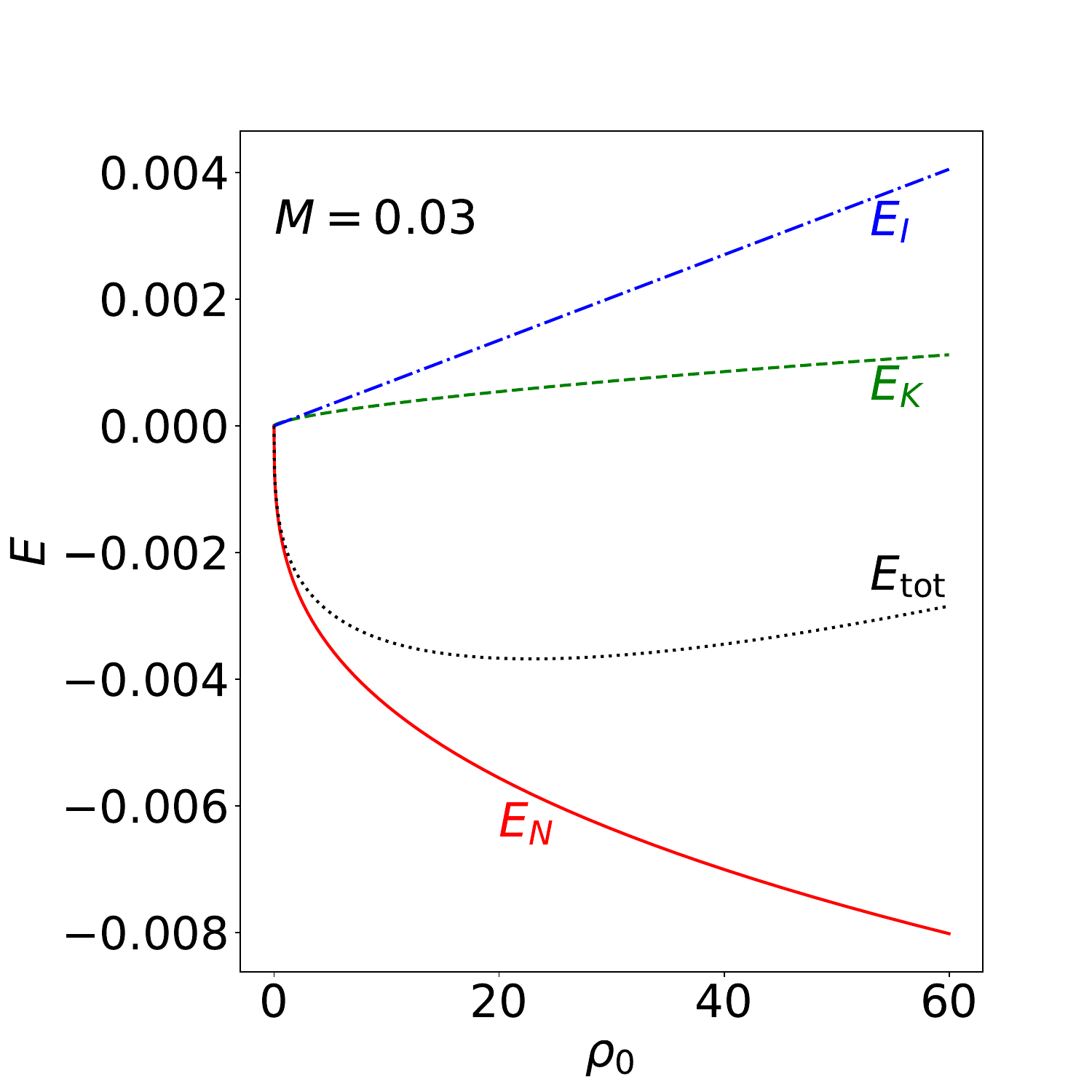}
\includegraphics[height=4.1cm,width=0.235\textwidth]{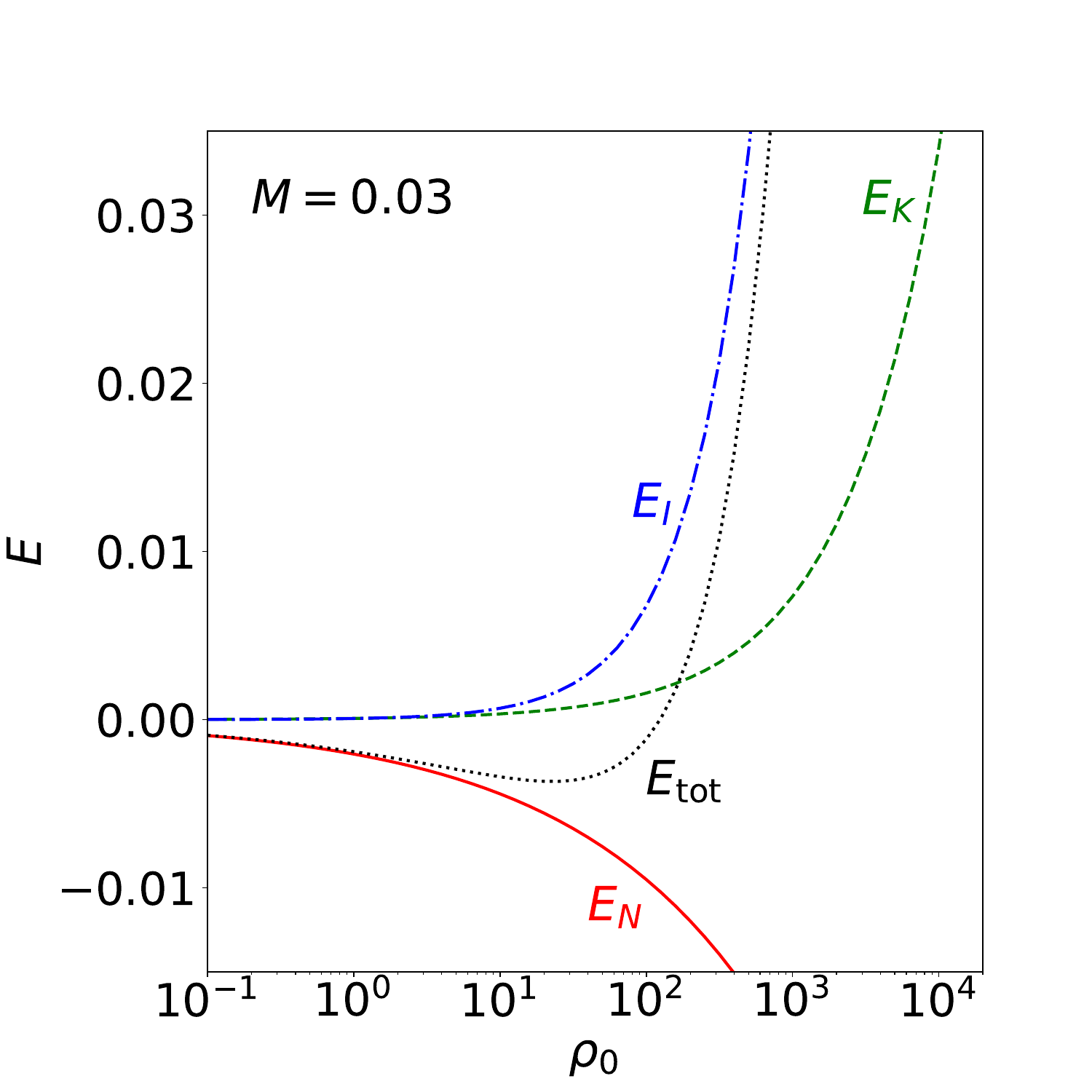}\\
\includegraphics[height=4.1cm,width=0.235\textwidth]{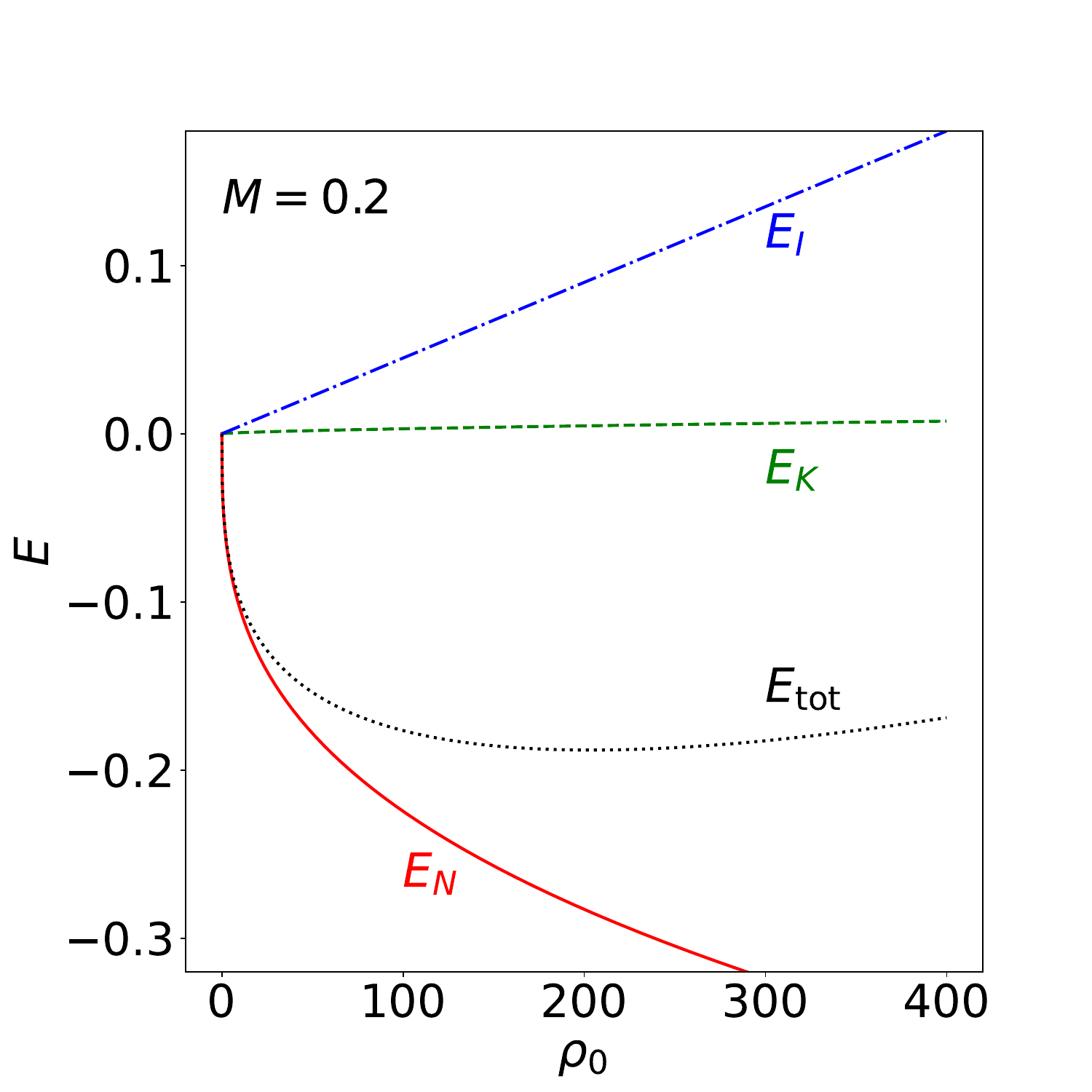}
\includegraphics[height=4.1cm,width=0.235\textwidth]{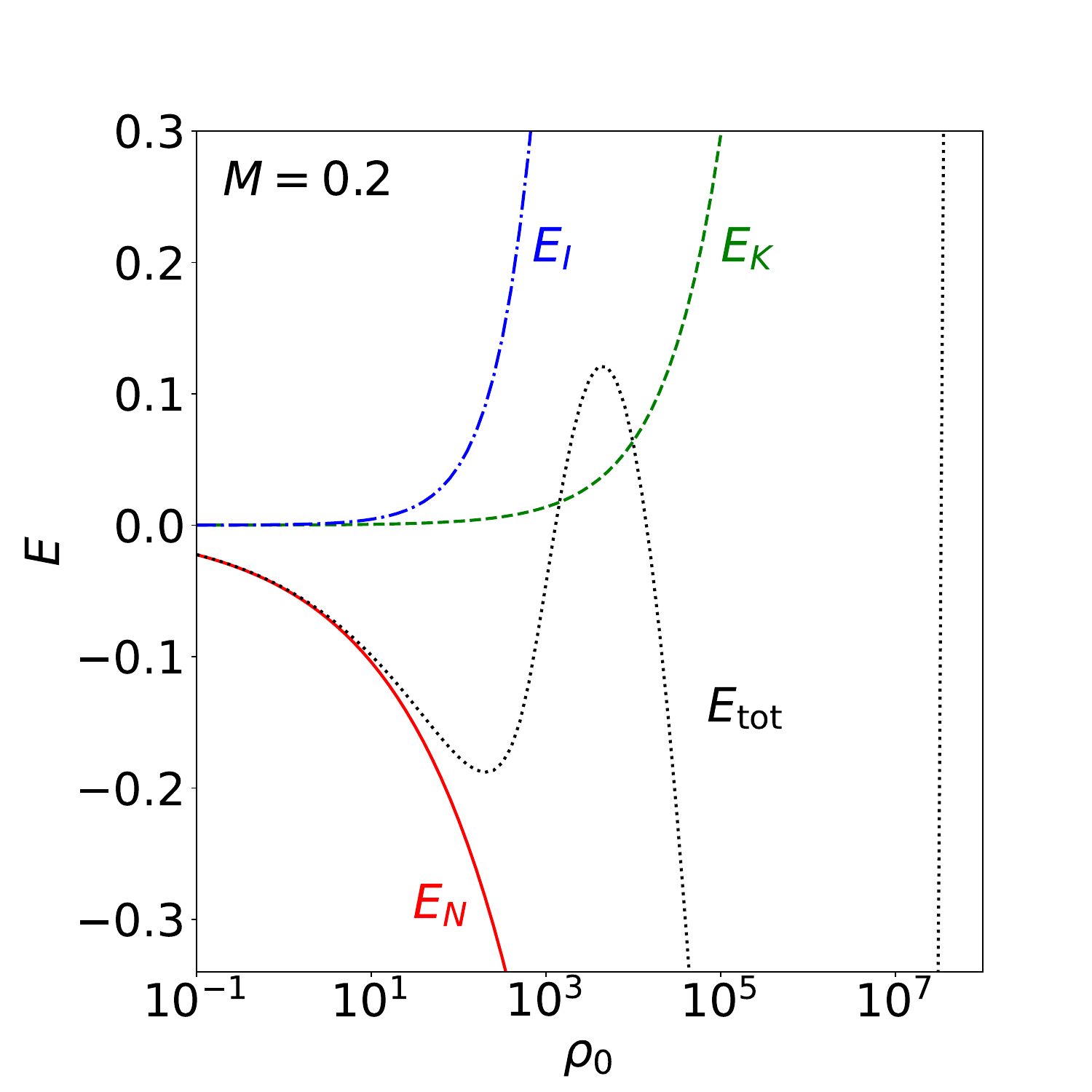}
\caption{
[$R_{\rm TF}=0.1$, $\rho_c=500$.]
Energy of a Gaussian density profile.}
\label{fig:R0p1-rhoc-500-Gauss}
\end{figure}

We show the case $\rho_c=500$ in Fig.~\ref{fig:R0p1-rhoc-500-Gauss}.
We can see that even to form a relatively low mass TF soliton, $M_{\rm TF} = 0.03$,
the central density $\rho_0$ must already be about 20. This is in contrast with the case studied
in Fig.~\ref{fig:R0p5-rhoc-100-Gauss}, where a TF soliton that makes up half of the mass of the
system has a central density as low as $\rho_0 \simeq 5$.
This explains why it takes much more time to form the small TF soliton in the simulation
of Fig.~\ref{fig:R0p1-rhoc-500} than in the simulation of Fig.~\ref{fig:R0p5-rhoc-100}.
In the case of Fig.~\ref{fig:R0p1-rhoc-500}, one must first wait for the slow growth of a density peak
among many random fluctuations to reach $\rho_0 \gtrsim 20$ to form a sufficiently massive and dense
TF soliton that can resist the interference with the other density fluctuations.
As seen in Fig.~\ref{fig:R0p1-rhoc-500}, this has not been achieved yet at $t=160$.
Then, we can see in the lower panels in Fig.~\ref{fig:R0p1-rhoc-500-Gauss} that
up to $M_{\rm TF}=0.2$, which is not reached by the end of our simulation, the TF soliton
is separated from the FDM soliton by a large energy barrier. This explains why it remains stable
and cannot collapse into a FDM soliton.

\subsubsection{Small density threshold, $\rho_c=0.5$}

\begin{figure}[ht]
\centering
\includegraphics[height=4.1cm,width=0.235\textwidth]{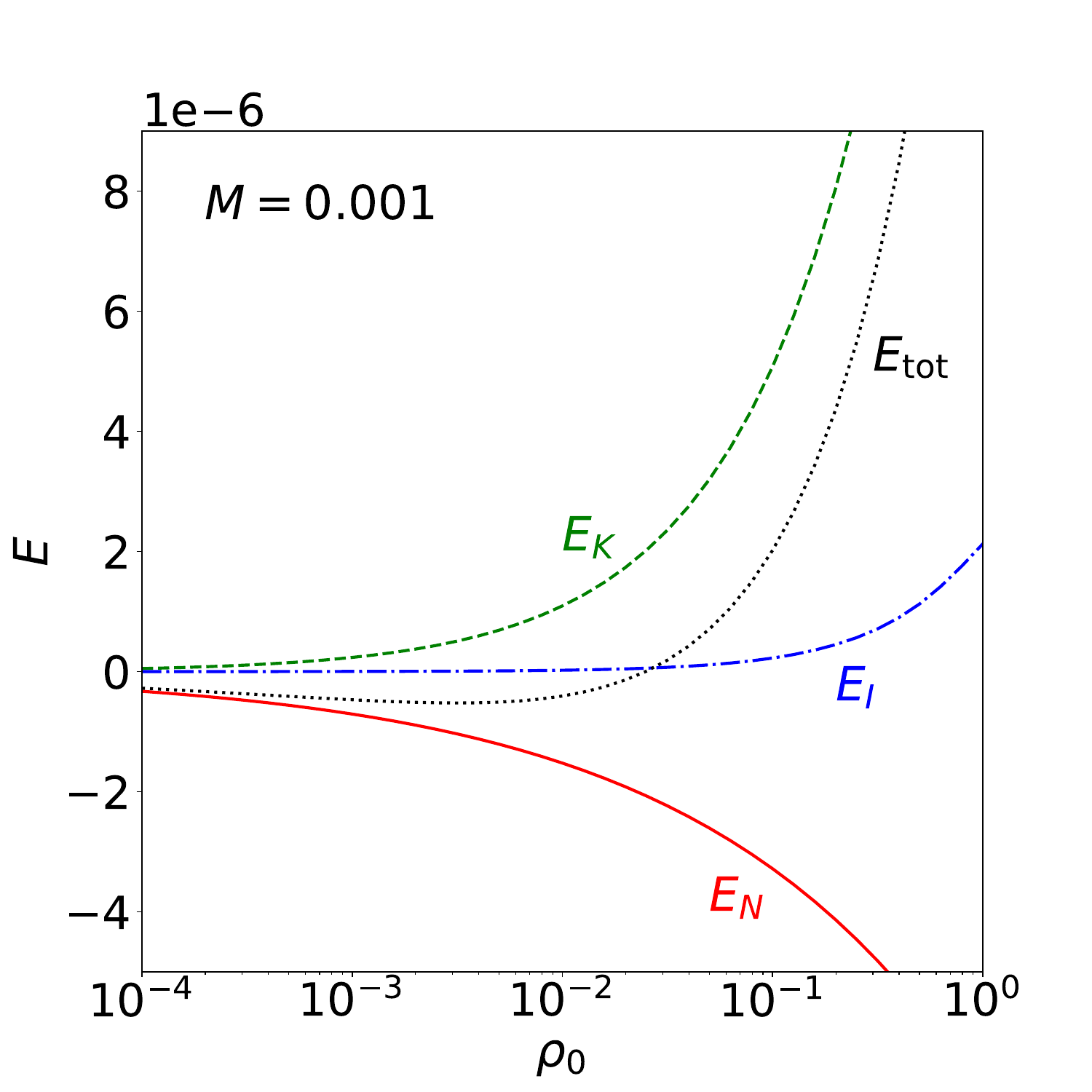}
\includegraphics[height=4.1cm,width=0.235\textwidth]{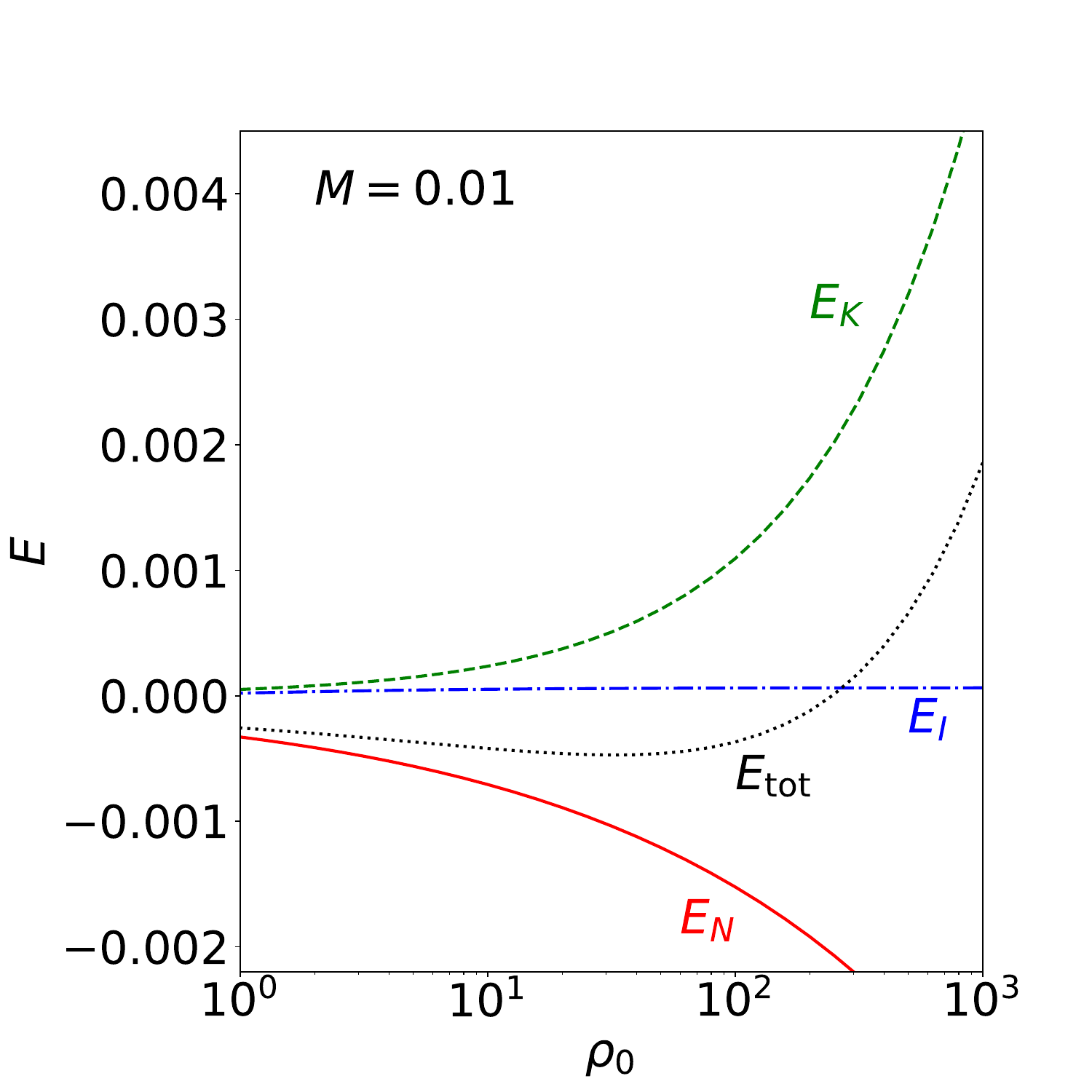}
\includegraphics[height=4.1cm,width=0.235\textwidth]{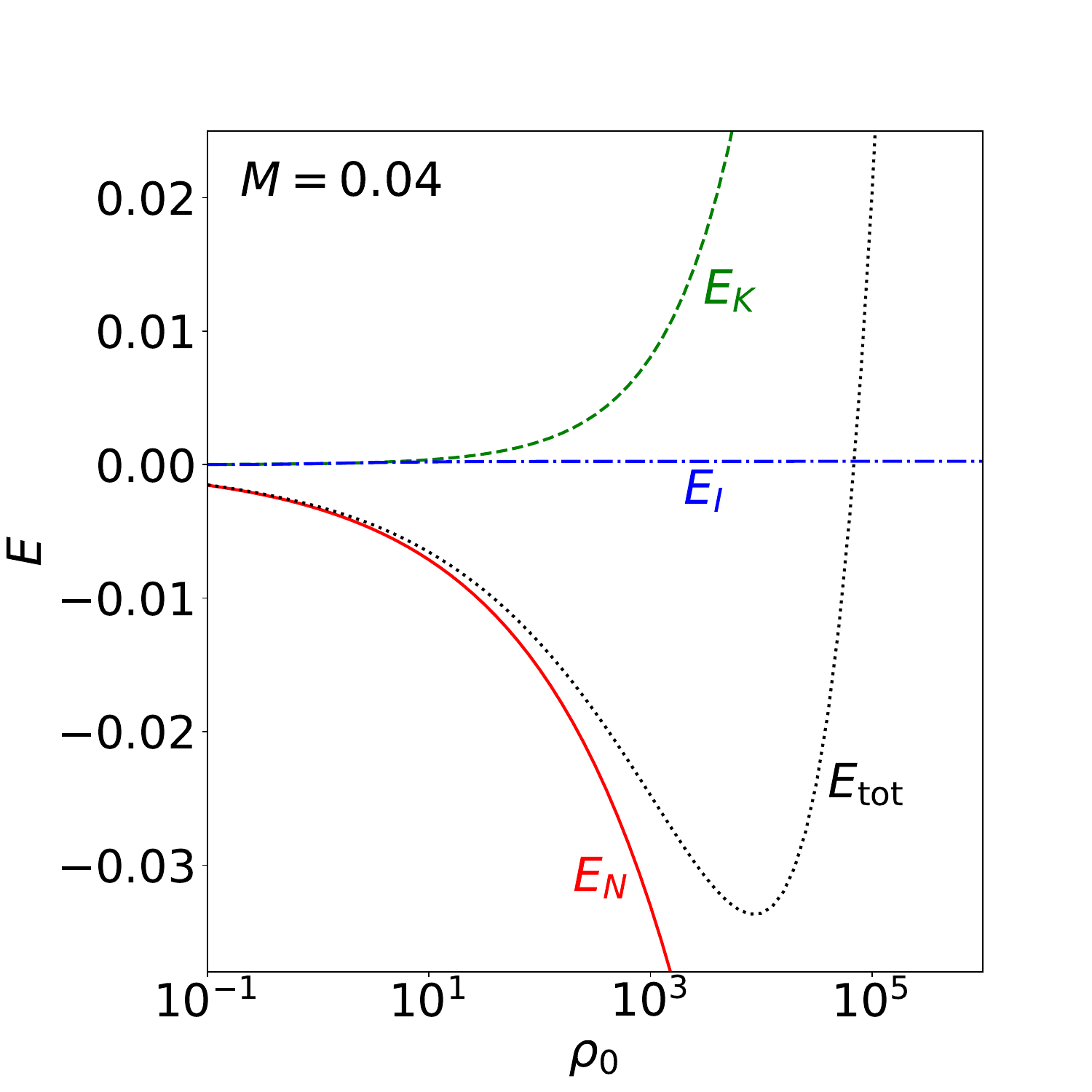}
\caption{
[$R_{\rm TF}=0.1$, $\rho_c=0.5$.]
Energy of a Gaussian density profile.}
\label{fig:R0p1-rhoc-0p5-Gauss}
\end{figure}

We show the case $\rho_c=0.5$ in Fig.~\ref{fig:R0p1-rhoc-0p5-Gauss}.
We can see that for a small mass $M_{\rm FDM}=0.001$, of the order of the central mass in the simulation
in Fig.~\ref{fig:R0p1-rhoc-0p5}, a FDM soliton would have a negligible density $\rho_0 \sim 0.01$
and energy $|E| \sim 10^{-7}$. This means such a state is irrelevant and hidden by the random fluctuations
of the system. For $M_{\rm FDM}=0.01$, the energy of the FDM soliton is still small, $|E| \sim 10^{-3}$,
but the central density is already large, $\rho_0 \gtrsim 30$.
This means that it is difficult for the system to settle into such a FDM soliton, as it must reach
rather large densities, associated with narrow peaks that have a small mass and energy and are thus
easily destroyed by random fluctuations.
This explains why it takes a very long time for the numerical simulation in Fig.~\ref{fig:R0p1-rhoc-0p5} 
to form a FDM soliton. It takes a mass of the order of $M_{\rm FDM} \simeq 0.04$, with a central density
$\rho_0 \sim 10^4$ to build a distinct soliton. This agrees with the FDM soliton formed after $t \sim 3000$
in the numerical simulation.

\subsubsection{Intermediate density threshold, $\rho_c=80$}

\begin{figure}[ht]
\centering
\includegraphics[height=4.1cm,width=0.235\textwidth]{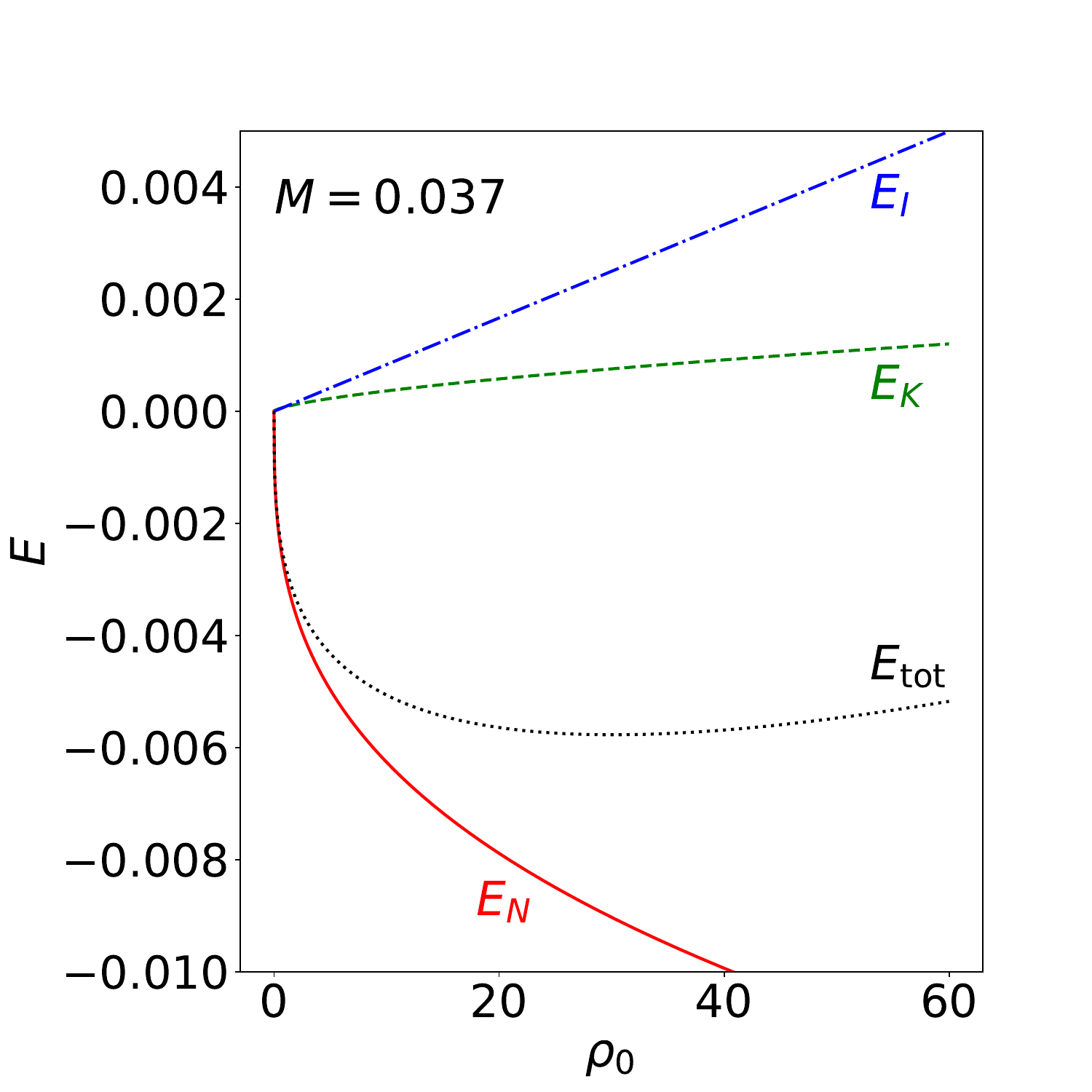}
\includegraphics[height=4.1cm,width=0.235\textwidth]{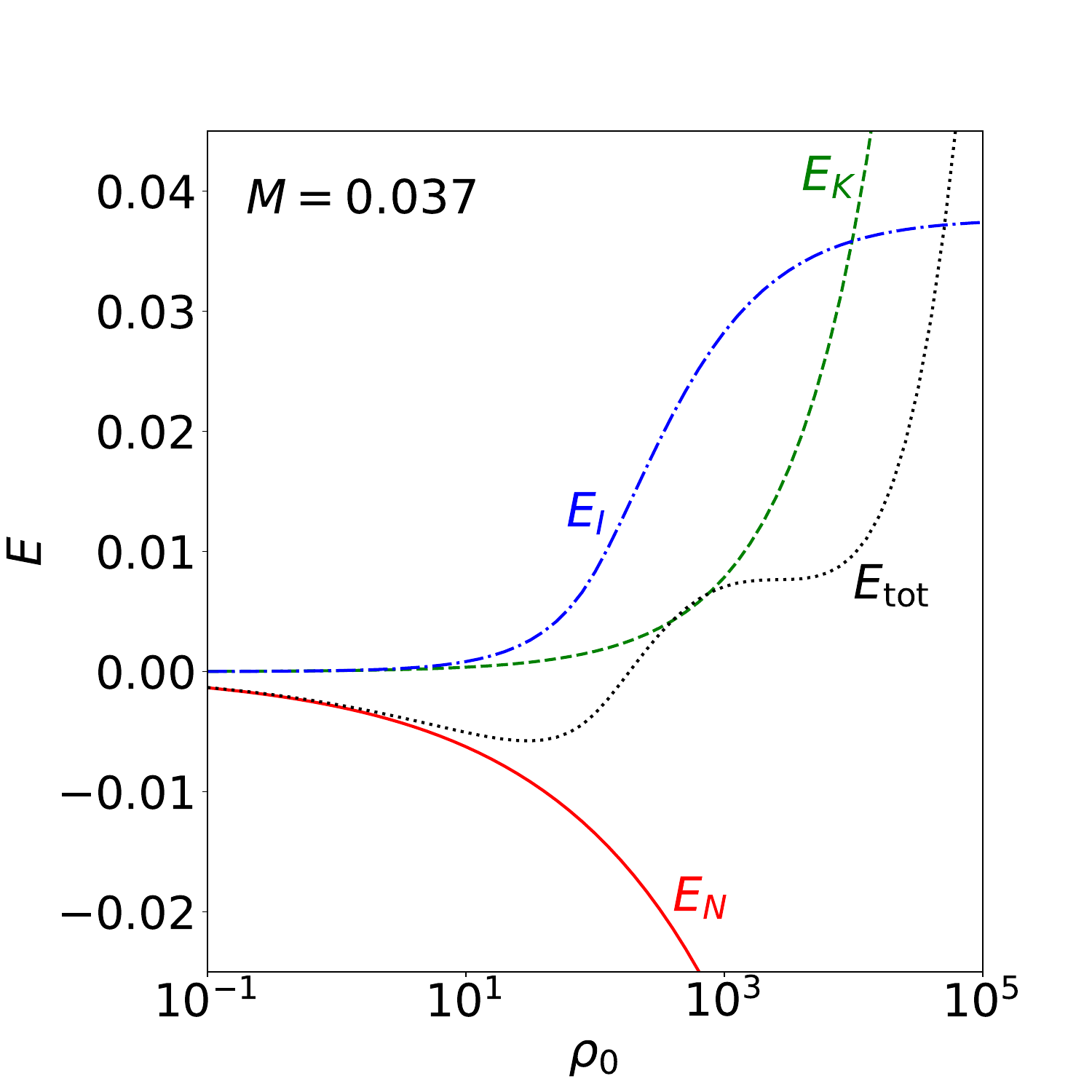}\\
\includegraphics[height=4.1cm,width=0.235\textwidth]{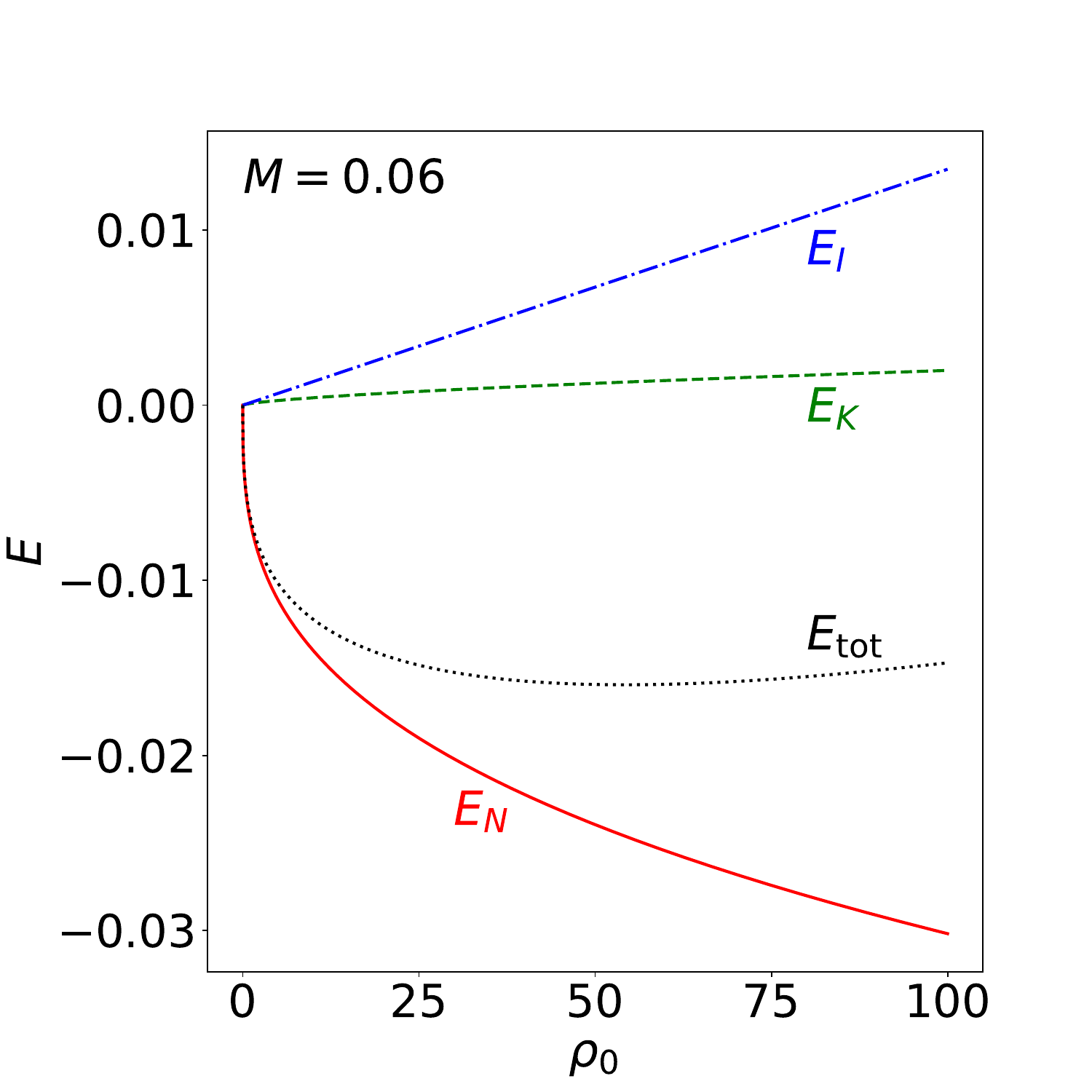}
\includegraphics[height=4.1cm,width=0.235\textwidth]{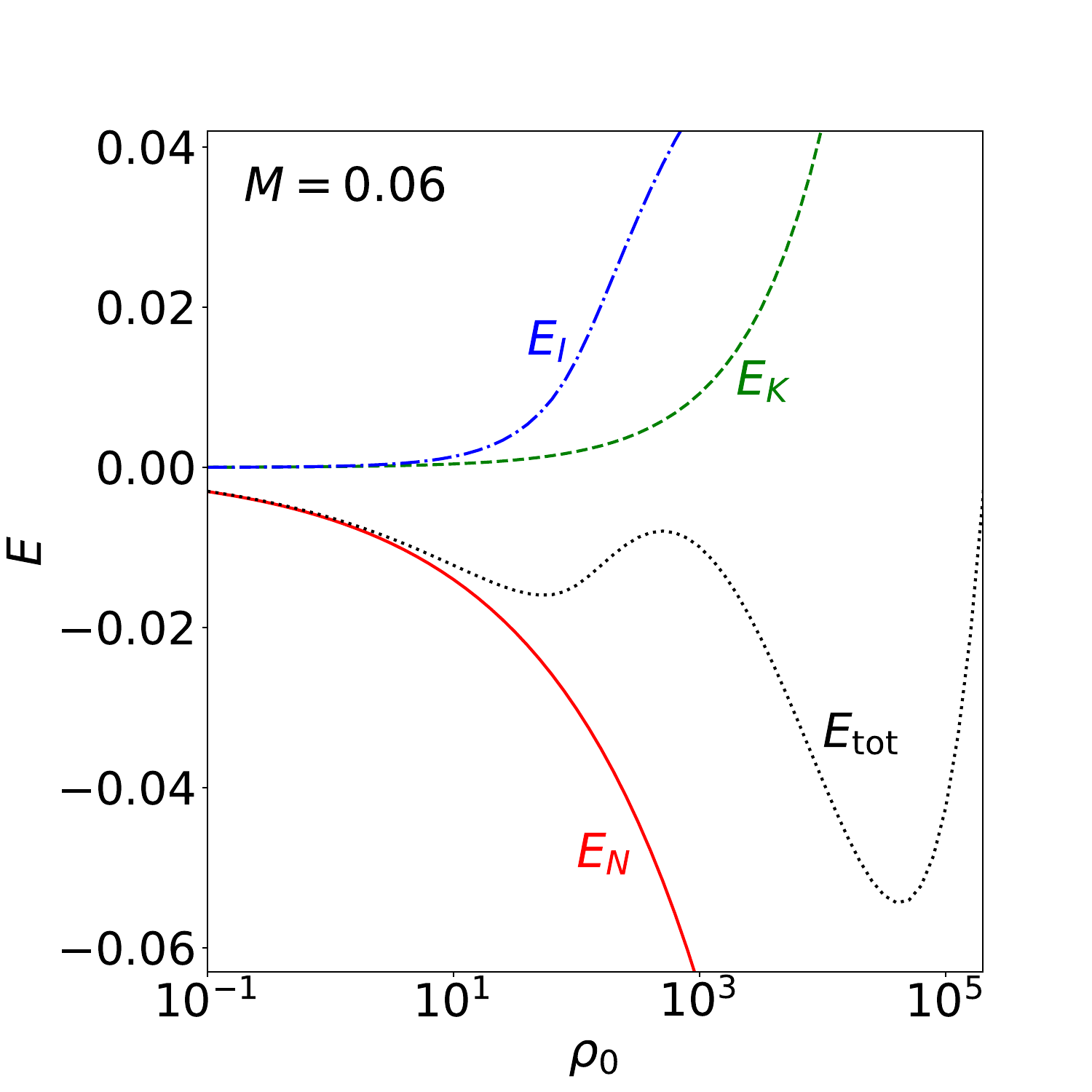}\\
\includegraphics[height=4.1cm,width=0.235\textwidth]{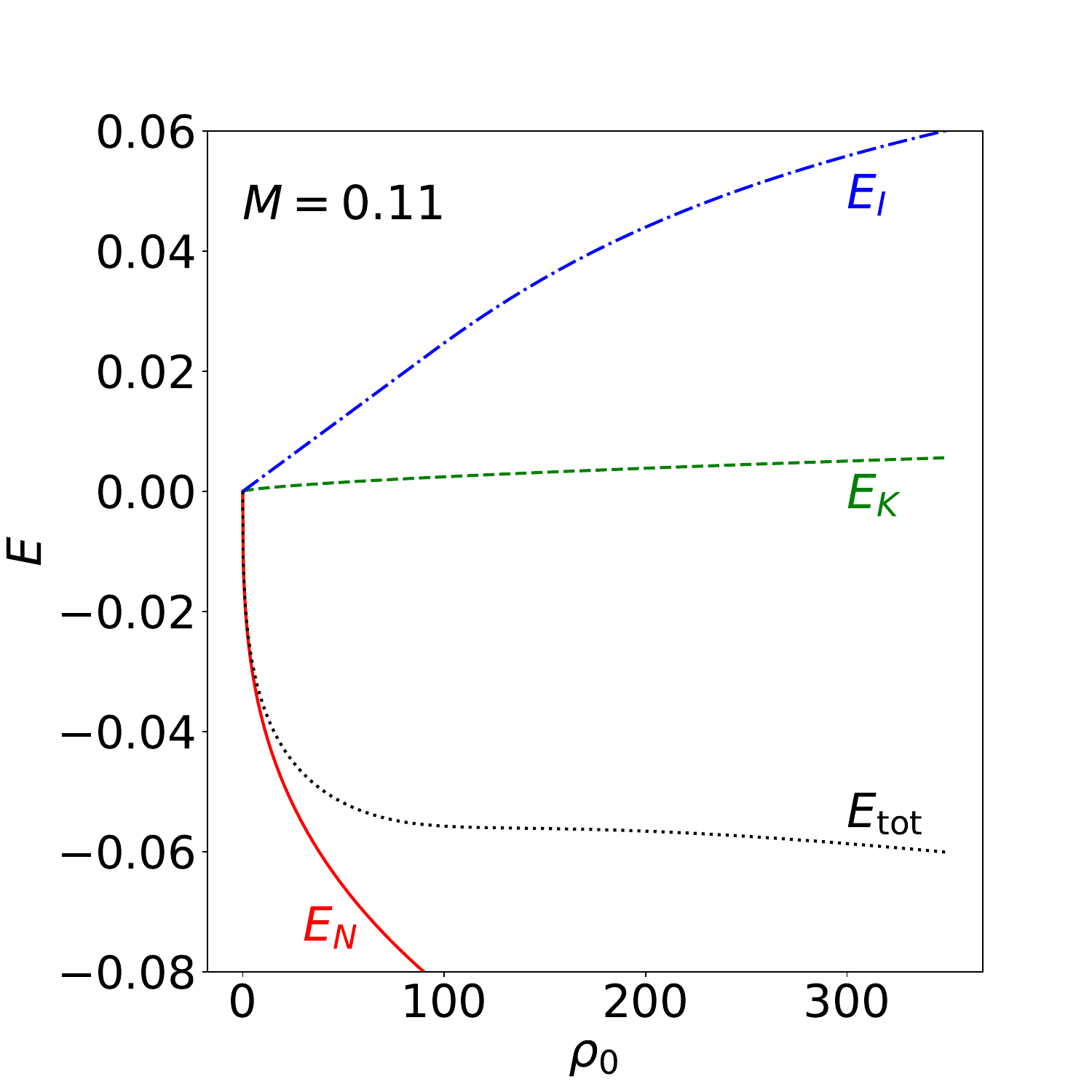}
\includegraphics[height=4.1cm,width=0.235\textwidth]{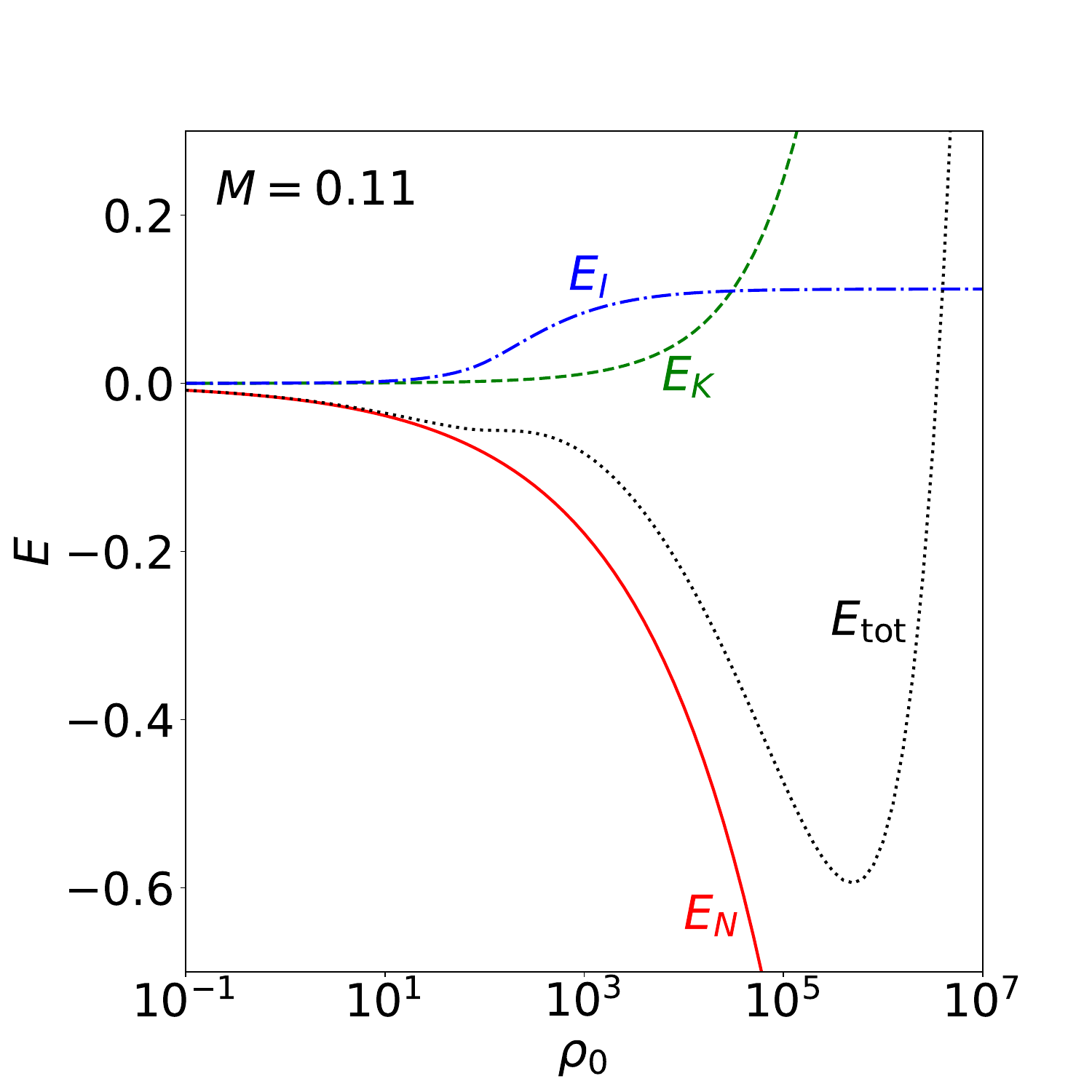}\\
\caption{
[$R_{\rm TF}=0.1$, $\rho_c=80$.]
Energy of a Gaussian density profile.}
\label{fig:R0p1-rhoc-80-Gauss}
\end{figure}

We show in Fig.~\ref{fig:R0p1-rhoc-80-Gauss} the intermediate case $\rho_c=80$.
We can see that until $M \lesssim 0.037$ there is only
one energy minimun, at a moderate density. This is the TF solution (\ref{eq:rho0-TF}),
which disappears at $M \simeq 0.11$ to leave only the new high-density FDM minimum.
This agrees with the simulation displayed in Fig.~\ref{fig:R0p1-rhoc-80}, where we find that the
TF soliton disappears at $t \simeq 200$, when it has reached a mass $M_{\rm TF} \simeq 0.1$,
and a FDM soliton forms with a mass $M_{\rm FDM} \simeq 0.06$.
As seen in the bottom left panel of Fig.~\ref{fig:R0p1-rhoc-80-Gauss}, when it becomes unstable the TF
soliton has an energy $E_{\rm TF} \simeq -0.055$. If we assume as in Sec.~\ref{sec:Gaussian-R0p5-rhoc3}
that the new FDM soliton has the same energy, the remaining matter having a zero energy as for
barely unbound particles, we can see in the middle right panel of Fig.~\ref{fig:R0p1-rhoc-80-Gauss}
that this leads to a mass $M_{\rm FDM} \simeq 0.06$ for the new FDM soliton.
This is close to the simulation result in Fig.~\ref{fig:R0p1-rhoc-80}.
In other words, because around the transition the energy levels of the TF and FDM solitons are not
too distant, a large fraction of the TF soliton is trapped in the new FDM soliton.

This is in contrast with the case studied in Figs.~\ref{fig:R0p5-rhoc-3} and \ref{fig:R0p5-rhoc-3-Gauss},
where the FDM soliton only makes up about $10\%$ of the previous TF soliton, and we argued
in Fig.~\ref{fig:R0p5-rhoc-3-Gauss} that this is due to the large difference between the energy levels
of these two configurations around the transition.
In other words, in the case studied in Figs.~\ref{fig:R0p5-rhoc-3} and \ref{fig:R0p5-rhoc-3-Gauss},
the much higher energy of the initial TF soliton meant that most of its matter was unbound in the
gravitational potential well of the newly formed FDM soliton, which could only trap a small fraction of
the mass.
In contrast, in the case shown in Figs.~\ref{fig:R0p1-rhoc-80} and \ref{fig:R0p1-rhoc-80-Gauss}
the similar energy levels of the two configurations mean that the FDM soliton is able to capture
a large fraction of the collapsing TF soliton.

\section{Conclusion}\label{sec:conclusion-cosine}

In this paper, we investigated the formation and evolution of solitons in a scalar dark matter
scenario with a truncated scalar potential.
This model, described in Sec.~\ref{sec:simplified-potential}, is based on a simplification of the
bounded cosine potential (\ref{eq:V_i-cosine}) of axion monodromy models.

As in \citep{Garcia:2023abs}, our initial conditions correspond to a flat-core halo,
with characteristic size, mass and density of the order of unity in dimensionless variables.
The initial wave function is expanded in the eigenmodes of the Schrödinger equation defined by
the associated gravitational potential.
In the semiclassical limit this corresponds to a virialized halo supported by its velocity dispersion.
However, the interference between the eigenmodes leads to large initial density fluctuations around
the mean profile, as seen in Eq.(\ref{eq:rho-variance}), that are a signature of these dark matter models.
In addition, if they are sufficiently large, the self-interactions can lead to the formation of
new hydrostatic equilibria.
The self-interactions are defined by the TF radius $R_{\rm TF}$ in (\ref{eq:Rsol-lambda}), which is
the radius of the solitons they can support, and the density threshold $\rho_c$ beyond which they vanish.

When $\rho_c$ is large, we recover the results of \citep{Garcia:2023abs}.
If $R_{\rm TF} = 0.5$ the system quickly forms a TF soliton, which contains about half of the system
mass and with a central density of the order of unity.
If $R_{\rm TF} = 0.1$, it takes a much longer time to form a TF soliton, after a long competition
between many density peaks. When the soliton is sufficiently massive to be unambiguously distinguished
from the other smaller density fluctuations, it contains less than $8\%$ of the total mass
but has a central density $\rho_0 \gtrsim 20$.

We explained that these behaviors can be understood from a Gaussian ansatz, which provides
the energy curve $E_{\rm tot}(\rho_0)$ as a function of the soliton central density
and is able to track both TF and FDM regimes. The solitons are then identified with the local
minima of the function $E_{\rm tot}(\rho_0)$.
This shows indeed that for $R_{\rm TF} = 0.5$ the TF soliton that makes up half of the total mass
has a density of the order of unity whereas for $R_{\rm TF} = 0.1$ even a soliton
that only makes up a few percent of the total mass requires a density $\rho_0 \gtrsim 20$.
Obviously, this is simply due to the smaller radius $R_{\rm TF} = 0.1$ as compared with the
system size $R \sim 1$, which implies a larger density for a given central mass.

When $\rho_c$ is small, no TF soliton appears in the numerical simulations.
In the case $R_{\rm TF} = 0.5$, after a long time where many small density fluctuations compete,
a density peak finally rises over the other fluctuations and forms a FDM soliton.
Although the self-interactions are weak, we find that they nevertheless
have a critical impact on the dynamics. Even though they cannot build an hydrostatic soliton,
they manage to increase the mass and the density within the radius $R_{\rm TF}=0.5$ at early times.
This helps the subsequent slow evolution with a gradual rise of competing density peaks
until the formation of a FDM soliton.
In contrast, in the case $R_{\rm TF} = 0.1$ the self-interactions have a spatial range
that is too small to provide any significant early boost.
Then, the dynamics proceed as in a pure FDM model, without any self-interactions, and
it takes a much longer time to eventually form a FDM soliton. 

This is a somewhat counter-intuitive behavior, as the self-interactions are associated with an additional
effective pressure in the hydrodynamical formulation and thus could be expected to decrease
density contrasts and hamper the formation of structures.
This shows that the Gross-Pitaevskii equation can display intricate behaviors and that
the hydrodynamical picture can be misleading in regimes where wave effects are important.

For intermediate values of the threshold $\rho_c$, the dynamics first proceed as for the large
$\rho_c$ case, with a fast ($R_{\rm TF}=0.5$) or slow ($R_{\rm TF}=0.1$) formation of a TF soliton.
Afterwards, as its central density reaches the threshold $\rho_c$ it can no longer be supported
by the self-interactions and it collapses to form a much smaller FDM soliton.
In the case $R_{\rm TF}=0.5$ the new FDM soliton only makes up $10\%$ of the previous TF soliton mass,
whereas in the case $R_{\rm TF}=0.1$ it captures most of the mass.
Again, this behavior can be understood from the Gaussian ansatz.
It shows that at the transition the energy (at fixed mass) of the FDM soliton is much lower than
that of the TF soliton, if $R_{\rm TF}=0.5$, whereas they have similar energies if $R_{\rm TF}=0.1$.
The large energy mismatch in the case $R_{\rm TF}=0.5$ implies that the TF soliton cannot transform
into the FDM soliton at fixed mass. The excess initial energy means that a large fraction of the
TF soliton mass is not bound to the new FDM gravitational potential well, which can only trap a small
fraction of the mass.
In the case $R_{\rm TF}=0.1$, the similar energy levels imply that the TF soliton can easily transform
into the FDM soliton, which is able to trap most of its mass.

In the case $R_{\rm TF}=0.5$, this two-step process accelerates the formation of the final FDM
soliton, as compared with the case where the self-interactions are negligible.
Indeed, in a first stage the system quickly forms a TF soliton. Then, this hydrostatic equilibrium
provides a favorable initial condition for the later collapse to the FDM soliton, which forms
much earlier and reaches a higher mass than in the case with negligible self-interactions.
In the case $R_{\rm TF}=0.1$ this two-step process is even more critical, as when the self-interactions
are fully negligible it takes a much longer time to form a FDM soliton.

Thus, we have shown in this paper how features in the self-interaction potential can lead
to interesting behaviors, with transitions between different types of equilibria.
They can modify the history of the formation of structures and speed up the building
of final solitons.
Moreover, even when they always remain subdominant, self-interactions can play a key role and
enable the formation of structures that would not form otherwise or would form on much longer
timescales.
We have also described how the solitons that can form and their transitions can be understood from
a simple Gaussian ansatz. However, to predict the timescales of secular processes, such as the slow
formation of a soliton or the slow growth of its mass, one needs other tools such as the kinetic theories
presented in \citep{Levkov:2018kau,Chan:2022bkz,Garcia:2023abs,Jain:2023ojg}.

\acknowledgments

R.G.G. was supported by the CEA NUMERICS program, which has received funding from the European
Union's Horizon 2020 research and innovation program under the Marie Sklodowska-Curie grant agreement
No 800945.
This work was granted access to the CCRT High-Performance Computing (HPC) facility under the Grant CCRT2024-valag awarded by the Fundamental Research Division (DRF) of CEA.

\bibliography{ref}

\begin{thebibliography}{47}
\expandafter\ifx\csname natexlab\endcsname\relax\def\natexlab#1{#1}\fi
\expandafter\ifx\csname bibnamefont\endcsname\relax
  \def\bibnamefont#1{#1}\fi
\expandafter\ifx\csname bibfnamefont\endcsname\relax
  \def\bibfnamefont#1{#1}\fi
\expandafter\ifx\csname citenamefont\endcsname\relax
  \def\citenamefont#1{#1}\fi
\expandafter\ifx\csname url\endcsname\relax
  \def\url#1{\texttt{#1}}\fi
\expandafter\ifx\csname urlprefix\endcsname\relax\def\urlprefix{URL }\fi
\providecommand{\bibinfo}[2]{#2}
\providecommand{\eprint}[2][]{\url{#2}}

\bibitem[{\citenamefont{Jungman et~al.}(1996)\citenamefont{Jungman,
  Kamionkowski, and Griest}}]{Jungman:1995df}
\bibinfo{author}{\bibfnamefont{G.}~\bibnamefont{Jungman}},
  \bibinfo{author}{\bibfnamefont{M.}~\bibnamefont{Kamionkowski}},
  \bibnamefont{and} \bibinfo{author}{\bibfnamefont{K.}~\bibnamefont{Griest}},
  \bibinfo{journal}{Phys. Rept.} \textbf{\bibinfo{volume}{267}},
  \bibinfo{pages}{195} (\bibinfo{year}{1996}), \eprint{hep-ph/9506380}.

\bibitem[{\citenamefont{Drees et~al.}(2005)\citenamefont{Drees, Godbole, and
  Roy}}]{Drees:2004jm}
\bibinfo{author}{\bibfnamefont{M.}~\bibnamefont{Drees}},
  \bibinfo{author}{\bibfnamefont{R.}~\bibnamefont{Godbole}}, \bibnamefont{and}
  \bibinfo{author}{\bibfnamefont{P.}~\bibnamefont{Roy}},
  \emph{\bibinfo{title}{Theory and Phenomenology of Sparticles}}
  (\bibinfo{publisher}{WORLD SCIENTIFIC}, \bibinfo{year}{2005}),
  \eprint{https://www.worldscientific.com/doi/pdf/10.1142/4001},
  \urlprefix\url{https://www.worldscientific.com/doi/abs/10.1142/4001}.

\bibitem[{\citenamefont{Steigman and Turner}(1985)}]{Steigman:1984ac}
\bibinfo{author}{\bibfnamefont{G.}~\bibnamefont{Steigman}} \bibnamefont{and}
  \bibinfo{author}{\bibfnamefont{M.~S.} \bibnamefont{Turner}},
  \bibinfo{journal}{Nucl. Phys. B} \textbf{\bibinfo{volume}{253}},
  \bibinfo{pages}{375} (\bibinfo{year}{1985}).

\bibitem[{\citenamefont{Schumann}(2019)}]{Schumann:2019eaa}
\bibinfo{author}{\bibfnamefont{M.}~\bibnamefont{Schumann}},
  \bibinfo{journal}{J. Phys. G} \textbf{\bibinfo{volume}{46}},
  \bibinfo{pages}{103003} (\bibinfo{year}{2019}), \eprint{1903.03026}.

\bibitem[{\citenamefont{Conrad}(2014)}]{Conrad:2014tla}
\bibinfo{author}{\bibfnamefont{J.}~\bibnamefont{Conrad}}, in
  \emph{\bibinfo{booktitle}{{Interplay between Particle and Astroparticle
  physics}}} (\bibinfo{year}{2014}), \eprint{1411.1925}.

\bibitem[{\citenamefont{Arcadi et~al.}(2018)\citenamefont{Arcadi, Dutra, Ghosh,
  Lindner, Mambrini, Pierre, Profumo, and Queiroz}}]{Arcadi:2017kky}
\bibinfo{author}{\bibfnamefont{G.}~\bibnamefont{Arcadi}},
  \bibinfo{author}{\bibfnamefont{M.}~\bibnamefont{Dutra}},
  \bibinfo{author}{\bibfnamefont{P.}~\bibnamefont{Ghosh}},
  \bibinfo{author}{\bibfnamefont{M.}~\bibnamefont{Lindner}},
  \bibinfo{author}{\bibfnamefont{Y.}~\bibnamefont{Mambrini}},
  \bibinfo{author}{\bibfnamefont{M.}~\bibnamefont{Pierre}},
  \bibinfo{author}{\bibfnamefont{S.}~\bibnamefont{Profumo}}, \bibnamefont{and}
  \bibinfo{author}{\bibfnamefont{F.~S.} \bibnamefont{Queiroz}},
  \bibinfo{journal}{Eur. Phys. J. C} \textbf{\bibinfo{volume}{78}},
  \bibinfo{pages}{203} (\bibinfo{year}{2018}), \eprint{1703.07364}.

\bibitem[{\citenamefont{Weinberg et~al.}(2015)\citenamefont{Weinberg, Bullock,
  Governato, Kuzio~de Naray, and Peter}}]{Weinberg:2013aya}
\bibinfo{author}{\bibfnamefont{D.~H.} \bibnamefont{Weinberg}},
  \bibinfo{author}{\bibfnamefont{J.~S.} \bibnamefont{Bullock}},
  \bibinfo{author}{\bibfnamefont{F.}~\bibnamefont{Governato}},
  \bibinfo{author}{\bibfnamefont{R.}~\bibnamefont{Kuzio~de Naray}},
  \bibnamefont{and} \bibinfo{author}{\bibfnamefont{A.~H.~G.}
  \bibnamefont{Peter}}, \bibinfo{journal}{Proc. Nat. Acad. Sci.}
  \textbf{\bibinfo{volume}{112}}, \bibinfo{pages}{12249}
  (\bibinfo{year}{2015}), \eprint{1306.0913}.

\bibitem[{\citenamefont{Del~Popolo and Le~Delliou}(2017)}]{DelPopolo:2016emo}
\bibinfo{author}{\bibfnamefont{A.}~\bibnamefont{Del~Popolo}} \bibnamefont{and}
  \bibinfo{author}{\bibfnamefont{M.}~\bibnamefont{Le~Delliou}},
  \bibinfo{journal}{Galaxies} \textbf{\bibinfo{volume}{5}}, \bibinfo{pages}{17}
  (\bibinfo{year}{2017}), \eprint{1606.07790}.

\bibitem[{\citenamefont{Nakama et~al.}(2017)\citenamefont{Nakama, Chluba, and
  Kamionkowski}}]{Nakama:2017ohe}
\bibinfo{author}{\bibfnamefont{T.}~\bibnamefont{Nakama}},
  \bibinfo{author}{\bibfnamefont{J.}~\bibnamefont{Chluba}}, \bibnamefont{and}
  \bibinfo{author}{\bibfnamefont{M.}~\bibnamefont{Kamionkowski}},
  \bibinfo{journal}{Phys. Rev. D} \textbf{\bibinfo{volume}{95}},
  \bibinfo{pages}{121302} (\bibinfo{year}{2017}), \eprint{1703.10559}.

\bibitem[{\citenamefont{Di~Luzio et~al.}(2020)\citenamefont{Di~Luzio,
  Giannotti, Nardi, and Visinelli}}]{DiLuzio:2020wdo}
\bibinfo{author}{\bibfnamefont{L.}~\bibnamefont{Di~Luzio}},
  \bibinfo{author}{\bibfnamefont{M.}~\bibnamefont{Giannotti}},
  \bibinfo{author}{\bibfnamefont{E.}~\bibnamefont{Nardi}}, \bibnamefont{and}
  \bibinfo{author}{\bibfnamefont{L.}~\bibnamefont{Visinelli}},
  \bibinfo{journal}{Phys. Rept.} \textbf{\bibinfo{volume}{870}},
  \bibinfo{pages}{1} (\bibinfo{year}{2020}), \eprint{2003.01100}.

\bibitem[{\citenamefont{Hu et~al.}(2000)\citenamefont{Hu, Barkana, and
  Gruzinov}}]{Hu:2000ke}
\bibinfo{author}{\bibfnamefont{W.}~\bibnamefont{Hu}},
  \bibinfo{author}{\bibfnamefont{R.}~\bibnamefont{Barkana}}, \bibnamefont{and}
  \bibinfo{author}{\bibfnamefont{A.}~\bibnamefont{Gruzinov}},
  \bibinfo{journal}{Phys. Rev. Lett.} \textbf{\bibinfo{volume}{85}},
  \bibinfo{pages}{1158} (\bibinfo{year}{2000}), \eprint{astro-ph/0003365}.

\bibitem[{\citenamefont{Hui et~al.}(2017)\citenamefont{Hui, Ostriker, Tremaine,
  and Witten}}]{Hui:2016ltb}
\bibinfo{author}{\bibfnamefont{L.}~\bibnamefont{Hui}},
  \bibinfo{author}{\bibfnamefont{J.~P.} \bibnamefont{Ostriker}},
  \bibinfo{author}{\bibfnamefont{S.}~\bibnamefont{Tremaine}}, \bibnamefont{and}
  \bibinfo{author}{\bibfnamefont{E.}~\bibnamefont{Witten}},
  \bibinfo{journal}{Phys. Rev. D} \textbf{\bibinfo{volume}{95}},
  \bibinfo{pages}{043541} (\bibinfo{year}{2017}), \eprint{1610.08297}.

\bibitem[{\citenamefont{Goodman}(2000)}]{Goodman:2000tg}
\bibinfo{author}{\bibfnamefont{J.}~\bibnamefont{Goodman}},
  \bibinfo{journal}{New Astron.} \textbf{\bibinfo{volume}{5}},
  \bibinfo{pages}{103} (\bibinfo{year}{2000}), \eprint{astro-ph/0003018}.

\bibitem[{\citenamefont{Matos et~al.}(2000)\citenamefont{Matos, Guzman, and
  Urena-Lopez}}]{Matos:1999et}
\bibinfo{author}{\bibfnamefont{T.}~\bibnamefont{Matos}},
  \bibinfo{author}{\bibfnamefont{F.~S.} \bibnamefont{Guzman}},
  \bibnamefont{and} \bibinfo{author}{\bibfnamefont{L.~A.}
  \bibnamefont{Urena-Lopez}}, \bibinfo{journal}{Class. Quant. Grav.}
  \textbf{\bibinfo{volume}{17}}, \bibinfo{pages}{1707} (\bibinfo{year}{2000}),
  \eprint{astro-ph/9908152}.

\bibitem[{\citenamefont{Schive et~al.}(2014)\citenamefont{Schive, Chiueh, and
  Broadhurst}}]{Schive:2014dra}
\bibinfo{author}{\bibfnamefont{H.-Y.} \bibnamefont{Schive}},
  \bibinfo{author}{\bibfnamefont{T.}~\bibnamefont{Chiueh}}, \bibnamefont{and}
  \bibinfo{author}{\bibfnamefont{T.}~\bibnamefont{Broadhurst}},
  \bibinfo{journal}{Nature Phys.} \textbf{\bibinfo{volume}{10}},
  \bibinfo{pages}{496} (\bibinfo{year}{2014}), \eprint{1406.6586}.

\bibitem[{\citenamefont{Ferreira}(2021)}]{Ferreira:2020fam}
\bibinfo{author}{\bibfnamefont{E.~G.~M.} \bibnamefont{Ferreira}},
  \bibinfo{journal}{Astron. Astrophys. Rev.} \textbf{\bibinfo{volume}{29}},
  \bibinfo{pages}{7} (\bibinfo{year}{2021}), \eprint{2005.03254}.

\bibitem[{\citenamefont{Li et~al.}(2014)\citenamefont{Li, Rindler-Daller, and
  Shapiro}}]{Li:2013nal}
\bibinfo{author}{\bibfnamefont{B.}~\bibnamefont{Li}},
  \bibinfo{author}{\bibfnamefont{T.}~\bibnamefont{Rindler-Daller}},
  \bibnamefont{and} \bibinfo{author}{\bibfnamefont{P.~R.}
  \bibnamefont{Shapiro}}, \bibinfo{journal}{Phys. Rev. D}
  \textbf{\bibinfo{volume}{89}}, \bibinfo{pages}{083536}
  (\bibinfo{year}{2014}), \eprint{1310.6061}.

\bibitem[{\citenamefont{Lee and Pang}(1992)}]{Lee:1991ax}
\bibinfo{author}{\bibfnamefont{T.~D.} \bibnamefont{Lee}} \bibnamefont{and}
  \bibinfo{author}{\bibfnamefont{Y.}~\bibnamefont{Pang}},
  \bibinfo{journal}{Phys. Rept.} \textbf{\bibinfo{volume}{221}},
  \bibinfo{pages}{251} (\bibinfo{year}{1992}).

\bibitem[{\citenamefont{Guth et~al.}(2015)\citenamefont{Guth, Hertzberg, and
  Prescod-Weinstein}}]{Guth:2014hsa}
\bibinfo{author}{\bibfnamefont{A.~H.} \bibnamefont{Guth}},
  \bibinfo{author}{\bibfnamefont{M.~P.} \bibnamefont{Hertzberg}},
  \bibnamefont{and}
  \bibinfo{author}{\bibfnamefont{C.}~\bibnamefont{Prescod-Weinstein}},
  \bibinfo{journal}{Phys. Rev. D} \textbf{\bibinfo{volume}{92}},
  \bibinfo{pages}{103513} (\bibinfo{year}{2015}), \eprint{1412.5930}.

\bibitem[{\citenamefont{Sikivie and Yang}(2009)}]{Sikivie:2009qn}
\bibinfo{author}{\bibfnamefont{P.}~\bibnamefont{Sikivie}} \bibnamefont{and}
  \bibinfo{author}{\bibfnamefont{Q.}~\bibnamefont{Yang}},
  \bibinfo{journal}{Phys. Rev. Lett.} \textbf{\bibinfo{volume}{103}},
  \bibinfo{pages}{111301} (\bibinfo{year}{2009}), \eprint{0901.1106}.

\bibitem[{\citenamefont{Marsh}(2016)}]{Marsh:2015xka}
\bibinfo{author}{\bibfnamefont{D.~J.~E.} \bibnamefont{Marsh}},
  \bibinfo{journal}{Phys. Rept.} \textbf{\bibinfo{volume}{643}},
  \bibinfo{pages}{1} (\bibinfo{year}{2016}), \eprint{1510.07633}.

\bibitem[{\citenamefont{Veltmaat et~al.}(2018)\citenamefont{Veltmaat, Niemeyer,
  and Schwabe}}]{Veltmaat:2018dfz}
\bibinfo{author}{\bibfnamefont{J.}~\bibnamefont{Veltmaat}},
  \bibinfo{author}{\bibfnamefont{J.~C.} \bibnamefont{Niemeyer}},
  \bibnamefont{and} \bibinfo{author}{\bibfnamefont{B.}~\bibnamefont{Schwabe}},
  \bibinfo{journal}{Phys. Rev. D} \textbf{\bibinfo{volume}{98}},
  \bibinfo{pages}{043509} (\bibinfo{year}{2018}), \eprint{1804.09647}.

\bibitem[{\citenamefont{Chavanis}(2011)}]{Chavanis:2011zi}
\bibinfo{author}{\bibfnamefont{P.-H.} \bibnamefont{Chavanis}},
  \bibinfo{journal}{Phys. Rev. D} \textbf{\bibinfo{volume}{84}},
  \bibinfo{pages}{043531} (\bibinfo{year}{2011}), \eprint{1103.2050}.

\bibitem[{\citenamefont{Chavanis}(2018)}]{Chavanis:2017loo}
\bibinfo{author}{\bibfnamefont{P.-H.} \bibnamefont{Chavanis}},
  \bibinfo{journal}{Phys. Rev. D} \textbf{\bibinfo{volume}{98}},
  \bibinfo{pages}{023009} (\bibinfo{year}{2018}), \eprint{1710.06268}.

\bibitem[{\citenamefont{Dawoodbhoy et~al.}(2021)\citenamefont{Dawoodbhoy,
  Shapiro, and Rindler-Daller}}]{Dawoodbhoy:2021beb}
\bibinfo{author}{\bibfnamefont{T.}~\bibnamefont{Dawoodbhoy}},
  \bibinfo{author}{\bibfnamefont{P.~R.} \bibnamefont{Shapiro}},
  \bibnamefont{and}
  \bibinfo{author}{\bibfnamefont{T.}~\bibnamefont{Rindler-Daller}},
  \bibinfo{journal}{Mon. Not. Roy. Astron. Soc.}
  \textbf{\bibinfo{volume}{506}}, \bibinfo{pages}{2418} (\bibinfo{year}{2021}),
  \eprint{2104.07043}.

\bibitem[{\citenamefont{Shapiro et~al.}(2021)\citenamefont{Shapiro, Dawoodbhoy,
  and Rindler-Daller}}]{Shapiro:2021hjp}
\bibinfo{author}{\bibfnamefont{P.~R.} \bibnamefont{Shapiro}},
  \bibinfo{author}{\bibfnamefont{T.}~\bibnamefont{Dawoodbhoy}},
  \bibnamefont{and}
  \bibinfo{author}{\bibfnamefont{T.}~\bibnamefont{Rindler-Daller}},
  \bibinfo{journal}{Mon. Not. Roy. Astron. Soc.}
  \textbf{\bibinfo{volume}{509}}, \bibinfo{pages}{145} (\bibinfo{year}{2021}),
  \eprint{2106.13244}.

\bibitem[{\citenamefont{Peccei and Quinn}(1977)}]{Peccei:1977hh}
\bibinfo{author}{\bibfnamefont{R.~D.} \bibnamefont{Peccei}} \bibnamefont{and}
  \bibinfo{author}{\bibfnamefont{H.~R.} \bibnamefont{Quinn}},
  \bibinfo{journal}{Phys. Rev. Lett.} \textbf{\bibinfo{volume}{38}},
  \bibinfo{pages}{1440} (\bibinfo{year}{1977}).

\bibitem[{\citenamefont{Weinberg}(1978)}]{Weinberg:1977ma}
\bibinfo{author}{\bibfnamefont{S.}~\bibnamefont{Weinberg}},
  \bibinfo{journal}{Phys. Rev. Lett.} \textbf{\bibinfo{volume}{40}},
  \bibinfo{pages}{223} (\bibinfo{year}{1978}).

\bibitem[{\citenamefont{Wilczek}(1978)}]{Wilczek:1977pj}
\bibinfo{author}{\bibfnamefont{F.}~\bibnamefont{Wilczek}},
  \bibinfo{journal}{Phys. Rev. Lett.} \textbf{\bibinfo{volume}{40}},
  \bibinfo{pages}{279} (\bibinfo{year}{1978}).

\bibitem[{\citenamefont{Kim}(1987)}]{Kim:1986ax}
\bibinfo{author}{\bibfnamefont{J.~E.} \bibnamefont{Kim}},
  \bibinfo{journal}{Phys. Rept.} \textbf{\bibinfo{volume}{150}},
  \bibinfo{pages}{1} (\bibinfo{year}{1987}).

\bibitem[{\citenamefont{Silverstein and Westphal}(2008)}]{Silverstein:2008sg}
\bibinfo{author}{\bibfnamefont{E.}~\bibnamefont{Silverstein}} \bibnamefont{and}
  \bibinfo{author}{\bibfnamefont{A.}~\bibnamefont{Westphal}},
  \bibinfo{journal}{Phys. Rev. D} \textbf{\bibinfo{volume}{78}},
  \bibinfo{pages}{106003} (\bibinfo{year}{2008}), \eprint{0803.3085}.

\bibitem[{\citenamefont{McAllister et~al.}(2010)\citenamefont{McAllister,
  Silverstein, and Westphal}}]{McAllister:2008hb}
\bibinfo{author}{\bibfnamefont{L.}~\bibnamefont{McAllister}},
  \bibinfo{author}{\bibfnamefont{E.}~\bibnamefont{Silverstein}},
  \bibnamefont{and} \bibinfo{author}{\bibfnamefont{A.}~\bibnamefont{Westphal}},
  \bibinfo{journal}{Phys. Rev. D} \textbf{\bibinfo{volume}{82}},
  \bibinfo{pages}{046003} (\bibinfo{year}{2010}), \eprint{0808.0706}.

\bibitem[{\citenamefont{Brax et~al.}(2019)\citenamefont{Brax, Cembranos, and
  Valageas}}]{Brax:2019fzb}
\bibinfo{author}{\bibfnamefont{P.}~\bibnamefont{Brax}},
  \bibinfo{author}{\bibfnamefont{J.~A.~R.} \bibnamefont{Cembranos}},
  \bibnamefont{and} \bibinfo{author}{\bibfnamefont{P.}~\bibnamefont{Valageas}},
  \bibinfo{journal}{Phys. Rev. D} \textbf{\bibinfo{volume}{100}},
  \bibinfo{pages}{023526} (\bibinfo{year}{2019}), \eprint{1906.00730}.

\bibitem[{\citenamefont{Madelung}(1927)}]{Madelung:1927ksh}
\bibinfo{author}{\bibfnamefont{E.}~\bibnamefont{Madelung}},
  \bibinfo{journal}{Z. Phys.} \textbf{\bibinfo{volume}{40}},
  \bibinfo{pages}{322} (\bibinfo{year}{1927}).

\bibitem[{\citenamefont{Chavanis and Delfini}(2011)}]{Chavanis:2011zm}
\bibinfo{author}{\bibfnamefont{P.~H.} \bibnamefont{Chavanis}} \bibnamefont{and}
  \bibinfo{author}{\bibfnamefont{L.}~\bibnamefont{Delfini}},
  \bibinfo{journal}{Phys. Rev. D} \textbf{\bibinfo{volume}{84}},
  \bibinfo{pages}{043532} (\bibinfo{year}{2011}), \eprint{1103.2054}.

\bibitem[{\citenamefont{Harko}(2011)}]{Harko:2011jy}
\bibinfo{author}{\bibfnamefont{T.}~\bibnamefont{Harko}}, \bibinfo{journal}{Mon.
  Not. Roy. Astron. Soc.} \textbf{\bibinfo{volume}{413}}, \bibinfo{pages}{3095}
  (\bibinfo{year}{2011}), \eprint{1101.3655}.

\bibitem[{\citenamefont{Brax et~al.}(2020)\citenamefont{Brax, Cembranos, and
  Valageas}}]{Brax:2020oye}
\bibinfo{author}{\bibfnamefont{P.}~\bibnamefont{Brax}},
  \bibinfo{author}{\bibfnamefont{J.~A.~R.} \bibnamefont{Cembranos}},
  \bibnamefont{and} \bibinfo{author}{\bibfnamefont{P.}~\bibnamefont{Valageas}},
  \bibinfo{journal}{Phys. Rev. D} \textbf{\bibinfo{volume}{102}},
  \bibinfo{pages}{083012} (\bibinfo{year}{2020}), \eprint{2007.04638}.

\bibitem[{\citenamefont{Lin et~al.}(2018)\citenamefont{Lin, Schive, Wong, and
  Chiueh}}]{Lin:2018whl}
\bibinfo{author}{\bibfnamefont{S.-C.} \bibnamefont{Lin}},
  \bibinfo{author}{\bibfnamefont{H.-Y.} \bibnamefont{Schive}},
  \bibinfo{author}{\bibfnamefont{S.-K.} \bibnamefont{Wong}}, \bibnamefont{and}
  \bibinfo{author}{\bibfnamefont{T.}~\bibnamefont{Chiueh}},
  \bibinfo{journal}{Phys. Rev. D} \textbf{\bibinfo{volume}{97}},
  \bibinfo{pages}{103523} (\bibinfo{year}{2018}), \eprint{1801.02320}.

\bibitem[{\citenamefont{Yavetz et~al.}(2022)\citenamefont{Yavetz, Li, and
  Hui}}]{Yavetz:2021pbc}
\bibinfo{author}{\bibfnamefont{T.~D.} \bibnamefont{Yavetz}},
  \bibinfo{author}{\bibfnamefont{X.}~\bibnamefont{Li}}, \bibnamefont{and}
  \bibinfo{author}{\bibfnamefont{L.}~\bibnamefont{Hui}},
  \bibinfo{journal}{Phys. Rev. D} \textbf{\bibinfo{volume}{105}},
  \bibinfo{pages}{023512} (\bibinfo{year}{2022}), \eprint{2109.06125}.

\bibitem[{\citenamefont{{Binney} and {Tremaine}}(2008)}]{Binney2008}
\bibinfo{author}{\bibfnamefont{J.}~\bibnamefont{{Binney}}} \bibnamefont{and}
  \bibinfo{author}{\bibfnamefont{S.}~\bibnamefont{{Tremaine}}},
  \emph{\bibinfo{title}{Galactic Dynamics: Second Edition}}
  (\bibinfo{publisher}{Princeton University Press}, \bibinfo{year}{2008}),
  \bibinfo{edition}{rev - revised, 2} ed.

\bibitem[{\citenamefont{{Pathria} and {Morris}}(1990)}]{1990JCoPh..87..108P}
\bibinfo{author}{\bibfnamefont{D.}~\bibnamefont{{Pathria}}} \bibnamefont{and}
  \bibinfo{author}{\bibfnamefont{J.~L.} \bibnamefont{{Morris}}},
  \bibinfo{journal}{Journal of Computational Physics}
  \textbf{\bibinfo{volume}{87}}, \bibinfo{pages}{108} (\bibinfo{year}{1990}).

\bibitem[{\citenamefont{{Zhang} and {Hayee}}(2008)}]{2008JLwT...26..302Z}
\bibinfo{author}{\bibfnamefont{Q.}~\bibnamefont{{Zhang}}} \bibnamefont{and}
  \bibinfo{author}{\bibfnamefont{M.~I.} \bibnamefont{{Hayee}}},
  \bibinfo{journal}{Journal of Lightwave Technology}
  \textbf{\bibinfo{volume}{26}}, \bibinfo{pages}{302} (\bibinfo{year}{2008}).

\bibitem[{\citenamefont{{Edwards} et~al.}(2018)\citenamefont{{Edwards},
  {Kendall}, {Hotchkiss}, and {Easther}}}]{2018JCAP...10..027E}
\bibinfo{author}{\bibfnamefont{F.}~\bibnamefont{{Edwards}}},
  \bibinfo{author}{\bibfnamefont{E.}~\bibnamefont{{Kendall}}},
  \bibinfo{author}{\bibfnamefont{S.}~\bibnamefont{{Hotchkiss}}},
  \bibnamefont{and}
  \bibinfo{author}{\bibfnamefont{R.}~\bibnamefont{{Easther}}},
  \bibinfo{journal}{JCAP} \textbf{\bibinfo{volume}{2018}}, \bibinfo{eid}{027}
  (\bibinfo{year}{2018}), \eprint{1807.04037}.

\bibitem[{\citenamefont{Garc\'\i{}a et~al.}(2024)\citenamefont{Garc\'\i{}a,
  Brax, and Valageas}}]{Garcia:2023abs}
\bibinfo{author}{\bibfnamefont{R.~G.} \bibnamefont{Garc\'\i{}a}},
  \bibinfo{author}{\bibfnamefont{P.}~\bibnamefont{Brax}}, \bibnamefont{and}
  \bibinfo{author}{\bibfnamefont{P.}~\bibnamefont{Valageas}},
  \bibinfo{journal}{Phys. Rev. D} \textbf{\bibinfo{volume}{109}},
  \bibinfo{pages}{043516} (\bibinfo{year}{2024}), \eprint{2304.10221}.

\bibitem[{\citenamefont{Levkov et~al.}(2018)\citenamefont{Levkov, Panin, and
  Tkachev}}]{Levkov:2018kau}
\bibinfo{author}{\bibfnamefont{D.~G.} \bibnamefont{Levkov}},
  \bibinfo{author}{\bibfnamefont{A.~G.} \bibnamefont{Panin}}, \bibnamefont{and}
  \bibinfo{author}{\bibfnamefont{I.~I.} \bibnamefont{Tkachev}},
  \bibinfo{journal}{Phys. Rev. Lett.} \textbf{\bibinfo{volume}{121}},
  \bibinfo{pages}{151301} (\bibinfo{year}{2018}), \eprint{1804.05857}.

\bibitem[{\citenamefont{Chan et~al.}(2024)\citenamefont{Chan, Sibiryakov, and
  Xue}}]{Chan:2022bkz}
\bibinfo{author}{\bibfnamefont{J.~H.-H.} \bibnamefont{Chan}},
  \bibinfo{author}{\bibfnamefont{S.}~\bibnamefont{Sibiryakov}},
  \bibnamefont{and} \bibinfo{author}{\bibfnamefont{W.}~\bibnamefont{Xue}},
  \bibinfo{journal}{JHEP} \textbf{\bibinfo{volume}{01}}, \bibinfo{pages}{071}
  (\bibinfo{year}{2024}), \eprint{2207.04057}.

\bibitem[{\citenamefont{Jain et~al.}(2023)\citenamefont{Jain, Amin, Thomas, and
  Wanichwecharungruang}}]{Jain:2023ojg}
\bibinfo{author}{\bibfnamefont{M.}~\bibnamefont{Jain}},
  \bibinfo{author}{\bibfnamefont{M.~A.} \bibnamefont{Amin}},
  \bibinfo{author}{\bibfnamefont{J.}~\bibnamefont{Thomas}}, \bibnamefont{and}
  \bibinfo{author}{\bibfnamefont{W.}~\bibnamefont{Wanichwecharungruang}},
  \bibinfo{journal}{Phys. Rev. D} \textbf{\bibinfo{volume}{108}},
  \bibinfo{pages}{043535} (\bibinfo{year}{2023}), \eprint{2304.01985}.

\end{thebibliography}

\end{document}